%% file: diss.tex
\newif\ifPDFLaTeX
\expandafter\ifx\csname pdfoutput\endcsname\relax\else\PDFLaTeXtrue\fi

\ifPDFLaTeX
  \documentclass[12pt,a4paper,chapters]{flex}
  \usepackage{type1cm}
  \usepackage[pdftex,colorlinks]{hyperref} 
  \usepackage[pdftex]{graphicx,feynmp,emp}
  \usepackage{thophys,mcite}
  \usepackage[latin1]{inputenc} 
  \usepackage{array,amsmath,amssymb,amscd,acronym,units}  
  \empaddtoTeX{\usepackage{amsmath,amssymb,units}}
  \empaddtoprelude{input graph;}
  \empaddtoprelude{input boxes;}
  \empaddtoTeX{\usepackage{type1cm}}
\else
  \documentclass[a4paper,chapters]{book}           %%% @SUBMIT@
  \usepackage{thophys,thohacks,mcite}
  \providecommand{\preprintno}[1]{\relax}
  \usepackage[T1]{fontenc}
  \usepackage[latin1]{inputenc}
  \usepackage{pstricks,shadow}
  \usepackage[monochrome]{color}
  \usepackage{graphicx,array,amsmath,amssymb,amscd,feynmp,emp,acronym,units}
  \empaddtoTeX{\usepackage{amsmath,amssymb,units}}
  \empaddtoprelude{input graph;}
  \empaddtoprelude{input boxes;}
\fi
\allowdisplaybreaks

\makeatletter
\makeatletter
  \def\fps@figure{t}
  \def\fps@table{b}
  \long\def\@makecaption#1#2{%
    \vskip\abovecaptionskip
    \sbox\@tempboxa{#1: \textit{#2}}%
    \ifdim\wd\@tempboxa>\hsize
      #1: \textit{#2}\par
    \else
      \global\@minipagefalse
      \hb@xt@\hsize{\hfil\box\@tempboxa\hfil}%
    \fi
    \vskip\belowcaptionskip}
\makeatother
\makeatletter
\newenvironment{cutequote@}[1]%
 {\def\cutequoteref{#1}%
  \begin{raggedleft}%
    \samepage%
    \leftskip.3\textwidth plus.6\textwidth%
    \bgroup\itshape}%
 {\\\egroup[\cutequoteref]\\
  \end{raggedleft}
  \vspace*{\baselineskip}
  \@afterheading}
 {\begin{cutequote@}{\textsc{#1}}}
 {\end{cutequote@}}
\newenvironment{cutequote*}[2]%
 {\begin{cutequote@}{\textsc{#1:} \textit{#2}}}
 {\end{cutequote@}}
\makeatother
 {\begin{list}{}%
   {\setlength{\leftmargin}{3em}%
    \setlength{\rightmargin}{3em}%
    \setlength{\itemindent}{1em}%
    \setlength{\listparindent}{0pt}%
    \settowidth{\labelwidth}{\textbf{#1:}}%
    }}%
 {\end{list}}
\newcommand{\PreserveBackslash}[1]{\let\temp=\\#1\let\\=\temp}

\newcommand{\ii}{\mathrm{i}}

\renewcommand{\Re}{\text{Re}}
\renewcommand{\Im}{\text{Im}}
{\def\$#1: #2 ${#2}}
\begin{document}
\thispagestyle{empty}
%\vspace*{\baselineskip}

\begin{center}
{\LARGE\bf

\mbox{Supersymmetry } \\
\vspace*{2mm} of Scattering Amplitudes \\
\vspace*{2mm} and Green Functions  \\
\vspace*{2mm} in Perturbation Theory \\
}
\vspace*{20mm}
\large
%%%%%%%Dem
Vom
Fachbereich Physik\\
der Technischen Universit\"at Darmstadt\\
\vspace*{8mm}
zur Erlangung des Grades\\
eines Doktors der Naturwissenschaften\\(Dr.~rer.~nat.)\\
\vspace*{8mm}
%%%%%%%vorgelegte
genehmigte
Dissertation von\\
\vspace*{2mm}
{\bf Dipl.-Phys.~J\"urgen~Reuter}\\
\vspace*{2mm}
aus Frankfurt am Main\\
%%%%%%%5\vspace*{50mm}
\vspace*{14mm}
Referent: Prof.~Dr.~P.~Manakos  \\
Korreferent: Prof.~Dr.~N.~Grewe \\
\hfill{}\\
\hfill{}\\
\vspace*{5mm}
Tag der Einreichung: 24.05.2002 \\
Tag der Pr\"ufung: 10.07.2002 \\
\hfill{}\\
%\hfill{}\\
\vspace*{9mm}
Darmstadt 2002\\
D 17
\end{center}
\pagebreak
\let\PBS=\PreserveBackslash   
\bibliographystyle{prsty-mod}%%%{phaip}
%\bibliographystyle{prsty-sort}%%%{phaip}
%
%\title{Supersymmetry of Scattering Amplitudes and Green Functions in
%  Perturbation Theory}  
%
%\author{%
%  Jürgen Reuter\\ 
%  \hfil\\
%  Technische Universität Darmstadt\\
%  Schlo\ss gartenstr.~9 \\
%  D-64289 Darmstadt, Germany\\
%  \hfil\\
%  \texttt{<reuter@hep.tu-darmstadt.de>}}
%
%%\preprintno{\hfil}
%\date{IKDA~2000/??\\April 2000\\\RCSId}
%
%%%%%%%%%%%%%%%%%%%%%%%%%%%%%%%%%%%%%%%%%%%%%%%%%%%%%%%%%%%%%%%%%%%%%%%%
%\maketitle
%\frontmatter
\vspace*{\stretch1}    

\hspace{30pt}\parbox{10cm}{\raggedright
{\em

  {\bf Cerro de la Estrella} 

  \vspace{1cm}

  \begin{tabular}{l}
    Aqu\'i los antiguos recib\'ian al fuego \\
    Aqu\'i el fuego creaba al mundo \\
    Al mediod\'ia las piedras se abren como frutos \\
    El agua abre los p\'arpados \\
    La luz resbala por la piel del d\'ia \\
    Gota inmensa donde el tiempo se refleja y se sacia [$\ldots$] 
  \end{tabular} }}
 \vspace{20pt}

\hspace{170pt}\parbox{7cm}{\raggedright
Octavio Paz, {\em "Libertad Bajo Palabra"}}
 
\vspace*{\stretch2}   
%\begin{abstract} 
%  \ldots
%\end{abstract}
%
%%%%%%%%%%%%%%%%%%%%%%%%%%%%%%%%%%%%%%%%%%%%%%%%%%%%%%%%%%%%%%%%%%%%%%%
\tableofcontents
%\mainmatter        
%%%%%%%%%%%%%%%%%%%%%%%%%%%%%%%%%%%%%%%%%%%%%%%%%%%%%%%%%%%%%%%%%%%%%%%
\begin{empfile}
\begin{fmffile}{\jobname pics}
\fmfset{arrow_ang}{10}
\fmfset{curly_len}{2mm}
\fmfset{wiggly_len}{3mm}     
\fmfcmd{vardef middir (expr p, ang) =
    dir (angle direction length(p)/2 of p + ang)
  enddef;}
\fmfcmd{style_def arrow_left expr p =
    shrink (.7);
      cfill (arrow p shifted (4thick * middir (p, 90)));
    endshrink
  enddef;}
\fmfcmd{style_def arrow_right expr p =
    shrink (.7);
      cfill (arrow p shifted (4thick * middir (p, -90)));
    endshrink
  enddef;}
\fmfcmd{style_def warrow_left expr p =
    shrink (.7);
      cfill (arrow p shifted (8thick * middir (p, 90)));
    endshrink
  enddef;}
\fmfcmd{style_def warrow_right expr p =
    shrink (.7);
      cfill (arrow p shifted (8thick * middir (p, -90)));
    endshrink
  enddef;}
\fmfcmd{style_def susy_ghost expr p =
      draw (zigzag p);
      cfill (arrow p);
  enddef;}
\newcommand{\threeexternal}[3]{%
  \fmfsurround{d1,e1,d2,e2,d3,e3}%
  \fmfv{label=$#1$,label.ang=0}{e1}%
  \fmfv{label=$#2$,label.ang=180}{e2}%
  \fmfv{label=$#3$,label.ang=0}{e3}}
\newcommand{\Threeexternal}[3]{%
  \fmfsurround{d1,e1,d3,e3,d2,e2}%
  \fmfv{label=$#1$,label.ang=0}{e1}%
  \fmfv{label=$#2$,label.ang=0}{e2}%
  \fmfv{label=$#3$,label.ang=180}{e3}}         
\newcommand{\Fourexternal}[4]{%
  \fmfsurround{d2,e2,d1,e1,d4,e4,d3,e3}%
  \fmfv{label=$#1$,label.ang=180}{e1}%
  \fmfv{label=$#2$,label.ang=0}{e2}%
  \fmfv{label=$#3$,label.ang=0}{e3}%
  \fmfv{label=$#4$,label.ang=180}{e4}}
\newcommand{\Fiveexternal}[5]{%
  \fmfsurround{d2,e2,d1,e1,d5,e5,d4,e4,d3,e3}%
  \fmfv{label=$#1$,label.ang=180}{e1}%
  \fmfv{label=$#2$,label.ang=0}{e2}%
  \fmfv{label=$#3$,label.ang=0}{e3}%
  \fmfv{label=$#4$,label.ang=0}{e4}%
  \fmfv{label=$#5$,label.ang=180}{e5}}
\newcommand{\twoincoming}{%
    \fmfdot{v}%
    \fmffreeze%
    \fmf{warrow_right}{e1,v}%
    \fmf{warrow_right}{e2,v}%
    \fmf{warrow_right}{v,e3}}
\newcommand{\threeincoming}{%
    \fmfdot{v}%
    \fmffreeze%
    \fmf{warrow_right}{e1,v}%
    \fmf{warrow_right}{e2,v}%
    \fmf{warrow_right}{e3,v}}
\newcommand{\threeoutgoing}{%
    \fmfdot{v}%
    \fmffreeze%
    \fmf{warrow_right}{v,e1}%
    \fmf{warrow_right}{v,e2}%
    \fmf{warrow_right}{v,e3}}
\newcommand{\fouroutgoing}{%
    \threeoutgoing%
    \fmf{warrow_right}{v,e4}}
\newcommand{\fiveoutgoing}{%
    \fouroutgoing%
    \fmf{warrow_right}{v,e5}}     
%%%%%%%%%%%%%%%%%%%%%%%%%%%%%%%%%%%%%%%%%%%%%%%%%%%%%%%%%%%%%%%%%%%%%%%

\include{chap0}

\include{chap1}

\part{SUSY Ward identities for asymptotic states}

\include{chap2}

\include{chap3}
\include{chap4}
\include{chap5}

\part{SUSY Ward identities via the current}

\include{chap6}
\include{chap7}

\include{chap8}

\part{SUSY Slavnov-Taylor identities}

\include{chap10}

\include{chap11}

%\include{chap9} %%% sugra-kapitel 

\part{Implementation, Summary and Outlook}

\include{chap12}

\include{chap13}

%\cite{Glashow:1961:SM,*Weinberg:1967:SM,*Salam:1968:SM}

%%%%%%%%%%%%%%%%%%%%%%%%%%%%%%%%%%%%%%%%%%%%%%%%%%%%%%%%%%%%%%%%%%%%%%%%
\appendix

\include{appen_bas}

\include{appen_mssm}

\include{appen1}
\include{appen_sym}
\chapter{Summary of models}
\include{appen_wz}
\include{appen_toy1}
\include{appen_or}
\include{appen_sagt}

\include{appen_modelsym}
%%%%%%%%%%%%%%%%%%%%%%%%%%%%%%%%%%%%%%%%%%%%%%%%%%%%%%%%%%%%%%%%%%%%%%%%

%%%%%%%%%%%%%%%%%%%%%%%%%%%%%%%%%%%%%%%%%%%%%%%%%%%%%%%%%%%%%%%%%%%%%%%%
\bibliography{diss}
%%%%%%%%%%%%%%%%%%%%%%%%%%%%%%%%%%%%%%%%%%%%%%%%%%%%%%%%%%%%%%%%%%%%%%%%
\include{dank}

%%%%%%%%%%%%%%%%%%%%%%%%%%%%%%%%%%%%%%%%%%%%%%%%%%%%%%%%%%%%%%%%%%%%%%%%
%%% \begin{acronym}
  \acrodef{QCD}{Quantenchromodynamik} 
  \acrodef{QED}{Quantenelectrodynamik} 
  \acrodef{QFT}{Quantenfeldtheorie} 
  \acrodef{SM}{Standardmodell}
%%% \end{acronym}

%%%%%%%%%%%%%%%%%%%%%%%%%%%%%%%%%%%%%%%%%%%%%%%%%%%%%%%%%%%%%%%%%%%%%%%%
\end{fmffile}    
\end{empfile}
\end{document}

%% file: chap0.tex
\chapter{Introduction and Motivation}

Although it is a theory of the utmost accuracy and success, 
the Standard Model (SM) of elementary particle physics cannot
describe Nature up to arbitrarily high energy scales and therefore is
not the last answer on our way in uncovering Nature's secrets. Today we look
upon the SM as merely an effective field theory which is
described by a local, causal quantum field theory up to an energy
scale yet unknown, but assumed to lie at about $10^{15}$ GeV. Though all
experimental data available today are in perfect agreement with the
description of Nature by the Standard Model, there are some loose 
ends in the framework of the SM from which we mention just one, the so
called {\em naturalness} or {\em hierarchy} problem. If the breaking
of the electroweak gauge symmetry is provided by an elementary scalar getting
a vacuum expectation value, the mass of that scalar, the Higgs boson,
should be of the order of the electroweak breaking scale. Typically, the
radiative corrections to the mass square of a scalar are proportional
to the square of the energy scale at which its quantum field theory is
embedded in a more fundamental theory, candidates for which being the
Planck scale, a GUT or a string scale of the order given above or
higher. This is not the case for fermions which receive only
logarithmic corrections. An immense fine tuning for the bare mass of
the Higgs scalar at the scale of the more fundamental theory is therefore
necessary to cancel the quadratic contributions from the
renormalization group flow. If we did not have these cancellations, the
``natural'' mass square of the Higgs scalar at the electroweak
breaking scale would be of the order of the square of the high scale;
this is called the naturalness problem. The hierarchy problem means the
sheer existence of the vast differences between the two energy
scales. 

%%%%%%%%%%%%%%%%%%%%%%%%%%%%%%%%%%%%%%%%%%%%%%%%%%%%%%%%%%%%%%%%%%%%%%

A possible solution of the naturalness problem serves as the
strongest motivation for supersymmetry. Supersymmetry is a symmetry
which interchanges bosons and fermions and could therefore naturally
explain the existence of light scalars. In the supersymmetric limit
each fermion loop contributing to the quadratically divergent Higgs
self-energy is accompanied by a scalar loop with the opposite sign.
Furthermore the coupling constants are required to be equal by supersymmetry,
hence the quadratic divergence cancels out and only the logarithmic
survives. As a second motivation we may mention gauge coupling unification
which is compatible with current data only in supersymmetric
extensions of the Standard Model but not in the SM itself. Hence, in
spite of technicolour models -- theories where the Higgs is a composite
object -- and models with extra dimensions (whether ``large'' or not)
as competitors, supersymmetric extensions of the SM are the most
widely accepted of the hypothesized models beyond the Standard Model.

After the first supersymmetric models had been established in the
early 1970s \cite{Wess/Bagger:SUSY:text}, phenomenology started and
supersymmetric extensions of the SM have been constructed, e.g.~see   
the reviews given in \cite{Haber/Kane:1985:SUSY},
\cite{Weinberg:QFTv3:Text}. The simplest of these extensions is called
the {\em Minimal Supersymmetric Standard Model} (MSSM), where the predicate
``minimal'' stands for minimal field content: Each SM field is embedded 
into a superfield where the SM fermions are accompanied by
scalars, the gauge bosons by fermions, called gauginos, and the Higgs bosons
also by fermionic superpartners. Moreover, the constraint of being
supersymmetric forces the existence of at least two Higgs superfields, one
with hypercharge $+1$ and one with hypercharge $-1$, to give mass to the up-
as well as down-type fermions; the appearance of two Higgs doublets is
necessary also to avoid anomalies. 

Therefore the prediction of supersymmetry
is the existence of superpartners for all yet known SM particles. Since they 
are constrained by SUSY to have the same masses as the SM particles
but have not been observed yet, supersymmetry has to be broken. Until
today the mechanism of supersymmetry breaking is unknown,
so we parameterize our ignorance by the most general explicit breaking
of supersymmetry, the so called {\em soft breaking terms}. They are
motivated by the fact that SUSY has to be broken by a whatsoever
mechanism at a high scale, producing these explicit breaking terms by
the renormalization group evolution of all relevant operators
compatible with all symmetries. Though SUSY is a very simple concept
and an enormously powerful symmetry, in addition to the huge number of
particles, these soft breaking terms make the MSSM tremendously
complicated as all particles which are by their quantum numbers
allowed to mix really do mix. Also the pure number of free parameters
in the MSSM becomes one order of magnitude higher as in the SM, namely
124 \cite{Haber:1997:SUSY}, or even, in a more general version, 178
\cite{Weinberg:QFTv3:Text}, \cite{Reuter:2000:SUSY}.  

%%%%%%%%%%%%%%%%%%%%%%%%%%%%%%%%%%%%%%%%%%%%%%%%%%%%%%%%%%%%%%%%%%%%%%%

Another issue is the incredible number of vertices considering all
Feynman rules of the MSSM (cf.~tables \ref{mssmvert1}, \ref{mssmvert2},
\ref{mssmvert3}, \ref{mssmvert4}) and the sometimes very complex structure of
the coupling constants, \cite{Simonsen:1995:SEWT}, \cite{Kuroda:1999:MSSM},
and \cite{Reuter:2000:SUSY}. There are some simplifying assumptions for
the structure of the coefficients of the soft breaking terms
(e.g.~{\em flavour alignment} or {\em universality}) which are
motivated by supergravity 
embeddings of the MSSM, but need not be fulfilled. One can steer a middle
course as a compromise for the model: as general as possible, but as
simple as necessary. We choose coefficients which are diagonal in
generation space (actually, the generation mixings must be very small not to
contradict the experimental thresholds for violation of the separate lepton
numbers $L_e$, $L_\mu$, $L_\tau$) but the diagonal elements need not be
equal in contrast to the prejudice given by the universality
constraint. The number of vertices in tables \ref{mssmvert1},
\ref{mssmvert2}, \ref{mssmvert3} and \ref{mssmvert4} has been
estimated under this assumption, but even as this is not the most complex
\begin{table}
\begin{center}
\begin{tabular}{|l|r|r|r|}\hline
        Process & $\#$ \begin{tabular}{c} {\em O'Mega}\\
        fusions\end{tabular} & $\#$ Propagators & $\#$ 
        Diagrams \\\hline 
        $e^+e^-\to \tilde{\chi}^0_1 \tilde{\chi}^0_2$ & 24 & 8 & 8 \\ 
	$e^+ e^-\to \tilde{e}_1^+ \tilde{e}_1^-$ & 27 & 9 & 9 \\
	$e^+ e^- \to \tilde{u}_1\tilde{u}_1^*\tilde{u}_1\tilde{u}_1^*$
        & 346 & 41 & 660 \\ 
        $e^+e^-\to e^+e^-\tilde{\chi}_1^0\tilde{\chi}_2^0$ & 610 & 60 & 1,552
        \\
        $e^+e^-\to \tilde{\chi}^0_1 \tilde{\chi}^0_2 \tilde{\chi}^0_3
        \tilde{\chi}_4^0$ & 782 & 66 & 2,208  \\ 
	$e^+ e^- \to \tilde{e}^+_1 \tilde{e}_1^- \tilde{u}_1
        \tilde{u}_1^* \tilde{u}_1 \tilde{u}_1^*$ & 4,002 & 153 &
        141,486 \\ 
        $e^+e^-\to e^+e^-\mu^+\mu^-\tilde{\chi}^0_1 \tilde{\chi}^0_2$ &  4,389
        & 172 & 239,518 \\
        $e^+e^-\to e^+e^-\tilde{\chi}^0_1 \tilde{\chi}^0_2 \tilde{\chi}^0_3
        \tilde{\chi}_4^0$ & 11,870 & 280 & 1,056,810 \\
        $e^+e^-\to \tilde{\chi}^0_1 \tilde{\chi}^0_1 \tilde{\chi}^0_2
        \tilde{\chi}^0_2 \tilde{\chi}^0_3 \tilde{\chi}_4^0$ & 17,075 &
        322 & 2,191,845 \\ 
        $e^+e^-\to e^+ e^- \mu^+ \mu^- u \bar{u} 
        \tilde{\chi}_1^0 \tilde{\chi}^0_2$ & 23,272 & 434 &
        50,285,616 \\   
        $e^+e^-\to \tilde{\chi}^0_1 \tilde{\chi}^0_1 \tilde{\chi}^0_2
        \tilde{\chi}^0_2 \tilde{\chi}^0_3 \tilde{\chi}^0_3
        \tilde{\chi}_4^0 \tilde{\chi}^0_4$ & 273,950 & 1,370 &
        470,267,024 \\\hline
\end{tabular}
\end{center}
\caption{\label{processes} Juxtaposition of the number of Feynman diagrams and
        of {\em O'Mega} fusions for some MSSM processes at a linear
        collider. By fusions we mean the fundamental calculational
        steps for constructing the amplitudes in {\em O'Mega}.}
\end{table}
of the ``minimal'' MSSMs, it has a discouraging number of more than
four thousand vertices.  

\vspace{3mm}
  
Today's generation of running and planned colliders (Tevatron, LHC,
and TESLA) will bring the 
decision which way Nature has chosen for electroweak symmetry breaking
(cf.~e.g.~\cite{Espinosa/Gunion:1999:NOLOSE}). But even if a Higgs boson is  
detected at one of the world's huge colliders in the next years, it will not
be easy to determine whether it is a ``standard'', a minimal supersymmetric,
a next to minimal supersymmetric one \cite{Higgs:1990:Hunter},
\cite{Higgs:1993:1}, \cite{Higgs:1997:2}, or something else. For this,
extensive knowledge about the alternatives to the SM must be available, and
besides the ubiquitous radiative corrections (within the SM, the MSSM
and other models), it is indispensable to calculate tree level
processes with up to eight particles in the final state, as in
highly energetic processes ($10^2 - 10^3$ GeV for the colliders above)
the final states are very complex. (The interest in eight final particles comes
from the desire to study $WW\to WW$ scattering, the inclusion of the
$WWWW$-vertex in eight-fermion production processes, production of
$t\bar{t}$-pairs and their decays as well as the production of superpartners
and SUSY cascade decays.) Of course, such calculations with
$10^4 - 10^8$ participating Feynman diagrams have to be done
automatically by matrix element generators like {\em O'Mega}
\cite{Ohl:2000:Omega}. Alternative models to the SM have
therefore to be incorporated into such matrix element generators 
as the SM was. The goal for the next years will be to compare
possibly found experimental deviations from the SM 
predictions with the theoretical results from alternative models like
the MSSM.  

As it soon becomes clear, the work is not done by simply writing a
model file for the MSSM to incorporate it in an matrix element
generator like {\em O'Mega}. Since the complexity of the model grows
immensely from the SM to the MSSM (compare tables
\ref{smvert1}-\ref{smvert2} with tables
\ref{mssmvert1}-\ref{mssmvert4}) it is inevitable to check the
consistency of such models like the MSSM. This is necessary for making
sure that all parameters (masses, coupling constants, widths, etc.)
are compatible with each other, to debug computer programs (model
files, numerical function library, etc.), and not to forget, to have
the numerical stability under control. Symmetry principles which have
always been strong concepts in physical theories provide such tests
for consistency checks here. The MSSM like the SM has its $SU(3)_C
\times SU(2)_L \times U(1)_Y$ gauge symmetry as a 
powerful tool for those checks; what is often used is the independence
of all physical results from the gauge parameter $\xi$ in general
$R_\xi$ gauges. Our aim is to make use of the Ward, or better, the
Slavnov-Taylor identities of the gauge symmetry
\cite{Kugo:Eichtheorie}, \cite{Kugo/Ojima:1979:quartet},
\cite{Weinberg:QFTv1:Text}, \cite{Weinberg:QFTv2:Text}. Both kinds of 
identities originate from the quantum generalization of the 
symmetry principle of the classical field theory, the first expressing
current conservation and being only valuable in the case
of global symmetries, the latter stemming from the BRST symmetry left
over after gauge fixing. 

In supersymmetric field theories we can, of course, use
supersymmetry as the underlying symmetry, and there, as long as we are not
concerned with local supersymmetry (supergravity), we are able to
employ Ward identities. As we will see for supersymmetric gauge
theories it is indispensable -- even at tree level -- to use the
Slavnov-Taylor identities. The stringency of the consistency checks is
also a drawback: the relations mentioned as vehicles for those tests
are quite complicated and involve a number of sophisticated
techniques. As a first and fundamental step, extensive knowledge about
how Ward- and Slavnov-Taylor identities for supersymmetric (gauge)
field theories work {\em analytically} (in perturbation theory) has to
be gained to use such identities for numerical checks. This will be
the concern for the major part of this thesis; first of all, the
investigation of the applicability (on-shell and/or off-shell, what
kind of method for which model) of the several kinds of methods to  be
presented here, furthermore -- and even more important -- to
understand the way the cancellations happen in these identities. The
latter point is inevitable in deciding which expressions to use in
numerical checks: expressions adjusted to the technical nature of cancellations
are likely to be numerically more stable than those which are not. A
third and last issue then is to transfer these analytical expressions
to the matrix element generator and perform numerical checks. Since it
is not possible to produce reliable theoretical predictions for future
experiments without having powerful 
consistency checks at hand, and since such consistency checks cannot
be under (numerical) control without a deeper understanding of how
they work analytically, the original intention of this work has
changed: from a purely phenomenological issue at the beginning -- to implement
realistic supersymmetric models as alternatives to the Standard Model
into the matrix element generator {\em 
O'Mega} -- to a more theoretical one -- to develop
stringent tests as consistency checks for these models and to
understand their fine points in detail. We hope to have convinced the
reader that the latter is the {\em sine qua non} for the first. Thus
the main part of this thesis is concerned with analytical perturbative
calculations of three different kinds of identities within several
models, to our knowledge never been done before. Let us
briefly summarize the content of this thesis.

%%%%%%%%%%%%%%%%%%%%%%%%%%%%%%%%%%%%%%%%%%%%%%%%%%%%%%%%%%%%%%%%%%%%%%

\section{Structure and Content}

After a short introduction to supersymmetry transformations, 
the main text is divided into four parts, the first showing a method to gain
on-shell Ward identities for supersymmetric field theories originally
invented in the late 1970s by Grisaru, Pendleton and van Nieuwenhuizen
but as far as we know this method has never been used diagrammatically. We
investigate that kind of Supersymmetric Ward Identities (SWI) for the
Wess-Zumino model and a more complex toy model to uncover some new
effects. As this formalism relies on the annihilation of the vacuum by
the supercharge, it does not work for spontaneously broken
supersymmetry. We provide an example within the framework of the
O'Raifeartaigh model. 

The second part is concerned with SWI constructed from Green functions
with one current insertion and contracted with the momentum brought
into the Green function by the current. At tree level these identities
are fulfilled on-shell and off-shell. For the latter the SWI are more
complicated due to the contributions of several ``contact terms''  and
provide more stringent tests than the on-shell identities. Examples are
calculated for the Wess-Zumino model, the toy model from part one and
for the O'Raifeartaigh model, as the supersymmetric current is
still conserved for spontaneously broken SUSY. It will be shown that
this method does not work for supersymmetric gauge theories. The
explanation of this phenomenon then blends over to the next part. 

There we introduce the BRST formalism for supersymmetric theories
where supersymmetry as a global symmetry is quantized with the help
of constant ghosts, \cite{White:1992:BRST},
\cite{Sibold/etal:2000:brst}. In order not to cloud the intricacies 
by a huge amount of fields and diagrams, we construct the simplest
possible supersymmetric Abelian toy model. We summarize the BRST
transformations with inclusion of supersymmetry and translations and
show several examples of supersymmetric Slavnov-Taylor identities in
that toy model and also in supersymmetric QCD.   

In the last part we discuss the problems concerned with the
implementation of supersymmetric models and the consistency checks
mentioned above. Connected with supersymmetric field theories is
the appearence of Majorana fermions -- real fermions -- which are their own
antiparticles. The solution of how to let the matrix element
generator evaluate the signs coming from Fermi statistics without 
expanding the Feynman diagrams is presented based on ideas in
\cite{Denner/etal:1992:feynmanrules}. Furthermore it is presented
there how one- and
two-point vertices arising together with the BRST formalism can be
handled within {\em O'Mega}, though their topologies are not compatible
with the way the amplitudes are built by {\em O'Mega}. It is
demonstrated that Slavnov-Taylor identities for gauge symmetries and
supersymmetry can be done within the same framework. Finally we will
give an outlook of what remains to be done in that field, possible
generalizations and improvements.

%%% Local Variables: 
%%% mode: latex
%%% TeX-master: t
%%% End: 

%% file: chap1.tex
%%% Local Variables: 
%%% mode: latex
%%% TeX-master: diss
%%% End: 

\chapter{SUSY Transformations}

\section{Classical transformations}

First of all, we want to summarize the supersymmetry transformations of
classical fields; as a general reference for the basics of supersymmetry we
mention the book of Julius Wess and Jonathan Bagger, {\em Supersymmetry and
Supergravity} \cite{Wess/Bagger:SUSY:text}. By contraction with a fermionic
(i.e. Grassmann odd) spinor transformation parameter we make the supercharges
bosonic 
\begin{equation}
  \label{eq:suplad}
  Q (\xi) \equiv \xi Q + \bar{\xi} \bar{Q}
\end{equation}
The component fields of a chiral multiplet, the scalar field $\phi$, the 
Weyl-spinor field $\psi$ and the scalar auxiliary field $F$ with dimension 
two undergo the following transformations generated by the supercharge $Q(\xi)$
(cf.~the appendix as well)
\begin{equation}
  \begin{aligned}
    \delta_\xi \phi & = \sqrt{2} \xi \psi \\
    \delta_\xi \psi & = - \ii \sqrt{2} \sigma^\mu \bar{\xi} \partial_\mu
    \phi + \sqrt{2} \xi F \\
    \delta_\xi F &  = - \ii \sqrt{2} \bar{\xi} \bar{\sigma}^\mu
    \partial_\mu \psi 
  \end{aligned}
\end{equation}
Compared to the book of Wess/Bagger the relative signs in the last two 
transformations have their origin in the different convention for
the metric used by Wess/Bagger. This causes differences in the definition of
the  4-vector of the Pauli matrices. 

Because $Q(\xi)$ is real (Hermitean as a generator for quantum fields), the
transformation properties of a field imply the properties of the complex
conjugated field. One simply has to define:
\begin{equation}
  \label{eq:defkomtrans}
  \left( \delta_\xi \Psi \right)^* = \delta_\xi \Psi^* \quad ,
\end{equation}
This is the natural choice for a real generator. The relation will still be
fulfilled in the quantized calculus. 

Better suited for our aim -- application of SUSY transformations in a
phenomenological particle physics context -- will be a formulation of
the transformation rules with bispinors. Therefore we reformulate 
the transformations given above in this formalism. We also split the
lowest and the highest components of the superfields into their scalar
and pseudoscalar parts, called ``chiral''. This will prove useful later. 
\begin{equation}
  \label{eq:chiralfeld}
  \phi = \dfrac{1}{\sqrt{2}} \left( A + \ii B \right) , \qquad F =
  \dfrac{1}{\sqrt{2}} \left( \mathcal{F} - \ii \mathcal{G} \right) .
\end{equation}
The resulting transformations are: 
\begin{equation}
  \label{eq:trafobispin}
  \boxed{
    \begin{aligned}
      \delta_\xi A & = \overline{\xi} \psi, \\
      \delta_\xi B & = \ii \overline{\xi} \gamma^5 \psi, \\
      \delta_\xi \psi & = - \ii \fmslash{\partial} \left( A + \ii
      \gamma^5 B \right) \xi + \left( \mathcal{F} + \ii \gamma^5
  \mathcal{G} \right) \xi , \\ 
      \delta_\xi \mathcal{F} & = - \ii \, \overline{\xi}
  \fmslash{\partial} \psi , \\ 
      \delta_\xi \mathcal{G} & = - \overline{\xi} \fmslash{\partial}
  \gamma^5 \psi . 
    \end{aligned}}
\end{equation}
In this list all spinors are understood as bispinors. For the translation of
the ``fundamental'' component fields to the ``chiral'' fields we refer
to section \ref{sec:auxprob}. 

%%%%%%%%%%%%%%%%%%%%%%%%%%%%%%%%%%%%%%%%%%%%%%%%%%%%%%%%%%%%%%%%%%%

\section{SUSY transformations in Hilbert space}

The following discussion should prevent the confusion with factors $\ii$ and 
signs when talking about SUSY transformations on the classical level and in 
the context of quantum field theory. Classically we review the results of the
last section:
\begin{equation}
  \label{eq:klasstrans}
  \begin{aligned}
    \delta_\xi \phi \equiv \left( \xi Q + \bar{\xi} \bar{Q} \right), \qquad
    \left( \xi Q + \bar{\xi} \bar{Q} \right)^* = \xi Q + \bar{\xi} \bar{Q} \\
    \delta_\xi \phi^* = \left( \delta_\xi \phi \right)^* = \left( \xi Q +
    \bar{\xi} \bar{Q} \right) \phi^* \quad ,
  \end{aligned}
\end{equation}
wherein $\phi$ could be a field of any geometrical character and any Grassmann
parity.  

In the quantum theory the transformation is represented by a unitary operator,
which is created by exponentiation of the supercharge -- now a Hermitean
generator -- multiplied with $\ii$:
\begin{equation}
  \label{eq:quanttrans}
  \left[ \ii Q(\xi) , \phi \right] = \delta_\xi \phi
\end{equation}
Again $\phi$ is a field (operator) of arbitrary geometrical character and
Grassmann parity. 
Moreover, $\delta_\xi \phi$ is the transformation of the classical fields 
incorporated into Hilbert space, i.e.~the classical term, in which the fields
have been replaced by operators acting in the Hilbert space. For the Hermitean
adjoint one finds: 
\begin{equation}
  \label{eq:adjquant}
  \begin{aligned}
    \left[ \ii Q(\xi) , \phi \right]^\dagger & = \left[ \ii Q(\xi) ,
    \phi^\dagger \right] \\
    & = \left(\delta_\xi \phi \right)^\dagger \\
    & \Longrightarrow \boxed{ \left[ \ii Q(\xi) , \phi^\dagger \right] = 
    \left( \delta_\xi \phi \right)^\dagger = \delta_\xi \phi^\dagger}
  \end{aligned}
\end{equation}
There is no subtlety in dealing with fermionic fields here because the rule for
reversing the order of Grassmann odd parameters classically is translated
to the rule for reversing the order of field operators when Hermitean
adjoined -- no matter whether they are fermionic or bosonic. But one still
has to take into account that Grassmann odd classical parameters like $\xi$
and fermionic field operators have to be reversed in order when Hermitean
adjoined. 

Finally there is a simple rule for the embedding of the classical
transformations into the quantum theory: Replace left multiplication
with $Q(\xi)$ by application of the commutator with $\ii Q(\xi)$. 

%%%%%%%%%%%%%%%%%%%%%%%%%%%%%%%%%%%%%%%%%%%%%%%%%%%%%%%%%%%%%%%%%%%%

\section[Problems with auxiliary fields]{General problems with auxiliary
  fields in supersymmetric field theories}\label{sec:auxprob}

As we will see, there is a possibility to implement SUSY Ward identities for
theories with exact supersymmetry and an {\em S}-matrix invariant under SUSY
transformations, by examining the transformation properties of the creation and
annihilation operators of {\em in} and {\em out} states. For the extraction of
the relations between amplitudes provided by supersymmetry, (in this
ansatz) asymptotic fields (cf., for example, Kugo,
\cite{Kugo:Eichtheorie}) have to be taken into account. The only
important parts of the asymptotic fields are the one-particle poles, so
we only have to keep those terms in the equations of motion of the auxiliary
fields $F$ and $D$ which stem from the bilinear parts of the superpotential.

For example in the Wess-Zumino model we have:
\begin{equation}
  \begin{aligned}
    F & = - m \phi^* - \dfrac{1}{2} \lambda (\phi^*)^2 \\
        & = - \dfrac{m}{\sqrt{2}} \left( A - \ii B \right) -
        \dfrac{\lambda}{4} \left( A - \ii B \right)^2 \\
        & \stackrel{!}{=} \dfrac{1}{\sqrt{2}} \left( \mathcal{F} - \ii
    \mathcal{G} \right)  
  \end{aligned}
\end{equation}
Out of this we obtain the equations of motion for the auxiliary fields:
\begin{equation}
  \label{eq:beweghilfs}
    \boxed{
      \begin{aligned}
        \mathcal{F} & = - m A - \dfrac{\lambda}{2 \sqrt{2}} \left( A^2
    - B^2 \right) \\ 
        \mathcal{G} & = - m B - \dfrac{\lambda}{\sqrt{2}} A B
      \end{aligned} }
\end{equation}

Off-shell there is no distinction possible between fields and auxiliary 
fields. The auxiliary fields are necessary to preserve the lemma stating
that the number of bosonic and fermionic degrees of freedom has to be equal.
For physical processes (with fields on the mass shell) one has to insert the 
equations of motion for the auxiliary fields. For the derivation of the
$S$-matrix via the LSZ reduction formula all one-particle poles have to be
accounted for. This implies further that in the equations of motion only the
one-particle poles have to be kept. In the MSSM these poles 
exclusively appear in  the mass terms (soft SUSY breaking terms) and the
bilinear Higgs term, the latter also generating masses. 

%%%%%%%%%%%%%%%%%%%%%%%%%%%%%%%%%%%%%%%%%%%%%%%%%%%%%%%%%%%%%%%%%%

\section{SUSY transformations of quantum fields}

Finally, we are able to write down the SUSY transformations in Hilbert space
for the chiral superfield:

\begin{equation}
  \label{eq:trafototal}
  \boxed{
  \begin{aligned} 
      \left[ \ii Q(\xi) , A \right] & = \overline{\xi} \psi, \\ 
      \left[ \ii Q(\xi) , B \right] & = \ii \overline{\xi} \gamma^5 \psi, \\
      \left[ \ii Q(\xi) , \psi \right] & = - \ii
      \fmslash{\partial} \left( A + \ii \gamma^5 B \right) \xi +
  \left( \mathcal{F} + \ii \gamma^5 \mathcal{G} \right) \xi
      , \\ \left[ \ii Q(\xi) , \mathcal{F} \right] & = - \ii \overline{\xi}
      \fmslash{\partial} \psi , \\ 
      \left[ \ii Q(\xi) , \mathcal{G} \right] & = -  \overline{\xi}
      \fmslash{\partial} \gamma^5 \psi     
  \end{aligned}}
\end{equation}

Taking into account only the one-particle poles, e.g. in the Wess-Zumino
model, yields:

\begin{equation}
  \label{eq:trafototal2}
  \boxed{
  \begin{aligned} 
      \left[ \ii Q(\xi) , A \right] & = \overline{\xi} \psi, \\ 
      \left[ \ii Q(\xi) , B \right] & = \ii \overline{\xi} \gamma^5 \psi, \\
      \left[ \ii Q(\xi) , \psi \right] & = - \left( \ii
      \fmslash{\partial} + m \right) \left( A + \ii \gamma^5 B \right) \xi 
  \end{aligned}}
\end{equation}

%%%%%%%%%%%%%%%%%%%%%%%%%%%%%%%%%%%%%%%%%%%%%%%%%%%%%%%%%%%%%%%%%%%%%

%%% Local Variables: 
%%% mode: latex
%%% TeX-master: "diss"
%%% End: 

%% file: chap2.tex
%%% Local Variables: 
%%% mode: latex
%%% TeX-master: t
%%% End: 

\chapter[SWI for asymptotic fields]{SUSY Ward Identities [SWI] for asymptotic
  fields} \label{chap_swi_on}

\section{Consequences of SUSY for $S$-matrix elements}\label{susy_rela}

In supersymmetric field theories supersymmetry is a symmetry of the theory,
meaning that the $S$-operator commutes with the supercharges: $\left[ Q , S
\right] = 0$. Later on we will see that in supersymmetric gauge theories the
gauge fixing required for quantization breaks supersymmetry, with the
result that the supercharge no longer commutes with the $S$-operator on
the complete Hilbert space but only with the $S$-operator on the cohomology of
the supercharge \cite{Sibold/etal:2000:brst}. The $S$-operator maps the
Hilbert space basis of asymptotic {\em in} states onto the one of the
asymptotic {\em out} states. Therefore we immediately conclude that the {\em
in} and {\em out} creation and annihilation operators have the same algebra, 
i.e.~commutation relations with the supercharge $Q$. Remember that we
are dealing at the moment with exact supersymmetry, so the vacuum is
invariant under SUSY transformations and must be annihilated by the
supercharge:  
\begin{equation}
  \label{eq:vakinv}
  Q \Ket{0} = 0.
\end{equation}

At this point we mention some common grounds and some differences of
supersymmetry and BRST symmetry. Both have in common that they are fermionic
generators of global symmetries of the theory (we do not treat
supergravity and local supersymmetry here) so there are some similarities
between them. BRST transformations leave many more states of Hilbert
space invariant (namely all physical states) than supersymmetry under
which only the vacuum (and perhaps soliton solutions) are
invariant. So for constructing relations between amplitudes of
different processes we are (in case of supersymmetry) left 
with on-shell relations between $S$-matrix amplitudes whereas in BRST
identities different off-shell Green functions can be compared. Later
on we will bring SUSY and BRST together and derive the most general
identities for supersymmetric gauge theories. 

For the derivation of SWIs the following relation is the basic ingredient to
start with:  
\begin{equation}
  \label{eq:entschgl}
  \boxed{ 
    \begin{aligned}
      0 = & \; \Vev{ \left[ Q , a_1^{\text{out}} \ldots a_n^{\text{out}}
      a_1^{\dagger \: \text{in}} \ldots a_m^{\dagger \: \text{in}} \right] }
      \\ = & \; \sum_i \Vev{ a_1^{\text{out}} \ldots \left[ Q,
      a_i^{\text{out}} \right] \ldots } + \sum_j \Vev{ a_1^{\text{out}} \ldots
      \left[ Q , a_j^{\dagger \: \text{in}} \right] \ldots }
    \end{aligned}}
\end{equation}
It follows, of course, from the invariance of the vacuum under SUSY
transformations. So starting with a string of creation operators differing
in spin by half a unit from the spin of the annihilation operators we get a
sum of amplitudes for different processes where all incoming and outgoing
particles are SUSY transformed successively. The creation and annihilation
operators needed in the SWI of that kind have to be extracted from the field
operators. An explanation for the way this is done will be given in
the next section. 

%%%%%%%%%%%%%%%%%%%%%%%%%%%%%%%%%%%%%%%%%%%%%%%%%%%%%%%%%%%%%%%%%%%%%%%%%%

\section{Projecting out creation and annihilation operators}

In this section we only summarize the inverse Fourier transformations by which
the creation and annihilation operators of excitations of a scalar or
fermionic quantum field can be projected out with, following these
prescriptions:  

\begin{equation}
  \label{eq:proj}
  \boxed{
    \begin{aligned}
      a (k) & = \ii \int d^3 \vec{x} \; e^{\ii k x}
      \stackrel{\leftrightarrow}{\partial}_t \phi (x) \\ & \\
      b (k,\sigma) & = \int d^3 \vec{x}  \,\; \overline{u} (k, \sigma) \,
      \gamma^0 \psi (x) e^{\ii k x} \\ & \\
      d^{\dagger} (k, \sigma) & = \int d^3 \vec{x} \,\; \overline{v} (k,
      \sigma) \, \gamma^0 \psi (x) e^{- \ii k x}
    \end{aligned}}
\end{equation}
In the first line we made use of the famous abbreviation:
\[ \left( a \stackrel{\leftrightarrow}{\partial}_\mu b \right) \equiv
a (\partial_\mu b) - (\partial_\mu a) b .\] In the case of Majorana spinor
fields, which are important in supersymmetric field theories, the last two
equations are identical. The verification of (\ref{eq:proj}) can be found in
appendix \ref{proofinvfourier}.

%%%%%%%%%%%%%%%%%%%%%%%%%%%%%%%%%%%%%%%%%%%%%%%%%%%%%%%%%%%%%%%%%%%%%

\section{Transformations of creation and annihilation operators}

As was discussed in the first section of this chapter for the derivation of
the SWIs we need the SUSY transformation properties of the creation and
annihilation operators. To derive them we go back to the so called ``chiral''
fields, $\phi$ and $\phi^*$, which are now called 
$\phi_-$ and $\phi_+$. We write down their definitions again:
\begin{equation}
  \label{eq:defchirfeld}
  \boxed{ \phi_\pm \equiv \dfrac{1}{\sqrt{2}} \Bigl( A \mp \ii B \Bigr) }
\end{equation}
At this point, there is a difference in the choice of sign compared to
the work of Grisaru, Pendleton and van Nieuwenhuizen
\cite{Grisaru/Pendleton:1977:smatrix}. 

Now we are -- by the use of the SUSY transformations of the quantum fields and
projecting the creation and annihilation operators out of the field
operators -- able to get the SUSY transformations of the ladder
operators. First we discuss the transformations of creation and
annihilation operators of the ``chiral'' scalar fields $\phi_\pm$, for
which the notation $a^{(\dagger)} (k, \sigma), \sigma \equiv \pm$ is:  
\begin{align}
    \left[ Q(\xi) , a (k, \sigma) \right] = & \; \ii \int d^3 \vec{x} \;
    e^{\ii k x} \stackrel{\leftrightarrow}{\partial}_0 \left[ Q(\xi) ,
    \phi_\sigma (x) \right] \notag\\ 
    = & \; - \dfrac{1}{\sqrt{2}} \int d^3 \vec{x} \; e^{\ii k x}
    \stackrel{\leftrightarrow}{\partial}_0 \Bigl( \overline{\xi} ( 1 + \sigma
    \gamma^5 ) \psi \Bigr) \notag\\
    = & \; \dfrac{\ii}{\sqrt{2}} \overline{\xi} ( 1 + \sigma \gamma^5 )
    \sum_\tau b (k, \tau) u (k, \tau)  
\end{align}
We find the transformation law
\begin{equation}
  \label{eq:transskalerz}
  \boxed{
  \left[ Q(\xi) , a (k, \sigma) \right] = \dfrac{\ii}{\sqrt{2}}
  \overline{\xi} ( 1 + \sigma \gamma^5 ) \sum_\tau b (k, \tau) u (k, \tau) }
\end{equation}

Consider a massless theory, where the spinors $u(k, \tau)$ und $v (k, \tau)$
are eigenstates of the matrix $\gamma^5$. We end up with the concise 
result:  
\begin{equation}
  \label{eq:masselosboson}
  \left[ Q (\xi) , a (k, \sigma) \right] = \sqrt{2} \ii \, 
  \overline{\xi} u (k, \sigma)  \, b (k, \sigma)
\end{equation}

Now we derive the transformation properties for the fermionic annihilation
operators: 
\begin{align}
    \left[ Q(\xi) , b (k, \sigma) \right] = & \; \int d^3 \vec{x} \;
    \overline{u} (k, \sigma) \gamma^0 \left[ Q(\xi) , \psi (x) \right] e^{\ii
    k x} \notag\\ = & \;
    - \ii \overline{u} (k, \sigma) \left( a_A (k) + \ii \gamma^5 a_B
    (k) \right) \xi,  \label{eq:qxib}
\end{align}
where we have used the spinor $\overline{u}$'s equation of motion: 
\begin{equation}
  \label{eq:dirac1}
  \overline{u} (p, \sigma) \left( \fmslash{p} - m \right) = 0.
\end{equation}

When using the chiral fields instead of the scalar and pseudoscalar
ones, it follows: 
\begin{equation}
  \label{eq:transfermerz}
  \boxed{ \left[ Q(\xi) , b (k, \sigma) \right] = - \dfrac{\ii}{\sqrt{2}}
  \sum_\tau \left( \overline{u} (k, \sigma) \left( 1 - \tau \gamma^5 \right)
  \xi \right) a (k, \tau) }
\end{equation}
In the massless case the bispinor is again an eigenstate of the chiral
projectors, so we find: 
\begin{equation}
  \label{eq:masselosfermerz}
    \left[ Q(\xi) , b (k, \sigma) \right] = - \sqrt{2} \ii \, 
    \overline{u} (k, \sigma) \xi a (k, \sigma) .
\end{equation}
We will derive the latter result in a more general context following
the discussion of Grisaru and Pendleton
\cite{Grisaru/Pendleton:1977:smatrix} in section \ref{sec:genderiv}. 

%%%%%%%%%%%%%%%%%%%%%%%%%%%%%%%%%%%%%%%%%%%%%%%%%%%%%%%%%%%%%%%

\section[Anticommutativity, Grassmann numbers and Generators]{Anticommutativity, Grassmann numbers and \\ Generators}

There is a subtlety which may easily be overlooked, but without it, it is
not possible to derive the SUSY transformations of the asymptotic creation
operators. 

For the quantization of field theories including fermions, Grassmann fields
are being used, i.e.~spinor fields whose components are Grassmann odd. This is
necessary to fulfill the demands of the fermions having Fermi-Dirac
statistics. Consider SUSY transformations which contain Grassmann odd constant 
spinors (as $\xi$ above). Those parameters must {\em anti}commute with
the Fermi fields. Consequently, spinor products normally being skew become
symmetric between Fermi fields or between a Fermi field and such a Grassmann
odd parameter (There are two signs when interchanging the two spinors in the
product, one which causes the skewness of the product, namely the contraction
direction of the spinor indices, but also another one from anticommuting the
Grassmann numbers (cf.~the appendix and \cite{Reuter:2000:SUSY})). In
quantizing such a theory, the anticommutativity must be maintained when going
from the classical Fermi fields to the field operators. Because -- with the
exception of the creation and annihilation operators (about which one could be
tempted to assume that they only are responsible for the anticommutativity of
fermions on Hilbert space) -- there are only commuting terms in the field
operators, we have to deduce that the creation and annihilation operators
for fermions remain Grassmann odd with respect to ``classical'' Grassmann
numbers. This means
\begin{equation}
  \label{eq:weitreichendefolgen}
  \left\{ \xi , b(k, \sigma) \right\} = \left\{ \xi, b^\dagger (k, \sigma)
  \right\} = \left\{ \xi, d(k, \sigma) \right\} = \left\{ \xi , d^\dagger (k,
  \sigma) \right\} = 0 ,
\end{equation}
which has noteworthy technical consequences. 

What happens after taking the Hermitean adjoint of an equation like
(\ref{eq:transskalerz})? The left hand side yields:
\begin{equation}
  \Bigl( \left[ Q(\xi) , a(k, \sigma) \right] \Bigr)^\dagger =
   - \left[ Q(\xi) , a^\dagger (k, \sigma) \right] 
\end{equation}
Again we used the Hermiticity of $Q(\xi)$:
\begin{equation}
  Q(\xi)^\dagger = \Bigl( \xi Q + \bar{\xi} \bar{Q} \Bigr)^\dagger = \bar{\xi}
  \bar{Q} + \xi Q = Q(\xi) 
\end{equation}

On the right hand side of (\ref{eq:transskalerz}) it has been taken into
account that a Hermitean adjoint for operators includes complex conjugation of 
ordinary numbers and Grassmann numbers. The order of Grassmann numbers has to
be reversed in complex conjugation:
\begin{equation}
  (g_1 g_2 \ldots g_n)^* = g_n^* \ldots g_2^* g_1^* \qquad g_i \;\;
  \text{Grassmann odd}
\end{equation}
One therefore gets:
\begin{equation}
  \begin{aligned}
    \biggl( \dfrac{\ii}{\sqrt{2}} \overline{\xi} \left( 1 + \sigma \gamma^5
    \right) \sum_\tau u (k, \tau) b (k, \tau) \biggr)^\dagger = & \; -
    \dfrac{\ii}{\sqrt{2}} \sum_\tau b^\dagger (k, \tau) u^\dagger (k, \tau)
    \left( 1 + \sigma \gamma^5 \right) \gamma^0 \xi \\ = & \; 
    - \dfrac{\ii}{\sqrt{2}} \sum_\tau b^\dagger (k, \tau) \overline{u} (k,
    \tau) \left( 1 - \sigma \gamma^5 \right) \xi \\ = & \; 
    + \dfrac{\ii}{\sqrt{2}} \sum_\tau \overline{u} (k, \tau) \left( 1 - \sigma
    \gamma^5 \right) \xi \: b^\dagger (k, \tau)  
  \end{aligned}
\end{equation}
In the last line we used (\ref{eq:weitreichendefolgen}). This finally produces 
the relation:
\begin{equation}
  \label{eq:weitreichendefolgen2}
  \boxed{
  \left[ Q(\xi) , a^\dagger (k, \sigma) \right] = - \dfrac{\ii}{\sqrt{2}}
  \sum_\tau \overline{u} (k, \tau) \left( 1 - \sigma \gamma^5 \right) \xi \:
  b^\dagger (k, \tau)} \quad .
\end{equation}
Altogether there are three signs: One due to the Hermitean adjoint of the
commutator, one by complex conjugation of the explicit factor $\ii$ and a third
one due to the anticommutativity of Fermi field operators and Grassmann
numbers. 

Another important difficulty about signs, related to the anticommutativity
of Fermi field operators and Grassmann numbers, will be discussed in 
chapter \ref{chap:toy}. 

%%%%%%%%%%%%%%%%%%%%%%%%%%%%%%%%%%%%%%%%%%%%%%%%%%%%%%%%%%%%%%%%

\section{General derivation of the transformations}\label{sec:genderiv}

When translating the identities of that kind introduced in the first
section of this chapter into the graphical language of Feynman
diagrams, we discover several subtleties concerning signs (a trade
mark of supersymmetry), which seem to be confusing at the first
sight. We discuss these specialties using an 
example with two incoming and two outgoing particles. Here we have two
{\em in} creation operators and two {\em out} annihilation
operators. With the abbreviation $c_\sigma (k_i)$ instead of $c (k_i,
\sigma)$ for $c \equiv a, b$ this SWI reads:  
\begin{align}
    0 = & \; \Vev{ \left[ Q(\xi) , a_-^{\text{out}} (k_3) b_+^{\text{out}}
    (k_4) a_-^{\dagger \: \text{in}} (k_1) a_-^{\dagger \: \text{in}} (k_2) 
    \right] } \notag\\ = & \; \; \; \; 
    \Vev{ \left[ Q(\xi) , a_-^{\text{out}} (k_3) \right] b_+^{\text{out}} (k_4)
    a_-^{\dagger \: \text{in}} (k_1) a_-^{\dagger \: \text{in}} (k_2)} \notag\\ &
    \; + \Vev{ a_-^{\text{out}} (k_3) \left[ Q(\xi) , b_+^{\text{out}} (k_4)
    \right] a_-^{\dagger \: \text{in}} (k_1) a_-^{\dagger \: \text{in}} (k_2)}
    \notag\\ & \; + \Vev{ a_-^{\text{out}} (k_3) b_+^{\text{out}} (k_4) \left[
    Q(\xi) , a_-^{\dagger \: \text{in}} (k_1) \right] a_-^{\dagger \:
    \text{in}} (k_2)} \notag\\ & \; + \Vev{ a_-^{\text{out}} (k_3)
    b_+^{\text{out}} 
    (k_4) a_-^{\dagger \: \text{in}} (k_1) \left[ Q(\xi) , a_-^{\dagger \:
    \text{in}} (k_2) \right] }     
\end{align}
With the help of the relations (\ref{eq:transskalerz}),
(\ref{eq:transfermerz}) and (\ref{eq:weitreichendefolgen2}) this can
be transformed into:
\begin{align}
    0 = & \; \; \; \; \dfrac{\ii}{\sqrt{2}} \sum_\sigma \Vev{ \left(
    \overline{\xi} {\cal P}_L u (k_3, \sigma) \right) b_\sigma^{\text{out}}
    (k_3) b_+^{\text{out}} (k_4) a_-^{\dagger \: \text{in}} (k_1)
    a_-^{\dagger \: \text{in}} (k_2)} \notag\\ & \; -
    \dfrac{\ii}{\sqrt{2}} \Vev{
    a_-^{\text{out}} (k_3) \left( \overline{u} (k_4, +) {\cal P}_L \xi \right)
    a_+^{\text{out}} (k_4) a_-^{\dagger \: \text{in}} (k_1) a_-^{\dagger \:
    \text{in}} (k_2)} \notag\\ & \; - \dfrac{\ii}{\sqrt{2}} \Vev{
    a_-^{\text{ out}}
    (k_3) \left( \overline{u} (k_4, +) {\cal P}_R \xi \right) a_-^{\text{out}}
    (k_4) a_-^{\dagger \: \text{in}} (k_1) a_-^{\dagger \: \text{in}} (k_2)}
    \notag\\ & \; - \dfrac{\ii}{\sqrt{2}} \sum_\sigma \Vev{
    a_-^{\text{out}} (k_3) 
    b_+^{\text{out}} (k_4) \left( \overline{u} (k_1, \sigma) {\cal P}_R \xi
    \right) b_\sigma^{\dagger \: \text{in}} (k_1) a_-^{\dagger \: \text{in}}
    (k_2)} \notag\\ & \; - \dfrac{\ii}{\sqrt{2}} \sum_\sigma \Vev{
    a_-^{\text{ out}}
    (k_3) b_+^{\text{out}} (k_4) a_-^{\dagger \: \text{in}} (k_1) \left(
    \overline{u} (k_2, \sigma) {\cal P}_R \xi \right) b_\sigma^{\dagger \:
    \text{in}} (k_2) }   
\end{align}
The sum in (\ref{eq:transfermerz}) has been split
up so that there are five terms now. To separate the spinor bilinears
produced by the SUSY transformations from the $S$-matrix
elements, we bring all these factors to the utmost left. Be aware of
picking up a sign in the last two lines by anticommuting the Grassmann
odd spinor bilinear and the fermionic annihilator. One ends up with
\begin{align}
    0 = & \; \; \; \; \dfrac{\ii}{\sqrt{2}} \sum_\sigma \left( \overline{\xi}
    {\cal P}_L u (k_3, \sigma) \right)\Vev{  b_\sigma^{\text{out}} (k_3)
    b_+^{\text{out}} (k_4) a_-^{\dagger \: \text{in}} (k_1) a_-^{\dagger \:
    \text{in}} (k_2)} \notag\\ & \; - \dfrac{\ii}{\sqrt{2}} \left( \overline{u}
    (k_4, +) {\cal P}_L \xi \right) \Vev{ a_-^{\text{out}} (k_3)
    a_+^{\text{out}} (k_4) a_-^{\dagger \: \text{in}} (k_1) a_-^{\dagger \: 
    \text{in}} (k_2)} \notag\\ & \; - \dfrac{\ii}{\sqrt{2}} \left( \overline{u}
    (k_4, +) {\cal P}_R \xi \right) \Vev{ a_-^{\text{out}} (k_3)
    a_-^{\text{out}} (k_4) a_-^{\dagger \: \text{in}} (k_1) a_-^{\dagger \:
    \text{in}} (k_2)} \notag\\ & \; + \dfrac{\ii}{\sqrt{2}} \sum_\sigma \left(
    \overline{u} (k_1, \sigma) {\cal P}_R \xi \right) \Vev{ a_-^{\text{out}}
    (k_3)  b_+^{\text{out}} (k_4)  b_\sigma^{\dagger \: \text{in}} (k_1)
    a_-^{\dagger \: \text{in}} (k_2)} \notag\\ & \; + \dfrac{\ii}{\sqrt{2}}
    \sum_\sigma \left( \overline{u} (k_2, \sigma) {\cal P}_R \xi \right)\Vev{
    a_-^{\text{out}} (k_3) b_+^{\text{out}} (k_4) a_-^{\dagger \: \text{in}}
    (k_1)  b_\sigma^{\dagger \: \text{in}} (k_2) }   
\end{align}
 
There is yet another source for producing signs, but it can only arise
in the context of Dirac fermions -- i.e. charged
fermions. Anticommutation of fermionic annihilators and/or creators
due to the Wick theorem is the origin of these additional sign
factors; we will go into the details in chapter \ref{chap:toy}, which deals
with a model in which Dirac fermions appear.   

Now we want to revisit part of a general derivation of the SWIs in the
formalism originally written down by M.T. Grisaru and H.N. Pendleton
used to derive helicity selection rules in gravitino--graviton scattering 
\cite{Grisaru/Pendleton:1977:SUSY_scattering}. Because the
supercharges commute with the momentum operator and change the
particles' spin by half a unit, we can derive the following relations
for the {\em in} annihilators of particles with spin $j$ and chirality
$\sigma$ in a supersymmetric theory:
\begin{equation}
  \label{eq:allgrel0}
  \begin{aligned}
    \left[ Q(\xi) , a_j (k, \sigma) \right] = & \; \Delta_j (\xi, k, \sigma)
    \cdot a_{j - \frac{1}{2}} (k, \sigma) , \\
    \left[ Q(\xi) , a_{j - \frac{1}{2}} (k, \sigma) \right] = & \; \Delta_{j -
    \frac{1}{2}} (\xi, k, \sigma) \cdot a_j (k, \sigma).
  \end{aligned}
\end{equation}
The momentum operator has the form:
\begin{equation}
  \label{eq:impulsop}
  P^\mu = \sum_\sigma \int d^3 \vec{p} \, p^\mu \Bigl( a_j^\dagger (p, \sigma)
  a_j (p, \sigma) + a_{j - \frac{1}{2}}^\dagger (p, \sigma) a_{j -
  \frac{1}{2}} (p, \sigma) \Bigr)  \: .
\end{equation}
From the fact that the supercharge and the momentum
operator commute, an equation for the two unknown functions $\Delta_j$,
$\Delta_{j - \frac{1}{2}}$ on the right hand side can be deduced 
\begin{equation}
  \begin{aligned}
    \left[ Q(\xi) , P^\mu \right] = & \; \sum_\sigma \int d^3 \vec{p} \, p^\mu 
    \biggl( a_j^\dagger (p, \sigma) \left[ Q(\xi) , a_j (p, \sigma) \right] +
    \left[ Q(\xi) , a_j^\dagger (p, \sigma) \right] a_j (p, \sigma) \\ &
    \;\; + a_{j-\frac{1}{2}}^\dagger (p, \sigma) \left[ Q(\xi) ,
    a_{j - \frac{1}{2}} (p, \sigma) \right] + \left[ Q(\xi) , a_{j -
    \frac{1}{2}}^\dagger (p, \sigma) \right] a_{j - \frac{1}{2}} (p, \sigma)
    \biggr) \\ = & \; \sum_\sigma \int d^3 \vec{p} \, p^\mu \biggl(
    a_j^\dagger (p, \sigma) a_{j - \frac{1}{2}} (p, \sigma) \left( \Delta_j
    (\xi,p,\sigma) - \Delta_{j - \frac{1}{2}}^* (\xi, p, \sigma) \right) \\ &
    \qquad + a_{j - \frac{1}{2}}^\dagger (p, \sigma) a_j (p, \sigma) \left(
    \Delta_{j - \frac{1}{2}} (\xi, p, \sigma) - \Delta_j^* (\xi, p, \sigma)
    \right) \biggr) \\ \stackrel{!}{=} & \; 0 \\ \Longrightarrow & \;
    \Delta_{j - \frac{1}{2}} (\xi, p, \sigma) \, = \, \Delta_j^* (\xi, p,
    \sigma) 
  \end{aligned}
\end{equation}
Defining $\Delta_j \equiv \Delta$ (\ref{eq:allgrel0}) reads
\begin{equation}
  \label{eq:allgrel}
  \boxed{
  \begin{aligned}
    \left[ Q(\xi) , a_j (k, \sigma) \right] = & \; \Delta (\xi, k, \sigma)
    \cdot a_{j - \frac{1}{2}} (k, \sigma) , \\
    \left[ Q(\xi) , a_{j - \frac{1}{2}} (k, \sigma) \right] = & \; \Delta^*
    (\xi, k, \sigma) \cdot a_j (k, \sigma)
  \end{aligned}} \;\; ,
\end{equation}
to be compared with (\ref{eq:masselosboson}) and (\ref{eq:masselosfermerz}). 

More relations can be gained from the Jacobi identity:
\begin{equation}
  \label{eq:jacob}
  \left[ \left[ Q(\xi) , Q(\zeta) \right] , a_j (k, \sigma) \right] + \left[
  \left[ Q(\zeta) , a_j (k, \sigma) \right] , Q(\xi) \right] + \left[ \left[
  a_j (k, \sigma) , Q(\xi) \right] , Q(\zeta) \right] = 0
\end{equation}
This implies the equation:
\begin{equation}
  \label{eq:relatimpdelta}
  \Delta (\zeta, k, \sigma) \cdot \Delta^* (\xi, k, \sigma) - \Delta (\xi, k,
  \sigma) \cdot \Delta^* (\zeta, k, \sigma) = 2 \overline{\xi} \fmslash{k}
  \zeta \quad . 
\end{equation}

As is shown in \cite{Grisaru/Pendleton:1977:SUSY_scattering}, the
explicit form of these functions can be found in the context of
special models. In the last section we derived them directly by
projecting out the annihilators from the field operators. In a general
model this procedure can become arbitrarily complicated, especially if
one has a nondiagonal metric on the space of states or if unphysical
modes are involved.   

%%%%%%%%%%%%%%%%%%%%%%%%%%%%%%%%%%%%%%%%%%%%%%%%%%%%%%%%%%%%%%%%%%%%%

%%% Local Variables: 
%%% mode: latex
%%% TeX-master: t
%%% End: 

%% file: chap3.tex
\chapter{The Wess-Zumino Model}\label{WZ} 

We want to test the SUSY Ward identities of the kind derived in the
last chapter for the Wess-Zumino (WZ) model. This is the
simplest supersymmetric field theoretic model with just one superfield
but the most general renormalizable superpotential. For details
about the model, the particle content and the Feynman rules see 
appendix \ref{appen_wz}.  
 
%%%%%%%%%%%%%%%%%%%%%%%%%%%%%%%%%%%%%%%%%%%%%%%%%%%%%%%%%%

\section{SWI for the WZ model}

We can use the formula (\ref{eq:entschgl}) derived in the last chapter
to check SWI in the WZ model. The starting point -- similar to the
derivation of the Slavnov-Taylor identities -- is a string of field
operators with half integer spin, which only by application of the
symmetry generator (here the supercharge), becomes a physically
possible (in particular non-vanishing) amplitude. First, we have to
translate the formulae from the previous chapter to the physical
fields of the WZ model -- by this we mean the real and imaginary part
of the complex scalar field $\phi$ or the scalar and pseudoscalar part,
respectively. 

To get the transformation properties of annihilators and creators of
the real part $A$ of the complex scalar
field $\phi$ one has to set the term proportional to $\gamma^5$ in
equation (\ref{eq:transskalerz}) equal to zero and to multiply the
result by $\sqrt{2}$. For the imaginary
part $B$ one has to set the term proportional to unity equal to zero, to
set $\sigma$ equal to one and multiply the result by a factor
$\sqrt{2} \ii$. This results in:
\begin{align}
  \left[ Q(\xi) , a_A (k) \right] & = \; \ii \sum_\sigma \overline{\xi} u
  (k, \sigma) b (k, \sigma) \label{trafovern1WZ} \\ & \notag \\
  \left[ Q(\xi) , a_B (k) \right] & = \; - \sum_\sigma \overline{\xi} \gamma^5
  u (k, \sigma) b (k, \sigma) \label{trafovern2WZ} \quad .
\end{align}
For the transformation law of the fermion annihilator it suffices 
to use (\ref{eq:qxib}), 
\begin{equation}
  \label{eq:trafovernWZ}
  \left[ Q(\xi) , b (k, \sigma) \right] = \; - \ii \overline{u} (k,
  \sigma) \Bigl( a_A (k) + \ii \gamma^5 a_B (k) \Bigr) \xi \quad .
\end{equation}

As an example, we take a transformation of a product of an {\em in} creation
operator for one $A$ and one $B$ field, and {\em out} annihilators for
an $A$ field and a Majorana fermion of positive helicity. We therefore
write relation (\ref{eq:entschgl}) in the form
\begin{equation}
  \label{eq:beispiel1}
  \begin{aligned}
    0 \stackrel{!}{=} & \; \Vev{ \left[ Q(\xi) , a_A^{\text{out}}
  (k_3) b^{\text{out}} (k_4, 
    +) a_A^{\dagger \: \text{in}} (k_1) a_B^{\dagger \: \text{in}} (k_2)
    \right] } \\ = & \quad \; \Vev{ a_A^{\text{out}} (k_3) b^{\text{out}}
    (k_4, +) a_A^{\dagger \: \text{in}} (k_1) \left[ Q(\xi) , a_B^{\dagger \:
    \text{in}} (k_2) \right] } \\ & \; + \Vev{ a_A^{\text{out}} (k_3)
    b^{\text{out}} (k_4, +)  \left[ Q(\xi) , a_A^{\dagger \: \text{in}} (k_1)
    \right] a_B^{\dagger \: \text{in}} (k_2) } \\ & \; + \Vev{
    a_A^{\text{out}} (k_3) \left[ Q(\xi) , b^{\text{out}} (k_4, +) \right]
    a_A^{\dagger \: \text{in}} (k_1) a_B^{\dagger \: \text{in}} (k_2) } \\ &
    \; + \Vev{ \left[ Q(\xi) , a_A^{\text{out}} (k_3) \right] b^{\text{out}}
    (k_4, +) a_A^{\dagger \: \text{in}} (k_1) a_B^{\dagger \: \text{in}} (k_2)
    } 
  \end{aligned}
\end{equation}
This seems to relate the amplitudes of four different physical
processes. But as the transformation of a fermionic annihilator
produces a linear combination of annihilators for the scalar and
pseudoscalar fields, $A$ and $B$, respectively, we get indeed five
different processes (here we adopt the convention that processes only
differing by the helicity of a fermion are counted as {\em one} process). 
\begin{equation}
  \label{eq:beispiel2}
  \begin{aligned}
    0 \stackrel{!}{=} & \; - \sum_\sigma \overline{u} (k_2, \sigma)
  \gamma^5 \xi 
    \Vev{a_A^{\text{out}} (k_3) b^{\text{out}} (k_4, +) a_A^{\dagger \:
    \text{in}} (k_1) b^{\dagger \: \text{in}} (k_2, \sigma) } \\ & \; + \ii
    \sum_\sigma \overline{u} (k_1, \sigma) \xi \Vev{a_A^{\text{out}} (k_3)
    b^{\text{out}} (k_4, +) b^{\dagger \: \text{in}} (k_1, \sigma)
    a_B^{\dagger \: \text{in}} (k_2) } \\ & \; - \ii \; \overline{u} (k_4,
    +) \xi \Vev{a_A^{\text{out}} (k_3) a_A^{\text{out}} (k_4) a_A^{\dagger \: 
    \text{in}} (k_1) a_B^{\dagger \: \text{in}} (k_2) } \\ & \; + \overline{u}
    (k_4, +) \gamma^5 \xi \Vev{a_A^{\text{out}} (k_3) a_B^{\text{out}} (k_4)
    a_A^{\dagger \: \text{in}} (k_1) a_B^{\dagger \: \text{in}} (k_2) } \\ &
    \; + \ii \sum_\sigma \overline{\xi} u (k_3, \sigma)
    \Vev{b^{\text{out}} (k_3, \sigma) b^{\text{out}} (k_4, +) a_A^{\dagger \:
    \text{in}} (k_1) a_B^{\dagger \: \text{in}} (k_2) }     
  \end{aligned}
\end{equation}
Note the double sign arising in the last two lines -- as explained in
section \ref{susy_rela} -- coming from a relative sign between the
transformation properties of a creation and an annihilation operator
and one from equation (\ref{eq:weitreichendefolgen2}). With the help
of the relation for $S$-matrix elements and amplitudes, which e.g.~can
be read off from \cite{Peskin/Schroeder:QFT:Text}, p.~105,
\begin{multline}
  \label{eq:relasmatrix}
  \Braket{q_1 \ldots q_n | S | p_1 \ldots p_m }_{\text{conn.}} = \\ \ii {\cal
  M} (p_1 , \ldots , p_m \longrightarrow q_1 , \ldots , q_n ) \: (2\pi)^4
  \delta^4 \biggl(\sum_{i=1}^m p_i \, - \, \sum_{j=1}^n q_j \biggr)
  \quad , 
\end{multline}
equation (\ref{eq:beispiel2}) can immediately be transferred into
Feynman diagrams (omitting the overall factor $\ii$ and also the delta
function for global momentum conservation): 
\begin{equation}
  \label{eq:beispiel3}
  \begin{aligned}
    0 \stackrel{!}{=} & \; - \sum_\sigma \overline{u} (k_2, \sigma)
  \gamma^5 \xi \cdot {\cal 
    M} (A (k_1) \Psi (k_2, \sigma) \longrightarrow A (k_3) \Psi (k_4, +) ) \\
    & \; + \ii \sum_\sigma \overline{u} (k_1, \sigma) \xi \cdot {\cal M}
    (\Psi (k_1, \sigma) B (k_2) \longrightarrow A (k_3) \Psi (k_4, +) ) \\ &
    \; - \ii \overline{u} (k_4, +) \xi \cdot {\cal M} (A (k_1) B (k_2)
    \longrightarrow A (k_3) A (k_4) ) \\ & \; + \overline{u} (k_4, +) \gamma^5
    \xi \cdot {\cal M} (A (k_1) B (k_2) \longrightarrow A (k_3) B (k_4) ) \\ &
    \; + \ii \sum_\sigma \overline{\xi} u (k_3, \sigma) \cdot {\cal M} (A
    (k_1) B (k_2) \longrightarrow \Psi (k_3, \sigma) \Psi (k_4, +) ) \quad .
  \end{aligned}
\end{equation}
Diagrammatically we can write down the following expression:

\begin{equation}
  \label{eq:switest1}
  \begin{aligned}
  0 = & \; - \sum_\sigma \overline{u} (k_2,\sigma) \gamma^5 \xi \cdot \left\{
  \parbox{2cm}{\begin{fmfchar*}(20,15) 
    \fmfleft{i1,i2}
    \fmfright{o1,o2}
    \fmf{plain}{i1,v1}
    \fmf{plain}{v1,o1}
    \fmf{dashes}{i2,v2}
    \fmf{dashes}{v2,o2}
    \fmf{dashes}{v1,v2}
    \fmfdot{v1,v2}
  \end{fmfchar*}} + \parbox{2cm}{\begin{fmfchar*}(20,15) 
    \fmfleft{i1,i2}
    \fmfright{o1,o2}
    \fmf{plain}{i1,v1}
    \fmf{plain}{v1,v2}
    \fmf{dashes}{i2,v1}
    \fmf{dashes}{v2,o2}
    \fmf{plain}{v2,o1}
    \fmfdot{v1,v2}
  \end{fmfchar*}} + \parbox{2cm}{\begin{fmfchar*}(20,15) 
    \fmfleft{i1,i2}
    \fmfright{o1,o2}
    \fmf{plain}{i1,v1}
    \fmf{plain}{v1,v2}
    \fmf{phantom}{i2,v1}
    \fmf{phantom}{v2,o2}
    \fmf{plain}{v2,o1}
    \fmfdot{v1,v2}
    \fmffreeze
    \fmf{dashes}{i2,v2}
    \fmf{dashes}{o2,v1}
  \end{fmfchar*}} \right\}
 \\ & \\ & \; + \ii \sum_\sigma \overline{u} (k_1, \sigma) \xi \cdot
  \left\{ \parbox{2cm}{\begin{fmfchar*}(20,15)
    \fmfleft{i1,i2}
    \fmfright{o1,o2}
    \fmf{plain}{i1,v1}
    \fmf{plain}{v1,o1}
    \fmf{dbl_dashes}{i2,v2}
    \fmf{dashes}{v2,o2}
    \fmf{dbl_dashes}{v1,v2}
    \fmfdot{v1,v2}
  \end{fmfchar*}} + \parbox{2cm}{\begin{fmfchar*}(20,15)
    \fmfleft{i1,i2}
    \fmfright{o1,o2}
    \fmf{plain}{i1,v1}
    \fmf{plain}{v1,v2}
    \fmf{dbl_dashes}{i2,v1}
    \fmf{dashes}{v2,o2}
    \fmf{plain}{v2,o1}
    \fmfdot{v1,v2}
  \end{fmfchar*}} +   \parbox{2cm}{\begin{fmfchar*}(20,15)
    \fmfleft{i1,i2}
    \fmfright{o1,o2}
    \fmf{plain}{i1,v1}
    \fmf{plain}{v1,v2}
    \fmf{phantom}{i2,v1}
    \fmf{phantom}{v2,o2}
    \fmf{plain}{v2,o1}
    \fmfdot{v1,v2}
    \fmffreeze
    \fmf{dashes}{v1,o2}
    \fmf{dbl_dashes}{v2,i2}
  \end{fmfchar*}} \right\} \; + \; 0 \\ & \\ & \;  + \overline{u} (k_4, +)
    \gamma^5 \xi \cdot 
  \left\{   \parbox{2cm}{\begin{fmfchar*}(20,15)
    \fmfleft{i1,i2}
    \fmfright{o1,o2}
    \fmf{dashes}{i1,v1}
    \fmf{dashes}{v1,o1}
    \fmf{dbl_dashes}{i2,v2}
    \fmf{dbl_dashes}{v2,o2}
    \fmf{dashes}{v1,v2}
    \fmfdot{v1,v2}
  \end{fmfchar*}} + \parbox{2cm}{\begin{fmfchar*}(20,15)
    \fmfleft{i1,i2}
    \fmfright{o1,o2}
    \fmf{dashes}{i1,v1}
    \fmf{dbl_dashes}{i2,v1}
    \fmf{dbl_dashes}{v1,v2}
    \fmf{dashes}{v2,o1}
    \fmf{dbl_dashes}{v2,o2}
    \fmfdot{v1,v2}
  \end{fmfchar*}} + \parbox{2cm}{\begin{fmfchar*}(20,15)
    \fmfleft{i1,i2}
    \fmfright{o1,o2}
    \fmf{dashes}{i1,v1}
    \fmf{phantom}{i2,v1}
    \fmf{dbl_dashes}{v1,v2}
    \fmf{dashes}{v2,o1}
    \fmf{phantom}{v2,o2}
    \fmfdot{v1,v2}
    \fmffreeze
    \fmf{dbl_dashes}{i2,v2}
    \fmf{dbl_dashes}{v1,o2}
  \end{fmfchar*}} + \parbox{2cm}{\begin{fmfchar*}(20,15)
    \fmfleft{i1,i2}
    \fmfright{o1,o2}
    \fmf{dashes}{i1,v}
    \fmf{dashes}{v,o1}
    \fmf{dbl_dashes}{i2,v}
    \fmf{dbl_dashes}{v,o2}
    \fmfdot{v}
  \end{fmfchar*}} \right\} \\ & \\ & \; + \ii \sum_\sigma \overline{\xi} u 
    (k_3, \sigma) \cdot \left\{ \parbox{2cm}{\begin{fmfchar*}(20,15)
    \fmfleft{i1,i2}
    \fmfright{o1,o2}
    \fmf{dashes}{i1,v1}
    \fmf{plain}{v2,o1}
    \fmf{dbl_dashes}{i2,v1}
    \fmf{dbl_dashes}{v1,v2}
    \fmf{plain}{v2,o2}
    \fmfdot{v1,v2}
  \end{fmfchar*}} + \parbox{2cm}{\begin{fmfchar*}(20,15)
    \fmfleft{i1,i2}
    \fmfright{o1,o2}
    \fmf{dashes}{i1,v1}
    \fmf{plain}{v1,o1}
    \fmf{dbl_dashes}{i2,v2}
    \fmf{plain}{v1,v2}
    \fmf{plain}{v2,o2}
    \fmfdot{v1,v2}
  \end{fmfchar*}} + \parbox{2cm}{\begin{fmfchar*}(20,15)
    \fmfleft{i1,i2}
    \fmfright{o1,o2}
    \fmf{phantom}{i2,v2}
    \fmf{phantom}{i1,v1}
    \fmf{plain}{v1,v2}
    \fmf{plain}{v1,o1}
    \fmf{plain}{v2,o2}
    \fmffreeze
    \fmf{dbl_dashes}{i2,v1}
    \fmf{dashes}{i1,v2}
    \fmfdot{v1,v2}
  \end{fmfchar*}} \right\}
  \end{aligned}
\end{equation}

For the calculation of the amplitudes it is useful to introduce the
Mandelstam variables,
\begin{align}
  s = & \; \left( k_1 + k_2 \right)^2 = \left( k_3 + k_4 \right)^2 \; ,
  \\ 
  t = & \; \left( k_3 - k_1 \right)^2 = \left( k_4 - k_2 \right)^2 \; ,
  \\
  u = & \; \left( k_4 - k_1 \right)^2 = \left( k_3 - k_2 \right)^2 \; .
\end{align}

The explicit analytical expressions for diagrams in which only scalar
(or pseudoscalar) particles are involved are easily found and work in
the same manner as in $\phi^4$ theory or the Standard Model. For the
diagrams with Majorana fermions the Feynman rules for
general fermions worked out by Denner et
al. \cite{Denner/etal:1992:feynmanrules} are needed.    

The terms in braces yield the following analytical expressions, in the
first line of (\ref{eq:switest1})
\begin{equation}
  - \dfrac{\ii \lambda^2}{2} \: \overline{u} (k_4, +) \left( \dfrac{3 m}{t
    - m^2} + \dfrac{\fmslash{k}_1 + \fmslash{k}_2 + m}{s - m^2} +
  \dfrac{\fmslash{k}_2 - \fmslash{k}_3 + m}{u - m^2} \right) u
    (k_2, \sigma) \quad ,
\end{equation}
in the second line
\begin{equation}
  - \dfrac{\lambda^2}{2} \: \overline{u} (k_4, +) \left( \dfrac{ m
    \gamma^5}{u - m^2} + \dfrac{(\fmslash{k}_1 + \fmslash{k}_2 + m)\gamma^5}{s
    - m^2} + \dfrac{\gamma^5 (\fmslash{k}_1 - \fmslash{k}_3 + m)}{t - m^2}
    \right) u (k_1, \sigma) \quad. 
\end{equation}
The diagrams in the third line add up to
\begin{equation}
  - \dfrac{\ii \lambda^2}{2} \left( \dfrac{3 m^2}{t - m^2} + \dfrac{m^2}{s
    - m^2} + \dfrac{m^2}{u - m^2} + 1 \right) \quad ,
\end{equation}
and finally in the last line of (\ref{eq:switest1}):
\begin{equation}
  - \dfrac{\lambda^2}{2} \: \overline{u} (k_4 , +) \left( \dfrac{ m
    \gamma^5}{s - m^2} + \dfrac{\gamma^5 (\fmslash{k}_4 - \fmslash{k}_2 +
    m)}{t - m^2} + \dfrac{(\fmslash{k}_4 - \fmslash{k}_1 + m) \gamma^5}{u -
    m^2} \right) v (k_3, \sigma) \quad .
\end{equation}

It proves to be more convenient for further simplification -- remember
that we still have to multiply the prefactors from equation
(\ref{eq:switest1}) -- to modify the analytical expression for the
diagrams in the last line. To apply the spin summation formula 
\begin{equation}
  \sum_\sigma u (p, \sigma) \overline{u} (p, \sigma) = \fmslash{p} + m
\end{equation}
we reverse the calculational direction of the Majorana fermion line for
the last process. How this works is explained in detail in
\cite{Denner/etal:1992:feynmanrules}. The result looks like
\begin{equation}
  + \dfrac{\lambda^2}{2} \: \overline{u} (k_3, \sigma) \left( \dfrac{m
    \gamma^5}{s - m^2} + \dfrac{(\fmslash{k}_3 - \fmslash{k}_1 + m)
    \gamma^5}{t - m^2} + \dfrac{\gamma^5 (\fmslash{k}_3 - \fmslash{k}_2 +
    m)}{u - m^2} \right) v (k_4, +) \quad ,
\end{equation}
with the change in sign coming from the antisymmetry of the charge
conjugation matrix. There are no additional signs from the vertices because
all couplings are scalar or pseudoscalar (cf. again
\cite{Denner/etal:1992:feynmanrules}). It is important to keep track
of the momenta's signs in the fermion propagators.   

Equation (\ref{eq:switest1}) now has the form:
\begin{equation}
  \label{eq:switest2}
  \begin{aligned}
   {} & 0 \stackrel{!}{=}  \\
   & \; \dfrac{\ii \lambda^2}{2} \: \overline{u} (k_4, +) \left(
    \dfrac{3 m}{t - m^2} + \dfrac{\fmslash{k}_1 + \fmslash{k}_2 + m}{s - m^2}
    + \dfrac{\fmslash{k}_2 - \fmslash{k}_3 + m}{u - m^2} \right)
    (\fmslash{k}_2 + m) \gamma^5 \xi \\
    & \; - \dfrac{\ii \lambda^2}{2} \overline{u} (k_4, +) \left( \dfrac{ m 
    \gamma^5}{u - m^2} + \dfrac{(\fmslash{k}_1 + \fmslash{k}_2 + m)\gamma^5}{s 
    - m^2} + \dfrac{\gamma^5 (\fmslash{k}_1 - \fmslash{k}_3 + m)}{t - m^2}
    \right) (\fmslash{k}_1 + m) \xi \\
    & \; - \dfrac{\ii \lambda^2}{2} \: \overline{u} (k_4, +) \left(
    \dfrac{3 m^2}{t - m^2} + \dfrac{m^2}{s - m^2} + \dfrac{m^2}{u - m^2} + 1
    \right) \gamma^5 \xi \\ & \; + \dfrac{\ii \lambda^2}{2} \overline{\xi}
    (\fmslash{k}_3 + m) \left( \dfrac{m \gamma^5}{s - m^2} +
    \dfrac{(\fmslash{k}_3 - \fmslash{k}_1 + m) \gamma^5}{t - m^2} +
    \dfrac{\gamma^5 (\fmslash{k}_3 - \fmslash{k}_2 + m)}{u - m^2} \right) v
    (k_4, +)
  \end{aligned}
\end{equation}
We divide everything by the common factor $\frac{\ii
\lambda^2}{2}$. To achieve the same structure for all four
contributions we reverse the fermion line in the last process a second
time to arrive at
\begin{equation}
  \label{eq:switest3}
  \begin{aligned}
   {} & 0 \stackrel{!}{=}  \\
   & \; \overline{u} (k_4, +) \left(
    \dfrac{3 m}{t - m^2} + \dfrac{\fmslash{k}_1 + \fmslash{k}_2 + m}{s - m^2}
    + \dfrac{\fmslash{k}_2 - \fmslash{k}_3 + m}{u - m^2} \right)
    (\fmslash{k}_2 + m) \gamma^5 \xi \\
    & \; - \overline{u} (k_4, +) \left( \dfrac{ m
    \gamma^5}{u - m^2} + \dfrac{(\fmslash{k}_1 + \fmslash{k}_2 + m)\gamma^5}{s 
    - m^2} + \dfrac{\gamma^5 (\fmslash{k}_1 - \fmslash{k}_3 + m)}{t - m^2}
    \right) (\fmslash{k}_1 + m) \xi \\
    & \; - \overline{u} (k_4, +) \left(
    \dfrac{3 m^2}{t - m^2} + \dfrac{m^2}{s - m^2} + \dfrac{m^2}{u - m^2} + 1
    \right) \gamma^5 \xi \\ & \; - \overline{u} (k_4, +)
    \left( \dfrac{m \gamma^5}{s - m^2} + \dfrac{\gamma^5 (\fmslash{k}_4 -
    \fmslash{k}_2 + m)}{t - m^2} + \dfrac{(\fmslash{k}_4 - \fmslash{k}_1 + m)
   \gamma^5}{u - m^2} \right) (- \fmslash{k}_3 + m) \xi  
  \end{aligned}
\end{equation} 
The terms proportional to $m^2$ in the first and third row cancel and
we are left with:
\begin{align}
  0 \stackrel{!}{=} & \; \overline{u} (k_4, +) \Biggl[ \dfrac{3 m
  \fmslash{k}_2}{t - m^2} + 
  \dfrac{(\fmslash{k}_1 + \fmslash{k}_2) (\fmslash{k}_2 + m) + m
  \fmslash{k}_2}{s - m^2} + \dfrac{(\fmslash{k}_2 - \fmslash{k}_3)
  (\fmslash{k}_2 + m) + m \fmslash{k}_2}{u - m^2} \notag \\ & \qquad \qquad \;
  + \dfrac{m (\fmslash{k}_1 - m)}{u - m^2} + \dfrac{(\fmslash{k}_1 +
  \fmslash{k}_2 + m) (\fmslash{k}_1 - m)}{s - m^2}  \notag \\ & \qquad\qquad
  \; - \dfrac{(\fmslash{k}_1 - \fmslash{k}_3 - m) (\fmslash{k}_1 - m)}{t -
  m^2} - 1- \dfrac{m (\fmslash{k}_3 + m)}{s - m^2} \notag \\ & \qquad\qquad\;
  + \dfrac{(\fmslash{k}_4 - \fmslash{k}_2 - m) (\fmslash{k}_3 + m)}{t - m^2} - 
  \dfrac{(\fmslash{k}_4 - \fmslash{k}_1 + m) (\fmslash{k}_3 + m)}{u - m^2}
  \Biggr] \gamma^5 \xi   
\end{align}
Considering the terms proportional to $(t - m^2)^{-1}$ and applying the
Dirac equation, 
\begin{equation}
  \label{eq:dirac2}
  \overline{u} (k_4 , +) \left( \fmslash{k}_4 - m \right) = 0 \quad ,
\end{equation}
and momentum conservation 
\begin{equation}
  \label{eq:impulserhalt}
  k_1 + k_2 = k_3 + k_4 \quad ,
\end{equation}
one gets
\begin{align}
  & \; (t - m^2)^{-1} \Bigl[ 3 m \fmslash{k}_2 + \fmslash{k}_2 (\fmslash{k}_1
  - m) - \fmslash{k}_2 (\fmslash{k}_3 + m) \Bigr] \notag \\ = & \; 
  (t - m^2)^{-1} \Bigl[ m \fmslash{k}_2 + \fmslash{k}_2 ( \fmslash{k}_1 -
  \fmslash{k}_3 ) \Bigr] \; \; = \; \; (t - m^2)^{-1} \Bigl[ \fmslash{k}_4
  \fmslash{k}_2 + \fmslash{k}_2 \fmslash{k}_4 - m^2 \Bigr] \notag \\ = & \; (t
  - m^2)^{-1} \Bigl[ 2 (k_2 k_4) - m^2 \Bigr] \; \; = \; \; (t - m^2)^{-1} ( -
  t + m^2 ) \notag \\ = & \; - 1
\end{align}
The terms proportional to $(s - m^2)^{-1}$ add up to
\begin{align}
  & \; (s - m^2)^{-1} \Bigl[ \fmslash{k}_1 \fmslash{k}_2 + m^2 + m
  (\fmslash{k}_1 + \fmslash{k}_2) + m \fmslash{k}_2 + \fmslash{k}_2
  \fmslash{k}_1 - m \fmslash{k}_2 - m \fmslash{k}_3 - m^2 \Bigr] \notag \\ = &
  \; (s - m^2)^{-1} \Bigl[ \fmslash{k}_1 \fmslash{k}_2 + \fmslash{k}_2
  \fmslash{k}_1 + m ( \fmslash{k}_1 + \fmslash{k}_2 - \fmslash{k}_3 ) \Bigr]
  \notag \\ = & \; (s - m^2)^{-1} \Bigl[ 2 (k_1 k_2) + m^2 \Bigr] \; \; = \;
  \; (s - m^2)^{-1} (s - m^2) \notag \\ = & \; + 1 ,
\end{align}
while the remaining $u$ terms yield:
\begin{align}
  & \; (u - m^2)^{-1} \Bigl[ m \fmslash{k}_2 + m^2 - m \fmslash{k}_3 -
  \fmslash{k}_3 \fmslash{k}_2 + m \fmslash{k}_2 + m \fmslash{k}_1 - m^2 -
  \fmslash{k}_2 \fmslash{k}_3 - m \fmslash{k}_2 \Bigr] \notag \\ = & \; (u -
  m^2)^{-1} \Bigl[ - \fmslash{k}_2 \fmslash{k}_3 - \fmslash{k}_3 \fmslash{k}_2
  + m (\fmslash{k}_1 + \fmslash{k}_2 - \fmslash{k}_3) \Bigr] \notag \\ = & \; 
  (u - m^2)^{-1} \Bigl[ - 2 (k_2 k_3) + m^2 \Bigr] \; \; = \; \; (u -
  m^2)^{-1} (u - m^2) \notag \\ = & \; + 1 \qquad .  
\end{align}
So finally all terms add up to zero and the SWI is fulfilled. 

%%%%%%%%%%%%%%%%%%%%%%%%%%%%%%%%%%%%%%%%%%%%%%%%%%%%%%%%%%%%

\section{Jacobi identities for the WZ model}

An important possibility to test the consistency of the SWI themselves is to
check whether the Jacobi identities for the appearing operators, i.e.~the
supercharge and the annihilation and creation operators for the particles, are
valid. 

In the sequel we frequently will use the properties of Grassmann odd
bilinears under the exchange of the two spinors. These can e.g.~be found in
\cite{Reuter:2000:SUSY} (cf.~also appendix \ref{majo}):
\begin{equation}
  \label{eq:symmbispimajo}
  \overline{\eta} \Gamma \xi = \; \left\{ 
    \begin{array}{ll} + \overline{\xi} \Gamma \eta & \text{für} \; \; \Gamma =
      1, \gamma^5, \gamma^5 \gamma^\mu \\ - \overline{\xi} \Gamma \eta &
      \text{für} \; \; \Gamma = \gamma^\mu , \left[ \gamma^\mu , \gamma^\nu
      \right]  \end{array} \right. 
\end{equation}

There is no complication in proving the Jacobi identities for the
scalar annihilation operators: 
\begin{equation}
  \label{eq:jacidwz1}
  - \Bigl[ \left[ Q(\xi) , Q(\eta) \right] , a_A (k) \Bigr] \stackrel{!}{=}
    \Bigl[ \left[ 
    Q(\eta) , a_A (k) \right] , Q(\xi) \Bigr] + \Bigl[ \left[ a_A (k) ,
    Q(\xi) \right] , Q(\eta) \Bigr]  
\end{equation}
For the left hand side we have
\begin{equation*}
  \text{LHS}  \; (\ref{eq:jacidwz1}) \; = - \left[ 2 \overline{\xi} \fmslash{P}
    \eta , a_A (k) \right] = + 2 \left( \overline{\xi} \fmslash{k} \eta
    \right) a_A (k) \; \; . 
\end{equation*}
The right hand side results in 
\begin{equation*}
  \begin{aligned}
  \text{RHS} \; (\ref{eq:jacidwz1}) \; = & \; - \ii \sum_\sigma
  \overline{\eta} u (k, \sigma) \left[ Q(\xi) , b(k,\sigma) \right] - \left(
  \xi \leftrightarrow \eta \right) \\ = & \; - \sum_\sigma \overline{\eta} u
  (k, \sigma) \, \overline{u} (k, \sigma) \Bigl( a_A (k) + \ii \gamma^5 a_B
  (k) \Bigr) \xi - \left(\xi \leftrightarrow \eta \right) \\
  = & \; - \overline{\eta} \left( \fmslash{k} + m \right) \Bigl( a_A (k) + \ii
  \gamma^5 a_B (k) \Bigr) \xi - \left( \xi \leftrightarrow \eta \right) \\
  = & \; - (\overline{\eta} \fmslash{k} \xi) a_A (k) + (\overline{\xi}
  \fmslash{k} \eta) a_A (k) \; \; = \; \;  2 (\overline{\xi} \fmslash{k} \eta)
  a_A (k) \qquad \surd 
  \end{aligned}
\end{equation*}
The calculation for the annihilator of the pseudoscalar particle $B$
is analogous, the only difference being the appearance of $\gamma^5$,
which lets the parts containing $a_A$ vanish and those with $a_B$
remain. 
\begin{equation}
  \label{eq:jacidwz2}
  - \Bigl[ \left[ Q(\xi) , Q(\eta) \right] , a_B (k) \Bigr] = \Bigl[ \left[
    Q(\eta) , a_B (k) \right] , Q(\xi) \Bigr] + \Bigl[ \left[ a_B (k) ,
    Q(\xi) \right] , Q(\eta) \Bigr]  
\end{equation}
\begin{equation*}
  \text{LHS}  \; (\ref{eq:jacidwz2}) \; = - \left[ 2 \overline{\xi} \fmslash{P}
    \eta , a_B (k) \right] = + 2 \left( \overline{\xi} \fmslash{k} \eta
    \right) a_B (k)  
\end{equation*}
\begin{equation*}
  \begin{aligned}
  \text{RHS} \; (\ref{eq:jacidwz2}) \; = & \; + \sum_\sigma \overline{\eta}
  \gamma^5 u (k, \sigma) \left[ Q(\xi) , b (k, \sigma) \right] - (\xi
  \leftrightarrow \eta) \\ = & \; - \ii \sum_\sigma \overline{\eta} \gamma^5 u
  (k, \sigma) \, \overline{u} (k, \sigma) \Bigl( a_A (k) + \ii \gamma^5 a_B
  (k) \Bigr) \xi - (\xi \leftrightarrow \eta) \\ = & \; 
  - \ii \overline{\eta} \gamma^5 \left( \fmslash{k} + m \right) \Bigl( a_A (k)
  + \ii \gamma^5 a_B (k) \Bigr) \xi - (\xi \leftrightarrow \eta) \\ = & \; -
  (\overline{\eta} \fmslash{k} \xi) a_B (k) + (\overline{\xi} \fmslash{k}
  \eta) a_B (k) \;\; = \;\; 2 (\overline{\xi} \fmslash{k} \eta) a_B (k) \qquad
  \surd 
  \end{aligned}
\end{equation*}

A more complicated task is the calculation of the Jacobi identity for
the fermion annihilators. We are forced to use the Fierz transformations,
the Gordon identity and all other formulae for spinors needed before.
First of all the Jacobi identity has, of course, the same form as
usual:
\begin{equation}
  \label{eq:jacidwz3}
  - \Bigl[ \left[ Q(\xi) , Q(\eta) \right] , b (k, \sigma) \Bigr]
    \stackrel{!}{=}  \Bigl[
    \left[ Q(\eta) , b (k, \sigma) \right] , Q(\xi) \Bigr] + \Bigl[ \left[ b
    (k, \sigma) , Q(\xi) \right] , Q(\eta) \Bigr]  
\end{equation}
For the momentum operator on the left hand side one has to insert only the
part of the particle number operators of the fermions, which yields
\begin{equation*}
  \text{LHS}  \; (\ref{eq:jacidwz3}) \; = - \left[ 2 \overline{\xi} \fmslash{P}
    \eta , b (k, \sigma) \right] = + 2 \left( \overline{\xi} \fmslash{k} \eta
    \right) b (k, \sigma) \quad .  
\end{equation*}
The right hand side can be manipulated in the following way:
\begin{equation*}
  \begin{aligned}
    \text{RHS} \; (\ref{eq:jacidwz3}) \; = & \; + \ii \overline{u} (k, \sigma)
    \Bigl( \left[ Q(\xi) , a_A (k) \right] + \ii \gamma^5 \left[ Q(\xi) , a_B
    (k) \right] \Bigr) \eta - (\xi \leftrightarrow \eta) \\ = & \; 
    - \sum_\tau (\overline{u} (k, \sigma) \eta) (\overline{\xi} u (k, \tau) )
    b (k, \tau) \\ & \; + \sum_\tau (\overline{u} (k, \sigma) \gamma^5 \eta) ( 
    \overline{\xi} \gamma^5 u (k, \tau) ) b (k, \tau) - (\xi \leftrightarrow
    \eta)  
  \end{aligned}
\end{equation*}

To calculate these products of spinor bilinears we have to use the Fierz
identities to be found in appendix \ref{appen_fierz} as well as
e.g.~in \cite{Itzykson/Zuber:1980:textbook}. For arbitrary commuting spinors
$\lambda_i, i=1,\ldots,4$ we therefore introduce these 
abbreviations: 
\begin{equation}
  \begin{aligned}
    s (4,2;3,1) = & \; (\overline{\lambda}_4 \lambda_2) \,
    (\overline{\lambda}_3 \lambda_1) \\
    v (4,2;3,1) = & \; (\overline{\lambda}_4 \gamma^\mu \lambda_2) \,
    (\overline{\lambda}_3 \gamma_\mu \lambda_1) \\
    t (4,2;3,1) = & \; \dfrac{1}{2} (\overline{\lambda}_4
    \sigma^{\mu\nu} \lambda_2) \, 
    (\overline{\lambda}_3 \sigma_{\mu\nu} \lambda_1) \\
    a (4,2;3,1) = & \; (\overline{\lambda}_4 \gamma^5 \gamma^\mu \lambda_2)\,
    (\overline{\lambda}_3 \gamma_\mu \gamma^5 \lambda_1) \\
    p (4,2;3,1) = & \; (\overline{\lambda}_4 \gamma^5 \lambda_2) \,
    (\overline{\lambda}_3 \gamma^5 \lambda_1)
  \end{aligned}
\end{equation}

The scalar and pseudoscalar combinations (take care of the sign which has to
be accounted for in case of spinors 2 and 3 being Grassmann odd!) give us the
following relations:
\begin{equation}
  \label{eq:relafierz1}
  s (4,2;3,1) = - \dfrac{1}{4} \Bigl( s (4,1;3,2) + v (4,1;3,2) + t (4,1;3,2)
  + a (4,1;3,2) + p (4,1;3,2) \Bigr) 
\end{equation}
\begin{equation}
  \label{eq:relafierz2}
  p (4,2;3,1) = - \dfrac{1}{4} \Bigl( s (4,1;3,2) - v (4,1;3,2) + t (4,1;3,2)
  - a (4,1;3,2) + p (4,1;3,2) \Bigr)
\end{equation}

Due to equation (\ref{eq:symmbispimajo}) the scalar, the pseudoscalar and the 
pseudovector are symmetric under interchange of the two Grassmann odd spinors,
hence after subtracting the ``exchange'' term $(\xi \leftrightarrow
\eta)$ these contributions vanish. The scalar and pseudoscalar combination
appear on the right hand side of equation (\ref{eq:jacidwz3}) with 
different signs, so the tensorial part of the equation cancels. Only
the vector contribution remains four times (scalar/pseudoscalar and a factor
two by adding the ``exchange'' term), so we have
\begin{equation}
  \label{eq:endlich}
    \text{RHS} \; (\ref{eq:jacidwz3}) \; = \; + \sum_\tau \left( \overline{u}
    (k, \sigma) \gamma^\mu u (k, \tau) \right) \left( \overline{\xi} \gamma_\mu
    \eta \right) b (k, \tau)   
\end{equation}
Finally the Gordon identity
(cf.~e.g.~\cite{Itzykson/Zuber:1980:textbook}, eq. (2.54)) 
\begin{equation}
  \label{eq:gordon}
  \overline{u} (p, \sigma) \gamma^\mu u (p', \tau) = \dfrac{1}{2 m}
  \overline{u} (p, \sigma) \Bigl( (p + p')^\mu + \ii \sigma^{\mu\nu} (p -
  p')_\nu \Bigr) u (p', \tau)
\end{equation}
for identical momenta $p = p' \equiv k$ is used, that is why the
second term vanishes. With the normalization of the Dirac spinors 
\begin{equation}
  \overline{u} (k, \sigma) u (k, \tau) = 2 m \, \delta_{\sigma\tau}
\end{equation}
the polarization sum over $\tau$ collapses and we end up with the desired
result 
\begin{equation}
  \label{eq:jacidwz4}
    \text{RHS} \; (\ref{eq:jacidwz3}) \; = \; + 2 \, \left( \overline{\xi}
    \fmslash{k} \eta \right) b (k, \sigma) \; . \qquad \surd  
\end{equation}

%%% Local Variables: 
%%% mode: latex
%%% TeX-master: "swi"
%%% End: 

%% file: chap4.tex
%%% Local Variables: 
%%% mode: latex
%%% TeX-master: "diss"
%%% End: 

\chapter{A toy model}\label{chap:toy}

\section{General remarks}

To study the effects stemming from mixings of component fields from different
superfields -- independent of the difficulty of spontaneous breakdown of
supersymmetry as in the O'Raifeartaigh model -- we consider another toy
model. It consists of two superfields, a mass term and a trilinear coupling. 
Like for the WZ model we summarize details about the model and the
derivation of the Feynman rules in appendix \ref{appen_toy1}. 

%%%%%%%%%%%%%%%%%%%%%%%%%%%%%%%%%%%%%%%%%%%%%%%%%%%%%%%%%%%%%%%%%%%%%%%

\section{SUSY transformations of Dirac spinors}

The main difference between this toy model and the WZ model is the problem of
diagonalizing the mass terms which arise by the existence of more
than one (at least two as here) superfields. By fusing a left- and a
righthanded Weyl spinor from different superfields (not connected through
Hermitean adjoint) a Dirac bispinor has been constructed. Moreover there is the
problem of ``clashing arrows'' in Feynman diagrams, i.e.~vertices with
apparently incompatible 
directions of the fermion lines. More accurately this means the appearance of
two fermions or two antifermions attached to a vertex in such models. This
may happen if quadratic terms of superfields, whose fermionic components are
combined into Dirac spinors, appear in the trilinear part of the 
superpotential. Another possibility is within the kinetic terms of the vector
superfields in the Lagrangean density of supersymmetric gauge theories
if their fermionic components are combined into Dirac fermions
together with the Weyl components of chiral matter 
superfields, as is the case for the charginos in the MSSM. 

First of all we want to derive the SUSY transformations of the scalar
annihilators, in analogy to the calculations in chapter
\ref{chap_swi_on}. The mode expansions of the {\em charged} scalar
fields -- the scalar component fields of the second superfield -- are
as follows  
\begin{equation}
  \label{eq:zerleggeladskal}
  \begin{aligned}
    \phi (x) = & \; \int \dfrac{d^3 \vec{p}}{(2\pi)^3 2E} \biggl( a_- (p) e^{-
    \ii p x} + a_+^\dagger (p) e^{+ \ii p x} \biggr) \\
    \phi^* (x) = & \; \int \dfrac{d^3 \vec{p}}{(2\pi)^3 2E} \biggl( a_+ (p)
    e^{- \ii p x} + a_-^\dagger (p) e^{+ \ii p x} \biggr)  
  \end{aligned}
\end{equation}
Analogously, the projection onto the two different annihilators
results in 
\begin{equation}
  \label{eq:projverngelad}
  a_- (k) = \ii \int d^3 \vec{x} \: e^{\ii k x}
  \stackrel{\leftrightarrow}{\partial}_t \phi (x) , \qquad  a_+ (k) = \ii
  \int d^3 \vec{x} \: \stackrel{\leftrightarrow}{\partial}_t \phi^* (x) 
\end{equation}
This enables us to write down the transformation laws of the annihilators.
\begin{equation*}
  \begin{aligned}
    \left[ Q(\xi) , a_+ (k) \right] = & \; \ii \int d^3 \vec{x} \: e^{\ii k x}
    \stackrel{\leftrightarrow}{\partial}_t  \left[ Q(\xi) , \phi^* (x) \right]
    \\ = & \; - \sqrt{2} \int d^3 \vec{x} \: e^{\ii k x}
    \stackrel{\leftrightarrow}{\partial}_t  \Bigl( \overline{\xi} {\cal P}_R
    \chi_2 (x) \Bigr) 
  \end{aligned}
\end{equation*}
Here and in the sequel $\chi_1$ and $\chi_2$ are the Majorana bispinors which
could be built of the fermionic component fields of the first and
the second superfield,
\begin{equation}
  \label{eq:majo1majo2}
  \chi_1 = \begin{pmatrix} \psi_1 \\ \bar{\psi}_1 \end{pmatrix} , \qquad
  \chi_2 = \begin{pmatrix} \psi_2 \\ \bar{\psi}_2 \end{pmatrix} .
\end{equation}
With the definition of the Dirac field (\ref{eq:diracspiel}) we are able to
express the righthanded Majorana field in terms of the Dirac field:
\begin{equation}
  \label{eq:diracmajoumform}
  \Psi = \begin{pmatrix} \psi_1 \\ \bar{\psi}_2 \end{pmatrix} = {\cal P}_L
  \chi_1 + {\cal P}_R \chi_2 \quad \Longrightarrow \quad {\cal P}_R \chi_2 =
  {\cal P}_R \Psi 
\end{equation} 
Inserting this in the above equation and performing a calculation in
the same manner as in chapter \ref{chap_swi_on} one finally gets the relation
\begin{equation}
  \label{eq:trafoaplusspiel}
  \boxed{ \left[ Q(\xi) , a_+ (k) \right] = \ii \sqrt{2} \sum_\sigma \Bigl(
  \overline{\xi} {\cal P}_R u (k, \sigma) \Bigr) \: b (k, \sigma) }
\end{equation}
Trying to proceed analogously for the annihilator $a_- (k)$ 
reveals a problem,  
\begin{equation*}
  \begin{aligned}
    \left[ Q(\xi) , a_- (k) \right] = & \; \ii \int d^3 \vec{x} \: e^{\ii k x}
    \stackrel{\leftrightarrow}{\partial}_t  \left[ Q(\xi) , \phi (x) \right]
    \\ = & \; - \sqrt{2} \int d^3 \vec{x} \: e^{\ii k x}
    \stackrel{\leftrightarrow}{\partial}_t  \Bigl( \overline{\xi} {\cal P}_L
    \chi_2 (x) \Bigr) \quad ,  
  \end{aligned} 
\end{equation*}
which consists of an impossibility -- at first look -- to express the
lefthanded Majorana field built of the spinor components of the second
superfield in terms of the components of the Dirac field. The solution
is to pass over to the charge conjugated Dirac field,
\begin{equation}
  \label{eq:diracmajoumform2}
  \Psi^c \equiv {\cal C} \overline{\Psi}^T = \begin{pmatrix} \psi_2 \\
  \bar{\psi}_1 \end{pmatrix} \quad \Longrightarrow \quad {\cal P}_L \chi_2 =
  {\cal P}_L \Psi^c \; \; , 
\end{equation}
with the charge conjugation matrix ${\cal C}$. Remembering the mode
expansion of the charge conjugated field operator,
\begin{equation}
  \label{eq:modentw}
  \Psi^c (x) = \int \dfrac{d^3 \vec{p}}{(2\pi)^3 2 E} \sum_\sigma \Bigl( u (p,
  \sigma) d (p, \sigma) e^{- \ii p x} + v (p, \sigma) b^\dagger (p, \sigma)
  e^{\ii p x} \Bigr) ,
\end{equation}
the result for the SUSY transformation of the antifermion annihilator
is found: 
\begin{equation}
  \label{eq:trafoaminusspiel}
  \boxed{ \left[ Q(\xi) , a_- (k) \right] = \ii \sqrt{2} \sum_\sigma \Bigl(
  \overline{\xi} {\cal P}_L u (k, \sigma) \Bigr) \: d (k, \sigma) }
\end{equation}

How to project the annihilation operators out of the scalar component
fields is well known by now:
\begin{equation}
  \label{eq:trafo_susy_ab}
  \begin{aligned}
    a_A (k) = & \; \ii \int d^3 \vec{x} \: e^{\ii k x}
    \stackrel{\leftrightarrow}{\partial}_t A (x) \\
    a_B (k) = & \; \ii \int d^3 \vec{x} \: e^{\ii k x}
    \stackrel{\leftrightarrow}{\partial}_t \: B (x) .
  \end{aligned}
\end{equation}
The derivation of the transformation laws is at first identical to those of
the annihilators $a_+ (k)$ and $a_- (k)$:
\begin{equation*}
  \begin{aligned}
    \left[ Q(\xi) , a_A (k) \right] = & \; \ii \int d^3 \vec{x} \: e^{\ii k x}
    \stackrel{\leftrightarrow}{\partial}_t \left[ Q(\xi) , A (x) \right] \\ =
    & \;  - \int d^3 \vec{x} \: e^{\ii k x}
    \stackrel{\leftrightarrow}{\partial}_t \left( \overline{\xi} \chi_1
    \right) \\ & \\  
    \left[ Q(\xi) , a_B (k) \right] = & \; \ii \int d^3 \vec{x} \: e^{\ii k x}
    \stackrel{\leftrightarrow}{\partial}_t \left[ Q(\xi) , B (x) \right] \\ =
    & \;  - \ii \int d^3 \vec{x} \: e^{\ii k x}
    \stackrel{\leftrightarrow}{\partial}_t \left( \overline{\xi} \gamma^5 
    \chi_1 \right)     
  \end{aligned}
\end{equation*}
The difference to the scalar fields of the second superfield is, that now the
whole Majorana spinor fields and not only the left- or righthanded
parts are present. In
consequence, the Dirac spinor field and its charge conjugate both appear in the
transformation laws for  $a_A (k)$ and $a_B (k)$ according to the expansion 
\begin{equation}
  \chi_1 = \begin{pmatrix} \psi_1 \\ \bar{\psi}_1 \end{pmatrix} =
  \begin{pmatrix} \psi_1 \\ 0 \end{pmatrix} + \begin{pmatrix} 0 \\
  \bar{\psi}_1 \end{pmatrix} = {\cal P}_L \Psi + {\cal P}_R \Psi^c
\end{equation}
After inserting the above we arrive at the final form of the
transformation laws for $a_A (k)$ and $a_B (k)$, which yield linear
combinations of the Dirac fermion's particle and antiparticle
annihilation operators:  
\begin{equation}
  \label{eq:trafoaskalarspiel}
  \boxed{
  \begin{aligned}
    \left[ Q(\xi) , a_A (k) \right] = & \; \ii \sum_\sigma \biggl(
    \left( \overline{\xi} {\cal P}_L u (k, \sigma) \right) b (k, \sigma) +
    \left( \overline{\xi} {\cal P}_R u (k, \sigma) \right) d (k, \sigma)
    \biggr) \\
    \left[ Q(\xi) , a_B (k) \right] = & \; \sum_\sigma \biggl(
    \left( \overline{\xi} {\cal P}_L u (k, \sigma) \right) b (k, \sigma) -
    \left( \overline{\xi} {\cal P}_R u (k, \sigma) \right) d (k, \sigma)
    \biggr)    
  \end{aligned}}
\end{equation}

{\bf{\em Remark:}} If the annihilators $b (k, \sigma)$ and $d (k,
\sigma)$ are identical we have a real, i.e. a Majorana fermion and the
equations (\ref{eq:trafoaskalarspiel}) are reduced to the relations
(\ref{trafovern1WZ}) and (\ref{trafovern2WZ}). For the chiral scalar fields 
$\phi$ and $\phi^*$ the same is true if we identify $b$ and $d$ and form
the linear combinations $(\sqrt{2})^{-1} \left( \phi +
\phi^* \right)$ and $(\ii \sqrt{2})^{-1} \left( \phi - \phi^* \right)$,
respectively. Hence the generalization of the Wess-Zumino model for Dirac
fermions is consistent. 

\vspace{1cm}

Deriving the SUSY transformations of the fermionic annihilators is more
complicated. We must be aware of the fact that the Dirac bispinor
field is composed from the Weyl spinor field $\psi_1$ as its
lefthanded component and from the Weyl spinor field $\bar{\psi}_2$ as
its righthanded component. Only these two chiral fields appear (we did not 
construct a Majorana bispinor field of   
the component fields $\psi_{1/2}$ and $\bar{\psi}_{1/2}$ from the
first chiral superfield or from the second superfield, respectively)
which means that here we only have to consider the transformations of the
components of the leftchiral superfield $\hat{\Phi}_1$ and the
rightchiral superfield $\hat{\Phi}_2^\dagger$ and not of their
Hermitean adjoints. Everything is consistent and chirality is
conserved. Going back to the roots, the transformation laws are:
\begin{equation}
  \label{eq:zunaechsttrafo1}
  \begin{aligned}
    \left[ Q(\xi) , {\cal P}_L \chi_1 \right] = {\cal P}_L \left[ Q(\xi) ,
    \chi_1 \right] = & \; - \ii {\cal P}_L (\ii \fmslash{\partial}) \left( A +
    \ii \gamma^5 B \right) \xi + \ii \sqrt{2} \, F_1 {\cal P}_L \xi \\
    \rightarrow & \; - \ii (\ii \fmslash{\partial}) {\cal P}_R \left( A + \ii
    \gamma^5 B \right) \xi - \ii m \sqrt{2} \phi^* {\cal P}_L \xi \\ = & \; -
    \ii (\ii \fmslash{\partial}) \left( A + \ii B \right) {\cal P}_R \xi - \ii
    m \sqrt{2} \phi^* {\cal P}_L \xi  
  \end{aligned}
\end{equation}
In the second line we inserted the equation of motion for the auxiliary
field $F_1$ and took the one-particle pole for the asymptotic fields
of the theory. By the same method one gets for the righthanded fermion
field of the second superfield
\begin{equation}
  \label{eq:zunaechsttrafo2}
  \begin{aligned}
    \left[ Q(\xi) , {\cal P}_R \chi_2 \right] = {\cal P}_R \left[ Q(\xi) ,
    \chi_2 \right] = & \; - \ii {\cal P}_R (\ii \fmslash{\partial}) \sqrt{2}
    \left( {\cal P}_R \phi + {\cal P}_L \phi^* \right) \xi + \ii \sqrt{2} \,
    F_2^* {\cal P}_R \xi \\ \rightarrow & \; - \ii (\ii \fmslash{\partial})
    \sqrt{2} {\cal P}_L \left( {\cal P}_R \phi + {\cal P}_L \phi^* \right) \xi
    - \ii m \left( A + \ii B \right) {\cal P}_R \xi \\ = & \; -
    \ii (\ii \fmslash{\partial}) \sqrt{2} \phi^* {\cal P}_L \xi - \ii
    m  \left( A + \ii B \right) {\cal P}_R \xi  \quad .
  \end{aligned}
\end{equation}
Here, for inserting the poles of the asymptotic fields into the
equation of motion for the auxiliary field $F_2$, it is important to
note that for the SUSY transformation of the lefthanded Weyl spinor
field the auxiliary field is multiplied with the lefthanded Grassmann
spinor $\xi$, whereas for the transformation of the righthanded Weyl
spinor field we have the complex conjugated auxiliary field multiplied
by the righthanded Grassmann spinor $\bar{\xi}$ (cf.~chapter 1, and
\cite{Reuter:2000:SUSY}, \cite{Weinberg:QFTv3:Text},
\cite{Wess/Bagger:SUSY:text}).  

Combining the two transformation laws (\ref{eq:zunaechsttrafo1}) and
(\ref{eq:zunaechsttrafo2}) one reaches 
\begin{equation}
  \label{eq:trafokomplexfermion}
  \begin{aligned}
    \left[ Q(\xi) , \Psi \right] = & \; {\cal P}_L \left[ Q(\xi) , \chi_1
    \right] + {\cal P}_R \left[ Q(\xi) , \chi_2 \right] \\ = & \; -\ii (\ii 
    \fmslash{\partial} + m) \left( A + \ii B \right) {\cal P}_R \xi - \ii (\ii 
    \fmslash{\partial} + m) \sqrt{2} \, \phi^* \, {\cal P}_L \xi
  \end{aligned}
\end{equation}

With the help of the equations (\ref{eq:proj}) from chapter
\ref{chap_swi_on} we are 
able to deduce the SUSY transformations of the asymptotic annihilation
operators (and as a by-product also those for the creation
operators). The calculations are analogous to those in (\ref{eq:qxib})
so that the positive-frequency part (the one with the annihilators) remains.
\begin{equation*}
  \begin{aligned}
    \left[ Q(\xi) , b (k, \sigma) \right] = & \; \int d^3 \vec{x} \,
    \overline{u} (k, \sigma) \gamma^0 e^{\ii k x} \left[ Q(\xi) , \Psi (x)
    \right] \\ = & \; - \ii \int d^3 \vec{x} \, \overline{u} (k, \sigma)
    \gamma^0 e^{\ii k x}  (\ii \fmslash{\partial} + m) \left( A + \ii B
    \right) {\cal P}_R \xi \\ & \; - \ii \int d^3 \vec{x} \, \overline{u} (k,
    \sigma) \gamma^0 e^{\ii k x} (\ii \fmslash{\partial} + m) \sqrt{2} \,
    \phi^* \, {\cal P}_L \xi  
  \end{aligned}
\end{equation*}
This implies:
\begin{equation}
  \label{eq:trafoferm1spiel}
  \boxed{ \left[ Q(\xi) , b (k, \sigma) \right] = - \ii \overline{u} (k,
  \sigma) \Bigl( a_A (k) {\cal P}_R + \ii a_B (k) {\cal P}_R + \sqrt{2} a_+
  (k) {\cal P}_L \Bigr) \xi}
\end{equation}

Finally, we reconsider in detail the calculation for the antifermion
creator on which originally is projected, wherein we use the
notation $k = (E, \vec{k})$ und $\tilde{k} = (E, - \vec{k})$:
\begin{align*}
    \left[ Q(\xi) , d^\dagger (k, \sigma) \right] = & \; \int d^3 \vec{x} \,
    \overline{v} (k,\sigma) \gamma^0 e^{- \ii k x} \left[ Q(\xi) , \Psi (x)
    \right] \\ = & \; - \ii \int d^3 \vec{x} \, \overline{v} (k,\sigma)
    \gamma^0 e^{- \ii k x} (\ii \fmslash{\partial} + m) \left( A + \ii B
    \right) {\cal P}_R \xi \\ & \; - \ii \int d^3 \vec{x} \, \overline{v} (k,
    \sigma) \gamma^0 e^{- \ii k x} (\ii \fmslash{\partial} + m) \sqrt{2} \,
    \phi^* \, {\cal P}_L \xi  \\ = & \; - \ii \int \dfrac{d^3 \vec{x} \, d^3
    \vec{p}}{(2\pi)^3 2E} \overline{v} (k, \sigma) \gamma^0 \Bigl( \left(
    \fmslash{p} + m \right) \left( a_A (p) + \ii a_B (p) \right) e^{- \ii (k + 
    p) x} \\ & \qquad \qquad \qquad - \left( \fmslash{p} - m \right) \left(
    a_A^\dagger (p) + \ii a_B^\dagger (p) \right) e^{\ii (p - k) x} \Bigr)
    {\cal P}_R \xi \\ & \; - \ii \sqrt{2} \int \dfrac{d^3 \vec{x} \, d^3
    \vec{p}}{(2\pi)^3 2E} \overline{v} (k, \sigma) \gamma^0 \Bigl( \left(
    \fmslash{p} + m \right) a_+ (p) e^{- \ii (k + p) x} \\ & \qquad \qquad
    \qquad - \left( \fmslash{p} - m \right) a_-^\dagger (p) e^{\ii (k - p) x}
    \Bigr) {\cal P}_L \xi \\ = & \; - \dfrac{\ii}{2 E} \overline{v} (k,
    \sigma) \gamma^0 \Bigl( \left( \fmslash{\tilde{k}} + m \right) \left( a_A
    (\tilde{k}) + \ii a_B (\tilde{k}) \right) \\ & \qquad \qquad \qquad -
    \left( \fmslash{k} - m \right) \left( a_A^\dagger (k) + \ii a_B^\dagger
    (k) \right) \Bigr) {\cal P}_R \xi \\ & \; - \dfrac{\sqrt{2} \ii}{2 E}
    \overline{v} (k, \sigma) \gamma^0 \Bigl( \left( \fmslash{\tilde{k}} + m
    \right) a_+ (\tilde{k}) - \left( \fmslash{k} - m \right) a_-^\dagger (k)
    \Bigr) {\cal P}_L \xi \\ = & \; 
    - \dfrac{\ii}{2 E} \overline{v} (k, \sigma) \Bigl( \left( \fmslash{k} + m
    \right) \gamma^0 \left( a_A (\tilde{k}) + \ii a_B (\tilde{k}) \right) \\ &
    \qquad \qquad \qquad + \left( - 2 E \gamma^0 + \fmslash{k} + m \right)
    \gamma^0 \left( a_A^\dagger (k) + \ii a_B^\dagger (k) \right) \Bigr) {\cal
    P}_R \xi \\ & \; - \dfrac{\sqrt{2} \ii}{2 E} \overline{v} (k, \sigma)
    \Bigl( \left( \fmslash{k} + m \right) \gamma^0 a_+ (\tilde{k}) + \left( -
    2 E \gamma^0 + \fmslash{k} + m \right) \gamma^0 a_-^\dagger (k) 
    \Bigr) {\cal P}_L \xi \\ = & \; + \ii \overline{v} (k, \sigma) \Bigl(
    a_A^\dagger(k) {\cal P}_R + \ii a_B^\dagger (k) {\cal P}_R + \sqrt{2}
    a_-^\dagger (k) {\cal P}_L \Bigr) \xi
\end{align*}
In the last line we used the Dirac equation in the form $\overline{v}
(k, \sigma) \left( \fmslash{k} + m \right) = 0$. Complex conjugation
changes this result into
\begin{equation*}
  \left[ Q(\xi) , d (k, \sigma) \right] = \; + \ii \overline{\xi} \Bigl(
  a_A(k) {\cal P}_L - \ii a_B (k) {\cal P}_L + \sqrt{2} a_- (k) {\cal P}_R
  \Bigr) v (k, \sigma) .
\end{equation*}
``Reversing the calculational direction of the fermion line'' with
respect to the Feynman rules \cite{Denner/etal:1992:feynmanrules} (this way of
speaking originates from changing the calculational directions of fermion
lines in diagrams and refers to the property of fermion bilinears summarized
in appendix \ref{majo}) gives rise to the final result:

\begin{equation}
  \label{eq:trafoferm2spiel}
  \boxed{ \left[ Q(\xi) , d (k, \sigma) \right] = - \ii \overline{u} (k,
  \sigma) \Bigl( a_A (k) {\cal P}_L - \ii a_B (k) {\cal P}_L + \sqrt{2} a_-
  (k) {\cal P}_R \Bigr) \xi}
\end{equation}

%%%%%%%%%%%%%%%%%%%%%%%%%%%%%%%%%%%%%%%%%%%%%%%%%%%%%%%%%%%%%%%%%%%%

\section{A cross-check: Jacobi identities}

The Jacobi identities for this toy model are mostly in complete
analogy to the Jacobi identities for the WZ model, but there are some
fine points which have to be handled carefully. So we show the
calculations in detail here. 

The Jacobi identity has the standard structure:
\begin{equation}
  \label{eq:jacobispiel_a}
  - \Bigl[ \left[ Q(\xi), Q(\eta) \right] , a_A (k) \Bigr] = \Bigl[ \left[
    Q(\eta) , a_A (k) \right] , Q(\xi) \Bigr] + \Bigl[ \left[ a_A (k) ,
    Q(\xi) \right] , Q(\eta) \Bigr]
\end{equation}
Up to now it is well known how to manipulate the left hand side
\begin{equation}
  \text{LHS} \; (\ref{eq:jacobispiel_a}) = + 2 \left( \overline{\xi}
  \fmslash{k} \eta \right) \, a_A (k) 
\end{equation}
There are more steps to take on the right hand side compared to the case of
the WZ model and they are a little bit more complex, too, 
\begin{align*}
    \text{RHS} \; (\ref{eq:jacobispiel_a}) = & \; \; \; \ii \sum_\sigma 
    \left( \overline{\eta} {\cal P}_L u (k, \sigma) \right) \left[ b (k,
    \sigma) , Q(\xi) \right] \\ & \; + \ii \sum_\sigma \left( \overline{\eta}
    {\cal P}_R u (k, \sigma) \right) \left[ d (k, \sigma) , Q(\xi) \right] -
    (\xi \leftrightarrow \eta) \\ = & \; - \overline{\eta}
    {\cal P}_L \left( \fmslash{k} + m \right) \Bigl( a_A (k)
    {\cal P}_R + \ii a_B (k) {\cal P}_R + \sqrt{2} {\cal P}_L a_+ (k) \Bigr)
    \xi \\ & \; - \overline{\eta} {\cal P}_R \left( \fmslash{k} +
    m \right) \Bigl( a_A (k) {\cal P}_L - \ii a_B (k) {\cal P}_L + \sqrt{2}
    {\cal P}_R a_- (k) \Bigr) \xi - (\xi \leftrightarrow \eta) \\ = & \; -
    \left( \overline{\eta} {\cal P}_L \fmslash{k} \xi \right) a_A (k) - \ii
    \left( \overline{\eta} {\cal P}_L \fmslash{k} \xi \right) a_B (k) -
    \sqrt{2} \, m \left( \overline{\eta} {\cal P}_L \xi \right) a_+
    (k) \\ & \; - 
    \left( \overline{\eta} {\cal P}_R \fmslash{k} \xi \right) a_A (k) + \ii
    \left( \overline{\eta} {\cal P}_R \fmslash{k} \xi \right) a_B (k) -
    \sqrt{2} \, m \left( \overline{\eta} {\cal P}_R \xi \right) a_+ (k) - (\xi
    \leftrightarrow \eta) \\ = & \; + 2 \left( \overline{\xi} \fmslash{k} \eta
    \right) a_A (k) \quad \surd
\end{align*}
In the second equation we used the polarization sum for the Dirac
spinors $u (k, \sigma)$, in the third equation the anticommutativity
of $\gamma^5$ with the other gamma matrices and finally, in the fourth
equation, we made use of the identity (\ref{eq:symmbispimajo}), which,
after subtracting the term $(\xi \leftrightarrow \eta)$, forces the
scalar and pseudoscalar parts to vanish so that only the vector
contribution with the annihilator $a (k, \sigma)$ remains. 

The calculation for the annihilation operator of the pseudoscalar
particle, $a_B (k)$, is almost completely analogous. 

What about the annihilators of the chiral scalar fields, i.e. the
component fields from the second supermultiplet? The difference
lies only in the commutator of the supercharge with the
annihilator now producing either the fermion or the antifermion
annihilator. In particular,
\begin{equation}
  \label{eq:jacobispiel_plus}
  - \Bigl[ \left[ Q(\xi) , Q(\eta) \right] , a_+ (k) \Bigr] = \Bigl[ \left[
    Q(\eta) , a_+ (k) \right] , Q(\xi) \Bigr] + \Bigl[ \left[ a_+ (k) , Q(\xi)
    \right] , Q(\eta) \Bigr],
\end{equation}
\begin{equation}
  \label{eq:jacobispiel_minus}
  - \Bigl[ \left[ Q(\xi) , Q(\eta) \right] , a_- (k) \Bigr] = \Bigl[ \left[
    Q(\eta) , a_- (k) \right] , Q(\xi) \Bigr] + \Bigl[ \left[ a_- (k) , Q(\xi)
    \right] , Q(\eta) \Bigr] .
\end{equation}
The left hand sides look as usual,
\begin{equation*}
  \text{LHS} \; (\ref{eq:jacobispiel_plus}) = + 2 \left( \overline{\xi}
  \fmslash{k} \eta \right) a_+ (k) , \qquad \text{LHS} \;
  (\ref{eq:jacobispiel_minus}) = + 2 \left( \overline{\xi} \fmslash{k} \eta
  \right) a_- (k) \quad .
\end{equation*}
No problems show up for the right hand sides:
\begin{equation*}
  \begin{aligned}
    \text{RHS} \; (\ref{eq:jacobispiel_plus}) = & \; \ii \sum_\sigma \left(
    \overline{\eta} {\cal P}_R u (k, \sigma) \right) \left[ b (k, \sigma) ,
    Q(\xi) \right] - (\xi \leftrightarrow \eta) \\ = & \; - \sqrt{2}
    \overline{\eta} {\cal P}_R \left( \fmslash{k} + m \right) \Bigl( a_A (k)
    {\cal P}_R + \ii a_B (k) {\cal P}_R + \sqrt{2} a_+ (k) {\cal P}_L \Bigr)
    \xi - (\xi \leftrightarrow \eta) \\ = & \; - \sqrt{2} \, m \left(
    \overline{\eta} {\cal P}_R \xi \right) a_A (k) - \sqrt{2} \, \ii m \left(
    \overline{\eta} {\cal P}_R \xi \right) a_B (k) \\ & \; - 2 \left(
    \overline{\eta} {\cal P}_R \fmslash{k} \xi \right) a_+ (k) - (\xi
    \leftrightarrow \eta) \\ = & \; + 2 \left( \overline{\xi} \fmslash{k} \eta
    \right) a_+ (k) \quad \surd  \qquad ,   
  \end{aligned}
\end{equation*}
\begin{equation*}
  \begin{aligned}
    \text{RHS} \; (\ref{eq:jacobispiel_minus}) = & \; \ii \sum_\sigma \left(
    \overline{\eta} {\cal P}_L u (k, \sigma) \right) \left[ d (k, \sigma) ,
    Q(\xi) \right] - (\xi \leftrightarrow \eta) \\ = & \; - \sqrt{2}
    \overline{\eta} {\cal P}_L \left( \fmslash{k} + m \right) \Bigl( a_A (k)
    {\cal P}_L - \ii a_B (k) {\cal P}_L + \sqrt{2} a_- (k) {\cal P}_R \Bigr)
    \xi - (\xi \leftrightarrow \eta) \\ = & \; - \sqrt{2} \, m \left(
    \overline{\eta} {\cal P}_L \xi \right) a_A (k) + \sqrt{2} \, \ii m \left(
    \overline{\eta} {\cal P}_L \xi \right) a_B (k) \\ & \; - 2 \left(
    \overline{\eta} {\cal P}_L \fmslash{k} \xi \right) a_- (k) - (\xi
    \leftrightarrow \eta) \\ = & \; + 2 \left( \overline{\xi} \fmslash{k} \eta
    \right) a_- (k) \quad \surd \qquad ,   
  \end{aligned}
\end{equation*}
where the last line again follows from (\ref{eq:symmbispimajo}).

There is nothing new about the Jacobi identities of the
fermion annihilation operators, but for the sake of completeness, we
list the calculations here, too. Again we have the standard structure
\begin{equation}
  \label{eq:jacobispiel_ferm1}
  - \Bigl[ \left[ Q(\xi) , Q(\eta) \right] , b (k, \sigma) \Bigr] = \Bigl[
    \left[ Q(\eta) , b (k, \sigma) \right] , Q(\xi) \Bigr] + \Bigl[ \left[ b
    (k, \sigma) , Q(\xi) \right] , Q(\eta) \Bigr] 
\end{equation}
\begin{equation}
  \label{eq:jacobispiel_ferm2}
  - \Bigl[ \left[ Q(\xi) , Q(\eta) \right] , d (k, \sigma) \Bigr] = \Bigl[
    \left[ Q(\eta) , d (k, \sigma) \right] , Q(\xi) \Bigr] + \Bigl[ \left[ d
    (k, \sigma) , Q(\xi) \right] , Q(\eta) \Bigr] 
\end{equation}
The left hand sides are:
\begin{equation*}
  \text{LHS} \; (\ref{eq:jacobispiel_ferm1}) = + 2 \left( \overline{\xi}
  \fmslash{k} \eta \right) b (k, \sigma) \quad ,
\end{equation*}
\begin{equation*}
  \text{LHS} \; (\ref{eq:jacobispiel_ferm2}) = + 2 \left( \overline{\xi}
  \fmslash{k} \eta \right) d (k, \sigma) \quad .
\end{equation*}
For the right hand side we find
\begin{equation*}
  \begin{aligned}
    \text{RHS} \; (\ref{eq:jacobispiel_ferm1}) = & \; - \ii \overline{u} (k,
    \sigma) \Bigl( \left[ a_A (k) , Q(\xi) \right] {\cal P}_R + \ii \left[ a_B
    (k) , Q(\xi) \right] {\cal P}_R \\ & \; \qquad + \sqrt{2} \left[ a_+ (k) ,
    Q(\xi) \right] {\cal P}_L \Bigr)- (\xi \leftrightarrow \eta) \\ = & \; -
    \sum_\tau \left( \overline{u} (k, \sigma) {\cal P}_R \eta \right) \Bigl(
    \overline{\xi} {\cal P}_L u (k, \tau) b (k, \tau) + \overline{\xi} {\cal
    P}_R u (k, \tau) d (k, \tau) \Bigr) \\ & \; 
    - \sum_\tau \left( \overline{u} (k, \sigma)
    {\cal P}_R \eta \right) \Bigl( \overline{\xi} {\cal P}_L u (k, \tau) b (k,
    \tau) - \overline{\xi} {\cal P}_R u (k, \tau) d (k, \tau) \Bigr) \\ & \; -
    2 \sum_\tau \left( \overline{u} (k, \sigma) {\cal P}_L \eta \right) \left(
    \overline{\xi} {\cal P}_R u (k, \tau) \right) b (k, \tau)  \quad - \quad
    (\xi \leftrightarrow \eta) 
  \end{aligned}
\end{equation*}

Obviously the contributions of the antifermion annihilators cancel
out. In this calculation, by multiplying out the chiral spinor
bilinears, one gets the same scalar and pseudoscalar terms as for the
Jacobi identity for the fermion annihilator of the WZ model
(\ref{eq:jacidwz3}), so we can use that earlier result. 
\begin{equation*}
  \begin{aligned}
    \text{RHS} \; (\ref{eq:jacobispiel_ferm1}) =  & \; - 2 \sum_\tau \Bigl(
    \left( \overline{u} (k, \sigma) {\cal P}_R \eta \right) \left(
    \overline{\xi} {\cal P}_L u (k, \tau) \right) \\ & \; \qquad + \left(
    \overline{u} (k, \sigma) {\cal P}_L \eta \right) \left( \overline{\xi}
    {\cal P}_R u (k, \tau) \right) \Bigr) b (k, \tau) - (\xi 
    \leftrightarrow \eta) \\ = & \; - \sum_\tau \left( \overline{u} (k,
    \sigma) \eta \right) \left( \overline{\xi} u (k, \tau) \right) b (k, \tau)
    \\ & \; + \sum_\tau \left( \overline{u} (k, \sigma) \gamma^5 \eta \right)
    \left( \overline{\xi} \gamma^5 u (k, \tau) \right) b (k, \tau) - (\xi
    \leftrightarrow \eta) \\ = & \; + 2 \left( \overline{\xi} \fmslash{k} \eta
    \right) \, b (k, \sigma) \quad  \surd 
  \end{aligned}
\end{equation*}
The calculation for $d(k, \sigma)$ is analogous. 

%%%%%%%%%%%%%%%%%%%%%%%%%%%%%%%%%%%%%%%%%%%%%%%%%%%%%%%%%%%%%%%%%%%%%%%%%%

\section{Wick theorem and plenty of signs}\label{sec:plentysigns}

Another point of utmost importance appears whenever charged fermions
come into play: We have to take care of relative signs between
amplitudes belonging to different processes in the same SWI. This is due to the
Wick theorem, with the signs stemming from disentangling the contractions of
the interaction operators of Yukawa type $\overline{\Psi} \Psi
\phi$. To illuminate this further, we want to show an example considering the
SWI:  
\begin{equation}
  \label{eq:SWIspiel}
  0 = \Vev{ \left[ Q(\xi) , a^{\text{out}}_A (k_3) d^{\text{out}} (k_4, +)
  a_A^{\text{in} \: \dagger} (k_1) a_A^{\text{in} \: \dagger} (k_2) \right] }
\end{equation}
This produces a relation between the following processes of
the diagrammatical form (for the vertices and propagators see appendix
\ref{appen_toy1}): 
\begin{align}
    0 = & \; \boxed{(-1)\cdot} \: \ii \sum_\sigma \left( \overline{\xi} {\cal 
    P}_L u (k_3,\sigma) \right) \cdot \left\{
  \parbox{2cm}{\begin{fmfchar*}(20,15)
    \fmfleft{i1,i2}
    \fmfright{o1,o2}
    \fmf{dashes}{i1,v1}
    \fmf{dashes}{i2,v1}
    \fmf{fermion}{o2,v2}
    \fmf{fermion}{v2,o1}
    \fmf{dashes}{v1,v2}
    \fmfdot{v1,v2}
  \end{fmfchar*}} + \parbox{2cm}{\begin{fmfchar*}(20,15)
    \fmfleft{i1,i2}
    \fmfright{o1,o2}
    \fmf{dashes}{i2,v2}
    \fmf{dashes}{i1,v1}
    \fmf{fermion}{o2,v2}
    \fmf{fermion}{v2,v1}
    \fmf{fermion}{v1,o1}
    \fmfdot{v1,v2}
  \end{fmfchar*}} + \parbox{2cm}{\begin{fmfchar*}(20,15)
    \fmfleft{i1,i2}
    \fmfright{o1,o2}
    \fmf{phantom}{i2,v2}
    \fmf{phantom}{i1,v1}
    \fmf{fermion}{o2,v2}
    \fmf{fermion}{v2,v1}
    \fmf{fermion}{v1,o1}
    \fmfdot{v1,v2}
    \fmffreeze
    \fmf{dashes}{i2,v1}
    \fmf{dashes}{i1,v2}
  \end{fmfchar*}} \right\} 
 \notag\\ & \notag\\ & \; - \ii \overline{u} (k_4, +)
    {\cal P}_L \xi \cdot                             
  \left\{   \parbox{1.8cm}{\begin{fmfchar*}(18,13)
    \fmfleft{i1,i2}
    \fmfright{o1,o2}
    \fmf{dashes}{i1,v1}
    \fmf{dashes}{v1,o1}
    \fmf{dashes}{i2,v2}
    \fmf{dashes}{v2,o2}
    \fmf{dashes}{v1,v2}
    \fmfdot{v1,v2}
  \end{fmfchar*}} + \parbox{1.8cm}{\begin{fmfchar*}(18,13)
    \fmfleft{i1,i2}
    \fmfright{o1,o2}
    \fmf{dashes}{i1,v1}
    \fmf{dashes}{i2,v1}
    \fmf{dashes}{v1,v2}
    \fmf{dashes}{v2,o1}
    \fmf{dashes}{v2,o2}
    \fmfdot{v1,v2}
  \end{fmfchar*}} + \parbox{1.8cm}{\begin{fmfchar*}(18,13)
    \fmfleft{i1,i2}
    \fmfright{o1,o2}
    \fmf{dashes}{i1,v1}
    \fmf{phantom}{i2,v1}
    \fmf{dashes}{v1,v2}
    \fmf{dashes}{v2,o1}
    \fmf{phantom}{v2,o2}
    \fmfdot{v1,v2}
    \fmffreeze
    \fmf{dashes}{i2,v2}
    \fmf{dashes}{v1,o2}
  \end{fmfchar*}} + \parbox{1.8cm}{\begin{fmfchar*}(18,13)
    \fmfleft{i1,i2}
    \fmfright{o1,o2}
    \fmf{dashes}{i1,v}
    \fmf{dashes}{v,o1}
    \fmf{dashes}{i2,v}
    \fmf{dashes}{v,o2}
    \fmfdot{v}
  \end{fmfchar*}} \right\} 
 \notag\\ & \notag\\ & \; + \: \boxed{(-1)\cdot} \: \ii \sum_\sigma \left( \overline{u}
 (k_1, \sigma) {\cal P}_L \xi \right) \cdot \left\{
 \parbox{2cm}{\begin{fmfchar*}(20,15) 
    \fmfleft{i1,i2}
    \fmfright{o1,o2}
    \fmf{fermion}{o2,v2}
    \fmf{fermion}{v2,i2}
    \fmf{dashes}{i1,v1}
    \fmf{dashes}{o1,v1}
    \fmf{dashes}{v1,v2}
    \fmfdot{v1,v2}
  \end{fmfchar*}} + \parbox{2cm}{\begin{fmfchar*}(20,15)
    \fmfleft{i1,i2}
    \fmfright{o1,o2}
    \fmf{fermion}{o2,v2}
    \fmf{dashes}{i1,v1}
    \fmf{fermion}{v1,i2}
    \fmf{dashes}{o1,v2}
    \fmf{fermion}{v2,v1}
    \fmfdot{v1,v2}
  \end{fmfchar*}} + \parbox{2cm}{\begin{fmfchar*}(20,15)
    \fmfleft{i1,i2}
    \fmfright{o1,o2}
    \fmf{fermion}{o2,v2}
    \fmf{phantom}{i1,v1}
    \fmf{fermion}{v1,i2}
    \fmf{phantom}{o1,v2}
    \fmf{fermion}{v2,v1}
    \fmffreeze
    \fmf{dashes}{i1,v2}
    \fmf{dashes}{v1,o1}
    \fmfdot{v1,v2}
    \end{fmfchar*}} \right\} 
 \notag\\ & \notag\\ & \; + \: \boxed{(-1)\cdot} \: \ii \sum_\sigma \left( \overline{u}
 (k_2, \sigma) {\cal P}_L \xi \right) \cdot \left\{
 \parbox{2cm}{\begin{fmfchar*}(20,15) 
    \fmfleft{i1,i2}
    \fmfright{o1,o2}
    \fmf{fermion}{o2,v2}
    \fmf{phantom}{v2,i2}
    \fmf{dashes}{o1,v1}
    \fmf{dashes}{v1,v2}
    \fmf{phantom}{i1,v1}
    \fmffreeze
    \fmf{fermion}{v2,i1}
    \fmf{dashes}{v1,i2}
    \fmfdot{v1,v2}
  \end{fmfchar*}} + \parbox{2cm}{\begin{fmfchar*}(20,15)
    \fmfleft{i1,i2}
    \fmfright{o1,o2}
    \fmf{fermion}{o2,v2}
    \fmf{dashes}{i2,v1}
    \fmf{fermion}{v1,i1}
    \fmf{dashes}{o1,v2}
    \fmf{fermion}{v2,v1}
    \fmfdot{v1,v2}
  \end{fmfchar*}} + \parbox{2cm}{\begin{fmfchar*}(20,15)
    \fmfleft{i1,i2}
    \fmfright{o1,o2}
    \fmf{fermion}{o2,v2}
    \fmf{dashes}{v2,i2}
    \fmf{fermion}{v1,i1}
    \fmf{dashes}{o1,v1}
    \fmf{fermion}{v2,v1}
    \fmfdot{v1,v2} \end{fmfchar*}} \right\} \notag \\ & \notag \\
    \label{eq:SWIspiel2} 
\end{align}   

Here we have omitted several processes giving vanishing contributions,
$A A \rightarrow A B$, $A A \rightarrow A \phi^{(*)}$, $A A \rightarrow
\overline{\Psi} \, \overline{\Psi}$ and $A \Psi \rightarrow A
\overline{\Psi}$. At first glance, the signs in boxes might seem totally
arbitrary, but can be verified by the Wick theorem. Before proving this 
statement we show that without these signs the SWI would indeed not be valid. 

The calculation for the SWI is principally analogous to similar calculations
in chapter \ref{WZ} done within the WZ model. Thus we may omit the details
here. No difficulties arise as we can switch directly from analytical
Feynman rules to diagrams. We use the polarization sum of Dirac
spinors and the change of sign, but not of chirality, when ``reversing'' a
fermion line, \cite{Denner/etal:1992:feynmanrules},
\begin{equation}
  \label{eq:umdrehen}
  - \ii \left( \overline{u} (k_4, +) {\cal P}_L \xi \right) = + \ii \left(
    \overline{\xi} {\cal P}_L v (k_4, +) \right) \qquad . 
\end{equation}

The first process $A (k_1) A(k_2) \rightarrow \Psi (k_3, \sigma)
\overline{\Psi} (k_4, +)$ yields, after multiplication with its prefactor and
performing the polarization sum,
\begin{multline}
  \label{eq:prozess_spiel1}
  - 2 g^2 \, \overline{\xi} {\cal P}_L \biggl( \dfrac{3 m \left( \fmslash{k}_3
    + m \right)}{s - m^2} + \dfrac{\left( \fmslash{k}_3 + m \right) \left(
    \fmslash{k}_3 - \fmslash{k}_2 + m \right)}{t - m^2} \\ + \dfrac{\left(
    \fmslash{k}_3 + m \right) \left( \fmslash{k}_3 - \fmslash{k}_1 + m
    \right)}{u - m^2} \biggr) v (k_4, +) \qquad .
\end{multline}

For the purely scalar process $A(k_1) A(k_2) \rightarrow A(k_3) A(k_4)$ we
have to ``reverse a fermion line'' (i.e. the spinorial prefactor) as mentioned
above  
\begin{equation}
  \label{eq:prozess_spiel2}
  + 6 g^2 \, \overline{\xi} {\cal P}_L \biggl(  \dfrac{3 m^2}{s - m^2} +
  \dfrac{3 m^2}{t - m^2} + \dfrac{3 m^2}{u - m^2} + 1 \biggr) v (k_4, +)
  \qquad .
\end{equation}

The scattering $A (k_1) \overline{\Psi} (k_2, \sigma) \rightarrow
A(k_3) \overline{\Psi} (k_4, +)$ of the scalar particle and the antifermion 
sums up to give the amplitude as follows:
\begin{multline}
  \label{eq:prozess_spiel3}
  - 2 g^2 \, \overline{\xi} {\cal P}_L \biggl( \dfrac{- 3 m \left(
    \fmslash{k}_2 - m \right)}{u - m^2} + \dfrac{\left( \fmslash{k}_2 - m
    \right) \left( \fmslash{k}_1 + \fmslash{k}_2 - m \right)}{s - m^2} \\ +
    \dfrac{\left( \fmslash{k}_2 - m \right) \left( \fmslash{k}_2 -
    \fmslash{k}_3 - m \right)}{t - m^2} \biggr) v(k_4, +) 
\end{multline}
With the help of the substitutions $k_1 \leftrightarrow k_2$ and $t
\leftrightarrow u$ we get the amplitude for the remaining fourth process
$\overline{\Psi} (k_1, \sigma) A(k_2) \rightarrow A(k_3) \overline{\Psi} (k_4,
+)$.  

Summing up the amplitudes of these four processes with the appropriate
prefactors gives zero. The calculation is totally identical to the
corresponding one done in the WZ model. Now it is obvious
that the three added signs are necessary for the SWI to be fulfilled. But
where do they come from?

Take a look at the first process as an $S$-matrix element:
\begin{equation}
  \label{eq:smatrix_spiel}
  \Vev{ b^{\text{out}} (k_3, \sigma) d^{\text{out}} (k_4, +) a^{\text{in} \:
  \dagger}_A (k_1) a^{\text{in} \: \dagger}_A (k_2) }
\end{equation}
When examining the three diagrams in the first line of (\ref{eq:SWIspiel2}),
the following expression arises, where we suppress the momentum and spin
arguments as well as the {\em in} and {\em out} labels, 
\begin{equation*}
  \vev{ \contracted[3mm]{}{b}{\;}{\contracted{}{d \;\bigl( }{\overline{\Psi}}
  {\Psi}{}} \contracted{}{A}{\bigr)\bigl(}{A} \contracted[2mm]{}{A}
  {\contracted{}{A}{\bigr)}{a}^\dagger{}} {a}^\dagger } = 
%  (-1) \cdot 
%  \vev{ \contracted[2mm]{}{b} {\contracted{\;}{d}{\bigl(}{\Psi}{}}
%  {\overline{\Psi}} \contracted{}{A}{\bigr)\bigl(}{A} \contracted[2mm]{}{A} 
%  {\contracted{}{A}{\bigr)}{a}^\dagger{}} {a}^\dagger } .
        (-1) \cdot 
  \vev{ \contracted{}{b}{}{\bigl(\overline{\Psi}}
  \contracted{}{\Psi}{}{d} 
  \contracted{}{A}{\bigr)\bigl(}{A} \contracted[2mm]{}{A} 
  {\contracted{}{A}{\bigr)}{a}^\dagger{}} {a}^\dagger } 
\end{equation*}
To disentangle the contraction lines we had to anticommute the fermion
annihilation operators. We used the conventional notations for contractions 
\begin{equation*}
  \begin{aligned}
    \contracted{}{A(x)}{}{A(y)} = & \; \int \dfrac{d^4 p}{(2\pi)^4}
    e^{- \ii p (x - y)} \dfrac{\ii}{p^2 - m^2 + \ii \epsilon} \\
    \Bra{0} \contracted{}{d (p, \sigma)}{}{\Psi} = & \; v (p, \sigma) \\
    \contracted{}{A}{\,}{\:a}^\dagger \Ket{0} = & \; 1 \\
    \ldots \quad & \; \; \ldots
  \end{aligned}
\end{equation*}
Using them, we can correctly convert the Feynman rules into analytical
expressions. By means of this anticommutation, a sign emerges. One is
easily convinced that the SWI with a fermion in the final state instead of an
antifermion does not need this anticommutation. Due to the reversed order of
the two fermion annihilation operators, no such sign arises in that
case. After a short calculation we find that the two other diagrams
contributing to the process considered above pick up signs by the same
mechanism, whereas this would not be the case if the two diagrams
contained the fermion instead of the antifermion annihilator in the 
$S$-matrix element.  

The structure of these signs can be understood with the help of
\cite{Denner/etal:1992:feynmanrules}, on top of page 4. From there we can read
off the sign of an $S$-matrix element to be $(-1)^{P+L+V}$, where $L$ is the
number of closed fermion loops, $P$ is the parity of the permutation of
asymptotic annihilators and creators after having disentangled the fermion
lines, and $V$ is the number of incoming and outgoing antifermions. We do only
deal with tree level diagrams here, so the number of loops always is zero and
no sign is produced by them. The signs stemming from the permutation of the
ladder operators are the same as those between the different
contributions from $s$- and $t$-channel in Bhabha scattering. We had already
taken them into account for the WZ model. While we had only Majorana fermions
there and could have contracted the field operators in an arbitrary way with
the ladder operators for external particles, the fact that we now have to
handle Dirac fermions and the sign problem connected with the existence of
antifermions discussed in \cite{Denner/etal:1992:feynmanrules} is a
new topic arising within our toy model. The signs in 
boxes in (\ref{eq:SWIspiel2}) are due to this effect.   

\vspace{2mm}

Because we need some additional techniques for
calculating an SWI for $(2\rightarrow 2)$-processes, we show a
detailed calculation here, starting with three fermions. In that case
vertices with ``clashing arrows'' will appear. This is examplified with  
\begin{equation}
  \label{eq:beispielinfty}
  0 \stackrel{!}{=} \Vev{ \left[ Q(\xi) , a_A^{\text{out}} (k_3)
  b^{\text{out}} (k_4,+) b^{\text{in} \: \dagger} (k_1, +) d^{\text{in} \:
  \dagger} (k_2, -) \right] } \quad .
\end{equation}
For the first process, $\Psi (k_1, +) \overline{\Psi} (k_2, -) \rightarrow
\overline{\Psi} (k_3, \sigma) \Psi(k_4,+)$, five diagrams contribute,
\begin{equation}
  \label{eq:extensivesbeispiel1}
    \parbox{2cm}{\begin{fmfchar*}(20,15)
    \fmfleft{i1,i2}
    \fmfright{o1,o2}
    \fmf{fermion}{v1,i2}
    \fmf{fermion}{i1,v1}
    \fmf{fermion}{o2,v2}
    \fmf{fermion}{v2,o1}
    \fmf{dashes}{v1,v2}
    \fmfdot{v1,v2}
  \end{fmfchar*}} + 
    \parbox{2cm}{\begin{fmfchar*}(20,15)
    \fmfleft{i1,i2}
    \fmfright{o1,o2}
    \fmf{fermion}{v1,i2}
    \fmf{fermion}{i1,v1}
    \fmf{fermion}{o2,v2}
    \fmf{fermion}{v2,o1}
    \fmf{dbl_dashes}{v1,v2}
    \fmfdot{v1,v2}
  \end{fmfchar*}} +
    \parbox{2cm}{\begin{fmfchar*}(20,15)
    \fmfleft{i1,i2}
    \fmfright{o1,o2}
    \fmf{fermion}{v1,i2}
    \fmf{fermion}{i1,v2}
    \fmf{fermion}{v1,o2}
    \fmf{fermion}{o1,v2}
    \fmf{dots}{v1,v2}
    \fmfdot{v1,v2}
  \end{fmfchar*}} - 
    \parbox{2cm}{\begin{fmfchar*}(20,15)
    \fmfleft{i1,i2}
    \fmfright{o1,o2}
    \fmf{fermion}{v1,i2}
    \fmf{fermion}{i1,v2}
    \fmf{fermion}{o2,v1}
    \fmf{fermion}{v2,o1}
    \fmf{dashes}{v1,v2}
    \fmfdot{v1,v2}
  \end{fmfchar*}} -
    \parbox{2cm}{\begin{fmfchar*}(20,15)
    \fmfleft{i1,i2}
    \fmfright{o1,o2}
    \fmf{fermion}{v1,i2}
    \fmf{fermion}{i1,v2}
    \fmf{fermion}{o2,v1}
    \fmf{fermion}{v2,o1}
    \fmf{dbl_dashes}{v1,v2}
    \fmfdot{v1,v2}
  \end{fmfchar*}} 
\end{equation}
The relative sign of the third diagram (containing the ``clashing arrows'')
has to be determined carefully from the Wick theorem and depends on the
``position of the fermion lines relative to each other''. More signs possibly
arise here, depending on the calculational directions of the fermion lines as
explained in \cite{Denner/etal:1992:feynmanrules}; this can happen, if it
is necessary to anticommute the two fermion field operators in the
interaction terms. Nevertheless this is compensated (cf.~again
\cite{Denner/etal:1992:feynmanrules}) by additional signs produced at the
gamma matrices attached to the vertices, giving the same result. For the
last two diagrams the relative signs, too, stem from the Wick theorem and
can be understood as belonging to exchange diagrams in the same manner as for
Bhabha scattering. The positive sign of the third diagram can be seen as
belonging to a $u$-channel, as the $u$-channel diagram has a relative sign
with respect to the $t$-channel diagrams but not to the $s$-channel
diagrams as in quantum 
electrodynamics (Of course, without Feynman number violating vertices it is
not possible to have $s$-, $t$- {\em and} $u$-channel diagrams there). But the
global sign (which is indispensable for comparison with the other processes
contributing to the SWI) is only calculable with the Wick theorem. For more
complicated processes it is inevitable to use the Wick theorem to get
the correct signs. Fortunately, as will be discussed later, it is
possible to do this in a way compatible with the {\em O'Mega} factorization
procedure. 

The five diagrams give, together with all signs and after summing over
the spin~$\sigma$:
\begin{equation*}
  \begin{aligned}
    \text{1.~process, (\ref{eq:extensivesbeispiel1})} \; = & \; 4 g^2 \Biggl\{
    \dfrac{1}{s - m^2} \Bigl( m \left( \overline{v} (k_2, -) {\cal P}_L u
    (k_1, +) \right) \left( \overline{u} (k_4, +) {\cal P}_R \xi \right) \\ &
    \qquad \qquad - \left( \overline{v} (k_2, -) {\cal P}_R u (k_1, +) \right)
    \left( \overline{u} (k_4, +) \fmslash{k}_3 {\cal P}_R \xi \right) \Bigr)
    \\ & \qquad \; + \dfrac{1}{t - m^2} \Bigl( \left( \overline{u} (k_4, +)
    {\cal P}_R u (k_1, +) \right) \left( \overline{v} (k_2, -) \fmslash{k}_3
    {\cal P}_R \xi \right) \\ & \qquad \qquad - m \left( \overline{u} (k_4, +)
    {\cal P}_L u (k_1, +) \right) \left( \overline{v} (k_2, -) {\cal P}_R \xi
    \right) \Bigr) \\ & \qquad \; - \dfrac{1}{u - m^2} \left( \overline{u}
    (k_4, +) {\cal P}_R u (k_2, -) \right) \left( \overline{v} (k_1, +)
    \fmslash{k}_3 {\cal P}_R \xi \right) \Biggr\}  
  \end{aligned}
\end{equation*}

The second process is decomposed into the two separate parts 
$\Psi (k_1,+) \overline{\Psi} (k_2, -) \rightarrow A (k_3) A (k_4)$,
\begin{equation}
  \label{eq:extensivesbeispiel2}
    \parbox{2cm}{\begin{fmfchar*}(20,15)
    \fmfleft{i1,i2}
    \fmfright{o1,o2}
    \fmf{fermion}{v1,i2}
    \fmf{fermion}{i1,v1}
    \fmf{dashes}{o2,v2}
    \fmf{dashes}{v2,o1}
    \fmf{dashes}{v1,v2}
    \fmfdot{v1,v2}
  \end{fmfchar*}} + 
    \parbox{2cm}{\begin{fmfchar*}(20,15)
    \fmfleft{i1,i2}
    \fmfright{o1,o2}
    \fmf{fermion}{v1,i2}
    \fmf{fermion}{i1,v2}
    \fmf{fermion}{v2,v1}
    \fmf{dashes}{v2,o1}
    \fmf{dashes}{v1,o2}
    \fmfdot{v1,v2}
  \end{fmfchar*}} +  
    \parbox{2cm}{\begin{fmfchar*}(20,15)
    \fmfleft{i1,i2}
    \fmfright{o1,o2}
    \fmf{fermion}{v1,i2}
    \fmf{fermion}{i1,v2}
    \fmf{fermion}{v2,v1}
    \fmf{phantom}{v2,o1}
    \fmf{phantom}{v1,o2}
    \fmffreeze
    \fmf{dashes}{v1,o1}
    \fmf{dashes}{v2,o2}
    \fmfdot{v1,v2}
  \end{fmfchar*}}  
\end{equation}
as well as $\Psi (k_1, +) \overline{\Psi} (k_2, -) \rightarrow A (k_3) B
(k_4)$: 
\begin{equation}
  \label{eq:extensivesbeispiel3}
    \parbox{2cm}{\begin{fmfchar*}(20,15)
    \fmfleft{i1,i2}
    \fmfright{o1,o2}
    \fmf{fermion}{v1,i2}
    \fmf{fermion}{i1,v1}
    \fmf{dashes}{o2,v2}
    \fmf{dbl_dashes}{v2,o1}
    \fmf{dbl_dashes}{v1,v2}
    \fmfdot{v1,v2}
  \end{fmfchar*}} + 
    \parbox{2cm}{\begin{fmfchar*}(20,15)
    \fmfleft{i1,i2}
    \fmfright{o1,o2}
    \fmf{fermion}{v1,i2}
    \fmf{fermion}{i1,v2}
    \fmf{fermion}{v2,v1}
    \fmf{dashes}{v2,o1}
    \fmf{dbl_dashes}{v1,o2}
    \fmfdot{v1,v2}
  \end{fmfchar*}} +  
    \parbox{2cm}{\begin{fmfchar*}(20,15)
    \fmfleft{i1,i2}
    \fmfright{o1,o2}
    \fmf{fermion}{v1,i2}
    \fmf{fermion}{i1,v2}
    \fmf{fermion}{v2,v1}
    \fmf{phantom}{v2,o1}
    \fmf{phantom}{v1,o2}
    \fmffreeze
    \fmf{dashes}{v1,o1}
    \fmf{dbl_dashes}{v2,o2}
    \fmfdot{v1,v2}
  \end{fmfchar*}}  
\end{equation}
It is not difficult to derive the analytical expressions. For 
(\ref{eq:extensivesbeispiel2}) we get 
\begin{equation*}
  - 2 g^2 \, \overline{v} (k_2, -) \biggl( \dfrac{3 m}{s - m^2} +
    \dfrac{\fmslash{k}_1 - \fmslash{k}_3 + m}{u - m^2} + \dfrac{\fmslash{k}_1
    - \fmslash{k}_4 + m}{t - m^2} \biggr) u (k_1, +) \; \left( \overline{u}
    (k_4, +) {\cal P}_R \xi \right), 
\end{equation*}
and for (\ref{eq:extensivesbeispiel3}):
\begin{equation*}
  - 2 g^2 \, \overline{v} (k_2, -) \biggl( \dfrac{m}{s - m^2} -
  \dfrac{\fmslash{k}_1 - \fmslash{k}_3 - m}{u - m^2} + \dfrac{\fmslash{k}_1 -
  \fmslash{k}_4 + m}{t - m^2} \biggr) \gamma^5 u (k_1, +)  \; \left(
  \overline{u} (k_4, +) {\cal P}_R \xi \right) 
\end{equation*}

SUSY transforming the antifermion in the initial state again gives
rise to two different processes, $\Psi (k_1, +) A (k_2) \rightarrow A
(k_3) \Psi (k_4, +)$,
\begin{equation}
  \label{eq:extensivesbeispiel4}
    \parbox{2cm}{\begin{fmfchar*}(20,15)
    \fmfleft{i1,i2}
    \fmfright{o1,o2}
    \fmf{fermion}{i1,v1}
    \fmf{fermion}{v2,o1}
    \fmf{dashes}{o2,v2}
    \fmf{dashes}{v1,i2}
    \fmf{fermion}{v1,v2}
    \fmfdot{v1,v2}
  \end{fmfchar*}} + 
    \parbox{2cm}{\begin{fmfchar*}(20,15)
    \fmfleft{i1,i2}
    \fmfright{o1,o2}
    \fmf{fermion}{i1,v1}
    \fmf{fermion}{v2,o2}
    \fmf{dashes}{v2,i2}
    \fmf{dashes}{v1,o1}
    \fmf{fermion}{v1,v2}
    \fmfdot{v1,v2}
  \end{fmfchar*}} +  
    \parbox{2cm}{\begin{fmfchar*}(20,15)
    \fmfleft{i1,i2}
    \fmfright{o1,o2}
    \fmf{fermion}{i1,v1}
    \fmf{fermion}{v1,o1}
    \fmf{dashes}{i2,v2}
    \fmf{dashes}{v2,o2}
    \fmf{dashes}{v1,v2}
    \fmfdot{v1,v2}
  \end{fmfchar*}}  
\end{equation}  
and $\Psi (k_1, +) B (k_2) \rightarrow A (k_3) \Psi (k_4, +)$:
\begin{equation}
  \label{eq:extensivesbeispiel5}
    \parbox{2cm}{\begin{fmfchar*}(20,15)
    \fmfleft{i1,i2}
    \fmfright{o1,o2}
    \fmf{fermion}{i1,v1}
    \fmf{fermion}{v2,o1}
    \fmf{dashes}{o2,v2}
    \fmf{dbl_dashes}{v1,i2}
    \fmf{fermion}{v1,v2}
    \fmfdot{v1,v2}
  \end{fmfchar*}} + 
    \parbox{2cm}{\begin{fmfchar*}(20,15)
    \fmfleft{i1,i2}
    \fmfright{o1,o2}
    \fmf{fermion}{i1,v1}
    \fmf{fermion}{v1,v2}
    \fmf{fermion}{v2,o2}
    \fmf{dbl_dashes}{v2,i2}
    \fmf{dashes}{v1,o1}
    \fmfdot{v1,v2}
  \end{fmfchar*}} +  
    \parbox{2cm}{\begin{fmfchar*}(20,15)
    \fmfleft{i1,i2}
    \fmfright{o1,o2}
    \fmf{fermion}{i1,v1}
    \fmf{fermion}{v1,o1}
    \fmf{dbl_dashes}{v2,i2}
    \fmf{dashes}{v2,o2}
    \fmf{dbl_dashes}{v1,v2}
    \fmfdot{v1,v2}
  \end{fmfchar*}}  
\end{equation}  
The corresponding terms are:
\begin{equation*}
  2 g^2 \; \overline{u} (k_4, +) \biggl( \dfrac{3 m}{t - m^2} +
  \dfrac{\fmslash{k}_1 + \fmslash{k}_2 + m}{s - m^2} + \dfrac{\fmslash{k}_1
  - \fmslash{k}_3 + m}{u - m^2} \biggr) u (k_1, +) \; \left( \overline{v}
  (k_2, -) {\cal P}_R \xi \right) ,
\end{equation*}
\begin{equation*}
  2 g^2 \: \overline{u} (k_4, +) \biggl( \dfrac{m}{t - m^2} -
  \dfrac{\fmslash{k}_1 - \fmslash{k}_3 - m}{u - m^2} + \dfrac{\fmslash{k}_1
  + \fmslash{k}_2 + m}{s - m^2} \biggr) \gamma^5 u (k_1, +) \; \left(
  \overline{v} (k_2, -) {\cal P}_R \xi \right) .
\end{equation*}

There still remains to perform the SUSY transformation of the fermion
in the initial state with the diagrams of the process $\phi^* (k_1)
\overline{\Psi} (k_2, -) \rightarrow A (k_3) \Psi (k_4, +)$: 
\begin{equation}
  \label{eq:extensivesbeispiel6}
    \parbox{2cm}{\begin{fmfchar*}(20,15)
    \fmfleft{i1,i2}
    \fmfright{o1,o2}
    \fmf{fermion}{v1,i2}
    \fmf{fermion}{v2,o2}
    \fmf{dashes}{o1,v2}
    \fmf{dots}{v1,i1}
    \fmf{fermion}{v1,v2}
    \fmfdot{v1,v2}
  \end{fmfchar*}} +
    \parbox{2cm}{\begin{fmfchar*}(20,15)
    \fmfleft{i1,i2}
    \fmfright{o1,o2}
    \fmf{fermion}{v1,i2}
    \fmf{fermion}{v2,v1}
    \fmf{fermion}{v2,o1}
    \fmf{dots}{v2,i1}
    \fmf{dashes}{v1,o2}
    \fmfdot{v1,v2}
  \end{fmfchar*}} +  
    \parbox{2cm}{\begin{fmfchar*}(20,15)
    \fmfleft{i1,i2}
    \fmfright{o1,o2}
    \fmf{fermion}{v1,i2}
    \fmf{fermion}{v1,o2}
    \fmf{dots}{v2,i1}
    \fmf{dashes}{v2,o1}
    \fmf{dots}{v1,v2}
    \fmfdot{v1,v2}
  \end{fmfchar*}}  
\end{equation}  
The relative (and, again, the global sign) of the second diagram results from
the Wick theorem (here we only show the fermion contractions
explicitly): 
\begin{equation}
  (-1)^2 \cdot \vev{ \contracted{a \,}{b}{\,\bigl(}{\overline{\Psi}}{} 
    \contracted{}{\Psi^c}{\phi^* \bigr) \bigl(} {\overline{\Psi^c}}{} 
    \contracted{}{\Psi^c}{\,A \bigr) \, a^\dagger \,} {d^\dagger} {} }  
\end{equation}
The trick in this calculation is to disentangle the contractions by rewriting
the second interaction operator,
\begin{equation*}
  \overline{\Psi} \Psi A \equiv \overline{\Psi} \Gamma \Psi A = \left(
  \overline{\Psi} \Gamma \Psi A \right)^T = (-1) \cdot \Psi^T {\cal C}^{-1}
  \left( {\cal C} \Gamma {\cal C}^{-1} \right) \Psi^c A \equiv (-1)^2 \cdot
  \overline{\Psi^c} \Psi^c A \quad ,  
\end{equation*}
because in this model only scalar, pseudoscalar or chiral
scalar couplings appear that are invariant (i.e.~their gamma
matrices) under the charge conjugation transformation. One of the
additional signs is due to the anticommutation of the Fermi
field operators when transposing, the other stems from the relations
\begin{equation}
  {\cal C} \overline{\Psi}^T = \Psi^c, \qquad \Psi^T {\cal C}^{-1} = -
  \overline{\Psi^c} \quad . 
\end{equation}
The sum of the last three diagrams results in:
\begin{multline*}
  - 4 g^2 \cdot \overline{u} (k_4, +) \biggl( \dfrac{\fmslash{k}_1 +
    \fmslash{k}_2 + m}{s - m^2} {\cal P}_R - {\cal P}_R \dfrac{\fmslash{k}_1 -
    \fmslash{k}_4 - m}{t - m^2} \\ + \dfrac{2 m}{u - m^2} {\cal P}_R \biggr) u 
    (k_2, -) \left( \overline{v} (k_1, +) {\cal P}_R \xi \right) \qquad .
\end{multline*}

Now we sum up the contributions of the several processes of this SWI
separately for each of the reaction channels. A common
prefactor $2 g^2$ is suppressed in the following.
\begin{equation}
  \label{eq:skanal}
  \begin{aligned}
    {\text{{\bf s-channel}}} \, \propto & \quad 
    2 m \bigl( \overline{v} (k_2, -) {\cal P}_L u (k_1, +) \bigr) \bigl(
    \overline{u} (k_4, +) {\cal P}_R \xi \bigr) \\ & \; - 2 \bigl(
    \overline{v} (k_2, -) {\cal P}_R u (k_1, +) \bigr) \bigl( \overline{u}
    (k_4, +) \fmslash{k}_3 {\cal P}_R \xi \bigr) \\ & \; - 3 m \bigl(
    \overline{v}(k_2, -) u (k_1, +) \bigr) \bigl( \overline{u} (k_4, +) {\cal
    P}_R \xi \bigr) \\ & \; - m \bigl( \overline{v} (k_2, -) \gamma^5 u (k_1,
    +) \bigr) \bigl( \overline{u} (k_4, +) {\cal P}_R \xi \bigr) \\ & \; +
    \bigl( \overline{u} (k_4, +) (\fmslash{k}_1 + \fmslash{k}_2 + m) u(k_1,+)
    \bigr) \bigl( \overline{v} (k_2, -) {\cal P}_R \xi \bigr) \\ & \; + \bigl(
    \overline{u} (k_4, +) (\fmslash{k}_1 + \fmslash{k}_2 + m) \gamma^5 u (k_1,
    +) \bigr) \bigl( \overline{v} (k_2, -) {\cal P}_R \xi \bigr) \\ & \; -
    2 \bigl( \overline{u} (k_4, +) (\fmslash{k}_1 + \fmslash{k}_2 + m) {\cal
    P}_R u (k_2, -) \bigr) \bigl( \overline{v} (k_1, +) {\cal P}_R \xi
    \bigr) 
  \end{aligned}
\end{equation}
\begin{equation}
  \label{eq:tkanal}
  \begin{aligned}
    {\text{{\bf t-channel}}} \, \propto & \quad
    2 \bigl( \overline{u} (k_4, +) {\cal P}_R u (k_1, +) \bigr) \bigl(
    \overline{v} (k_2, -) \fmslash{k}_3 {\cal P}_R \xi \bigr) \\ & \; - 2 m
    \bigl( \overline{u} (k_4, +) {\cal P}_L u (k_1, +) \bigr) \bigl(
    \overline{v} (k_2, -) {\cal P}_R \xi \bigr) \\ & \; - \bigl( \overline{v}
    (k_2, -) (\fmslash{k}_1 - \fmslash{k}_4 + m) u(k_1, +) \bigr) \bigl(
    \overline{u} (k_4, +) {\cal P}_R \xi \bigr) \\ & \; - \bigl( \overline{v}
    (k_2, -) (\fmslash{k}_1 - \fmslash{k}_4 + m) \gamma^5 u (k_1, +) \bigr)
    \bigl( \overline{u} (k_4, +) {\cal P}_R \xi \bigr) \\ & \; + 3 m \bigl(
    \overline{u} (k_4, +) u (k_1, +) \bigr) \bigl( \overline{v} (k_2, -)
    {\cal P}_R \xi \bigr) \\ & \; + m \bigl( \overline{u} (k_4, +) \gamma^5 u
    (k_1, +) \bigr) \bigl( \overline{v} (k_2, -) {\cal P}_R \xi \bigr) \\ &
    \; + 2 \bigl( \overline{u} (k_4, +) {\cal P}_R (\fmslash{k}_1 -
    \fmslash{k}_4 - m) u (k_2, -) \bigr) \bigl( \overline{v} (k_1, +) {\cal
    P}_R \xi \bigr) 
  \end{aligned}
\end{equation}
\begin{equation}
  \label{eq:ukanal}
  \begin{aligned}
    {\text{{\bf u-channel}}} \, \propto & \; 
    - 2 \bigl( \overline{u} (k_4, +) {\cal P}_R u(k_2, -) \bigr) \bigl(
    \overline{v} (k_1, +) \fmslash{k}_3 {\cal P}_R \xi \bigr) \\ & \; - \bigl(
    \overline{v} (k_2, -) (\fmslash{k}_1 - \fmslash{k}_3 + m) u (k_1, +)
    \bigr) \bigl( \overline{u} (k_4, +) {\cal P}_R \xi \bigr) \\ & \; + \bigl(
    \overline{v} (k_2, -) (\fmslash{k}_1 - \fmslash{k}_3 - m) \gamma^5 u (k_1,
    +) \bigr) \bigl( \overline{u} (k_4, +) {\cal P}_R \xi \bigr) \\ & \; +
    \bigl( \overline{u} (k_4, +) (\fmslash{k}_1 - \fmslash{k}_3 + m) u (k_1,
    +) \bigr) \bigl( \overline{v} (k_2, -) {\cal P}_R \xi \bigr) \\ & \; -
    \bigl( \overline{u} (k_4, +) (\fmslash{k}_1 - \fmslash{k}_3 - m) \gamma^5
    u (k_1, +) \bigr) \bigl( \overline{v} (k_2, -) {\cal P}_R \xi \bigr) \\ &
    \; - 4 m \bigl( \overline{u} (k_4, +) {\cal P}_R u (k_2, -) \bigr) \bigl(
    \overline{v} (k_1, +) {\cal P}_R \xi \bigr)
  \end{aligned}
\end{equation}

The first, third and fourth line of (\ref{eq:skanal}) can be combined
to give 
\begin{equation*}
  - 4 m \bigl( \overline{v}_2 {\cal P}_R u_1 \bigr) \bigl( \overline{u}_4
    {\cal P}_R \xi \bigr) 
\end{equation*}
(in the sequel we abbreviate $u (k_1, +)$ by $u_1$ etc.). Adding the
second line from equation (\ref{eq:skanal}), we arrive at
\begin{equation}
  \label{eq:dummy1}
  - 2 \bigl( \overline{v}_2 {\cal P}_R u_1 \bigr) \bigl( \overline{u}_4 \left( 
    \fmslash{k}_3 + 2 m \right) {\cal P}_R \xi \bigr) \quad .
\end{equation}
Adding the fifth and sixth line of (\ref{eq:skanal}) yields
\begin{equation}
  \label{eq:dummy2}
  2 \bigl( \overline{v}_2 {\cal P}_R \xi \bigr) \bigl( \overline{u}_4 \left(
  \fmslash{k}_1 + \fmslash{k}_2 + m \right) {\cal P}_R u_1 \bigr) \quad .
\end{equation}
Applying the Fierz identities, we bring this expression and also the
term of the last line in (\ref{eq:skanal}) into the form of
(\ref{eq:dummy1}). In the following calculation we use the notation
$k_{12} \equiv k_1 + k_2$. The brackets indicate 
our fundamental spinors in spinor products of the Fierz identities. In
contrast to the Fierz identities used for checking the Jacobi
identities, there is no additional sign in here as there is only one
anticommuting spinor.
\begin{equation}
  \label{eq:fierz_dummy1}
  \begin{aligned}
    2 \bigl( \left[ \overline{u}_4 (\fmslash{k}_{12} + m) \right]
    {\cal P}_R u_1 \bigr) \bigl( \overline{v}_2 \left[ {\cal P}_R \xi \right]
    \bigr) = & \; + \dfrac{1}{2} \bigl( \overline{u}_4 (\fmslash{k}_{12} + m)
    {\cal P}_R \xi \bigr) \bigl( \overline{v}_2 {\cal P}_R u_1 \bigr) \\ & \;
    + \dfrac{1}{2} \bigl( \overline{u}_4 (\fmslash{k}_{12} + m) \gamma^5 {\cal
    P}_R \xi \bigr) \bigl( \overline{v}_2 \gamma^5 {\cal P}_R u_1 \bigr) \\ &
    \; + \dfrac{1}{2} \bigl( \overline{u}_4 (\fmslash{k}_{12} + m) \gamma^\mu
    {\cal P}_R \xi \bigr) \bigl( \overline{v}_2 \gamma_\mu {\cal P}_R u_1
    \bigr) \\ & \; + \dfrac{1}{2} \bigl( \overline{u}_4 (\fmslash{k}_{12} + m)
    \gamma^5 \gamma^\mu {\cal P}_R \xi \bigr) \bigl( \overline{v}_2 \gamma_\mu
    \gamma^5 {\cal P}_R u_1 \bigr) \\ & \; + \dfrac{1}{4} \bigl(
    \overline{u}_4 (\fmslash{k}_{12} + m) \sigma^{\mu\nu} {\cal P}_R \xi
    \bigr) \bigl( \overline{v}_2 \sigma_{\mu\nu} {\cal P}_R u_1 \bigr) 
  \end{aligned}
\end{equation}
By Fierzing, the last line of (\ref{eq:skanal}) can be written as
\begin{equation}
  \label{eq:fierz_dummy2}
  \begin{aligned}
    - 2 \bigl( \left[ \overline{u}_4 (\fmslash{k}_{12} + m) \right] {\cal P}_R
    u_2 \bigr) \bigl( \overline{v}_1 \left[ {\cal P}_R \xi \right] \bigr) = &
    \; - \dfrac{1}{2} \bigl( \overline{u}_4 (\fmslash{k}_{12} + m) {\cal P}_R
    \xi \bigr) \bigl( \overline{v}_1 {\cal P}_R u_2 \bigr) \\ & \; -
    \dfrac{1}{2} \bigl( \overline{u}_4 (\fmslash{k}_{12} + m) \gamma^5 {\cal
    P}_R \xi \bigr) \bigl( \overline{v}_1 \gamma^5 {\cal P}_R u_2 \bigr) \\ &
    \; - \dfrac{1}{2} \bigl( \overline{u}_4 (\fmslash{k}_{12} + m) \gamma^\mu
    {\cal P}_R \xi \bigr) \bigl( \overline{v}_1 \gamma_\mu {\cal P}_R u_2
    \bigr) \\ & \; - \dfrac{1}{2} \bigl( \overline{u}_4 (\fmslash{k}_{12} + m) 
    \gamma^5 \gamma^\mu {\cal P}_R \xi \bigr) \bigl( \overline{v}_1 \gamma_\mu
    \gamma^5 {\cal P}_R u_2 \bigr) \\ & \; - \dfrac{1}{4} \bigl(
    \overline{u}_4 (\fmslash{k}_{12} + m) \sigma^{\mu\nu} {\cal P}_R \xi
    \bigr) \bigl( \overline{v}_1 \sigma_{\mu\nu} {\cal P}_R u_2 \bigr) \quad
    . 
  \end{aligned} 
\end{equation}
To give the expressions a common structure we again use the rules of
\cite{Denner/etal:1992:feynmanrules} to ``turn round'' the second term
in parentheses on the right hand side of (\ref{eq:fierz_dummy2}):
\begin{equation}
  \label{eq:fierz_dummy3}
  \begin{aligned}
    (\ref{eq:fierz_dummy2}) \: = &
    \; + \dfrac{1}{2} \bigl( \overline{u}_4 (\fmslash{k}_{12} + m) {\cal P}_R
    \xi \bigr) \bigl( \overline{v}_2 {\cal P}_R u_1 \bigr) \\ & \; +
    \dfrac{1}{2} \bigl( \overline{u}_4 (\fmslash{k}_{12} + m) \gamma^5 {\cal
    P}_R \xi \bigr) \bigl( \overline{v}_2 \gamma^5 {\cal P}_R u_1 \bigr) \\ &
    \; - \dfrac{1}{2} \bigl( \overline{u}_4 (\fmslash{k}_{12} + m) \gamma^\mu
    {\cal P}_R \xi \bigr) \bigl( \overline{v}_2 \gamma_\mu {\cal P}_L u_1
    \bigr) \\ & \; + \dfrac{1}{2} \bigl( \overline{u}_4 (\fmslash{k}_{12} + m)
    \gamma^5 \gamma^\mu {\cal P}_R \xi \bigr) \bigl( \overline{v}_2 \gamma_\mu
    \gamma^5 {\cal P}_L u_1 \bigr) \\ & \; - \dfrac{1}{4} \bigl(
    \overline{u}_4 (\fmslash{k}_{12} + m) \sigma^{\mu\nu} {\cal P}_R \xi
    \bigr) \bigl( \overline{v}_2 \sigma_{\mu\nu} {\cal P}_R u_1 \bigr) 
  \end{aligned}
\end{equation}
When adding (\ref{eq:fierz_dummy1}) and (\ref{eq:fierz_dummy3}) the
tensor part vanishes. Absorbing the $\gamma^5$ matrices into the chiral
projectors the vector contributions in (\ref{eq:fierz_dummy1}) and
(\ref{eq:fierz_dummy3}) cancel the terms containing the axial vector, while
the scalar and pseudoscalar contributions can be combined to give:  
\begin{equation}
  \label{eq:dummy3}
  2 \bigl( \overline{u}_4 (\fmslash{k}_{12} + m) {\cal P}_R \xi \bigr) \bigl(
  \overline{v}_2 {\cal P}_R u_1 \bigr) \quad .
\end{equation}
Summing up (\ref{eq:dummy1}) and (\ref{eq:dummy3}) yields the
following result for the whole $s$-channel contributions
\begin{equation}
  \label{eq:skanal2}
  2 \bigl( \overline{u}_4 (\fmslash{k}_1 + \fmslash{k}_2 - \fmslash{k}_3 - m)
  {\cal P}_R \xi \bigr) \bigl( \overline{v}_2 {\cal P}_R u_1 \bigr) = 2 \bigl(
  \overline{u}_4 (\fmslash{k}_4 - m) {\cal P}_R \xi \bigr) \bigl(
  \overline{v}_2 {\cal P}_R u_1 \bigr) = 0 \quad .
\end{equation}

In the analytical expression for the $t$-channel diagrams
(\ref{eq:tkanal}), combining the first two as well as the fifth and
the sixth line gives
\begin{equation}
  \label{eq:dummy4}
  2 \bigl( \overline{v}_2 (\fmslash{k}_3 + 2m) {\cal P}_R \xi \bigr) \bigl(
  \overline{u}_4 {\cal P}_R u_1 \bigr) \quad . 
\end{equation}
On the other hand, the third and fourth line yield
\begin{equation}
  \label{eq:dummy5}
  - 2 \bigl( \overline{v}_2 (\fmslash{k}_1 - \fmslash{k}_4 + m) {\cal P}_R u_1
    \bigr) \bigl( \overline{u}_4 {\cal P}_R \xi \bigr) \quad .
\end{equation}
To perform the calculation in a more effective way, we manipulate the last
line in (\ref{eq:tkanal}), in particular we ``turn round'' the first
term in parentheses,
\begin{equation}
  \label{eq:dummy6}
 + 2 \bigl( \overline{v}_2 (\fmslash{k}_1 - \fmslash{k}_4 + m) {\cal P}_R v_4
   \bigr) \bigl( \overline{v}_1 {\cal P}_R \xi \bigr)  \quad .
\end{equation}
It also has to be Fierz transformed, together with (\ref{eq:dummy5}),
to get the same spinor structure as (\ref{eq:dummy4}). Again we use
the notation $k_{14} \equiv k_1 - k_4$, the brackets distinguishing the
spinors used as the fundamental ones in the Fierz identities. From 
(\ref{eq:dummy5}) we obtain
\begin{equation}
  \label{eq:fierz_dummy4}
  \begin{aligned}
    - 2 \bigl( \left[ \overline{v}_2 (\fmslash{k}_{14} + m) \right] {\cal P}_R 
    u_1 \bigr) \bigl( \overline{u}_4 \left[ {\cal P}_R \xi \right] \bigr) = &
    \; - \dfrac{1}{2} \bigl( \overline{v}_2 (\fmslash{k}_{14} + m) {\cal P}_R
    \xi \bigr) \bigl( \overline{u}_4 {\cal P}_R u_1 \bigl) \\ & \; -
    \dfrac{1}{2} \bigl( \overline{v}_2 (\fmslash{k}_{14} + m) \gamma^5 {\cal
    P}_R \xi \bigr) \bigl( \overline{u}_4 \gamma^5 {\cal P}_R u_1 \bigr) \\ &
    \; - \dfrac{1}{2} \bigl( \overline{v}_2 (\fmslash{k}_{14} + m) \gamma^\mu
    {\cal P}_R \xi \bigr) \bigl( \overline{u}_4 \gamma_\mu {\cal P}_R u_1
    \bigr) \\ & \; - \dfrac{1}{2} \bigl( \overline{v}_2 (\fmslash{k}_{14} + m)
    \gamma^5 \gamma^\mu {\cal P}_R \xi \bigr) \bigl( \overline{u}_4 \gamma_\mu
    \gamma^5 {\cal P}_R u_1 \bigr) \\ & \; - \dfrac{1}{4} \bigl(
    \overline{v}_2 (\fmslash{k}_{14} + m) \sigma^{\mu\nu} {\cal P}_R \xi
    \bigr) \bigl( \overline{u}_4 \sigma_{\mu\nu} {\cal P}_R u_1 \bigr)
  \end{aligned}
\end{equation}
For the Fierz transformation of (\ref{eq:dummy6}) we ``turn round''
the product containing the spinors $\overline{v}_1$ and $v_4$, getting
the spinors $\overline{u}_4$ and $u_1$.  
\begin{equation}
  \label{eq:fierz_dummy5}
  \begin{aligned}
     2 \bigl( \left[ \overline{v}_2 (\fmslash{k}_{14} + m) \right] {\cal P}_R
      v_4 \bigr) \bigl( \overline{v}_1 \left[ {\cal P}_R \xi \right] \bigr) =
      & \;  - \dfrac{1}{2} \bigl( \overline{v}_2 (\fmslash{k}_{14} + m) {\cal
      P}_R \xi \bigr) \bigl( \overline{u}_4 {\cal P}_R u_1 \bigr) \\ & \; -
      \dfrac{1}{2} \bigl( \overline{v}_2 (\fmslash{k}_{14} + m) \gamma^5 {\cal
      P}_R \xi \bigr) \bigl( \overline{u}_4 \gamma^5 {\cal P}_R u_1 \bigr) \\
      & \; + \dfrac{1}{2} \bigl( \overline{v}_2 (\fmslash{k}_{14} + m)
      \gamma^\mu {\cal P}_R \xi \bigr) \bigl( \overline{u}_4 \gamma_\mu {\cal
      P}_L u_1 \bigr) \\ & \; - \dfrac{1}{2} \bigl( \overline{v}_2
      (\fmslash{k}_{14} + m) \gamma^5 \gamma^\mu {\cal P}_R \xi \bigr) \bigl(
      \overline{u}_4 \gamma_\mu \gamma^5 {\cal P}_L u_1 \bigr) \\ & \; +
      \dfrac{1}{4} \bigl( \overline{v}_2 (\fmslash{k}_{14} + m)
      \sigma^{\mu\nu} {\cal P}_R \xi \bigr) \bigl( \overline{u}_4
      \sigma_{\mu\nu} {\cal P}_R u_1 \bigr)
  \end{aligned}
\end{equation}
As was the case for the $s$-channel, the tensor contributions to
(\ref{eq:fierz_dummy4}) and (\ref{eq:fierz_dummy5}) cancel out, while
in each equation the vector part again cancels the axial vector. The
scalar and pseudoscalar parts from both Fierz transformations give
\begin{equation}
  \label{eq:dummy7}
  + 2 \bigl( \overline{v}_2 (\fmslash{k}_4 - \fmslash{k}_1 - m) {\cal P}_R \xi
  \bigr) \bigl( \overline{u}_4 {\cal P}_R u_1 \bigr) \quad ,
\end{equation}
so finally the result for the $t$-channel is written as:
\begin{equation}
  \label{eq:tkanal2}
  2 \bigl( \overline{v}_2 (\fmslash{k}_3 + \fmslash{k}_4 - \fmslash{k}_1 + m)
  {\cal P}_R \xi \bigr) \bigl( \overline{u}_4 {\cal P}_R u_1 \bigr) = 2 \bigl(
  \overline{v}_2 (\fmslash{k}_2 + m) {\cal P}_R \xi \bigr) \bigl(
  \overline{u}_4 {\cal P}_R u_1 \bigr) = 0 
\end{equation}

The same calculation goes through for the $u$-channel, transferring
(\ref{eq:ukanal}): 
\begin{equation}
  \label{eq:ukanal2}
  \begin{aligned}
    (\ref{eq:ukanal}) = & \; - 2 \bigl( \overline{v}_1 (\fmslash{k}_3 + 2
    m) {\cal P}_R \xi \bigr) \bigl( \overline{u}_4 {\cal P}_R u_2 \bigr) \\ &
    \; + 2 \bigl( \overline{v}_1 (\fmslash{k}_2 - \fmslash{k}_4 + m) {\cal
    P}_R u_2 \bigr) \bigl( \overline{u}_4 {\cal P}_R \xi \bigr) \\ & \; - 2
    \bigl( \overline{v}_1 (\fmslash{k}_2 - \fmslash{k}_4 + m) {\cal P}_R v_4
    \bigr) \bigl( \overline{v}_2 {\cal P}_R \xi \bigr) 
  \end{aligned}
\end{equation}
The Fierz transformations of the last two lines (again we ``invert''
the products containing $\overline{v}_2$ and $v_4$ in the third line
and abbreviate $k_2 - k_4$ by $k_{24}$) are:
\begin{equation}
  \label{eq:fierz_dummy6}
  \begin{aligned}
    2 \bigl( \left[ \overline{v}_1 (\fmslash{k}_{24} + m) \right] {\cal P}_R
    u_2 \bigr) \bigl( \overline{u}_4 \left[ {\cal P}_R \xi \right] \bigr) = &
    \quad \dfrac{1}{2} \bigl( \overline{v}_1 (\fmslash{k}_{24} + m) {\cal P}_R
    \xi \bigr) \bigl( \overline{u}_4 {\cal P}_R u_2 \bigr) \\ & \; +
    \dfrac{1}{2} \bigl( \overline{v}_1 (\fmslash{k}_{24} + m) \gamma^5 {\cal
    P}_R \xi \bigr) \bigl( \overline{u}_4 \gamma^5 {\cal P}_R u_2 \bigr) \\ &
    \; + \dfrac{1}{2} \bigl( \overline{v}_1 (\fmslash{k}_{24} + m) \gamma^\mu
    {\cal P}_R \xi \bigr) \bigl( \overline{u}_4 \gamma_\mu {\cal P}_R u_2
    \bigr) \\ & \; + \dfrac{1}{2} \bigl( \overline{v}_1 (\fmslash{k}_{24} + m)
    \gamma^5 \gamma^\mu {\cal P}_R \xi \bigr) \bigl( \overline{u}_4 \gamma_\mu
    \gamma^5 {\cal P}_R u_2 \bigr) \\ & \; + \dfrac{1}{4} \bigl(
    \overline{v}_1 (\fmslash{k}_{24} + m) \sigma^{\mu\nu} {\cal P}_R \xi
    \bigr) \bigl( \overline{u}_4 \sigma_{\mu\nu} {\cal P}_R u_2 \bigr)  
  \end{aligned}
\end{equation}
\begin{equation}
  \label{eq:fierz_dummy7}
  \begin{aligned}
    - 2 \bigl( \left[ \overline{v}_1 (\fmslash{k}_{24} + m) \right] {\cal P}_R
    v_4 \bigr) \bigl( \overline{v}_2 \left[ {\cal P}_R \xi \right] \bigr) = &
    \quad \dfrac{1}{2} \bigl( \overline{v}_1 (\fmslash{k}_{24} + m) {\cal P}_R
    \xi \bigr) \bigl( \overline{u}_4 {\cal P}_R u_2 \bigr) \\ & \; +
    \dfrac{1}{2} \bigl( \overline{v}_1 (\fmslash{k}_{24} + m) \gamma^5 {\cal
    P}_R \xi \bigr) \bigl( \overline{u}_4 \gamma^5 {\cal P}_R u_2 \bigr) \\ &
    \; - \dfrac{1}{2} \bigl( \overline{v}_1 (\fmslash{k}_{24} + m) \gamma^\mu
    {\cal P}_R \xi \bigr) \bigl( \overline{v}_2 \gamma_\mu {\cal P}_R v_4
    \bigr) \\ & \; - \dfrac{1}{2} \bigl( \overline{v}_1 (\fmslash{k}_{24} + m)
    \gamma^5 \gamma^\mu {\cal P}_R \xi \bigr) \bigl( \overline{v}_2 \gamma_\mu
    \gamma^5 {\cal P}_R v_4 \bigr) \\ & \; - \dfrac{1}{4} \bigl(
    \overline{v}_1 (\fmslash{k}_{24} + m) \sigma^{\mu\nu} {\cal P}_R \xi
    \bigr) \bigl( \overline{u}_4 \sigma_{\mu\nu} {\cal P}_R u_2 \bigr)  
  \end{aligned}
\end{equation}
The vector contributions as well as the axial vector parts vanish 
separately for each process as in the $s$- and $t$-channels, while the
tensor contributions of (\ref{eq:fierz_dummy6}) and
(\ref{eq:fierz_dummy7}) cancel each other. The scalar and pseudoscalar
contributions are equal und sum up to  
\begin{equation}
  \label{eq:dummy8}
  2 \bigl( \overline{v}_1 (\fmslash{k}_{24} + m) {\cal P}_R \xi \bigr) \bigl(
  \overline{u}_4 {\cal P}_R u_2 \bigr) .
\end{equation}
Therefore the result of (\ref{eq:ukanal2}) is
\begin{equation}
  \label{eq:dummy9}
  2 \bigl( \overline{v}_1 (\fmslash{k}_2 - \fmslash{k}_3 - \fmslash{k}_4 - m)
  {\cal P}_R \xi \bigr) \bigl( \overline{u}_4 {\cal P}_R u_2 \bigr) = - 2
  \bigl( \overline{v}_1 (\fmslash{k}_1 + m) {\cal P}_R \xi \bigr) \bigl(
  \overline{u}_4 {\cal P}_R u_2 \bigr) = 0 \quad .
\end{equation}

So finally we can see that $s$-, $t$- and $u$-channel diagrams vanish
separately and we find the SWIs of ($2 \rightarrow 2$) processes
containing two as well as four fermions to be fulfilled. 

%%% Local Variables: 
%%% mode: latex
%%% TeX-master: "diss"
%%% End: 

%% file: chap5.tex
%%%%%%%%%%%%%%%%%%%%%%%%%%%%%%%%%%%%%%%%%%%%%%%%%%%%%%%%%%%%

\chapter{The O'Raifeartaigh model}

\section{Spontaneous breaking of Supersymmetry}

The simplest model in which supersymmetry is spontaneously broken is
the O'Raifear\-taigh model. To be more precise it is a whole class of models
(cf. \cite{Weinberg:QFTv3:Text}), the particular O'Raifeartaigh
model being only a special case. The particle content, some special
remarks and the Feynman rules of the O'Raifeartaigh model (from
hereon referred to as the OR model) are collected in the appendix. As
was proven by O'Raifeartaigh, at least three chiral
superfields are needed to make spontaneous supersymmetry breaking possible. 

This model offers the opportunity to examine what happens to the SWI
in the case of spontaneous breaking. Of course, the derivation of 
identity (\ref{eq:entschgl}) breaks down together with our symmetry
since the vacuum is no longer left invariant by the action of the
supercharge. But we want to show an example of an SWI, in the sense,
that we calculate a SWI as if (\ref{eq:entschgl}) were still valid and
take a look at the terms violating the SWI. The latter should turn out
to be proportional to the parameters of SUSY breaking. 

%%%%%%%%%%%%%%%%%%%%%%%%%%%%%%%%%%%%%%%%%%%%%%%%%%%%%%%%%%%%%%%%%%%%%

\section{Preliminaries to the O'Raifeartaigh model}

For the OR model as a spontaneously broken supersymmetric model the
relation 
\begin{equation}
  \label{eq:supladverni}
  Q \Ket{0} = 0
\end{equation}
is no longer fulfilled, but this had to be postulated to be
able to derive the SWI. This section will show what happens to the SWI if
we were to assume (\ref{eq:supladverni}) to be valid anyhow. 

There is a higher number of particles in the OR model than in
previously considered models. We gratefully make use of this fact
as the number of participating diagrams in an SWI shrinks
enormously with a growing variety of external particles. 
Unfortunately this advantage is partly lost since
up to three different scalar particles appear as a 
result of the SUSY transformations of fermionic annihilation and
creation operators. 

With the experience from last chapter's toy model we
can immediately write down the transformation laws of the annihilators
(and therefore also for the creators). 

First of all we want to introduce a common notation for all particles: The
annihilators of the scalars are denoted by $a_A$, $a_B$, $a^\phi_\pm$
and $a^\Phi_\pm$, the Majorana fermion's annihilator by $c$, while
the annihilators for the Dirac fermion are denoted by $b$ and $d$ as
usual. The creators are the Hermitean adjoints, of course.

As for the toy model, the fermionic partner of the scalar field
which is split into real and imaginary parts, gives the lefthanded
component of a Dirac fermion so we can directly take over the result
(\ref{eq:trafoaskalarspiel}):
\begin{equation}
  \label{eq:trafo_OR_1}
  \boxed{
    \begin{aligned}
      \left[ Q(\xi) , a_A (k) \right] = & \; \ii \sum_\sigma \biggl( \left(
      \overline{\xi} {\cal P}_L u (k, \sigma) \right) b (k, \sigma) + \left(
      \overline{\xi} {\cal P}_R u (k, \sigma) \right) d (k, \sigma) \biggr) \\
      \left[ Q(\xi) , a_B (k) \right] = & \; \sum_\sigma \biggl( \left(
      \overline{\xi} {\cal P}_L u (k, \sigma) \right) b (k, \sigma) - \left(
      \overline{\xi} {\cal P}_R u (k, \sigma) \right) d (k, \sigma) \biggr) 
    \end{aligned}}
\end{equation}

The fermionic partner for the complex scalar field from the third
superfield and its Hermitean adjoint are the righthanded component of
that Dirac spinor. Consequently we can maintain
(\ref{eq:trafoaplusspiel}) and (\ref{eq:trafoaminusspiel}),
\begin{equation}
  \label{eq:trafo_OR_2}
  \boxed{ \left[ Q(\xi) , a^\Phi_+ (k) \right] = \ii \sqrt{2} \sum_\sigma
  \Bigl( \overline{\xi} {\cal P}_R u (k, \sigma) \Bigr) \: b (k, \sigma) }
  \quad , 
\end{equation}
\begin{equation}
  \label{eq:trafo_OR_3}
  \boxed{ \left[ Q(\xi) , a^\Phi_- (k) \right] = \ii \sqrt{2} \sum_\sigma
  \Bigl( \overline{\xi} {\cal P}_L u (k, \sigma) \Bigr) \: d (k, \sigma) }
  \quad .
\end{equation}          
In the case of the scalar field $\phi$ -- the scalar component of the
first superfield and superpartner of the Goldstino -- we just have to
set the two annihilators $b$ and $d$ equal to the Majorana annihilator
$c$: 
\begin{equation}
  \label{eq:trafo_OR_4}
  \boxed{ \left[ Q(\xi) , a^\phi_+ (k) \right] = \ii \sqrt{2} \sum_\sigma
  \Bigl( \overline{\xi} {\cal P}_R u (k, \sigma) \Bigr) \: c (k, \sigma) }
  \quad , 
\end{equation}
\begin{equation}
  \label{eq:trafo_OR_5}
  \boxed{ \left[ Q(\xi) , a^\phi_- (k) \right] = \ii \sqrt{2} \sum_\sigma
  \Bigl( \overline{\xi} {\cal P}_L u (k, \sigma) \Bigr) \: c (k, \sigma) }
  \quad .
\end{equation} 
The transformations of the Dirac annihilators are analogous to
(\ref{eq:trafoferm1spiel}) and (\ref{eq:trafoferm2spiel}),
respectively:
\begin{equation}
  \label{eq:trafo_OR_6}
  \boxed{ \left[ Q(\xi) , b (k, \sigma) \right] = - \ii \overline{u} (k,
  \sigma) \Bigl( a_A (k) {\cal P}_R + \ii a_B (k) {\cal P}_R + \sqrt{2}
  a^\Phi_+ (k) {\cal P}_L \Bigr) \xi} \quad , 
\end{equation}    
\begin{equation}
  \label{eq:trafo_OR_7}
  \boxed{ \left[ Q(\xi) , d (k, \sigma) \right] = - \ii \overline{u} (k,
  \sigma) \Bigl( a_A (k) {\cal P}_L - \ii a_B (k) {\cal P}_L + \sqrt{2}
  a^\Phi_- (k) {\cal P}_R \Bigr) \xi} \quad .
\end{equation} 
For the first superfield we use equation (\ref{eq:transfermerz}) und
get 
\begin{equation}
  \label{eq:trafo_OR_8}
  \boxed{ \left[ Q(\xi) , c (k, \sigma) \right] = - \ii \sqrt{2} \:
  \overline{u} (k, \sigma) \Bigl( a_-^\phi {\cal P}_R + a_+^\phi {\cal P}_L
  \Bigr) \xi } 
\end{equation}

%%%%%%%%%%%%%%%%%%%%%%%%%%%%%%%%%%%%%%%%%%%%%%%%%%%%%%%%%%%%%%%%%%%%%

\section{Example for an SWI in the OR model}

As for the WZ model before, we want to construct an example for an
SWI. Again we start with a string of fields in which the spin of initial
and final states differ by half a unit. As mentioned above, we want to
make use of the greater variety of particles available in this model. 

Our choice for an example is the following:
\begin{equation}
  \label{eq:swiverletz}
  \begin{aligned}
    0 \neq & \; \Vev{ \left[ Q(\xi) , a_A (k_3) c (k_4, +) a_-^{\Phi \:
    \dagger} (k_1) a_+^{\phi \: \dagger} (k_2) \right] } \\ = & \quad \; \ii
    \sum_\sigma \left( \overline{\xi} {\cal P}_L u (k_3, \sigma) \right) \cdot
    {\cal M} \bigl(\Phi (k_1) \phi^* (k_2) \rightarrow \Psi (k_3, \sigma) \chi
    (k_4, +) \bigr) \\ & \; + \ii \sum_\sigma \left( \overline{\xi} {\cal P}_R
    u(k_3, \sigma) \right) \cdot {\cal M} \big(\Phi (k_1) \phi^* (k_2)
    \rightarrow \overline{\Psi} (k_3, \sigma) \chi (k_4, +) \bigr) \\ & \; +
    \ii \sqrt{2} \sum_\sigma \left( \overline{u} (k_1, \sigma) {\cal P}_L \xi
    \right) \cdot {\cal M} \bigl(\overline{\Psi} (k_1, \sigma) \phi^* (k_2)
    \rightarrow A (k_3) \chi (k_4, +) \bigr) 
  \end{aligned}
\end{equation}
The processes resulting from the SUSY transformations of the Majorana
fermion in the final state and the massless boson in the initial state
do not contribute. For the transformation of the remaining
particles we write down only the nonvanishing terms. The first process
with two diagrams
\begin{equation}
  \label{eq:swi_or1}
  \parbox{2cm}{\begin{fmfchar*}(20,15)
    \fmfleft{i1,i2}
    \fmfright{o1,o2}
    \fmf{dots}{i2,v1}
    \fmf{dbl_dots}{i1,v1}
    \fmf{fermion}{v2,o2}
    \fmf{plain}{v2,o1}
    \fmf{dashes}{v1,v2}
    \fmfdot{v1,v2}
  \end{fmfchar*}} + \parbox{2cm}{\begin{fmfchar*}(20,15)
    \fmfleft{i1,i2}
    \fmfright{o1,o2}
    \fmf{dots}{i2,v1}
    \fmf{dbl_dots}{i1,v1}
    \fmf{fermion}{v2,o2}
    \fmf{plain}{v2,o1}
    \fmf{dbl_dashes}{v1,v2}
    \fmfdot{v1,v2}
  \end{fmfchar*}} ,
\end{equation}
produces, after multiplication with the appropriate prefactor, the
analytical expression
\begin{equation}
  \label{eq:swi_or2}
  2 g^2 m \cdot \left( \overline{\xi} {\cal P}_L \fmslash{k}_3 v (k_4, +)
  \right) \cdot \left( \dfrac{1}{s - m^2 + 2 \lambda g} - \dfrac{1}{s - m^2 -
  2 \lambda g} \right) \qquad .
\end{equation}
The second process is analogous:
\begin{equation}
  \label{eq:swi_or3}
  \parbox{2cm}{\begin{fmfchar*}(20,15)
    \fmfleft{i1,i2}
    \fmfright{o1,o2}
    \fmf{dots}{i2,v1}
    \fmf{dbl_dots}{i1,v1}
    \fmf{fermion}{o2,v2}
    \fmf{plain}{v2,o1}
    \fmf{dashes}{v1,v2}
    \fmfdot{v1,v2}
  \end{fmfchar*}} + \parbox{2cm}{\begin{fmfchar*}(20,15)
    \fmfleft{i1,i2}
    \fmfright{o1,o2}
    \fmf{dots}{i2,v1}
    \fmf{dbl_dots}{i1,v1}
    \fmf{fermion}{o2,v2}
    \fmf{plain}{v2,o1}
    \fmf{dbl_dashes}{v1,v2}
    \fmfdot{v1,v2}
  \end{fmfchar*}} 
\end{equation}
The result is
\begin{equation}
  \label{eq:swi_or4}
  - 2 g^2 m \cdot \left( \overline{\xi} {\cal P}_R \fmslash{k}_3 v (k_4, +)
    \right) \cdot \left( \dfrac{1}{s - m^2 + 2 \lambda g} + \dfrac{1}{s - m^2
    - 2 \lambda g} \right) .
\end{equation}
There exists just one diagram for the third process,
\begin{equation}
  \label{eq:swi_or5}
    \parbox{2cm}{\begin{fmfchar*}(20,15)
    \fmfleft{i1,i2}
    \fmfright{o1,o2}
    \fmf{dots}{i2,v1}
    \fmf{fermion}{v1,i1}
    \fmf{plain}{o1,v2}
    \fmf{dashes}{v2,o2}
    \fmf{fermion}{v1,v2}
    \fmfdot{v1,v2}
  \end{fmfchar*}} \qquad .
\end{equation}
Here we have to keep an eye on the signs again, while having to apply the
Wick theorem. The resulting amplitude is
\begin{equation}
  \label{eq:swi_or6}
  + 4 g^2 m \cdot \left( \overline{\xi} {\cal P}_R (\fmslash{k}_1 +
  \fmslash{k}_2 ) v (k_4, +) \right) \cdot \dfrac{1}{s - m^2} =   + 4 g^2 m
  \cdot \left( \overline{\xi} {\cal P}_R \fmslash{k}_3 v (k_4, +) \right)
  \cdot \dfrac{1}{s - m^2}  
\end{equation}
Implicitly we used momentum conservation and the Dirac equation for
$v (k_4)$, which simply is $\fmslash{k}_4 v (k_4) = 0$ for the
Majorana fermion being the Goldstino. 

When choosing special transformation spinors $\xi$, we see that the
righthanded and lefthanded part of the identity must be fulfilled
separately. As is immediately seen the SWI is violated as we
expected from the beginning. Inspecting the
limit $\lambda \rightarrow 0$ shows that the contribution containing the
lefthanded chiral projector vanishes and the parts with the
righthanded chiral projectors cancel each other. This is
understandable by remembering that the parameter $\lambda$ controls
the spontaneous symmetry breaking of the OR model as it produces the
mass splitting between the particles of the second and the third
superfield, which came from diagonalizing the mass terms. 

This violation of the SWI of the type derived in
\cite{Grisaru/Pendleton:1977:smatrix} and
\cite{Grisaru/Pendleton:1977:SUSY_scattering} stems from the
noninvariance of the vacuum under SUSY transformations in
spontaneously broken SUSY theories. It can be avoided by using a
formalism based on the concept of a conserved Noether current for the
supersymmetry; this will be shown in the next part.

%%% Local Variables: 
%%% mode: latex
%%% TeX-master: "diss"
%%% End: 

%% file: chap6.tex
\chapter{The supersymmetric current and SWI} 

There are some inherent problems in the method of 
calculating SWIs the way presented in
the last part: It does not work for spontaneously broken supersymmetry
and is also only applicable for on-shell identities. To
develop stringent tests for supersymmetric field theories, it will
prove useful to consider off-shell identities as well, as much more of
the underlying physics is involved in such relations. In this
part we will first present how SWI can be implemented when using the
current of the supersymmetry and then show examples for the Wess-Zumino
model. To verify that this method is also valid for spontaneously
broken supersymmetry, we extend our calculations to the O'Raifeartaigh
model. Afterwards we turn to the combination of (global)
supersymmetry and gauge symmetries when examining currents in
supersymmetric Yang-Mills theories. This is important because
realistic models should, of course, incorporate at least the gauge
symmetries of the Standard Model. 

%%%%%%%%%%%%%%%%%%%%%%%%%%%%%%%%%%%%%%%%%%%%%%%%%%%%%%%%%%%%%%%%%%%%%

\section{Ward identities -- current vs.~external states}

In this section we describe the connection between the SWI in the
formalism derived in \cite{Grisaru/Pendleton:1977:smatrix} and
\cite{Grisaru/Pendleton:1977:SUSY_scattering} and similar relations
which can be obtained with the help of supersymmetric current conservation. The
name ``supersymmetric'' current is a bit 
misleading as this current is not invariant under SUSY
transformations. In fact, the current mentioned here is closely related
to a spinor component of a real superfield provided with an
additional vector index, called the supercurrent
(cf. \cite{Weinberg:QFTv3:Text}, \cite{Piguet/Sibold:1986:book}). The
scalar component of the supercurrent is the current of $R$ symmetry,
while the vector component is given by the energy-momentum tensor. The
supersymmetric current has the Lorentz transformation properties of a
vectorspinor. In a local version of supersymmetry -- supergravity -- the
corresponding gauge field is the gravitino. 

To derive this kind of SWI we write down a time-ordered
product of a string of field operators (appearing in the supersymmetric
model under consideration) with the operator insertion of the
supersymmetric current,
\begin{equation}
  \label{eq:zeitgeordnetprodstrom}
  \Greensfunc{ {\cal J}^\mu (x) {\cal O}_1 (y_1) {\cal O}_2 (y_2) \ldots {\cal
  O}_n (y_n)} 
\end{equation}
Taking the derivative of this expression with respect to $x^\mu$ (we
use the abbreviation $\partial_\mu^x \equiv \partial / \partial
x^\mu$), we get:
\begin{multline}
  \label{stromerhaltung1}
  \ii\partial_\mu^x \Greensfunc{ {\cal J}^\mu (x) {\cal O}_1 (y_1) \ldots {\cal
  O}_n (y_n)} \\ = \; \; \sum_{i = 1}^n \chi_i \Greensfunc{ {\cal O}_1 \ldots
  {\cal O}_{i-1} \left[ \ii {\cal J}^0 (x) , {\cal O}_i (y_i)
  \right]_{P_i} \delta (x^0 - y_i^0) \, {\cal O}_{i+1} \ldots {\cal
  O}_n } \\ + \ii \Greensfunc{ 
  \partial_\mu^x {\cal J}^\mu (x) {\cal O}_1 (y_1) \ldots {\cal O}_n (y_n) }
\end{multline}
Here $\chi_i$ has the meaning of a sign prefactor  
\begin{equation}
  \chi_i \equiv (- 1)^{\sum_{j=1}^{i-1} P_j} , 
\end{equation}
which arises by anticommuting the Grassmann odd current with Fermi
field operators. $P$ is the Grassmann parity of the fields, $1$ for
fermions and $0$ for bosons. In the same manner we have introduced the
graded commutator
\begin{equation}
  \label{eq:skommutator}
  \left[ A , B \right]_{P = 1} \equiv \left\{ A , B \right\} \, \text{for
  fermions}, \quad \left[ A , B \right]_{P = 0} \equiv \left[ A , B \right]
  \, \text{otherwise}
\end{equation}
as an anticommutator in the case of two fermionic operators and a
commutator in all other cases.

The last term in (\ref{stromerhaltung1}), which is created by applying the
derivative to the current, vanishes due to current
conservation. The terms with the graded commutators arise when acting
with the time derivative on the step functions in 
the time ordered product. We make use of the fact that the equal
time commutator (or anticommutator in the case of a fermionic
operator) of the zero component of the current with an operator (for
instance, the field operator of the fundamental fields of the theory)
equals the symmetry transformation (in our case the SUSY
transformation) of the considered field:
\begin{equation}
  \label{eq:SUSY_trafo_strom}
  \left[ \, \ii \overline{\xi} {\cal J}^0 (x) , {\cal O} (y) \right]
  \delta (x^0 - y^0) = \delta_\xi {\cal O} (y) \cdot \delta^4 (x - y) 
\end{equation}
With the help of this relation we can rewrite the right hand side of
(\ref{stromerhaltung1}). Furthermore we switch to momentum space and replace
the spacetime derivative acting on the left hand side of equation
(\ref{stromerhaltung1}) by the momentum $k_\mu$ which flows into the Green
function through the current operator insertion (so $-k^\mu = \sum_i
p^\mu_i$ is the sum over the incoming momenta of all other external legs). 
\begin{multline}
  \label{eq:stromerhaltung2}
  k_\mu \mathrm{F.T.} \Greensfunc{ \overline{\xi} {\cal J}^\mu (x)
  {\cal O}_1 (y_1) 
  \ldots {\cal O}_n (y_n)}        \\
  = \sum_{i = 1}^n \mathrm{F.T.} \Greensfunc{ {\cal O}_1 \ldots {\cal
  O}_{i-1} \left( \delta_\xi {\cal O}_i (y_i) \right) {\cal O}_{i+1}
  \ldots {\cal O}_n} \cdot \delta^4 (x - y_i)  
\end{multline}
In (\ref{eq:stromerhaltung2}) the supersymmetric current has been multiplied by
the SUSY transformation parameter $\xi$ and hence became a bosonic
operator. There are two consequences: we could forget about the
sign prefactor which was part of (\ref{stromerhaltung1}) and all
graded commutators became commutators. In (\ref{eq:SUSY_trafo_strom}) and 
(\ref{eq:stromerhaltung2}) we used the usual notation for the SUSY
transformations of the fields (with transformation parameter $\xi$). 

At tree level the identity (\ref{stromerhaltung1}) is valid for
linearly as well as nonlinearly realized symmetries both for on-shell and
off-shell processes (cf.~for instance the path integral derivation of
the Ward identities in \cite{diFrancesco:1996:CFT}). In the case of
nonlinearly realized symmetries, not only higher than quadratic terms will
appear in the current operator but also composite operators in the
transformations of the fields. To put the identity (\ref{stromerhaltung1}) on
the mass shell we have to apply the LSZ reduction formula
\cite{Kugo:Eichtheorie}, \cite{Weinberg:QFTv1:Text},
\cite{Peskin/Schroeder:QFT:Text} to all external legs except the current
itself, which remains unamputated: 
\begin{equation*}
  \label{eq:1pow}
  \parbox{5cm}{
  \hfil\\\hfil\\
  \begin{fmfgraph*}(32,20)
    \fmfleft{l1,l2,l3,l4}
    \fmfright{r1,r2,r3,r4,r5}
    \fmf{phantom}{l1,v,r5}
    \fmf{phantom}{l4,v,r1}
    \fmffreeze
    \fmf{plain}{l1,v} \fmf{plain}{l2,v} \fmf{plain}{l4,v}
    \fmf{dbl_plain}{v,r3} \fmf{plain}{v,r1} \fmf{plain}{r5,v}
    \fmfv{decor.shape=circle,decor.filled=30,decor.size=0.8cm}{v}
    \fmflabel{\begin{footnotesize}$\ii\int d^4x_1 f_1^* (\Box_1 +
    m_1^2)$\end{footnotesize}}{l1}
    \fmflabel{\begin{footnotesize}$\ii\int d^4x_2 f_2^* (\Box_2 +
    m_2^2)$\end{footnotesize}}{l2}
    \fmflabel{$\vdots$}{l3}
    \fmflabel{\begin{footnotesize}$\ii\int d^4x_n f_n^* (\Box_n +
    m_n^2)$\end{footnotesize}}{l4}
    \fmflabel{\begin{footnotesize}$\ii\int d^4x_{n+1} f_{n+1} (\Box_{n+1} +
    m_{n+1}^2)$\end{footnotesize}}{r1}
    \fmflabel{\begin{footnotesize}$\ii\int d^4x_{n+m} f_{n+m} (\Box_{n+m} +
    m_{n+m}^2)$\end{footnotesize}}{r5}
    \fmflabel{$\quad k_\mu\mathcal{J}^\mu$}{r3}
    \fmflabel{$\vdots$}{r2}
    \fmflabel{$\vdots$}{r4}
    \fmfv{decor.shape=circle,decor.filled=shaded,decor.size=4mm}{r3}
  \end{fmfgraph*}
  \hfil\\\hfil\\ }
\end{equation*}
We used the abbreviation $f_i \equiv e^{-\ii k_i x_i} / \sqrt{(2\pi)^3
2k_i^0}$.
For simplicity we denoted only the amputation procedure for bosons. 
The big grey blob stands for the process under
consideration (i.e.~the interaction operators needed to connect the
external fields in (\ref{stromerhaltung1})), while the smaller blob
will become our standard convention for a current insertion. On shell,
all the so called contact terms on the right hand side of equation
(\ref{stromerhaltung1}) vanish. This is seen by inspection of the
amputation procedure for those Green functions with the transformed
fields: Let the external particle corresponding to the $i$th field
have momentum $p_i$ on the left hand side, then on the right hand side
the particle corresponding to the transformed field has its momentum
increased by the momentum influx through the current $p_i + k$. For
the sake of simplicity, we show an example involving only scalar
fields:
\begin{equation}
        D_F(p_i)^{-1} \cdot D_F (p_i + k) = \dfrac{p_i^2 - m_i^2}{(p_i
        + k)^2 - m_i^2} 
\end{equation}
These two propagator factors do not cancel like all other propagators of
external particles do, hence when setting the external momenta $p_j,
j=1,2,\ldots$ on the mass shell, this yields zero for every term on the
right hand side. 

Another interesting phenomenon happens for spontaneously broken
symmetries, where a field gets a vacuum expectation value and is
therefore shifted by a constant. A term linear in the field appears 
in the current, or more precisely, a term proportional to the derivative
of the Goldstone boson. This contributes tadpole-like diagrams which,
if resummed, shift the appropriate poles of the fields according to
the mass splitting from the spontaneous symmetry breaking. Since
coupling constants and vacuum expectation values are combined to yield
masses of particles, there is a mixing of different orders in
perturbation theory contributing to the Ward identity. For
supersymmetric field theories the corresponding term in the current is
given by a gamma matrix times the derivative of the Goldstino
field. We will study this in detail in the O'Raifeartaigh model below. 

%%%%%%%%%%%%%%%%%%%%%%%%%%%%%%%%%%%%%%%%%%%%%%%%%%%%%%%%%%%%%%%%%%%%%%%%%

\section{Simplest example -- Wess-Zumino model}

Like any continous symmetry in a field theory, supersymmetry possesses
a conserved current whose charge is the generator of the symmetry
transformation. Supersymmetry is no symmetry of the Lagrangean density
but only of the action. It transforms the Lagrangean density
into a total derivative which vanishes upon integration over
spacetime. The following discussion is similar to that in 
\cite{Weinberg:QFTv3:Text}. If we assume that the change of the Lagrangean
density under a SUSY transformation takes on the form
\begin{equation}
  \label{eq:susytraflag}
  \delta_\xi {\cal L} = \overline{\xi} \partial_\mu K^\mu ,
\end{equation}
we can calculate the structure of $K^\mu$. We want to derive the
supersymmetric current for the WZ model as this is the simplest supersymmetric
model. In the Lagrangean density only the $D$-term of the kinetic part
and the $F$-terms from the superpotential appear. The SUSY
transformation of a $D$-term of an arbitrary superfield is given by
\cite{Weinberg:QFTv3:Text}, \cite{Reuter:2000:SUSY}~\footnote{The relative
  factor of $\ii$ between both references comes from the different conventions
  concerning the metric and hence the gamma matrices.}  
\begin{equation}
  \label{eq:trafodterm}
  \delta_\xi D = \overline{\xi} \gamma^5 \fmslash{\partial} \lambda \quad .
\end{equation}
Here $\lambda$ is a spinor being the $\theta^3$ coefficient
in a superspace expansion of a general superfield. We conclude, that for the
kinetic part of the WZ model Lagrangean density as a product of a right- and a
lefthanded chiral superfield, $\hat{\Phi}^\dagger \hat{\Phi}$, 
\begin{equation}
  \delta_\xi {\cal L}_{\text{kin}} = \delta_\xi \left[ \dfrac{1}{2}
  \hat{\Phi}^\dagger \hat{\Phi} \right]_D = \overline{\xi} \gamma^5
  \fmslash{\partial} \dfrac{1}{2} \left[ \hat{\Phi}^\dagger \hat{\Phi}
  \right]_\lambda  \;\;\; .
\end{equation}
The appropriate $\lambda$ can be read off from equation (26.2.24) in
\cite{Weinberg:QFTv3:Text} or, in our conventions, from equation (5.116) in
\cite{Reuter:2000:SUSY}, by taking into consideration that the
general superfield $\hat{\Phi}_1$ there is to be set to the right chiral
superfield 
$\hat{\Phi}^\dagger$ and the second superfield $\hat{\Phi}_2$ to the
Hermitean adjoint left chiral superfield $\hat{\Phi}$. This enables us
to make the following replacements (of course, the SUSY transformation
can be done by brute force in a component language but the superfield
formalism is much more elegant)~\footnote{In the appendix a detailed
derivation for the supersymmetric current in supersymmetric Yang-Mills
theories can be found.}:
\begin{equation}
  \begin{aligned}
    \lambda_1 \equiv & \; 0 \qquad & \lambda_2 \equiv & \; 0 \\ V_1^\mu \equiv
    & \; - \ii \partial^\mu \phi^* \qquad & V^\mu_2 \equiv & \; \ii
    \partial^\mu \phi  \\
    C_1 \equiv & \; \phi^* \qquad & \omega_1 \equiv & \; \ii \sqrt{2} \Psi_R
    \\ C_2 \equiv & \; \phi \qquad & \omega_2 \equiv & \; - \ii \sqrt{2}
    \Psi_L \\ N_1 \equiv & \; F^* \qquad & M_1 \equiv & \; - \ii
    F^* \\ N_2 \equiv & \; F \qquad & M_2 \equiv & \; \ii
    F   
  \end{aligned}
\end{equation}
The result is
\begin{equation}
  \label{eq:kkinetisch}
  K^\mu_{\text{kin}} = \dfrac{1}{\sqrt{2}} \gamma^\mu \biggl( \left(
  \fmslash{\partial} \phi \right) \Psi_R + \left( \fmslash{\partial} \phi^*
  \right) \Psi_L - \ii F \Psi_R - \ii F^* \Psi_L \biggr) \quad .
\end{equation}
From the transformation of the superpotential's $F$-terms we
write down the relation
\begin{equation}
  \delta_\xi {\cal L}_{\text{pot}} = \delta_\xi \left[ \dfrac{m}{2}
  \hat{\Phi}^2 + \dfrac{\lambda}{3!} \hat{\Phi}^3 \right]_F \text{+ h.c.} =
  - \ii \sqrt{2} \, \overline{\xi} {\cal P}_L \fmslash{\partial} \left[
  \dfrac{m}{2} \hat{\Phi}^2 + \dfrac{\lambda}{3!} \hat{\Phi}^3 \right]_\psi
  \text{+ h.c.}  
\end{equation}
The contribution from the potential becomes
\begin{equation}
  K^\mu_{\text{pot}} = - \ii \sqrt{2} \gamma^\mu \biggl( m \Psi_L \phi + m
  \Psi_R \phi^* + \dfrac{1}{2} \lambda \Psi_L \phi^2 + \dfrac{1}{2} \lambda
  \Psi_R (\phi^*)^2 \biggr)
\end{equation}
(NB: Herein $\lambda$ is the coupling constant of the WZ model,
not a spinor component of a superfield.) So altogether we get for this
contribution to the supersymmetric current
\begin{multline}
  \label{eq:supstromanteil}
  K^\mu = \dfrac{1}{\sqrt{2}} \gamma^\mu \biggl( \left(
  \fmslash{\partial} \phi \right) \Psi_R + \left( \fmslash{\partial} \phi^*
  \right) \Psi_L - \ii F \Psi_R - \ii F^* \Psi_L - 2 m \ii \Psi_L 
  \phi \\ - 2 m \ii \Psi_R \phi^* - \ii \lambda \Psi_L \phi^2 - \ii \lambda
  \Psi_R (\phi^*)^2 \biggr) 
\end{multline}
Inserting the definitions of the fields $A$, $B$, $\mathcal{F}$ and
$\mathcal{G}$ yields 
\begin{multline}
  \label{eq:supstromanteil2}
  K^\mu = \dfrac{1}{2} \gamma^\mu (\fmslash{\partial} A) \Psi - \dfrac{\ii}{2}
  \gamma^\mu \gamma^5 (\fmslash{\partial} B) \Psi - \dfrac{\ii}{2} \gamma^\mu
  \mathcal{F} \Psi - \dfrac{1}{2} \gamma^\mu \gamma^5 \mathcal{G} \Psi \\ -
  \ii m \gamma^\mu A 
  \Psi - m \gamma^\mu \gamma^5 B \Psi - \dfrac{\ii \lambda}{2 \sqrt{2}}
  \gamma^\mu \left( A^2 - B^2 \right) \Psi - \dfrac{\lambda}{\sqrt{2}}
  \gamma^\mu \gamma^5 A B \Psi 
\end{multline}

The so called Noether part of the supersymmetric current (by which the
current is given in the case of an invariant Lagrangean density) reads
\begin{equation}
  \label{eq:noetheranteil}
  \sum_{\text{all fields}} \dfrac{\partial_R {\cal L}}{\partial (\partial_\mu
  \Phi)} \delta_\xi \Phi = - \overline{\xi} N^\mu \quad .
\end{equation}
In the WZ models these derivatives are
\begin{equation}
    \dfrac{\partial_R {\cal L}}{\partial (\partial_\mu A)} = \partial^\mu A,
    \qquad \dfrac{\partial_R {\cal L}}{\partial (\partial_\mu B)} =
    \partial^\mu B, \qquad \dfrac{\partial_R {\cal L}}{\partial (\partial_\mu
    \Psi)} = \dfrac{\ii}{2} \overline{\Psi} \gamma^\mu ,
\end{equation}
while the SUSY transformations of the several fields are stated in
(\ref{eq:trafobispin}). The Noether part therefore is
\begin{equation}
  \label{eq:noetheranteil2}
  N^\mu = - \left( \partial^\mu A \right) \Psi - \ii \left( \partial^\mu B
  \right) \gamma^5 \Psi - \dfrac{1}{2} \left[ \fmslash{\partial} \left( A -
  \ii \gamma^5 B \right) \right] \gamma^\mu \Psi + \dfrac{\ii}{2} \left(
  \mathcal{F} + \ii \gamma^5 \mathcal{G} \right) \gamma^\mu \Psi 
\end{equation}
Adding the two parts (\ref{eq:supstromanteil2}) and
(\ref{eq:noetheranteil}) results in the supersymmetric current for the WZ
model
\begin{equation}
  \label{eq:strom_WZ}
  \boxed{\begin{aligned}
  {\cal J}^\mu = & \; K^\mu + N^\mu \\ = & \; \; \; \ii \left( (\ii
  \fmslash{\partial} - m) A \right) \gamma^\mu \Psi + \left( (\ii
  \fmslash{\partial} + m) B \right) \gamma^5 \gamma^\mu \Psi \\ & \; 
  - \dfrac{\ii \lambda}{2 \sqrt{2}} \gamma^\mu \left( A^2 - B^2 \right) \Psi - 
  \dfrac{\lambda}{\sqrt{2}} \gamma^\mu \gamma^5 A B \Psi   
  \end{aligned} }
\end{equation}

Now we can check -- even if it is a little bit cumbersome --
the current conservation explicitly.
\begin{equation*}
  \begin{aligned}
    \partial_\mu {\cal J}^\mu = & \; - (\Box A) \Psi - \ii m A
    (\fmslash{\partial} \Psi) - \underline{(\fmslash{\partial} A)
    (\fmslash{\partial} \Psi)} - \underline{\ii m (\fmslash{\partial} A) \Psi}
    - \ii (\Box B) \gamma^5 \Psi + m B \gamma^5 (\fmslash{\partial} \Psi) \\ &
    \; + \underline{\ii (\fmslash{\partial} B) \gamma^5 (\fmslash{\partial}
    \Psi)} - \underline{m (\fmslash{\partial} B) \gamma^5 \Psi} - \dfrac{\ii
    \lambda}{2 \sqrt{2}} \left( A^2 - B^2 \right) \fmslash{\partial} \Psi -
    \underline{\dfrac{\ii \lambda}{\sqrt{2}} (\fmslash{\partial} A) A \Psi} \\
    & \; + \underline{\dfrac{\ii \lambda}{\sqrt{2}} (\fmslash{\partial} B) B
    \Psi} + \dfrac{\lambda}{\sqrt{2}} \gamma^5 A B \fmslash{\partial} \Psi +
    \underline{\dfrac{\lambda}{\sqrt{2}} \gamma^5 (\fmslash{\partial} A) B
    \Psi} + \underline{\dfrac{\lambda}{\sqrt{2}} \gamma^5 A
    (\fmslash{\partial} B) \Psi} \\ = & \; \dfrac{\lambda}{2 \sqrt{2}}
    (\overline{\Psi} \Psi) \Psi - m \mathcal{F} \Psi -
    \dfrac{\lambda}{\sqrt{2}} A \mathcal{F} \Psi - \dfrac{\lambda}{\sqrt{2}} B
    \mathcal{G} \Psi + \dfrac{\lambda}{2 \sqrt{2}} (\overline{\Psi} \gamma^5
    \Psi) \gamma^5 \Psi - \ii m \mathcal{G} \gamma^5 \Psi \\ & 
    \; + \dfrac{\ii \lambda}{\sqrt{2}} B \mathcal{F} \gamma^5 \Psi - \dfrac{\ii
    \lambda}{\sqrt{2}} A \mathcal{G} \gamma^5 \Psi + \ii \mathcal{F}
    (\fmslash{\partial} \Psi) - \mathcal{G} 
    \gamma^5 (\fmslash{\partial} \Psi)  
  \end{aligned}
\end{equation*}
The underlined terms cancel due to the equation of motion of the
Majorana field $\Psi$. In the second equality the first eight terms stem from
the equations of motion for the scalar fields $A$ and $B$,
while the last two come from inserting the equations of motion for the
spinor field into the terms not underlined. The terms linear in $\mathcal{F}$
and $\mathcal{G}$ can be combined to give the equations of motion for the
Majorana field and we are left with the trilinear fermion
terms. Noting that third powers of Grassmann odd two component
spinors $(\psi \psi) \psi$ vanish, the calculation 
\begin{equation}
  \begin{aligned}
    \left( \overline{\Psi} \Psi \right) \Psi & = (\psi \psi + \bar{\psi}
    \bar{\psi}) \cdot \begin{pmatrix} \psi \\ \bar{\psi} \end{pmatrix} = 
    \begin{pmatrix} (\bar{\psi} \bar{\psi}) \psi \\ (\psi \psi) \bar{\psi} 
    \end{pmatrix} \\
    \left( \overline{\Psi} \gamma^5 \Psi \right) \gamma^5 \Psi & = (- \psi
    \psi + \bar{\psi} \bar{\psi}) \cdot \begin{pmatrix} - \psi \\ \bar{\psi}
    \end{pmatrix} =  \begin{pmatrix} - (\bar{\psi} \bar{\psi}) \psi \\ - (\psi
    \psi) \bar{\psi}  
    \end{pmatrix}    
  \end{aligned} ,
\end{equation}
shows the cancellation of the trilinear fermion terms. This finishes the proof
of the desired current conservation:
\begin{equation}
  \label{eq:stromerhaltung}
  \boxed{\partial_\mu {\cal J}^\mu = 0 }
\end{equation}

The current for a general model with an arbitrary number of superfields and
the proof for its conservation can be found in appendix
\ref{sec:currentgenmodel}. 

%%%%%%%%%%%%%%%%%%%%%%%%%%%%%%%%%%%%%%%%%%%%%%%%%%%%%%%%%%%%%%%%%%%%%%%%%

%%% Local Variables: 
%%% mode: latex
%%% TeX-master: "swi"
%%% End: 

%% file: chap7.tex
\chapter{SWI via the current} 

\section{Starting point: WZ model}

In this section we want to calculate supersymmetric Ward identities
(SWI) for the WZ model obtained with the help of the current as 
constructed in the previous chapter. The current for the WZ model is
given by (\ref{eq:strom_WZ}). We will show an example for an on-shell
identity with three external particles (SWI with two external
particles are just given by the propagators of the theory in the
contact terms and are rather trivial) as well as for an off-shell SWI with the
same number of external particles. 

For the on-shell example, where the contact terms are 
absent, we choose a $(2\rightarrow 1)$ process with two
incoming scalar particles $A$, one outgoing fermion $\Psi$ and a current
insertion, to which (in lowest order perturbation theory) four different
diagrams contribute:

\begin{equation*}
    \parbox{21mm}{%
      \begin{fmfgraph*}(20,20)
        \fmfleft{i1,i2}\fmfright{o1,o2}
        \fmf{plain}{o2,v2}
        \fmf{dashes}{i2,v2}
        \fmf{plain}{v1,v2}
        \fmf{dbl_plain}{o1,v1}
        \fmf{dashes}{i1,v1}
        \fmfdot{v1,v2}
        \fmfblob{.25w}{o1}
      \end{fmfgraph*}
        %\hfil\\
        }\qquad + \quad
    \parbox{21mm}{%
      \begin{fmfgraph*}(20,20)
        \fmfleft{i1,i2}\fmfright{o1,o2}
        \fmf{plain}{o2,v2}
        \fmf{phantom}{i2,v2}
        \fmf{plain}{v1,v2}
        \fmf{dbl_plain}{o1,v1}
        \fmf{phantom}{i1,v1}
        \fmffreeze
        \fmf{dashes}{i1,v2}
        \fmf{dashes}{i2,v1}
        \fmfdot{v1,v2}
        \fmfblob{.25w}{o1}
      \end{fmfgraph*}
        %\hfil\\
        }\qquad + \quad
    \parbox{21mm}{%
      \begin{fmfgraph*}(20,20)
        \fmfleft{i1,i2}\fmfright{o1,o2}
        \fmf{phantom}{o2,v2}
        \fmf{dashes}{i2,v2}
        \fmf{dashes}{v1,v2}
        \fmf{dbl_plain}{o1,v1}
        \fmf{phantom}{i1,v1}
        \fmffreeze
        \fmf{dashes}{i1,v2}
        \fmf{plain}{v1,o2}
        \fmfdot{v1,v2}
        \fmfblob{.25w}{o1}
      \end{fmfgraph*}
        %\hfil\\
        }\qquad + \quad    
    \parbox{21mm}{%
      \begin{fmfgraph*}(20,20)
        \fmfleft{i1,i2}\fmfright{o1,o2}
        \fmf{plain}{o2,v}
        \fmf{dashes}{i2,v}
        \fmf{dbl_plain}{o1,v}
        \fmf{dashes}{i1,v}
        \fmfdot{v}
        \fmfblob{.25w}{o1}
      \end{fmfgraph*}
        %\hfil\\
        }
\end{equation*}                                                   

\vspace{2pt}

The momenta of the incoming $A$s are denoted by $k_1$ and $k_2$ while
the outgoing Majorana fermion's momentum is $k'$. The analytical
expressions for the four diagrams are (from right to left):
\begin{subequations}
  \label{eq:swistrom_wz1}
\begin{align}
    (1) & \qquad - \dfrac{\ii \lambda}{\sqrt{2}} \gamma^\mu v (k') , \\
    (2) & \qquad + \dfrac{3 \ii m \lambda}{\sqrt{2} \, (s - m^2)} \left(
      \fmslash{k}_1 + \fmslash{k}_2 - m \right) \gamma^\mu v (k'), \\
    (3) & \qquad + \dfrac{\ii \lambda}{\sqrt{2} \, (t - m^2)} \left(
      \fmslash{k}_1 - m \right) \gamma^\mu \left( \fmslash{k}_2 - \fmslash{k}'
      + m \right) v (k'), \\
    (4) & \qquad + \dfrac{\ii \lambda}{\sqrt{2} \, (u - m^2)} \left(
      \fmslash{k}_2 - m \right) \gamma^\mu \left( \fmslash{k}_1 - \fmslash{k}'
      + m \right) v (k').
\end{align}
\end{subequations}
For this problem the Mandelstam variables are
\begin{equation*}
  s \equiv (k_1 + k_2)^2 , \qquad t \equiv (k_2 - k')^2, \qquad u \equiv (k_1
  - k')^2 . 
\end{equation*}
The verification of the SWI only needs the use of the Dirac equation
$\left( \fmslash{k}' + m \right) v (k') = 0$ and the relation
$\fmslash{k} \fmslash{k} = k^2$. Applying the 4-gradient to the above
matrix element produces the following sum that can be easily
confirmed to be zero:
\begin{align}
  \partial_\mu \Braket{\Psi| {\cal J}^\mu | A A} = & \;
  \dfrac{\lambda}{\sqrt{2}} \biggl[ \left( \fmslash{k}_1 + \fmslash{k}_2 + m
  \right) - \dfrac{3 m}{s - m^2} \left( \fmslash{k}_1 + \fmslash{k}_2 - m
  \right) \left( \fmslash{k}' - \fmslash{k}_1 - \fmslash{k}_2 \right) \notag
  \\ & \qquad \quad - \dfrac{\left( \fmslash{k}_1 - m \right) \left(
  \fmslash{k}' - \fmslash{k}_2 - \fmslash{k}_1 \right) \left( \fmslash{k}' -
  \fmslash{k}_2 - m \right)}{t - m^2} \notag \\ & \qquad \quad - \dfrac{\left( 
  \fmslash{k}_2 - m \right) \left( \fmslash{k}' - \fmslash{k}_1 -
  \fmslash{k}_2 \right) \left( \fmslash{k}' - \fmslash{k}_1 - m \right)}{u -
  m^2} \biggr] v (k') \notag \\ = & \; \dfrac{\lambda}{\sqrt{2}} \biggl[
  \left( \fmslash{k}_1 + \fmslash{k}_2 + m \right) - 3 m - \dfrac{\left(
  \fmslash{k}_1 - m \right) \left( t + m \fmslash{k}_1 \right)}{t - m^2}
  \notag \\ & \qquad \quad - \dfrac{\left( \fmslash{k}_1 - m \right) \left(
  \fmslash{k}_1 + m \right) \left( \fmslash{k}_2 - \fmslash{k}' \right)}{t -
  m^2} - \dfrac{\left(\fmslash{k}_2 - m \right) \left( u + m \fmslash{k}_2
  \right)}{u - m^2} \notag \\ & \qquad \quad - \dfrac{\left( \fmslash{k}_2 - m
  \right) \left( \fmslash{k}_2 + m \right) \left( \fmslash{k}_1 - \fmslash{k}'
  \right)}{u - m^2} \biggr] v (k') \notag \\ = & \; \dfrac{\lambda}{\sqrt{2}}
  \biggl[ m + \fmslash{k}_1 + \fmslash{k}_2 - 3 m - \fmslash{k}_2 + m -
  \fmslash{k}_1 + m \biggr] v (k') \; \; = \; \; 0 \quad \surd 
\end{align}

Concerning (nonlinear) transformations, on-shell only the 
one-particle pole contributes. But for off-shell Ward identities the 
nonlinear terms give nonvanishing contributions in contact terms. The
correct method to handle that difficulty is to define local operator
insertions for every nonlinear term appearing in the
transformations. 

As an example for an off-shell identity we take the insertion of an
$A$, a $B$ and a $\Psi$ field as the left hand side in
(\ref{eq:stromerhaltung2})  
\begin{multline}
  \mathrm{F.T.} \Braket{0|\mathrm{T}\overline{\mathcal{J}_\mu}(y) \xi A(x_1)
         B(x_2)\Psi(x_3)|0} = \\
  \frac{\mathrm{i}}{p_1^2-m^2}\frac{\mathrm{i}}{p_2^2-m^2}
  \dfrac{-\ii}{\fmslash{p}_3+m} \left( \mathrm{F.T.} \Braket{0|\mathrm{T}
  \overline{\mathcal{J}_\mu} (y)A(x_1)B(x_2)\Psi(x_3)|0}_{\text{amp.}}
  \right) \xi, 
\end{multline}
where $\text{F.T.}$ stands for the Fourier transform. Compared to the on-shell
identity we just changed one scalar into a pseudoscalar. As 
this is an off-shell identity we need not to distinguish incoming and
outgoing particles. The nonvanishing contributions to the contact
terms for this SWI are:
\begin{subequations}
   \begin{align}
     \mathrm{F.T.} \Braket{0|\mathrm{T} \overline{\xi} \Psi(x_1)
       B(x_2)\Psi(x_3)|0} &= \dfrac{-\lambda}{\sqrt{2}} 
       \dfrac{\ii}{p_2^2 - m^2} \dfrac{-\ii}{\fmslash{p}_3+m} \gamma^5 
       \dfrac{\ii}{\fmslash{p}_1 +\fmslash{k}-m} \xi 
   \end{align}
   \begin{align}
     \mathrm{F.T.} \Braket{0|\mathrm{T}A(x_1)(\ii\overline{\xi} \gamma^5
       \Psi (x_2))\Psi(x_3)|0}
        &= \dfrac{\lambda}{\sqrt{2}} \dfrac{\ii}{p_1^2-m^2} \dfrac{
        -\ii}{\fmslash{p}_3+m} \dfrac{\ii}{\fmslash{p}_2 + \fmslash{k} - m}
        \gamma^5 \xi 
    \end{align}
    \begin{multline}
     \mathrm{F.T.} \Braket{0|\mathrm{T}A(x_1)B(x_2)(-\ii) \left( \ii 
     \fmslash{\partial}_{x_3} + m \right) B(x_3) \gamma^5 \xi|0} =  \\
     \dfrac{-m \lambda}{\sqrt{2}} \dfrac{\ii}{p_1^2-m^2} \dfrac{\ii}{
     p_2^2-m^2} \dfrac{\ii}{(p_3+k)^2-m^2} \left( -\fmslash{p}_3 - 
     \fmslash{k} + m \right) \gamma^5 \xi 
   \end{multline}
   \begin{align}
      \dfrac{-\ii\lambda}{\sqrt{2}} \mathrm{F.T.}
     \Braket{0|\mathrm{T}A(x_1)B(x_2) (A B) (x_3) \gamma^5 \xi|0} &= 
      \dfrac{-\ii\lambda}{\sqrt{2}} \dfrac{\ii}{p_1^2-m^2}
     \dfrac{\ii}{p^2_2-m^2} \gamma^5 \xi 
   \end{align}
\end{subequations}

To evaluate the 4-point function with the current
insertion we rewrite the current $\overline{\xi} \mathcal{J}_\mu$ as
$\overline{\mathcal{J}_\mu} \xi$, which is identical due to the
Majorana properties of the current and the transformation parameter:
\begin{align}
\overline{\xi} \mathcal{J}_\mu = & \; \overline{\xi}
\biggl\{ \ii \left( (\ii \fmslash{\partial} - m) A \right) 
    \gamma_\mu \Psi + \left( (\ii \fmslash{\partial} + m) B \right) 
    \gamma^5 \gamma_\mu \Psi \notag \\ &\qquad\qquad\qquad  
     - \dfrac{\ii \lambda}{2 \sqrt{2}} \gamma_\mu \left( A^2 - B^2 \right) 
    \Psi - \dfrac{\lambda}{\sqrt{2}} \gamma_\mu \gamma^5 A B \Psi
    \biggr\} \notag 
  \\ = & \; \biggl\{ \overline{\Psi} \gamma_\mu \ii \left( \ii 
  \fmslash{\partial} + m\right) A  - \overline{\Psi} \gamma_\mu 
  \left( \ii \fmslash{\partial} + m\right) B \gamma^5  \notag \\ & 
  \qquad\qquad\qquad + \dfrac{\ii\lambda}{2\sqrt{2}} \overline{\Psi}
  \gamma_\mu \left(
  A^2 - B^2 \right)  - \dfrac{\lambda}{\sqrt{2}} \overline{\Psi} 
  \gamma_\mu \gamma^5 A B \biggr\} \xi
\end{align}
This brings the propagator of the (matter) fermion to the left.
Again there are four diagrams for the Green
function with current insertion:
\begin{equation}
    \parbox{21mm}{%
      \begin{fmfgraph*}(20,20)
        \fmfleft{i1,i2}\fmfright{o1,o2}
        \fmf{plain}{o2,v2}
        \fmf{dbl_dashes}{i2,v2}
        \fmf{plain}{v1,v2}
        \fmf{dbl_plain}{i1,v1}
        \fmf{dashes}{o1,v1}
        \fmfdot{v1,v2}
        \fmfblob{.25w}{i1}
      \end{fmfgraph*}}\qquad + \quad
    \parbox{21mm}{%
      \begin{fmfgraph*}(20,20)
        \fmfleft{i1,i2}\fmfright{o1,o2}
        \fmf{plain}{o2,v2}
        \fmf{dashes}{i2,v2}
        \fmf{plain}{v1,v2}
        \fmf{dbl_plain}{i1,v1}
        \fmf{dbl_dashes}{o1,v1}
        \fmfdot{v1,v2}
        \fmfblob{.25w}{i1}
      \end{fmfgraph*}}\qquad + \quad
    \parbox{21mm}{%
      \begin{fmfgraph*}(20,20)
        \fmfleft{i1,i2}\fmfright{o1,o2}
        \fmf{dashes}{o2,v2}
        \fmf{dbl_dashes}{i2,v2}
        \fmf{dbl_dashes}{v1,v2}
        \fmf{dbl_plain}{i1,v1}
        \fmf{plain}{o1,v1}
        \fmfdot{v1,v2}
        \fmfblob{.25w}{i1}
      \end{fmfgraph*}}\qquad + \quad
    \parbox{21mm}{%
      \begin{fmfgraph*}(20,20)
        \fmfleft{i1,i2}\fmfright{o1,o2}
        \fmf{plain}{o2,v}
        \fmf{dashes}{i2,v}
        \fmf{dbl_plain}{i1,v}
        \fmf{dbl_dashes}{o1,v}
        \fmfdot{v}
        \fmfblob{.25w}{i1}
      \end{fmfgraph*}}\qquad\quad
\end{equation}

For the sign of the fermion propagator one has to take care of the
momentum flow.

\begin{multline}
  \mathrm{F.T.} \Braket{0|\mathrm{T}\overline{\mathcal{J}_\mu}(y)
        A(x_1)B(x_2)\Psi(x_3)|0}_{\text{amp.}} \xi =
    - \dfrac{\ii \lambda}{\sqrt{2}} \gamma^5 \dfrac{\ii}{\fmslash{p}_1 +
    \fmslash{k}-m} \gamma_\mu \left( \fmslash{p}_1 + m \right) \xi \\
    + \dfrac{\ii\lambda}{\sqrt{2}} \dfrac{\ii}{\fmslash{p}_2 + \fmslash{k}
    -m} \gamma_\mu \left( \fmslash{p}_2 + m\right) \gamma^5 \xi 
    - \dfrac{\lambda}{\sqrt{2}} \gamma_\mu \gamma^5 \xi 
    \\
    - \dfrac{\ii m \lambda}{\sqrt{2}} \dfrac{\ii}{(p_3 + k)^2 - m^2} 
    \gamma_\mu \left( \fmslash{p}_3 + \fmslash{k} - m \right) \gamma^5 \xi
\end{multline}
Dotting the momentum $k_\mu=-(p_1+p_2+p_3)_\mu$ into this expression
yields 
\begin{align*}
  & \; \dfrac{\ii}{\fmslash{p}_3+m} \dfrac{1}{(p_1^2-m^2)(p_2^2-m^2)} \;
    k^\mu \mathrm{F.T.} \Braket{0|\mathrm{T}
    \overline{\mathcal{J}_\mu}(y) A(x_1)B(x_2)
     \Psi(x_3)|0}_{\text{amp.}} \xi \notag 
\end{align*}    
\begin{align}
    = & \; \dfrac{\ii\lambda}{\sqrt{2}} 
    \dfrac{1}{\fmslash{p}_3+m} \dfrac{1}{(p_1^2-m^2)(p_2^2-m^2)} 
    \cdot \biggl\{
    \left( \fmslash{p}_1 + \fmslash{p}_2 +
    \fmslash{p}_3 \right) \notag \\ & \;
    - \dfrac{1}{\fmslash{p}_2 + 
    \fmslash{p}_3 - m} \left( \fmslash{p}_1 + \fmslash{p}_2 + \fmslash{p}_3
    \right) \left( \fmslash{p}_1 - m \right) \notag \\ & \;
    - \dfrac{1}{\fmslash{p}_1 + \fmslash{p}_3
    +m} \left( \fmslash{p}_1 + \fmslash{p}_2 + \fmslash{p}_3 \right) \left(
    \fmslash{p}_2 + m\right) \notag \\ & \;
    + m \dfrac{1}{(p_1 + p_2)^2 - m^2} 
    \left( \fmslash{p}_1 + \fmslash{p}_2 + \fmslash{p}_3 \right) \left(
    \fmslash{p}_1 + \fmslash{p}_2 + m \right) \biggr\} \gamma^5 \xi 
    \notag \\
    = & \;
    \dfrac{\ii\lambda}{\sqrt{2}} \dfrac{1}{\fmslash{p}_3+m} \dfrac{1}{
    (p_1^2-m^2)(p_2^2-m^2)} \left( \fmslash{p}_1 + \fmslash{p}_2 + 
    \fmslash{p}_3 \right) \gamma^5 \xi \notag \\ & \; 
    - \dfrac{\ii\lambda}{\sqrt{2}} \dfrac{1}{\fmslash{p}_3+m} \dfrac{1}{(p_1^2
    -m^2)(p_2^2-m^2)} \left( \fmslash{p}_1 - m \right) \gamma^5 \xi \notag
    \\ & \; - \dfrac{\ii\lambda}{\sqrt{2}} \dfrac{1}{\fmslash{p}_3+m} 
    \dfrac{1}{p_2^2-m^2} \dfrac{1}{\fmslash{p}_2 + \fmslash{p}_3-m} 
    \gamma^5 \xi \notag \\ & \; 
    - \dfrac{\ii\lambda}{\sqrt{2}} \dfrac{1}{\fmslash{p}_3+m} \dfrac{1}{(p_1^2
    -m^2)(p_2^2-m^2)} \left(\fmslash{p}_2+m\right) \gamma^5 \xi \notag \\
    & \; - \dfrac{\ii\lambda}{\sqrt{2}} \dfrac{1}{\fmslash{p}_3+m} \dfrac{1}{
    p_1^2-m^2} \dfrac{1}{\fmslash{p}_1+\fmslash{p}_3+m} \gamma^5 \xi \notag 
    \\ & \;
    + \dfrac{\ii m\lambda}{\sqrt{2}} \dfrac{1}{(p_1^2-m^2)(p_2^2-m^2)}
    \dfrac{1}{(p_1 + p_2)^2 - m^2} \left( \fmslash{p}_1 + \fmslash{p}_2 
    + m \right)\gamma^5 \xi \notag \\ & \; + \dfrac{\ii m \lambda}{\sqrt{2}} 
    \dfrac{1}{\fmslash{p}_3+m} \dfrac{1}{(p_1^2-m^2)(p_2^2-m^2)} \gamma^5
    \xi
\end{align}                                                          

The third, fifth and sixth term equal the ones from the linearly transformed
fields of the r.h.s.:
\begin{multline}
  - \dfrac{\ii\lambda}{\sqrt{2}} \biggl\{ \dfrac{1}{p_2^2-m^2}
   \dfrac{1}{\fmslash{ 
   p}_3+m} \dfrac{1}{\fmslash{p}_2 + \fmslash{p}_3 - m} + \dfrac{1}{p_1^2-
   m^2} \dfrac{1}{\fmslash{p}_3+m} \dfrac{1}{\fmslash{p}_1+\fmslash{p}_3
   +m} \\ - m \dfrac{1}{p_1^2-m^2} \dfrac{1}{p_2^2-m^2} \dfrac{1}{(p_1+p_2)^2
   -m^2} \left( \fmslash{p}_1 + \fmslash{p}_2 + m \right) \biggr\} \gamma^5
   \xi 
\end{multline}
The remaining terms add up to:
\begin{multline}
 - \dfrac{\ii \lambda}{\sqrt{2}} \dfrac{1}{(p_1^2-m^2)(p_2^2-m^2)} \dfrac{1}{
 \fmslash{p}_3 +m} \biggl\{ - \fmslash{p}_1 - \fmslash{p}_2 - \fmslash{p}_3
 + \fmslash{p}_1 - m + \fmslash{p}_2 + m - m 
 \biggr\} \gamma^5 \xi \\ = \dfrac{\ii \lambda}{\sqrt{2}} \dfrac{1}{(p_1^2-
 m^2)(p_2^2-m^2)} \gamma^5 \xi 
\end{multline}  
This equals the single term coming from the local operator insertion, 
so that the Ward identity is indeed fulfilled.

%%%%%%%%%%%%%%%%%%%%%%%%%%%%%%%%%%%%%%%%%%%%%%%%%%%%%%%%%%%%%%%%%%%%%%

\section{Currents and SWI in the O'Raifeartaigh model}

Taking the general formula (\ref{eq:supstromallg}) derived in appendix
\ref{appen:generalcurrent} we can derive the supersymmetric current
for the O'Raifeartaigh model (short: OR model). From the
superpotential in which the superfields have been substituted by their
scalar components    
\begin{equation}
  \label{eq:or_suppot_skalar}
  f (\phi_1, \phi_2, \phi_3) = \lambda \phi_1 + m \phi_2 \phi_3 + g \phi_1
  \phi_2^2 
\end{equation}
we can read off the derivatives with respect to the scalar fields
(there is no difference whether we take the mixings of the fields
into account first and take the derivatives afterwards or vice versa): 
\begin{alignat}{2}
  \dfrac{\partial f(\phi_1, \phi_2, \phi_3)}{\partial \phi_1} = & \; \lambda +
  g \phi_2^2  && =  \; \lambda + \dfrac{g}{2} \left( A^2 - B^2 + 2
  \ii A B \right) \\
  \dfrac{\partial f(\phi_1, \phi_2, \phi_3)}{\partial \phi_2} = & \; m \phi_3
  + 2 g \phi_1 \phi_2 && = \; m \Phi + \sqrt{2} g \left(
  A + \ii B \right) \\ 
  \dfrac{\partial f(\phi_1, \phi_2, \phi_3)}{\partial \phi_3} = & \; m \phi_2
  && = \; \dfrac{m}{\sqrt{2}} \left( A + \ii B \right) 
\end{alignat}

After inserting these derivatives and sorting the terms we get 
\begin{equation}
  \label{eq:or_superstrom}
  \begin{aligned}
    {\cal J}^\mu = & \; - \sqrt{2} (\fmslash{\partial} \phi) \gamma^\mu {\cal
    P}_R \chi - \sqrt{2} (\fmslash{\partial} \phi^*) \gamma^\mu {\cal P}_L
    \chi - \sqrt{2} \ii \lambda \gamma^\mu \chi + \ii {\cal P}_L \left( (\ii
    \fmslash{\partial} - m) A \right) \gamma^\mu \Psi \\ & \; + \ii {\cal P}_R
    \left( (\ii \fmslash{\partial} - m) A \right) \gamma^\mu \Psi^c + {\cal
    P}_L \left( (\ii \fmslash{\partial} - m) B \right) \gamma^\mu \Psi - {\cal
    P}_R \left( (\ii \fmslash{\partial} - m) B \right) \gamma^\mu \Psi^c \\ &
    \; + \ii \sqrt{2} {\cal P}_R \left( (\ii \fmslash{\partial} - m) \Phi
    \right) \gamma^\mu \Psi  + \ii \sqrt{2} {\cal P}_L \left( (\ii
    \fmslash{\partial} - m) \Phi^* \right) \gamma^\mu \Psi^c \\ & \; -
    \dfrac{\ii g}{\sqrt{2}} \left( A^2 - B^2 \right) \gamma^\mu \chi -
    \sqrt{2} g A B \gamma^\mu \gamma^5 \chi - 2 \ii g \gamma^\mu A \phi {\cal
    P}_L \Psi - 2 \ii g \gamma^\mu A \phi^* {\cal P}_R \Psi^c \\ & \; + 2 g
    \gamma^\mu B \phi {\cal P}_L \Psi - 2 g \gamma^\mu B \phi^* {\cal P}_R
    \Psi^c   
  \end{aligned}
\end{equation}

Let us start with a rather trivial example, which relates 2- and
3-point functions in lowest order perturbation theory. We consider the SWI
\begin{equation}
  \label{eq:2punkt3punkt}
  \begin{aligned}
    & k_\mu \, \mathrm{F.T.} \Greensfunc{ \left( \overline{\xi} {\cal J}^\mu
    \right) A (x_1) \Psi (x_2)} \\ & \stackrel{!}{=}
    \quad \mathrm{F.T.} \Greensfunc{ \Psi (x_2)
  \left( \overline{\Psi} (x_1) {\cal P}_R \xi \right)} \delta^4
  (x-x_1)  \\ & \qquad \qquad \; \; \; \; \, + 
  \mathrm{F.T.} \Greensfunc{A(x_1) \left( - \ii \fmslash{\partial} - m \right)
    A(x_2) {\cal P}_R \xi} \delta^4 (x-x_2) + {\cal O} (g)   
  \end{aligned}
\end{equation}
We have only kept those of the contact terms giving
nonvanishing contributions. The right hand side will be calculated
first; we adopt the convention that all momenta be incoming. The right hand
side is 
\begin{equation}
  \label{eq:rechteseite23}
  \text{RHS} \, (\ref{eq:2punkt3punkt}) = \dfrac{\ii (- \fmslash{k}_2
  + m)}{k_2^2 
  - m^2} {\cal P}_R \xi - \dfrac{\ii (\fmslash{k}_1 + m)}{k_1^2 - m^2 - 2
  \lambda g} {\cal P}_R \xi \; + \; {\cal O} (g)
\end{equation}
As mentioned earlier, for the calculation of the left hand side care
has to be taken about possible higher orders in perturbation theory
which may contribute to this SWI. In these diagrams the linear
part of the current will be coupled to the external particles via the
Goldstino, wherein the coupling constant combined with the parameter
for the spontaneous symmetry breaking $\lambda$ is responsible for the
mass splitting between the participating particles $A$ and
$\Psi$. This will prove important -- as we will see soon -- for
constructing the propagators with the correct poles. The pole of the
Goldstino at zero mass always cancels out of those diagrams against the
momentum influx from the current. Diagrammatically the left hand side looks
like ($k = k_1 + k_2$): 
\begin{equation}
  \label{eq:linkeseite23_1}
  \text{LHS} \, (\ref{eq:2punkt3punkt}) =  \qquad
    \parbox{2cm}{\begin{fmfchar*}(15,18) 
    \fmftop{t}
    \fmfbottom{b1,b2}
    \fmf{dots}{b1,v}
    \fmf{fermion}{v,b2}
    \fmf{dbl_plain,label=\begin{math}k_\mu {\cal J}^\mu
    \end{math}}{v,t} 
    \fmfdot{v}
    \fmfblob{.25w}{t}
  \end{fmfchar*}} + 
  \parbox{2cm}{\begin{fmfchar*}(15,18)
    \fmftop{t}
    \fmfbottom{b1,b2}
    \fmf{dots}{b1,v1}
    \fmf{fermion}{v1,b2}
    \fmf{plain}{v1,v2}
    \fmf{dbl_plain,label=\begin{math}k_\mu {\cal J}^\mu
    \end{math}}{v2,t} 
    \fmfdot{v1,v2}
    \fmfblob{.25w}{t}
  \end{fmfchar*}} 
\end{equation}
The analytical expression for the left hand side (\ref{eq:2punkt3punkt}) is
\begin{align}
    \text{LHS} \, (\ref{eq:2punkt3punkt}) = & \; \; \; \; \ii \partial_\mu^x
    \Greensfunc{\left( \overline{\xi} {\cal J}^\mu (x) \right) A (x_1) \Psi
    (x_2)}_{(0)} \notag \\ & \; + \ii \partial_\mu^x \cdot \ii \int d^4 y
    \Greensfunc{ \left( \overline{\xi} {\cal J}^\mu (x) \right) A (x_1) \Psi  
    (x_2) {\cal L}_{\text{int}} } + \ldots \notag \\ = & \; \; \; \; \ii
    \partial_\mu^x \Greensfunc{ A(x_1) \Psi(x_2) \ii \Bigl( \overline{\Psi}
    (x) \gamma^\mu \left[ (\ii \fmslash{\partial} + m) A (x) \right] {\cal
    P}_R \xi \Bigr) } \notag \\ & \; - \partial_\mu^x \int d^4 y
    \Greensfunc{\left( \sqrt{2} \ii \lambda \overline{\xi} \gamma^\mu \chi
    (x) \right) \left( \sqrt{2} g \overline{\Psi}(y) {\cal P}_R \chi (y) 
    \right) A(x_1) \Psi(x_2)} \notag \\ & \; + \text{higher orders}
    \notag \\
    \stackrel{\text{FT}}{\rightarrow} & \; \dfrac{\ii (- \fmslash{k}_2 +
    m)}{k_2^2 - m^2} \left( \fmslash{k}_1 + \fmslash{k}_2 \right)
    \dfrac{\fmslash{k}_1 + m}{k_1^2 - m^2 - 2 \lambda g} {\cal P}_R
    \xi \notag \\ &
    \; - \dfrac{\ii (-\fmslash{k}_2 + m)}{k_2^2 - m^2} \dfrac{2 \lambda
    g}{k_1^2 - m^2 - 2 \lambda g} {\cal P}_R \xi \; + \; {\cal O} (g)
    \notag \\
    = & \; \dfrac{\ii ( - \fmslash{k}_2 + m)}{k_2^2 - m^2} {\cal P}_R \xi 
    - \dfrac{\ii (\fmslash{k}_1 + m)}{k_1^2 - m^2 - 2 \lambda g} {\cal P}_R
    \xi \; + \; {\cal O} (g) \quad = \quad \text{RHS} \,
  (\ref{eq:2punkt3punkt}) 
    \quad \surd    \label{eq:linkeseite23_2}
\end{align}
The SWI is fulfilled. Amputating the external legs (except for the
current) by means of the LSZ reduction formula produces the on-shell
identity which is thence automatically fulfilled as well.

%%% Local Variables: 
%%% mode: latex
%%% TeX-master: "diss"
%%% End: 

%% file: chap8.tex
\chapter{Gauge theories and Supersymmetry} 

In gauge theories there appears a new phenomenon not met in the
previous chapters: the participation of (massless or massive) vector
bosons connected to the concept of gauge symmetry and gauge
transformations. These are indispensable ingredients for a
realistic field theoretic model describing elementary particle
phenomenology. The gauge principle, i.e.~the covariance of the fields under
local phase transformations, must in a supersymmetric field
theory be incorporated in a SUSY covariant manner. As shown in
\cite{Weinberg:QFTv3:Text} and \cite{Reuter:2000:SUSY}  the kinetic
terms with minimal coupling can be written down in a SUSY-covariant
form by introducing a vector superfield $\hat{V}$ (this is a real
superfield with $\hat{V}^\dagger = \hat{V}$), and making the
replacement 
\begin{equation}
\label{eq:kinlageich}
S_{\text{kin}} = \int d^4 x \, \dfrac{1}{2} \left[ \hat{\Phi}^\dagger
\hat{\Phi} \right]_D \longrightarrow \int d^4 x \, \dfrac{1}{2} \left[
\hat{\Phi}^\dagger e^{\pm c \hat{V}} \Phi \right]_D   .
\end{equation}
Therein $c$ is a normalization constant depending on the normalization
of the algebra of the gauge symmetry which is as changing from author
to author as the choice of sign. The sign of $c$ is related to the sign in
the gauge-covariant derivative,
\begin{equation}
D_\mu = \partial_\mu \pm \ii g \sum_a T^a A_\mu^a .
\end{equation}

The kinetic term for the gauge fields is produced with the help of
spinor superfields, chiral superfields equipped with an additional
spinor index. They are established by triply applying the
super-covariant derivative $\mathcal{D}$ to the vector superfield 
\begin{equation}
\hat{W}(x,\theta) = - \dfrac{1}{4} \left( \overline{\mathcal{D}} \mathcal{D}
\right) \mathcal{D} \: \hat{V} (x,\theta).
\end{equation}
Then the kinetic part of the gauge fields can be expressed as
\begin{equation}
 S_{\text{gauge}} = \frac{1}{2} \int d^4 x \, \Re \Bigl[ \sum_a
 \overline{W^a_R} W^a_L \Bigr]  .
\end{equation}

There is a high redundancy in the superfield formulation of
supersymmetric gauge theories. The new superfield $\hat{V}$
there contains a huge amount of unphysical degrees of freedom. But we can
get rid of them. The kinetic part (and the superpotential as well)
are not only invariant under SUSY and gauge transformations but also
under so called extended gauge transformations. These are gauge
transformations where the gauge parameter (usually a scalar spacetime
dependent parameter) is replaced by a complete superfield
$\hat{\Lambda}(x,\theta)$. We can use these transformations to gauge away
the superfluous degrees of freedom, three scalar and one spinor
component field so that only the gauge field, the gaugino and a scalar
field with canonical dimension two remain. This is called the
Wess-Zumino gauge. After having fixed the above mentioned components,
only the ordinary gauge transformations survive from the
extended gauge transformations. 

The Lagrangean density of the matter fields with minimal coupling therefore
has the structure:
\begin{multline}
\label{eq:lmateich}
\mathcal {L}_{\text{mat}} = \left( D_\mu \phi\right)^\dagger \left(
D^\mu \phi \right) + \dfrac{\ii}{2} \left( \overline{\Psi} \fmslash{D}
\Psi \right) + F^\dagger F - \sqrt{2} g \overline{\lambda^a} \phi^\dagger T^a
\Psi_L \\ - \sqrt{2} g \overline{\Psi_L} T^a \phi \lambda^a + g
\phi^\dagger T^a \phi D^a + \mathcal{W} (\phi, \Psi, F)
\end{multline}
Here $\mathcal{W} (\phi, \Psi, F)$ stands for the superpotential
parts of the matter Lagrangean density which are globally and locally
invariant under the gauge symmetry group. It does not contain any
derivatives of the fields. 

The kinetic terms of the gauge fields and gauginos are:
\begin{equation}
\label{eq:leicheich}
\mathcal{L}_{\text{gauge}} = - \dfrac{1}{4} F_{\mu\nu}^a F^{\mu\nu}_a +
\dfrac{\ii}{2} \overline{\lambda^a} \left( \fmslash{D} \lambda
\right)^a + \dfrac{1}{2} D^a D^a 
\end{equation}

Since it is consistent with the gauge symmetry, we may add a {\em
Fayet-Iliopoulos} term 
\begin{equation}
\label{lfieich}
\mathcal{L}_{\text{FI}} = \zeta^a D^a \qquad \text{with} \qquad
f^a_{bc} \zeta^a = 0 . 
\end{equation}
The last condition is necessary in the non-Abelian case for this term
to transform into a total derivative under SUSY. It forces the gauge
field part in the covariant derivative of the gauginos produced when
SUSY-transforming the auxiliary field to vanish.  

%%%%%%%%%%%%%%%%%%%%%%%%%%%%%%%%%%%%%%%%%%%%%%%%%%%%%%%%%%%%%%%%

\section{The de~Wit--Freedman transformations}

The Wess-Zumino supergauge fixing procedure destroys
invariance of the Lagrangean density under SUSY transformations as
well as under extended gauge transformations. When performing a SUSY
transformation the states gauged away in the WZ gauge are populated
again with the effect that the Lagrangean density is no longer WZ
gauged. This can be remedied by performing another extended gauge
transformation to newly reach WZ gauge. From last section's discussion 
this is understandable from the fact that SUSY and gauge
transformations are not completely orthogonal to each other. The
several transformations and their relations are displayed below:

\begin{equation*}
\boxed{
\begin{pspicture}(0,0)(6,3)
  \psline{->}(1.5,.5)(4.5,.5)
  \psline{->}(1.5,2.5)(4.5,2.5)
  \psline{<-}(1,1)(1,2)
  \psline{<-}(5,1)(5,2)
  \psline{->}(1.5,1)(4.5,2)
  %\psline(0,3)(6,3)    \psline(0,0)(6,0)
  \rput(1,.5){$\mathcal{L}^{WZ}$}       \rput(5,.5){$\mathcal{L}^{WZ}$}
  \rput(1,2.5){$\mathcal{L}$}           \rput(5,2.5){$\mathcal{L}$}
  \rput(3,.2){$\mathcal{T}_{\text{deW-F}}$}
  \rput(.5,1.5){$\mathcal{T}_{\text{ext.g.}}$}
  \rput(5.6,1.5){$\mathcal{T}_{\text{ext.g.}}$}
  \rput(2.3,1.6){$\mathcal{T}_{\text{SUSY}}$}
  \rput(3,2.8){$\mathcal{T}_{\text{SUSY}}$}
\end{pspicture}}
\end{equation*}                           
When performing a SUSY transformation and an extended gauge
transformation afterwards (for the details cf.~\cite{Weinberg:QFTv3:Text})
this results in a combined transformation called 
{\em de~Wit--Freedman transformation} which leaves the Lagrangean
density in WZ gauge invariant \cite{deWit/Freedman:1975:susyeich}. 
In de~Wit--Freedman transformations the spacetime derivatives are
replaced by gauge covariant derivatives; furthermore there are some
additional terms. So de~Wit--Freedman transformations are the gauge-covariant
version of the SUSY transformations. For supersymmetric Yang--Mills theories
they are (we put a tilde on them to
distinguish them from the ordinary supersymmetry transformations; for more
details see appendix \ref{dwf-appen}):

\begin{equation}
\boxed{
\begin{aligned}
 \tilde{\delta}_\xi \phi = & \; \sqrt{2} \left( \overline{\xi} \Psi_L
 \right), \\
 \tilde{\delta}_\xi \psi = & \; - \ii \sqrt{2} \gamma^\mu \left( (D_\mu
 \phi) {\cal P}_R +  (D_\mu \phi)^\dagger {\cal P}_L \right)
 \xi + \sqrt{2} (F {\cal P}_L + F^\dagger {\cal P}_R) \xi , \\ 
 \tilde{\delta}_\xi F = & \; - \ii \sqrt{2} \left( \overline{\xi}
 \fmslash{D} \Psi_L \right) + 2 g \overline{\xi} T^a \phi \lambda^a_R,
 \\ 
 \tilde{\delta}_\xi A_\mu^a = & \; - \left( \overline{\xi} \gamma_\mu
 \gamma_5 \lambda^a \right), \\
 \tilde{\delta}_\xi \lambda^a = & \; - \dfrac{\ii}{2} F_{\mu\nu}^a
 \gamma^\mu \gamma^\nu \gamma^5 \xi + D^a \xi , \\
 \tilde{\delta}_\xi D^a = & \; - \ii \overline{\xi} \left( \fmslash{D}
 \lambda \right)^a. 
\end{aligned}}
\end{equation}

%%%%%%%%%%%%%%%%%%%%%%%%%%%%%%%%%%%%%%%%%%%%%%%%%%%%%%%%%%%%%%%%%%%%%%

\section[The current in SYM theories]{The current in supersymmetric Yang--Mills theories}

Because it is a complicated and lengthy topic we postpone the
detailed derivation of the supersymmetric current for supersymmetric
Yang--Mills theories (SYM) to the appendix, \ref{sec:symcurrent}. We simply
state the result for the SUSY current in a supersymmetric Yang-Mills theory 
\begin{equation}
  \label{eq:strom_sym}
\boxed{
\begin{aligned}
  \mathcal{J}^\mu = & \; - \sqrt{2} \gamma^\nu \gamma^\mu (D_\nu \phi)^T
  \Psi_R - \sqrt{2} \gamma^\nu \gamma^\mu (D_\nu \phi)^\dagger \Psi_L - \ii
  \gamma^\mu \zeta^a \lambda^a \\ & \; + \dfrac{1}{2} \gamma^\alpha
  \gamma^\beta \gamma^\mu \gamma^5 F_{\alpha\beta}^a \lambda^a - \ii g
  \gamma^\mu \left( \phi^\dagger \vec{T} \phi \right) \cdot \lambda \\ & \; -
  \ii \sqrt{2} \gamma^\mu \left( \dfrac{\partial f (\phi)}{\partial \phi}
  \right)^T \Psi_L - \ii \sqrt{2} \gamma^\mu \left( \dfrac{\partial f
  (\phi)}{\partial \phi} \right)^\dagger \Psi_R 
\end{aligned}} \quad .
\end{equation}
It is conserved, 
\begin{equation}
        \boxed{ \partial_\mu \mathcal{J}^\mu = 0 } \quad, 
\end{equation}
as will also be proven in the appendix, \ref{sec:symcurrentcons}. 

%%%%%%%%%%%%%%%%%%%%%%%%%%%%%%%%%%%%%%%%%%%%%%%%%%%%%%%%%%%%%%%%%%%%%%%

\section{Comparison of the currents -- physical interpretation}

The use of the de~Wit--Freedman transformation is not mandatory
\cite{Weinberg:QFTv3:Text}. It is also possible to use the
``ordinary'' SUSY transformations to calculate the current. We do
want to show now that the current in SYM theories remains the same when 
using SUSY instead of de~Wit--Freedman transformations in its
derivation. The difference between both transformations shows up in
the auxiliary fields $F$, $F^\dagger$ and $D^a$, as well as in the
matter fermions. The two Noether parts -- calculated on the one hand
with the dWF transformation, on the other hand with the ordinary SUSY
transformation -- differ by a term
\begin{equation}
  \label{deltastromnoeth}
  \overline{\xi} \mathcal{J}^\mu_{\text{dWF}} - \overline{\xi}
  \mathcal{J}^\mu_{\text{ord.}}  = - \dfrac{\ii g}{\sqrt{2}}
  \left( \overline{\xi} \gamma^\nu \gamma^\mu \phi^T \vec{T}^T
  \vec{A}_\nu \Psi_R \right) + \dfrac{\ii g}{\sqrt{2}} \left(
  \overline{\xi} \gamma^\nu \gamma^\mu \phi^\dagger \vec{T}
  \vec{A}_\nu \Psi_L \right) .
\end{equation}
We now list all terms produced by a dWF transformation of the Lagrangean
density in addition to the SUSY transformation:
\begin{align}
F^\dagger F \longrightarrow & \;  - 2 g F^\dagger \left(
\overline{\xi} \vec{T} \phi \cdot \vec{\lambda}_R \right) - 2 g \left(
\overline{\xi} \phi^\dagger \vec{T} F \cdot \vec{\lambda}_L \right) +
\sqrt{2} g \left( \overline{\xi} F^\dagger \vec{T} \fmslash{\vec{A}}
\Psi_L \right) \notag \\ & \; - \sqrt{2} g \left( \overline{\xi} F^T \vec{T}^T 
\fmslash{\vec{A}} \Psi_R \right) \label{zusatz_ff} \\ \dfrac{\ii}{2} \left(
\overline{\Psi} 
\fmslash{D} \Psi \right) \longrightarrow & \; \dfrac{\ii g}{\sqrt{2}} \left(
\overline{\xi} \gamma^\mu \gamma^\nu (D_\nu \vec{T}^T \vec{A}_\mu \phi^T)
\Psi_R \right) - \dfrac{\ii g}{\sqrt{2}} \left( \overline{\xi} \gamma^\mu
\gamma^\nu (D_\nu \vec{T} \vec{A}_\mu \phi)^\dagger \Psi_L \right) \notag \\ &
\; + \dfrac{\ii g}{\sqrt{2}} \left( \overline{\xi} \gamma^\mu (\vec{T}
\vec{A}_\mu \phi^\dagger) \fmslash{D} \Psi_L \right) - \dfrac{\ii g}{\sqrt{2}}
\left( \overline{\xi} \gamma^\mu (\vec{T}^T \vec{A}_\mu \phi^T) \fmslash{D}
\Psi_R \right) \label{zusatz_psi} \\ - \sqrt{2} g \left(
\overline{\vec{\lambda}} \cdot 
\phi^\dagger \vec{T} \Psi_L \right) \longrightarrow & \; 2 g^2 \left(
\overline{\xi} \phi^\dagger \vec{T} (\vec{T} \fmslash{\vec{A}} \phi) \cdot
\vec{\lambda}_L \right) \label{zusatz_g1} \\ - \sqrt{2} g \left(
\overline{\Psi_L} \vec{T} \phi 
\cdot \vec{\lambda} \right) \longrightarrow & \; - 2 g^2 \left( \overline{\xi} 
\gamma^\mu (\vec{T} \vec{A}_\mu \phi^\dagger) \vec{T} \phi \cdot
\vec{\lambda}_R \right) \label{zusatz_g2} \\ g \left( \phi^\dagger \vec{T}
\phi \right) \cdot D 
\longrightarrow & \; \ii g^2 \left( \phi^\dagger T^a \phi \right) \left(
\overline{\xi} f^a_{\;bc} \fmslash{A}^b \lambda^c \right) \label{zusatz_d} \\
\mathcal{L}_{\text{gauge}} \longrightarrow & \; \ii D^a \left( \overline{\xi}
f^a_{\;bc} \fmslash{A}^b \lambda^c \right) \label{zusatz_eich} 
\end{align}
From these additional terms as many as possible are eliminated. The
first two terms out of (\ref{zusatz_ff}) vanish by the condition
(\ref{suppoteichinv}). From the last two equations (\ref{zusatz_d})
and (\ref{zusatz_eich}), the contributions cancel due to the equation
of motion for the auxiliary field $D^a$. Next we multiply the terms
with the covariant derivatives of the fermions in (\ref{zusatz_psi})
by a factor two and subtract them once. In the doubled expressions we
insert the equation of motion for the fermions; with the help of the
two identities derived from (\ref{suppoteichinv}) in the paragraph
below that equation we see, that the remaining terms from
(\ref{zusatz_ff}) cancel as well as the gaugino contributions
(\ref{zusatz_g1}) and (\ref{zusatz_g2}). We are left with 
\begin{equation}
  \begin{aligned}
   & \;  \dfrac{\ii g}{\sqrt{2}} \left( \overline{\xi} \gamma^\mu \gamma^\nu
   (D_\nu \vec{T}^T \vec{A}_\mu \phi^T) \Psi_R \right) - \dfrac{\ii
   g}{\sqrt{2}} \left( \overline{\xi} \gamma^\mu \gamma^\nu (D_\nu \vec{T}
   \vec{A}_\mu \phi)^\dagger \Psi_L \right) \\ & \;  
    - \dfrac{\ii g}{\sqrt{2}} \left( \overline{\xi} \gamma^\mu (\vec{T}
   \vec{A}_\mu \phi^\dagger) \fmslash{D} \Psi_L \right) + \dfrac{\ii
   g}{\sqrt{2}} \left( \overline{\xi} \gamma^\mu (\vec{T}^T \vec{A}_\mu
   \phi^T) \fmslash{D} \Psi_R \right)  .
  \end{aligned}
\end{equation}
All terms containing two gauge fields cancel each other so that the
remaining term is the following derivative:
\begin{equation}
  \partial_\nu \biggl[ \dfrac{\ii g}{\sqrt{2}} \left( \overline{\xi}
  \gamma^\mu \gamma^\nu (\vec{T}^T \vec{A}_\mu \phi^T) \Psi_R \right) -
  \dfrac{\ii g}{\sqrt{2}} \left( \overline{\xi} \gamma^\mu \gamma^\nu (\vec{T}
  \vec{A}_\mu \phi^\dagger) \Psi_L \right) \biggr] .
\end{equation}
This cancels exactly the contribution to the current from the Noether
part, (\ref{deltastromnoeth}), and both currents are equal. 

The fact that the two currents are identical can be interpreted physically in
the following way: The supersymmetric current of SYM may be derived in the
superfield formalism independent from any choice of supergauge. Since the
supercurrent (the superfield containing the supersymmetric current and the
energy-momentum tensor) is a scalar with respect to (extended) gauge
transformations, the supersymmetric current remains the same when expressed in
different supergauges. 

%%%%%%%%%%%%%%%%%%%%%%%%%%%%%%%%%%%%%%%%%%%%%%%%%%%%%%%%%%%%%%%%%%%%%%%%%

\section{SWI in an Abelian toy model}

To test our supersymmetric Ward identities (SWI) for a supersymmetric
gauge theory, we choose the simplest possible example, a model with one
matter superfield and a $U(1)$ gauge symmetry. This is not SQED -- the
supersymmetric extension of QED, since there is only a single
superfield. Gauge invariance then forces the superpotential to vanish,
so that all particles of our model are massless. Furthermore the matter
fermion is of Majorana type and the gauge field must couple to it as an
axial vector, as this is the only possibility for a gauge field
to have a nonvanishing coupling to a Majorana fermion. The whole gauge
superfield must then be axial as well and the model bears an 
anomaly, the supersymmetric extension of the axial vector anomaly. But
as long as we are only concerned with tree level processes we do not have
to care about anomalies -- they will only become important for higher
order calculations. The details of our Abelian toy model can again be
found in appendix \ref{sec:sagt}. Here we just quote the Lagrangean density
and the current (to avoid confusion with the scalar particle $A$ we
denote the gauge boson by $G_\mu$ in this model):
\begin{equation}
\boxed{
\begin{aligned}
  \mathcal{L} = \dfrac{1}{2} (\partial_\mu A) (\partial^\mu A) +
    \dfrac{1}{2} (\partial_\mu B) (\partial^\mu B) + \dfrac{\ii}{2}
    \overline{\Psi} \fmslash{\partial} \Psi - \dfrac{1}{4} F_{\mu\nu}
    F^{\mu\nu} + \dfrac{\ii}{2} \overline{\lambda} \fmslash{\partial}
    \lambda \\
    + e G_\mu \left( B \partial^\mu A - A \partial^\mu B
    \right) + \dfrac{e^2}{2} G_\mu G^\mu \left( A^2 + B^2 \right) - e
    \left( \overline{\Psi} \lambda \right) A \\
    - \ii e \left( \overline{\Psi} \gamma^5 \lambda \right) B -
    \dfrac{e}{2} \overline{\Psi} \fmslash{G} \gamma^5 \Psi - \dfrac{e^2}{8}
    \left( A^4 + B^4 + 2 A^2 B^2 \right)
\end{aligned}}
\end{equation}

\begin{equation}
\boxed{
\begin{aligned}
  \mathcal{J}^\mu = - (\fmslash{\partial} A) \gamma^\mu \Psi - \ii
  (\fmslash{\partial} B) \gamma^\mu \gamma^5 \Psi + \ii e A \fmslash{G}
  \gamma^\mu \gamma^5 \Psi - e B \fmslash{G}  \gamma^\mu \Psi \\   +
  \dfrac{1}{2} \lbrack \gamma^\alpha , \gamma^\beta \rbrack \gamma^\mu
  \gamma^5 (\partial_\alpha G_\beta) \lambda - \dfrac{\ii e}{2} \left( A^2
  + B^2 \right) \gamma^\mu \lambda
\end{aligned}}
\end{equation}      

We first show a simple example for an on-shell Ward identity:
\begin{multline}
  J_\mu(p_1,p_2) = \mathrm{F.T.} \Braket{0|\mathcal{J}_\mu(x)|A (p_1)
  \Psi(p_2)} \\ = \mathrm{F.T.}
  \Braket{0|\mathcal{J}_\mu(x)|A(p_1)\Psi(p_2)}_{(0)} 
  = - \fmslash{p}_1 \gamma_\mu u(p_2) 
\end{multline}
\begin{equation}
  (p_1+p_2)^\mu J_\mu(p_1,p_2) = - \fmslash{p}_1 \left( \fmslash{p}_1 +
    \fmslash{p}_2 \right) u(p_2) = 0 
\end{equation}
The second term in parentheses vanishes due to the Dirac equation
while $p_1^2$ is zero (all particles are massless in this model by
gauge invariance). 

We now discuss several examples of Ward identites calculated
off-shell. By amputation in the sense of LSZ reduction we can then
easily get back to the on-shell identities. Let us first -- as a warm-up --
examine an SWI with two fields beneath the current insertion written
down in the form (\ref{eq:stromerhaltung2}) 
\begin{multline}
        k^\mu \mathrm{F.T.} \Braket{0|
        \mathrm{T} \,\overline{\xi}\mathcal{J}_\mu(y)A(x_1)\Psi(x_2)|0} 
        \stackrel{!}{=} \mathrm{F.T.} \Braket{0|\mathrm{T} \left(\delta_\xi
        A(x_1)\right) \Psi(x_2)|0} \delta^4 (x_1-y) \\ \qquad + \mathrm{F.T.}
        \Braket{0|\mathrm{T}A(x_1) \left( \delta_\xi \Psi(x_2) \right)
        |0} \delta^4 (x_2-y) \\ = \mathrm{F.T.}
        \Braket{0|\mathrm{T}\Psi(x_2)\overline{\Psi}(x_1)\xi|0} \delta^4 (x_1
        - y)
     \\ + \mathrm{F.T.} \Braket{0|\mathrm{T}A(x_1) (-\ii
        \fmslash{\partial}A(x_2))\xi|0} \delta^4(x_2-y)
\end{multline}              
Graphically we denote the momentum influx by a dotted line. Then we 
have the following relation ($k + p_1 + p_2 = 0$ and all momenta
incoming)  
\begin{equation}
  \parbox{21mm}{\hfil\\\hfil\\%
    \begin{fmfgraph*}(20,15)
      \fmftop{t}
      \fmfbottom{b1,b2}
      \fmf{dashes}{b1,v}
      \fmf{plain}{b2,v}
      \fmf{dbl_plain,label=\begin{math}k_\mu {\cal J}^\mu
        \end{math}}{v,t}
      \fmfdot{v}
      \fmfblob{.25w}{t}
    \end{fmfgraph*}\hfil\\} \qquad \stackrel{!}{=} \qquad   
  \parbox{21mm}{\hfil\\\hfil\\%
    \begin{fmfgraph*}(20,15)
      \fmfleft{i,di}\fmfright{o,do}
      \fmftop{k}
      \fmf{plain,label=$p_2$,l.side=right}{o,i}
      \fmf{dots,left=.2,tension=0.5,label=$k$,l.side=left}{i,k}
    \end{fmfgraph*}\hfil\\} \quad  + \quad 
    \parbox{21mm}{\hfil\\\hfil\\%
    \begin{fmfgraph*}(20,15)
      \fmfleft{i,di}\fmfright{o,do}
      \fmftop{k}
      \fmf{dashes,label=$p_1$,l.side=left}{i,o}
      \fmf{dots,left=.2,tension=0.5,label=$k$,l.side=left}{k,o}
    \end{fmfgraph*}\hfil\\} 
\end{equation}
\begin{equation}
  - k^\mu \dfrac{\ii}{p_1^2} \dfrac{-\ii}{\fmslash{p}_2} \gamma_\mu 
  (-\ii\fmslash{p}_1) \xi = \dfrac{-\ii}{p_1^2}
  \dfrac{1}{\fmslash{p}_2} \left( \fmslash{p}_1 + \fmslash{p}_2
  \right) \fmslash{p}_1 \xi \stackrel{!}{=} 
  \left( \dfrac{-\ii}{\fmslash{p}_2} + \dfrac{-\ii
      \fmslash{p}_1}{p_1^2} \right) \xi  
\end{equation}
The SWI is fulfilled. 

\vspace{.5cm}            

Another SWI for a 2-point Green function will be calculated now to 
show a new effect.
\begin{equation}
  \parbox{21mm}{%
        \hfil\\
    \begin{fmfgraph*}(20,15)
      \fmftop{t}
      \fmfbottom{b1,b2}
      \fmf{photon}{b1,v}
      \fmf{plain}{b2,v}
      \fmf{dbl_plain,label=\begin{math}k_\mu {\cal J}^\mu
        \end{math}}{v,t}
      \fmffreeze \fmf{photon}{b2,v}
      \fmfdot{v}
      \fmfblob{.25w}{t}
    \end{fmfgraph*}
        \hfil} \qquad \stackrel{!}{=} \qquad 
  \parbox{21mm}{%
      \hfil\\
    \begin{fmfgraph*}(20,15)
      \fmfleft{i,di}\fmfright{o,do}
      \fmftop{k}
      \fmf{plain,label=$p_2$,l.side=left}{i,o}
      \fmf{dots,right=.2,tension=0.5,label=$k$,l.side=right}{k,i}
      \fmffreeze \fmf{photon}{i,o}
    \end{fmfgraph*}\\
     \hfil}\quad + \quad 
  \parbox{21mm}{%
      \hfil\\
    \begin{fmfgraph*}(20,15)
      \fmfleft{i,di}\fmfright{o,do}
      \fmftop{k}
      \fmf{photon,label=$p_1$,l.side=left}{i,o}
      \fmf{dots,left=.2,tension=0.5,label=$k$,l.side=left}{k,o}
    \end{fmfgraph*}\\
        \hfil}
\end{equation}
\begin{align}
  &  \mathrm{F.T.} \Braket{0|\mathrm{T} \left( \delta_\xi G_\nu 
    (x_1) \right) \lambda(x_2)|0} \delta^4(x_1-y) \notag 
  \\ & \qquad\qquad + \mathrm{F.T.} \Braket{0|\mathrm{T}G_\nu(x_1) \left(
    \delta_\xi 
    \lambda(x_2) \right) |0} \delta^4(x_2-y) \notag \\
  = \; &  - \mathrm{F.T.}
    \Braket{0|\mathrm{T}\lambda(x_2)\overline{\lambda}(x_1)\gamma_\nu\gamma^5\xi|0} \delta^4(x_1-y)
    \notag \\ & \qquad \qquad
  - \dfrac{\ii}{2} \mathrm{F.T.} \Braket{0|\mathrm{T}G_\nu(x_1)
    (\partial_\alpha^{x_2} G_\beta(x_2))\lbrack \gamma^\alpha ,
    \gamma^\beta \rbrack \gamma^5 \xi|0} \delta^4(x_2-y) \notag \\
  \stackrel{!}{=} \; & k^\mu \mathrm{F.T.}
  \Braket{0|\mathrm{T} \overline{\xi}
    \mathcal{J}_\mu(y)G_\nu(x_1)\lambda(x_2)|0} \notag 
  \\ = \; &
  \dfrac{1}{2} k^\mu \mathrm{F.T.}  \Braket{0|\mathrm{T}
    \lambda(x_2) \overline{\lambda}(y) \gamma^5 \gamma_\mu \lbrack
    \gamma^\alpha , \gamma^\beta \rbrack (\partial_\alpha^y G_\beta (y))
    G_\nu(x_1) \xi |0} \; \; , 
\end{align}              
i.e.
\begin{multline}
  \dfrac{\ii}{\fmslash{p}_2} \gamma_\nu \gamma^5 \xi - \dfrac{1}{2}
  \dfrac{\ii}{p_1^2} \lbrack -\fmslash{p}_1 , \gamma_\nu \rbrack
  \gamma^5 \xi  \\ \stackrel{!}{=}
  \dfrac{1}{2} (-1) (p_1^\mu + p_2^\mu) \dfrac{-\ii}{\fmslash{p}_2}
  \gamma^5 \gamma_\mu \lbrack \gamma^\alpha , \gamma^\beta \rbrack (-
  \ii p_{1,\alpha}) \dfrac{-\ii \eta_{\beta\nu}}{p_1^2} \xi
\end{multline}           
Multiplication by a factor $\ii$ and simplification yields the result
\begin{multline}
  \dfrac{-1}{\fmslash{p}_2} \gamma_\nu \gamma^5 \xi - \dfrac{1}{2}
  \dfrac{1}{p_1^2} \lbrack \fmslash{p}_1 , \gamma_\nu \rbrack
  \gamma^5 \xi  \\ \stackrel{!}{=}
  \dfrac{1}{2} \dfrac{1}{\fmslash{p}_2} \gamma^5 (\fmslash{p}_1 +
  \fmslash{p}_2) \lbrack \fmslash{p}_1 , \gamma_\nu \rbrack
  \dfrac{1}{p_1^2} \xi = - \dfrac{1}{2} \dfrac{1}{p_1^2} \lbrack
  \fmslash{p}_1 , \gamma_\nu \rbrack \gamma^5 \xi - \dfrac{1}{2} \gamma^5
  \dfrac{1}{\fmslash{p}_2} \fmslash{p}_1 \lbrack \fmslash{p}_1 , \gamma_\nu
  \rbrack \dfrac{1}{p_1^2} \xi \\
  = - \dfrac{1}{2} \dfrac{1}{p_1^2} \lbrack \fmslash{p}_1 , \gamma_\nu
  \rbrack \gamma^5 \xi - \dfrac{1}{\fmslash{p}_2} \gamma_\nu \gamma^5 \xi
  + \dfrac{1}{\fmslash{p}_2} \dfrac{\fmslash{p}_1}{p_1^2} p_{1,\nu}
  \gamma^5 \xi
\end{multline}
Astonishingly there is an additional term on the side of the current so that
{\em this SWI is not fulfilled off-shell.} But we notice that this
additional (the third) term is proportional to the momentum of the
gauge boson. Using LSZ reduction (multiplying with the inverse
propagator and with the polarization vector) and setting the particles
on the mass shell yields zero by gauge boson transversality. Hence,
{\em on-shell} the SWI {\em is} valid. 

\vspace{.5cm}    

Now we attack a more complicated example, a 3-point function. In formula
(\ref{eq:stromerhaltung2}) we choose -- together with the current insertion -- 
the fields $A(x_1)$, $G_\nu(x_2)$ and $\Psi(x_3)$. We get four nonvanishing
contributions from the SUSY transformations of these fields. 
\begin{subequations}
  \begin{align}
    \mathrm{F.T.} \Braket{0|\mathrm{T}\overline{\xi} \Psi(x_1)
      G_\nu(x_2)\Psi(x_3)|0} &=
    + \ii e \frac{(-\mathrm{i})}{p_2^2} \frac{\mathrm{i}}{\fmslash{p}_3}
    \gamma_\nu \gamma^5 \frac{\mathrm{i}}{\fmslash{p}_1 + \fmslash{k}}
    \xi \\
    \mathrm{F.T.} \Braket{0|\mathrm{T}A(x_1)(-\overline{\xi} \gamma_\nu
      \gamma^5 \lambda (x_2))\Psi(x_3)|0}
    &= - \ii e \frac{\mathrm{i}}{p_1^2}
    \frac{\mathrm{i}}{\fmslash{p}_3} \frac{\mathrm{i}}{\fmslash{p}_2 +
      \fmslash{k}} \gamma_\nu \gamma^5 \xi
  \end{align}
  \begin{multline}
    \mathrm{F.T.} \Braket{0|\mathrm{T}A(x_1)G_\nu(x_2) (-\gamma^5
      \fmslash{\partial} B(x_3)\xi)|0} =  \\
    + \ii e \frac{\mathrm{i}}{p_1^2} \frac{-\mathrm{i}}{p_2^2}
    (p_{3,\nu}-p_{1,_\nu}+k_\nu) \frac{\mathrm{i}}{(p_3+k)^2} \gamma^5
    (\fmslash{p}_3 - \fmslash{k}) \xi
  \end{multline}
  \begin{equation}
    \mathrm{F.T.} \Braket{0|\mathrm{T}A(x_1)G_\nu(x_2) (-\gamma^\mu
      (G_\mu A)(x_3) \gamma^5 \xi)|0} = \dfrac{\ii}{p_1^2} \dfrac{-\ii
      \eta_{\mu\nu}}{p_2^2} (- e) \gamma^\mu \gamma^5 \xi
  \end{equation}
\end{subequations}
The last term includes a composite operator insertion coming from the
nonlinearities in the current of the supersymmetry due to the
de~Wit-Freedman transformation (others stem from the elimination of
the auxiliary fields). For the first two processes one has to take
care of the sign of the rightmost fermion propagator whose
calculational direction is opposite to the momentum flow.

Now we have to calculate the 4-point function with current insertion;
there are four contributing diagrams (again we use the trick to
rewrite $\overline{\xi} \mathcal{J}_\mu$ as $\overline{\mathcal{J}_\mu} \xi$,
which brings the propagator of the matter fermion to the rightmost
position in the fermion line):
\begin{equation}
  \parbox{21mm}{%
    \begin{fmfgraph*}(20,20)
      \fmfleft{i1,i2}\fmfright{o1,o2}
      \fmf{dashes}{o2,v2}
      \fmf{photon}{i2,v2}
      \fmf{dbl_dashes}{v1,v2}
      \fmf{dbl_plain}{i1,v1}
      \fmf{plain}{o1,v1}
      \fmfdot{v1,v2}
      \fmfblob{.25w}{i1}
    \end{fmfgraph*}\hfil\\}\qquad + \quad
  \parbox{21mm}{%
    \begin{fmfgraph*}(20,20)
      \fmfleft{i1,i2}\fmfright{o1,o2}
      \fmf{plain}{o2,v2}
      \fmf{photon}{i2,v2}
      \fmf{plain}{v1,v2}
      \fmf{dbl_plain}{i1,v1}
      \fmf{dashes}{o1,v1}
      \fmfdot{v1,v2}
      \fmfblob{.25w}{i1}
    \end{fmfgraph*}\hfil\\}\qquad + \quad
  \parbox{21mm}{%
    \begin{fmfgraph*}(20,20)
      \fmfleft{i1,i2}\fmfright{o1,o2}
      \fmf{plain}{o2,v2}
      \fmf{dashes}{i2,v2}
      \fmf{plain}{v1,v2}
      \fmf{dbl_plain}{i1,v1}
      \fmf{photon}{o1,v1}
      \fmffreeze
      \fmf{photon}{v1,v2}
      \fmfdot{v1,v2}
      \fmfblob{.25w}{i1}
    \end{fmfgraph*}\hfil\\}\qquad + \quad
  \parbox{21mm}{%
    \begin{fmfgraph*}(20,20)
      \fmfleft{i1,i2}\fmfright{o1,o2}
      \fmf{plain}{o2,v}
      \fmf{photon}{i2,v}
      \fmf{dbl_plain}{i1,v}
      \fmf{dashes}{o1,v}
      \fmfdot{v}
      \fmfblob{.25w}{i1}
    \end{fmfgraph*}\hfil\\}\qquad\quad
\end{equation} 
\begin{multline}
  \mathrm{F.T.} \Braket{0|\mathrm{T}\overline{\mathcal{J}_\mu}(y) \xi A(x_1)
    G_\nu(x_2)\Psi(x_3)|0} = \\
  \frac{\mathrm{i}}{p_1^2}\frac{-\mathrm{i}}{p_2^2}
  \dfrac{-\ii}{\fmslash{p}_3} \left( \mathrm{F.T.} \Braket{0|\mathrm{T}
      \overline{\mathcal{J}_\mu} (y)A(x_1)G_\nu(x_2)\Psi(x_3)|0}_{\text{amp.}}
  \right) \xi
\end{multline}
Note the momentum flow for the sign of the fermion propagator. 
\begin{multline}
  \mathrm{F.T.} \Braket{0|\mathrm{T}\overline{\mathcal{J}_\mu}(y)
    A(x_1)G_\nu(x_2)\Psi(x_3)|0}_{\text{amp.}} \xi =
  - \ii e \gamma_\mu \gamma^5 \gamma_\nu \xi \\
  - \ii e \dfrac{\ii}{\fmslash{p}_2 + \fmslash{k}} \left(-\dfrac{1}{2}
  \right) (-\ii p_{2,\alpha}) \gamma_\mu \gamma^5 \lbrack \gamma^\alpha ,
  \gamma_\nu \rbrack
  \xi + \ii e \gamma_\nu \gamma^5 \dfrac{\ii}{\fmslash{p}_1 +
    \fmslash{k}} \gamma_\mu (-\ii \fmslash{p}_1) \xi \\
  + \dfrac{\ii}{(p_3 + k)^2} e \left( p_{1,\nu} - p_{3,\nu} - k_\nu \right)
  \ii \gamma_\mu \gamma^5 \ii \left( \fmslash{p}_3 + \fmslash{k} \right)
  \xi
\end{multline}
with $k_\mu=-(p_1+p_2+p_3)_\mu$ 
\begin{align}
  & \; \dfrac{-\ii}{\fmslash{p}_3} \dfrac{1}{p_1^2 p_2^2} \;
  k^\mu \mathrm{F.T.} \Braket{0|\mathrm{T}
    \overline{\mathcal{J}_\mu}(y) A(x_1)G_\nu(x_2)
    \Psi(x_3)|0}_{\text{amp.}} \xi \notag \\ =
  & \; - \dfrac{\ii}{\fmslash{p}_3} \dfrac{1}{p_1^2 p_2^2} \biggl\{
  + \ii e \left( \fmslash{p}_1 + \fmslash{p}_2 + \fmslash{p}_3 \right)
  \gamma^5 \gamma_\nu + \dfrac{\ii e}{2} \, \dfrac{1}{\fmslash{p}_1 +
    \fmslash{p}_3} \left( \fmslash{p}_1 + \fmslash{p}_2 + \fmslash{p}_3
  \right) \gamma^5 \lbrack \fmslash{p}_2 , \gamma_\nu \rbrack \notag
  \\ & \; \qquad - \ii e \left( p_{1,\nu} - p_{3,\nu} + k_\nu \right)
  \dfrac{1}{(p_3 - k)^2} \left( \fmslash{p}_1 + \fmslash{p}_2 +
    \fmslash{p}_3 \right) \gamma^5 \left( \fmslash{p}_1 + \fmslash{p}_2
  \right) \notag \\ & \; \qquad + \ii e \gamma_\nu \gamma^5
  \dfrac{1}{\fmslash{p}_2 + \fmslash{p}_3} \left( \fmslash{p}_1 +
    \fmslash{p}_2 + \fmslash{p}_3 \right) \fmslash{p}_1 \biggr\} \xi
  \notag \\ = & \;
  - \dfrac{e}{p_1^2 p_2^2} \gamma_\nu \gamma^5 \xi - \dfrac{e}
  {p_1^2 p_2^2} \dfrac{1}{\fmslash{p}_3} \left( \fmslash{p}_1 +
    \fmslash{p}_2 \right) \gamma_\nu \gamma^5 \xi
  \notag \\ & \;
  + \dfrac{e}{2} \dfrac{1}{p_1^2 p_2^2} \dfrac{1}{\fmslash{p}_3}
  \gamma^5 \lbrack \fmslash{p}_2 , \gamma_\nu \rbrack  \xi + \dfrac{
    e}{p_1^2} \dfrac{1}{\fmslash{p}_3} \dfrac{1}{\fmslash{p}_1 +
    \fmslash{p}_3} \gamma_\nu \gamma^5 \xi - \dfrac{e}{p_1^2 p_2^2}
  \dfrac{1}{\fmslash{p}_3} \dfrac{1}{\fmslash{p}_1 + \fmslash{p}_3}
  \fmslash{p}_2 p_{2,\nu} \gamma^5 \xi
  \notag \\ & \;
  + \dfrac{e}{p_1^2 p_2^2} \dfrac{1}{\fmslash{p}_3} \left( 2 p_{1,\nu}
    + p_{2,\nu} \right) \gamma^5 \xi - \dfrac{e}{p_1^2 p_2^2} \left(
    2 p_{1,\nu} + p_{2,\nu} \right) \dfrac{1}{(p_1 + p_2)^2} \gamma^5
  \left( \fmslash{p}_1 + \fmslash{p}_2 \right) \xi \notag
  \\ & \;
  + e \dfrac{1}{p_1^2 p_2^2} \dfrac{1}{\fmslash{p}_3} \gamma_\nu
  \gamma^5 \fmslash{p}_1 \xi + e \dfrac{1}{p_2^2}
  \dfrac{1}{\fmslash{p}_3} \gamma_\nu \gamma^5 \dfrac{1}{\fmslash{p}_2 +
    \fmslash{p}_3} \xi
\end{align}
The first term from the first line, each of the second terms from
the second and third line as well as the second term from the last
line yield the sum of the four contact terms with the SUSY transformed
fields, given by:
\begin{multline}
  e \dfrac{1}{p_2^2} \dfrac{1}{\fmslash{p}_3} \gamma_\nu \gamma^5
  \dfrac{1}{\fmslash{p}_2 + \fmslash{p}_3} \xi + e \dfrac{1}{p_1^2}
  \dfrac{1}{\fmslash{p}_3} \dfrac{1}{\fmslash{p}_1 + \fmslash{p}_3}
  \gamma_\nu \gamma^5 \xi - \dfrac{e}{p_1^2 p_2^2} \gamma_\nu \gamma^5
  \xi \\ - e \dfrac{1}{p_1^2} \dfrac{1}{p_2^2}
  \left( 2 p_{1,\nu} + p_{2,\nu} \right) \dfrac{1}{(p_1 + p_2)^2} \gamma^5
  (\fmslash{p}_1 + \fmslash{p}_2) \xi
\end{multline}
The remaining terms sum up to 
\begin{equation}
  - \dfrac{e}{p_1^2 p_2^2} \dfrac{1}{\fmslash{p}_3} \dfrac{1}{
    \fmslash{p}_1 + \fmslash{p}_3} \fmslash{p}_2 p_{2,\nu} \gamma^5 \xi
\end{equation}
Again the term violating the off-shell validity of the SWI is
proportional to the momentum of the gauge boson as observed for the
second of our 2-point function examples $\Greensfunc{\lbrack Q(\xi) ,
G_\mu(x_1)\lambda(x_2)\rbrack}$, so that it will not survive LSZ
reduction.  

One is tempted to say that it is the external gauge boson's ``fault'' but
this allegation is contradicted by another example not containing any gauge
boson,
\begin{multline}
 k^\mu \mathrm{F.T.} \Greensfunc{\overline{\xi}\mathcal{J}_\mu A(x_1) B(x_2)
 \lambda(x_3)} \\ \stackrel{?}{=} \mathrm{F.T.}
 \Greensfunc{(\overline{\xi} \Psi(x_1)) B(x_2) \lambda(x_3)} \delta^4(x_1-y)
 \qquad\qquad\qquad \\ + \mathrm{F.T.} \Greensfunc{A(x_1) (\ii \overline{\xi} 
 \gamma^5 \Psi(x_2)) \lambda(x_3)} \delta^4(x_2-y) \\ - \dfrac{\ii}{2}
 \mathrm{F.T.} \Greensfunc{A(x_1) B(x_2) \partial_\alpha G_\beta(x_3) \lbrack
 \gamma^\alpha , \gamma^\beta \rbrack \gamma^5 \xi} \delta^4(x_3-y)
\end{multline}
There are three diagrams contributing at tree level to the Green
function with current insertion:
\begin{equation}
  \parbox{21mm}{%
    \begin{fmfgraph*}(20,20)
      \fmfleft{i1,i2}\fmfright{o1,o2}
      \fmf{dbl_dashes}{o2,v2}
      \fmf{dashes}{i2,v2}
      \fmf{photon}{v1,v2}
      \fmf{dbl_plain}{i1,v1}
      \fmf{plain}{o1,v1} \fmffreeze \fmf{photon}{o1,v1}
      \fmfdot{v1,v2}
      \fmfblob{.25w}{i1}
    \end{fmfgraph*}}\qquad + \quad
  \parbox{21mm}{%
    \begin{fmfgraph*}(20,20)
      \fmfleft{i1,i2}\fmfright{o1,o2}
      \fmf{plain}{o2,v2}
      \fmf{dashes}{i2,v2}
      \fmf{plain}{v1,v2}
      \fmf{dbl_plain}{i1,v1}
      \fmf{dbl_dashes}{o1,v1} \fmffreeze
      \fmf{photon}{o2,v2} 
      \fmfdot{v1,v2}
      \fmfblob{.25w}{i1}
    \end{fmfgraph*}}\qquad + \quad
  \parbox{21mm}{%
    \begin{fmfgraph*}(20,20)
      \fmfleft{i1,i2}\fmfright{o1,o2}
      \fmf{plain}{o2,v2}
      \fmf{dbl_dashes}{i2,v2}
      \fmf{plain}{v1,v2}
      \fmf{dbl_plain}{i1,v1}
      \fmf{dashes}{o1,v1} \fmffreeze
      \fmffreeze
      \fmf{photon}{o2,v2} 
     \fmfdot{v1,v2}
      \fmfblob{.25w}{i1}
    \end{fmfgraph*}}
\end{equation} 

\vspace{5mm}

All momenta are understood as incoming ($k_\mu$ is the momentum
influx through the current). With the relation 
\begin{equation}
 \lbrack (\fmslash{p}_1 + \fmslash{p}_2) , (\fmslash{p}_1 -
 \fmslash{p}_2) \rbrack = - 2 \lbrack \fmslash{p}_1 , \fmslash{p}_2
 \rbrack 
\end{equation}
we can calculate the three diagrams (for simplicity we multiply the
relation by a factor $+\ii$)
\begin{multline*}
 -\ii (p_1 + p_2 + p_3)^\mu \left( - \dfrac{\ii e}{p_1^2 p_2^2
   \fmslash{p}_3} \right) \biggl\{ \dfrac{1}{(p_1+p_2)^2} \gamma_\mu
   \lbrack \fmslash{p}_1 , \fmslash{p}_2 \rbrack - 
   \dfrac{1}{\fmslash{p}_1 + \fmslash{p}_3} \gamma_\mu \fmslash{p}_2 +
   \dfrac{1}{\fmslash{p}_2 + \fmslash{p}_3} \gamma_\mu \fmslash{p}_1
   \biggr\} \gamma^5 \xi  \\ 
   = - \dfrac{e}{p_1^2 p_2^2 (p_1+p_2)^2} \lbrack \fmslash{p}_1 ,
   \fmslash{p}_2 \rbrack \gamma^5 \xi - \dfrac{e}{p_1^2 p_2^2
   (p_1+p_2)^2 \fmslash{p}_3} (\fmslash{p}_1 + \fmslash{p}_2) \lbrack
   \fmslash{p}_1 , \fmslash{p}_2 \rbrack \gamma^5 \xi \\ \qquad \quad + 
   \dfrac{e}{p_1^2 p_2^2 \fmslash{p}_3} \fmslash{p}_2 \gamma^5 \xi +
   \dfrac{e}{p_1^2 \fmslash{p}_3} \dfrac{1}{\fmslash{p}_1 +
   \fmslash{p}_3} \gamma^5 \xi - \dfrac{e}{p_1^2 p_2^2 \fmslash{p}_3}
   \fmslash{p}_1 \gamma^5 \xi - \dfrac{e}{p_2^2 \fmslash{p}_3}
   \dfrac{1}{\fmslash{p}_2 + \fmslash{p}_3} \gamma^5 \xi    
\end{multline*}
The first part of the identity 
\begin{equation}
(\fmslash{p}_1 + \fmslash{p}_2) \lbrack \fmslash{p}_1 , \fmslash{p}_2
\rbrack = - (p_1+p_2)^2 (\fmslash{p}_1 - \fmslash{p}_2) +
(\fmslash{p}_1 + \fmslash{p}_2) \left( p_1^2 - p_2^2 \right) ,
\end{equation}
inserted into the second term of the analytical expression for the current
Green function cancels its third and fifth term. It remains:
\begin{multline}
  \label{ward_abl}
  \ii k^\mu \, \mathrm{F.T.} \Greensfunc{\mathcal{J}_\mu A(x_1) B(x_2)
  \lambda(x_3)} = 
  - \dfrac{e}{p_1^2 p_2^2 (p_1+p_2)^2} \lbrack \fmslash{p}_1 ,
   \fmslash{p}_2 \rbrack \gamma^5 \xi \\ - \dfrac{e}{p_1^2 p_2^2 (p_1+p_2)^2
   \fmslash{p}_3} (\fmslash{p}_1 + \fmslash{p}_2) (p_1^2 - p_2^2) \gamma^5
   \xi + \dfrac{e}{p_1^2 \fmslash{p}_3} \dfrac{1}{\fmslash{p}_1 +
   \fmslash{p}_3} \gamma^5 \xi - \dfrac{e}{p_2^2 \fmslash{p}_3}
   \dfrac{1}{\fmslash{p}_2 + \fmslash{p}_3} \gamma^5 \xi     
\end{multline}

The three contact terms yield (again multiplied by $+\ii$)
\begin{equation}
  - \dfrac{e}{p_2^2 \fmslash{p}_3} \dfrac{1}{\fmslash{p}_2 + \fmslash{p}_3}
    \gamma^5 \xi + \dfrac{e}{p_1^2 \fmslash{p}_3} \dfrac{1}{\fmslash{p}_1 +
    \fmslash{p}_3} \gamma^5 \xi - \dfrac{e}{p_1^2 p_2^2 (p_1+p_2)^2} \lbrack
    \fmslash{p}_1 , \fmslash{p}_2 \rbrack \gamma^5 \xi .
\end{equation}
We see that the difference of the left and right hand sides of the SWI
is an additional term from the Green function with current insertion,
the second one on the right hand side of (\ref{ward_abl}), proportional to the
difference of the squared momenta of the scalar and pseudoscalar particles
which vanishes on-shell but need not to do so off-shell. 

\vspace{5mm}

As a final statement we can say that everything is fine when treating SWI
calculated with the current as on-shell identities. But there are two
obstacles: Ward-Takahashi identities for on-shell amplitudes do not provide
tests as stringent as do off-shell amplitudes since they only check the
current's couplings to external lines. Hence it would be more satisfying
to have also off-shell checks for the consistency at hand. That is
the practical point; but more disturbing is the theoretical aspect: Why are
the SWI not fulfilled off-shell? The deeper cause is that supersymmetry is no
longer a symmetry of the $S$-matrix for SUSY gauge theories. As will be
explained in the next part, the gauge-fixing procedure required for the
quantization of gauge theories is not compatible with SUSY, since it breaks
the invariance of the action under supersymmetry. This restricts
the validity of the SWI built with the current from the whole Hilbert space to
its physical part. From this it is clear, that in the case of supersymmetric
gauge theories the SWI presented in this part are only fulfilled for physical
on-shell states. The next part will bring a way to get rid of that obstacle.

%%% Local Variables: 
%%% mode: latex
%%% TeX-master: "diss"
%%% End: 

%% file: chap10.tex
\section*{}

The problem of supersymmetric gauge theories that Ward identities of
the supersymmetry are not valid off-shell, arises from the fact that the
supercharge does no longer commute with the $S$-operator, or in other
words, it is no longer constant in time, cf.~\cite{Sibold/etal:2000:brst}.   
As derived therein, the difference of the action of the SUSY charge
operator on the space of asymptotic $in$ and asymptotic $out$ states
is the BRST transformation of the derivative of the effective action
with respect to the ghost of the supersymmetry:
\begin{equation}
        \label{eq:susynonconserved}
        Q_{\text{out}} - Q_{\text{in}} = \ii \left[ Q_{\text{BRST}} ,
        \dfrac{\partial \Gamma_{\text{eff}}}{\partial
        \overline{\epsilon}} \right] \tag{III.1}
\end{equation}
This can be rewritten, in the language of
\cite{Piguet/Sibold:1986:book}, as a commutator of the SUSY charge with
the $S$-operator
\begin{equation}
        \lbrack Q_{\text{in}} , S \rbrack = -\ii \left[ Q_{\text{BRST}}
        , \dfrac{\partial \Gamma_{\text{eff}}}{\partial
        \overline{\epsilon}} \circ S \right] \quad , \tag{III.2}
\end{equation} 
where the symbol $\circ$ means that the derivative of the effective action has
to be understood as an operator insertion on the right hand side. The 
right hand side vanishes between physical states, so the SUSY charge --
if not conserved on Hilbert space -- is a conserved symmetry
operator on the cohomology of the BRST charge, the physical Hilbert
space.  

There are some remarks in order: Following the pioneering idea of
Peter L.~White \cite{White:1992:BRST} we enlarge the BRST formalism
not only to include supersymmetry transformations but also 
translations. As long as supergravity is not considered,
supersymmetry remains a global symmetry and hence the ghosts of
supersymmetry being {\em commuting} spinors are constants, as well as the
translation ghosts. This allows a filtration, a power series expansion of
functionals and also of Slavnov-Taylor identities with respect to the constant
ghosts. Since they are constant, we were able to take an ordinary derivative
in (\ref{eq:susynonconserved}) instead of a functional one. Using the
BRST formalism, the nonlinear representation of supersymmetry causes no
problems for renormalization any longer, but we are not interested in that
topic here. The crucial point being responsible for the
nonconservation of the SUSY charge, eq.~(\ref{eq:susynonconserved}),
is that gauge fixing, necessary for constructing a well-defined perturbation
theory for quantized gauge theories, does not take place in a SUSY invariant
manner and hence breaks supersymmetry -- fortunately only in the unphysical
sector of Hilbert space. Another noteworthy matter is the fact that the
enlarged BRST algebra initially closes only on-shell, which can be remedied by 
the Batalin-Vilkovisky formalism \cite{Weinberg:QFTv2:Text},
introducing quadratic terms for the sources of the non-linear parts of
the BRST transformations (those for the fermions), sometimes called
antifields. However, in lowest order perturbation theory we do not have to
take care of that subtlety. 

The important thing for us is that with the generalization of the BRST
formalism we have a possibility at hand to calculate off-shell
identities for supersymmetry -- the Slavnov-Taylor identities (STI) of
the generalized BRST algebra in lowest order perturbation theory. The
prize to pay for this is the proliferation of several kinds of ghosts
and BRST vertices. The details of this formalism and its application
to supersymmetric gauge theories will be the content of this
part. Therein we obtain strong insights into the structure of
supersymmetric STI. In particular, we will see that in the Abelian
case the Faddeev-Popov ghosts do no longer decouple from matter when
considering SUSY STI. An example for the non-Abelian case reveals the
details of the cancellations between the gauge and the SUSY parts of
the BRST transformations and shows that almost all ingredients for
a non-Abelian model are necessary to fulfill the STI there. 

%%%%%%%%%%%%%%%%%%%%%%%%%%%%%%%%%%%%%%%%%%%%%%%%%%%%%%%%%%%%%%%%%%%% 

\chapter{BRST formalism and SUSY transformations}

\section{Definitions of the ghosts}

In this section we try to solve the problem of the ghosts' properties 
concerning reality and statistics. We take account of the
existence of not only the ``gauge ghosts'' (Faddeev-Popov ghosts) but also of
ghosts for translations and SUSY transformations. This is necessary to achieve
closure of the algebra, as well as nilpotency for the BRST charge. We will
discuss this in detail in the sequel.   

Let us consider a pure gauge transformation with real gauge (transformation)
parameter $\theta^{a\:*}=\theta^a$, with $a$ being the index of the gauge
group. The ghost of a gauge symmetry can be derived by splitting a Grassmann
odd, constant parameter $\lambda$ from the gauge parameter; from the parameter
being, of course, itself spacetime dependent, remains a Grassmann odd
spacetime dependent field, the {\em (Faddeev-Popov) ghost}. As it is
not consistent to choose ghost and antighost as Hermitean adjoints of each
other the obvious alternative is to consider both as real, i.e.~Hermitean
fields. The ghost is an anticommuting scalar field which (as unphysical degree
of freedom) violates the spin-statistics theorem; this would be the case for
all ghosts. From the Hermiticity of the ghost it follows that the parameter
$\lambda$ is imaginary:
\begin{equation}
\mathbb{R} \ni \theta^a = \lambda c^a \qquad \left(\lambda c^a\right)^*
= c^a \lambda^* = - \lambda^* c^a \; \Rightarrow \; \boxed{\lambda^* = -
\lambda} 
\end{equation}

We have to proceed in the same manner for the SUSY transformation
parameter (at first, we use the two component notation) and establish
SUSY ghosts: transformation parameter $\xi^\alpha,
\bar{\xi}_{\dot{\alpha}} \longrightarrow\;$ SUSY ghosts
$\epsilon^\alpha, \bar{\epsilon}_{\dot{\alpha}}$.  
If we define 
\begin{equation}
\xi^\alpha = \lambda \epsilon^\alpha ,
\end{equation}
use the property $\left( \xi^\alpha \right)^* =
\bar{\xi}^{\dot{\alpha}}$ and require the same relation to hold
for the SUSY ghosts $\left(\epsilon^\alpha\right)^* =
\bar{\epsilon}^{\dot{\alpha}}$ (this is necessary for consistency as
there is an identical relation between the corresponding generators
$Q^\alpha$ and $\bar{Q}^{\dot{\alpha}}$ of their SUSY transformations), we
get the following identity:
\begin{equation}
(\xi^\alpha)^* = \left( \lambda \epsilon^\alpha \right)^* = \lambda^*
(\epsilon^\alpha)^* = - \lambda \bar{\epsilon}^{\dot{\alpha}}
\stackrel{!}{=} \bar{\xi}^{\dot{\alpha}},
\end{equation}
so altogether
\begin{equation}
\xi^\alpha = \lambda \epsilon^\alpha, \qquad\qquad\qquad
\bar{\epsilon}_{\dot{\alpha}} = - \lambda
\bar{\epsilon}_{\dot{\alpha}} .
\end{equation}
Introducing the bispinor notation now, 
\begin{equation}
\xi \equiv \begin{pmatrix} \xi_\alpha \\ \bar{\xi}^{\dot{\alpha}}
\end{pmatrix}, \qquad \qquad \epsilon \equiv \begin{pmatrix}
\epsilon_\alpha \\ \bar{\epsilon}^{\dot{\alpha}}
\end{pmatrix}, 
\end{equation}
we get the final result:
\begin{equation}
\boxed{\xi = - \lambda \gamma^5 \epsilon}
\end{equation}

For deriving an analogous relation for the translation ghosts we
note, that in general an infinitesimal translation of a function has
the form 
\begin{equation}
        \delta_a f(x) = a^\mu \partial_\mu f(x) .
\end{equation}
\cite{Sibold/etal:2000:brst} and \cite{Hollik/etal:1999:susyrenorm}
adopt the following conventions 
\begin{equation}
        a^\mu = \ii \lambda \omega^\mu
\end{equation}
for the connection between transformation parameter and translation ghost.
The translation (of course only as a global symmetry here) is a
bosonic symmetry like (ordinary) gauge symmetry, so the translation
ghost $\omega^\mu$ is a Grassmann odd vector. From the reality of the
transformation parameter $a^\mu$ we conclude with the help of 
\begin{equation}
\Bbb{R}^4 \ni a^\mu \; \Rightarrow \; \left(\ii \lambda \omega^\mu
\right)^* = - \ii \omega^{\mu\:*} \lambda^* = + \ii \lambda^*
\omega^{\mu\:*} = - \ii \lambda \omega^{\mu\:*} \stackrel{!}{=} \ii
\lambda \omega^\mu 
\end{equation}
\begin{equation}
 \Longrightarrow \boxed{\omega^{\mu\:*} = - \omega^\mu}
\end{equation}

We summarize the properties of all ghosts in the following table
($d_s$ is the unspecified dimension of the BRST charge). 

\vspace{2mm}

{\renewcommand{\arraystretch}{1.5}   
\begin{tabular}{|c|c|c|c|c|}\hline
  Ghost & Dim. & Grassmann P. & Charge & Ghost Number \\
  \hline\hline
  $c$          & $s$             & $1$ & $0$ & $+1$ \\\hline
  $\bar{c}$    & $2-d_s$           & $1$ & $0$ & $-1$ \\\hline
  $\epsilon$   & $d_s-\frac{1}{2}$ & $0$ & $0$ & $+1$ \\\hline
  $\omega^\mu$ & $d_s - 1$         & $1$ & $0$ & $+1$ \\\hline
\end{tabular}}

%%%%%%%%%%%%%%%%%%%%%%%%%%%%%%%%%%%%%%%%%%%%%%%%%%%%%%%%%%%%%%%%%%%%%%%%

\section{BRST symmetry in our Abelian toy model}\label{sec:abeliantoy}

\subsection{The model}

To illustrate the BRST formalism for supersymmetric gauge theories in
detail, we use the Abelian toy model invented in the last part in the
context of the supersymmetric current. The Lagrangean density, the
field content, the propagators, vertices, equations of motion, as well
as the SUSY transformations can be found in the appendix. 

%%%%%%%%%%%%%%%%%%%%%%%%%%%%%%%%%%%%%%%%%%%%%%%%%%%%%%%%%%%%%%%%%%%%%%%%%

\subsection{BRST transformations}

With the above given definitions for the ghosts and the summary of
transformations from the appendix we can immediately write down the
BRST transformations for our Abelian toy model 
\begin{subequations}
  \label{brsttransform}
  \begin{align}
    s A(x) &=\; - e c(x) B(x) - \overline{\epsilon} \gamma^5 \Psi (x) - \ii
    \omega^\nu \partial_\nu A(x)  \\
    s B(x) &=\; + e c(x) A(x) - \ii \overline{\epsilon} \Psi(x) - \ii
    \omega^\nu \partial_\nu B(x) \\
    s \Psi(x) &=\; - \ii e c(x) \gamma^5 \Psi(x) + \ii (\fmslash{\partial} -
    \ii e \fmslash{G} \gamma^5) (A(x) \gamma^5 + \ii B(x)) \epsilon - \ii
    \omega^\nu \partial_\nu \Psi(x) \\
    s \overline{\Psi}(x) &=\; \ii e \overline{\Psi}(x) \gamma^5 c(x) - \ii
    \overline{\epsilon} (\fmslash{\partial} + \ii e \gamma^5 \fmslash{G})
    (A(x)\gamma^5 - \ii B(x)) - \ii \omega^\nu\partial_\nu \overline{\Psi}(x)
    \\ 
    s \lambda(x) &=\; \dfrac{\ii}{2} F_{\alpha\beta} (x) \gamma^\alpha
    \gamma^\beta \epsilon + \dfrac{e}{2} \left( A^2 (x) + B^2(x) \right)
    \gamma^5 \epsilon - \ii \omega^\nu \partial_\nu \lambda(x) \\
    s \overline{\lambda} (x) &=\; - \dfrac{\ii}{2} \overline{\epsilon}
    F_{\alpha\beta} (x) \gamma^\alpha \gamma^\beta + \dfrac{e}{2}
    \overline{\epsilon} \gamma^5 \left( A^2 (x) + B^2 (x) \right) - \ii
    \omega^\nu \partial_\nu \overline{\lambda} (x) \\
    s G_\mu (x) &=\; \partial_\mu c(x) - \overline{\epsilon} \gamma_\mu
    \lambda(x) - \ii \omega^\nu \partial_\nu G_\mu(x) \\
    s c(x) &=\; \ii \overline{\epsilon} \gamma^\mu \epsilon G_\mu (x) - \ii
    \omega^\nu \partial_\nu c(x) \\
    s \overline{c}(x) &=\; \ii \tilde{B}(x) - \ii \omega^\nu \partial_\nu
    \overline{c}(x) \\
    s \tilde{B}(x) &=\; \overline{\epsilon} \gamma^\mu \epsilon \partial_\mu
    \overline{c}(x) - \ii \omega^\nu \partial_\nu \tilde{B}(x)\\ 
    s \epsilon &=\; 0 \\
    s \omega^\mu &=\; \overline{\epsilon} \gamma^\mu \epsilon
  \end{align}
\end{subequations}
We have denoted the Nakanishi-Lautrup field by $\tilde{B}$ to avoid confusion
with the pseudoscalar field. To derive the identities 
for adjoint fields we use that for bosonic fields $B$ and for
fermionic fields $F$ the relations 
\begin{equation}
  s B^\dagger = (s B)^\dagger, \qquad\qquad s F^\dagger = - (s F)^\dagger.
\end{equation}
hold. In the case of the adjoint spinors we have to take care of the
commutation properties of the several fields. The first part of
(\ref{brsttransform}) -- if present -- stems from the gauge
transformation, the second from the SUSY transformation, and the last
obviously from translation. The somewhat strange looking and
unmotivated transformations of the several ghosts are necessary for
the closure of the algebra (cf.~\cite{White:1992:BRST},
\cite{Sibold/etal:2000:brst}) and can be understood from examination
of the super-Poincar\'e algebra. 

It is not hard to check that the BRST transformation is nilpotent
except for the transformation of the fermion fields, where the square
of the BRST operator gives the equation of motion for them:
\begin{align}
  s^2 A = s^2 B = s^2 G_\mu = s^2 c = s^2 \overline{c} = s^2 \tilde{B} = s^2
  \epsilon = s^2 \omega_\mu = 0, \notag \\
  s^2 \Psi = - \dfrac{1}{2} (\overline{\epsilon} \gamma^\mu \epsilon)
  \gamma_\mu \dfrac{\delta\Gamma}{\delta\overline{\Psi}}, \qquad \qquad
  s^2\lambda =  - \dfrac{1}{4} (\overline{\epsilon} \gamma^\mu \epsilon)
  \gamma_\mu   \dfrac{\delta \Gamma}{\delta \overline{\lambda}} 
\end{align}
For the derivation of the identity for the matter fermion one has to
multiply use the Fierz identities. As the calculations are a little
bit intricate, we show one example for the matter fermion:
\begin{align}
s^2 \Psi =&\; - \ii e (sc) \gamma^5 \Psi + \ii e c \gamma^5 (s \Psi) +
\ii (\fmslash{\partial} - \ii e \fmslash{G} \gamma^5) \left( (s A)
\gamma^5 + \ii (s B) \right) \epsilon \notag\\ &\; + e\gamma^\mu
(sG_\mu) \gamma^5 \left( A \gamma^5 + \ii B \right) \epsilon - \ii (s
\omega^\nu) \partial_\nu \Psi + \ii \omega^\nu \partial_\nu (s \Psi)
\notag \\ =&\;e (\overline{\epsilon} \gamma^\mu\epsilon)G_\mu\gamma^5
\Psi - \underline{e \omega^\nu (\partial_\nu c) \gamma^5 \Psi} +
\underline{\underline{e^2 c^2 \Psi}} - \underline{e c \gamma^5
(\fmslash{\partial} - \ii e \fmslash{G} \gamma^5) \left( A \gamma^5 +
\ii B \right) \epsilon} \notag\\ &\;+ \underline{e c \gamma^5 \omega^\nu
\partial_\nu \Psi} + \underline{\ii(\fmslash{\partial} - \ii e
\fmslash{G} \gamma^5) \ii e c \gamma^5 \left( A \gamma^5 + \ii B
\right) \epsilon} \notag\\&\; + \underline{\ii (\fmslash{\partial} -
\ii e \fmslash{G} \gamma^5) (-\ii) \omega^\nu \partial_\nu \left( A
\gamma^5 + \ii B\right) \epsilon} + \ii (\fmslash{\partial} - \ii e
\fmslash{G} \gamma^5) \left[ (\overline{\epsilon} \Psi) -
(\overline{\epsilon} \gamma^5 \Psi) \gamma^5 \right] \epsilon \notag\\
&\; + \underline{e  (\fmslash{\partial} c) \gamma^5 \left( A \gamma^5
+ \ii B \right) \epsilon} - e \gamma^\mu (\overline{\epsilon}
\gamma_\mu \lambda) \gamma^5 \left( A \gamma^5 + \ii B \right)
\epsilon \notag \\&\; - \underline{\ii e \omega^\nu (\partial_\nu
\fmslash{G}) \gamma^5 \left( A \gamma^5 + \ii B \right) \epsilon} - \ii
(\overline{\epsilon} \gamma^\nu \epsilon) \partial_\nu \Psi +
\underline{e \omega^\nu \partial_\nu (c \gamma^5 \Psi)} \notag \\ &\;-
\underline{\omega^\nu \partial_\nu (\fmslash{\partial} - \ii e
\fmslash{G} \gamma^5) \left( A \gamma^5 + \ii B \right)\epsilon} +
\underline{\underline{ \omega^\mu \omega^\nu  \partial_\mu
\partial_\nu \Psi}} + \underline{\ii^2 e (\fmslash{\partial} c)
\gamma^5 \left( A \gamma^5 + \ii B \right) \epsilon}    
\end{align} 
Underlined terms cancel each other, while doubly underlined ones
vanish identically. Sorting and ordering the remaining terms we are
left with:
\begin{multline}
\label{nilpotenz}
s^2 \Psi =\; - \ii (\overline{\epsilon} \gamma^\mu \epsilon) \left(
\partial_\mu \Psi + \ii e G_\mu \gamma^5 \Psi \right) + \ii
(\fmslash{\partial} - \ii e \fmslash{G} \gamma^5) \left[
(\overline{\epsilon} \Psi) - (\overline{\epsilon} \gamma^5 \Psi)
\gamma^5 \right] \epsilon \\ - e \gamma^\mu (\overline{\epsilon}
\gamma_\mu \lambda) \gamma^5 \left( A \gamma^5 + \ii B \right)
\epsilon
\end{multline}
We use the following Fierz identities
\begin{subequations}
\begin{align}
(\overline{\epsilon} \Psi) \epsilon =&\; \dfrac{1}{8}
(\overline{\epsilon} \sigma_{\mu\nu} \epsilon) \sigma^{\mu\nu} \Psi +
\dfrac{1}{4} (\overline{\epsilon} \gamma^\mu \epsilon) \gamma_\mu \Psi \\
(\overline{\epsilon} \gamma^5 \Psi) \gamma^5 \Psi =&\; \dfrac{1}{8}
(\overline{\epsilon} \sigma_{\mu\nu} \epsilon) \sigma^{\mu\nu} \Psi -
\dfrac{1}{4} (\overline{\epsilon} \gamma^\mu \epsilon) \gamma_\mu \Psi .
\end{align}
\end{subequations}
For {\em commuting} Majorana spinors the scalar, pseudoscalar and
pseudovectorial combinations vanish due to the symmetry properties of
Majorana bilinears $\overline{\epsilon}\Gamma\epsilon$. We get for the second
term of (\ref{nilpotenz}):  
\begin{multline}
\ii (\fmslash{\partial} - \ii e \fmslash{G} \gamma^5) \left[
(\overline{\epsilon} \Psi) - (\overline{\epsilon} \gamma^5 \Psi)
\gamma^5 \right] \epsilon \\ = - \dfrac{1}{2} (\overline{\epsilon}
\gamma^\mu \epsilon) \ii \gamma_\mu (\fmslash{\partial} + \ii e
\fmslash{G} \gamma^5) \Psi + \ii (\overline{\epsilon} \gamma^\mu
\epsilon) \left( \partial_\mu + \ii e G_\mu \gamma^5 \Psi \right) 
\end{multline}
Here we have used the Dirac algebra; the second term cancels the first
from (\ref{nilpotenz}). When Fierzing the last term, only the vector
combination is nonvanishing and we can directly write down:
\begin{equation}
- e \gamma^\mu (\overline{\epsilon}
\gamma_\mu \lambda) \gamma^5 \left( A \gamma^5 + \ii B \right)
\epsilon = \dfrac{1}{2} (\overline{\epsilon} \gamma^\mu \epsilon) e
\gamma_\mu \left( A - \ii B \gamma^5 \right) \lambda 
\end{equation}
Summing up all terms gives the desired result.

%%%%%%%%%%%%%%%%%%%%%%%%%%%%%%%%%%%%%%%%%%%%%%%%%%%%%%%%%%%%%%%%%%%%%%%

\section{Gauge fixing and kinetic ghost term}

For the quantization of a gauge theory we have to carry out a gauge
fixing in the usual way.
\begin{equation}
  S_{\text{GF+FP}} = - \ii \int d^4 x \; s (\overline{c} F) = - \ii \int d^4 x
  \left[ (s \overline{c} ) F - \overline{c} (s F) \right]
\end{equation}
$F$ is the gauge fixing function:
\begin{equation}
  F = \partial_\mu G^\mu + \dfrac{\xi}{2} \tilde{B} .
\end{equation}
By $\xi$ we denote the gauge parameter, not to be confused with the
(anticommuting) SUSY transformation parameter. After applying the BRST
transformation the terms containing the translation ghost $\omega^\mu$
cancel out since the gauge fixing function is translation
invariant. It remains:
\begin{multline}
  \boxed{ S_{\text{GF+FP}} = \int d^4 x \; \left\{ \tilde{B} \partial_\mu
  G^\mu +  \dfrac{\xi}{2} \tilde{B}^2 + \ii \overline{c} \Box c - \ii
  \overline{c} (\overline{\epsilon} \fmslash{\partial} \lambda) + \ii
  \dfrac{\xi}{2} \overline{c} (\overline{\epsilon} \gamma^\mu \epsilon)
  \partial_\mu \overline{c} \right\} }
\end{multline}
Integrating out the Nakanishi-Lautrup field gives:
\begin{equation}
  S_{\text{GF+FP}} = \int d^4 x \; \left\{ - \dfrac{1}{2 \xi} (\partial_\mu
  G^\mu)^2 +  \ii \overline{c} \Box c - \ii
  \overline{c} (\overline{\epsilon} \fmslash{\partial} \lambda) + \ii
  \dfrac{\xi}{2} \overline{c} (\overline{\epsilon} \gamma^\mu \epsilon)
  \partial_\mu \overline{c} \right\} 
\end{equation}
The contributions with the derivative of the gauge field yield,
together with the terms from the kinetic part, the gauge boson
propagator in $R_\xi$ gauge:
\begin{align}
  \parbox{21mm}{%
    \begin{fmfgraph*}(20,5)
      \fmfleft{i}\fmfright{o}
      \fmflabel{$G_\mu(-p)$}{i}
      \fmflabel{$G_\nu(p)$}{o}           
      \fmf{photon}{i,o}
      \fmfdot{i,o}
    \end{fmfgraph*}}\qquad\quad
  &= \frac{- \mathrm{i}}{p^2+\mathrm{i}\epsilon} \left( \eta_{\mu\nu} -
  (1-\xi) \dfrac{p_\mu p_\nu}{p^2} \right)
\end{align}
Furthermore we get the ghost propagator:
\begin{align}
  \parbox{21mm}{%
    \begin{fmfgraph*}(20,5)
      \fmfleft{i}\fmfright{o}
      \fmflabel{$c(-p)$}{i}
      \fmflabel{$\overline{c}(p)$}{o}           
      \fmf{dots_arrow}{o,i}
      \fmfdot{i,o}
    \end{fmfgraph*}}\qquad\quad
  &= \dfrac{-1}{p^2 + \ii \epsilon}
\end{align}
But there also arise two new vertices containing the gauge antighosts
as well as the SUSY ghosts. Albeit being an Abelian model the ghosts
do not decouple from the (matter) fields. To say it sloppily, since
supersymmetry and gauge symmetry are not commuting subalgebras in the
de~Wit--Freedman description, the model becomes formally
non-Abelian. The two vertices are
\begin{subequations}
  \begin{align}
    \parbox{21mm}{%
      \hfil\\\hfil\\
      \begin{fmfgraph*}(20,15)
        \fmfleft{p1}\fmfright{p2,p3}
        \fmflabel{$\overline{c}(-p)$}{p1}
        \fmflabel{$\lambda(p)$}{p2}
        \fmflabel{$\overline{\epsilon}$}{p3}
        \fmf{dots_arrow}{v,p1}
        \fmf{susy_ghost}{p3,v}
        \fmf{plain}{p2,v}  \fmffreeze
        \fmf{photon}{p2,v}
        \fmfdot{v}
        \fmfv{decor.shape=square,decor.filled=full,decor.size=2mm}{p3} 
      \end{fmfgraph*}\\
      \hfil}\qquad\quad
    &= - \ii \fmslash{p} \\ & \notag \\ 
       \parbox{21mm}{%
         \hfil\\\hfil\\
         \begin{fmfgraph*}(20,20)
           \fmfleft{p1,p2}\fmfright{p3,p4}
           \fmflabel{$\epsilon$}{p1}
           \fmflabel{$\overline{c}(-p)$}{p2}
           \fmflabel{$\overline{\epsilon}$}{p3}
           \fmflabel{$\overline{c}(p)$}{p4}
           \fmf{dots_arrow}{v,p2}
           \fmf{dots_arrow}{v,p4}
           \fmf{susy_ghost}{p3,v}
           \fmf{susy_ghost}{p1,v}
           \fmfv{decor.shape=square,decor.filled=full,decor.size=2mm}{p1,p3}
           \fmfdot{v}
         \end{fmfgraph*}\\
         \hfil}\qquad\quad
           &= \xi \fmslash{p}
  \end{align} 
\end{subequations}   
For the four-ghost vertex there is a symmetry factor two for the gauge
antighost, but no symmetry factor for the SUSY ghost. This is because
the SUSY ghost is a constant, so what we do is merely a Taylor
expansion in the power of the SUSY ghosts where the factorials cancel
the symmetry factors. Only the gauge ghosts are propagating fields,
all other ghosts are simply constant insertions. The black box in the
Feynman diagrams should indicate that for the constance of the ghost
the line ends there. 

%%%%%%%%%%%%%%%%%%%%%%%%%%%%%%%%%%%%%%%%%%%%%%%%%%%%%%%%%%%%%%%%%%%%%%%%

\section[Slavnov-Taylor identities in the Abelian toy model]{Slavnov-Taylor identities in the Abelian toy \\ model}

In this section we want to review the Green functions from the last
section of the second part in the light of the BRST formalism and the
Slavnov-Taylor identities. There is no one-to-one correspondence
between the Ward identities of the last part and the Slavnov-Taylor
identities; but one can replace the current insertion from the SWI by
the BRST charge acting on the same string of fields as in the SWI. So
for the first example where the SWI had not been valid off-shell we
write down the STI~\footnote{All calculations in this chapter are done
in the Feynman gauge, $\xi\equiv 1$, but as is easily seen all
calculations go through analogously for a general $R_\xi$ gauge and
are therefore independent from the choice of gauge.}
\begin{equation}
\Greensfunc{ \left\{ Q_{\text{BRST}} , G_\nu(x_1) \lambda(x_2) \right\} }
= 0 \; \; \text{wegen} \; \; Q_{\text{BRST}} \Ket{0} = 0 .
\end{equation}
There are three (!) contributing diagrams:
\begin{equation}
       \parbox{21mm}{%
         \begin{fmfgraph*}(20,15)
           \fmftop{t1}\fmfbottom{b1,b2}
           \fmflabel{$x$}{b1}
           \fmflabel{$y$}{b2}
           \fmf{susy_ghost}{t1,b1}
           \fmfv{decor.shape=square,decor.filled=full,decor.size=2mm}{t1} 
           \fmfv{decor.shape=square,decor.filled=empty,decor.size=1.8mm}{b1} 
           \fmf{plain,label=$k \;\; \rightarrow$}{b1,b2}
         \end{fmfgraph*}
         \hfil}  \qquad + \qquad 
       \parbox{21mm}{%
         \begin{fmfgraph*}(20,15)
           \fmftop{t1}\fmfbottom{b1,b2}
           \fmflabel{$x$}{b1}
           \fmflabel{$y$}{b2}
           \fmf{photon,label=$k \;\; \rightarrow$}{b1,b2}
           \fmf{susy_ghost}{t1,b2}
           \fmfv{decor.shape=square,decor.filled=full,decor.size=2mm}{t1} 
           \fmfv{decor.shape=square,decor.filled=empty,decor.size=1.8mm}{b2} 
         \end{fmfgraph*}
         \hfil}  \qquad + \qquad 
       \parbox{21mm}{%
         \begin{fmfgraph*}(20,15)
           \fmftop{t1}\fmfbottom{b1,b2}
           \fmflabel{$x$}{b1}
           \fmflabel{$y$}{b2}
           \fmf{dots_arrow}{v,b1} 
           \fmf{plain,label=$k \searrow$,l.side=left}{v,b2}
           \fmf{susy_ghost}{t1,v}
           \fmffreeze
           \fmf{photon}{v,b2}
           \fmfv{decor.shape=square,decor.filled=full,decor.size=2mm}{t1} 
           \fmfv{decor.shape=square,decor.filled=empty,decor.size=1.8mm}{b1} 
           \fmfdot{v}
         \end{fmfgraph*}
         \hfil} 
\end{equation}

\vspace{5mm}

For the BRST vertices we use the notations of
\cite{Kugo:Eichtheorie}. The first diagram yields
\begin{equation}
  - \Greensfunc{(\overline{\epsilon} \gamma_\nu \lambda(x)) \lambda(y)} = +
    \Greensfunc{\lambda(y) (\overline{\lambda}(x) \gamma_\nu \epsilon)} =
    \dfrac{\ii}{\fmslash{k}} \gamma_\nu \epsilon.
\end{equation}
Several signs have to be accounted for: $(\overline{\epsilon} \gamma_\nu
\lambda) = + (\overline{\lambda} \gamma_\nu \epsilon)$, as $\epsilon$
is commuting; for the same reason: $(\overline{\lambda} \gamma_\nu
\epsilon) \lambda = - \lambda (\overline{\lambda} \gamma_\nu \epsilon) $. 

From the second diagram we get
\begin{equation}
  \dfrac{\ii}{2} \Greensfunc{G_\nu(x) F_{\alpha\beta}(y)
  \gamma^\alpha\gamma^\beta \epsilon} = \dfrac{\ii}{2}
  \dfrac{-\ii\eta_{\nu\beta}}{k^2} (-\ii k_\alpha) \lbrack \gamma^\alpha ,
  \gamma^\beta \rbrack \epsilon = - \dfrac{\ii}{2} \dfrac{1}{k^2} \lbrack
  \fmslash{k}, \gamma_\nu \rbrack \epsilon .
\end{equation}

After insertion of an interaction operator the third diagram
contributes
\begin{multline}
  \Greensfunc{\partial_\nu^x c(x) \lambda(y)} = - \Greensfunc{\partial_\nu^x
  c(x) \overline{c}(y) \lambda(y) (\overline{\lambda}(z)
  \stackrel{\leftarrow}{\fmslash{\partial}_z} \epsilon)} \\ = -
  \dfrac{-1}{k^2} (\ii k_\nu) \dfrac{\ii}{\fmslash{k}} (\ii \fmslash{k})
  \epsilon = - \dfrac{\ii k_\nu}{k^2} .
\end{multline}
Adding the three terms yields
\begin{equation}
 \dfrac{\ii}{\fmslash{k}} \gamma_\nu \epsilon - \dfrac{\ii}{2} \dfrac{1}{k^2}
  \lbrack \fmslash{k}, \gamma_\nu \rbrack \epsilon - \dfrac{\ii k_\nu}{k^2} = 
  \dfrac{\ii}{\fmslash{k}} \gamma_\nu \epsilon - \dfrac{\ii}{\fmslash{k}}
  \gamma_\nu \epsilon + \dfrac{\ii k_\nu}{k^2} - \dfrac{\ii k_\nu}{k^2} = 0.
\end{equation}

The STI is in contrast to the SWI fulfilled {\em off-shell} (and
therefore automatically on-shell). Note the crucial importance
of the term where the two ghosts couple to the gaugino. 

\vspace{2mm}

We now turn to the more complex examples. At first, 
\begin{align}
  0 = &\; \Greensfunc{ \left\{ Q_{\text{BRST}} , A(x_1) G_\nu(x_2) \Psi(x_3)
  \right\}} \notag \\
   = & \; - e \Greensfunc{c(x_1)B(x_1) G_\nu(x_2)
  \Psi (x_3)} + \Greensfunc{(\overline{\Psi}(x_1) \gamma^5 \epsilon)
  G_\nu(x_2) \Psi(x_3)} \notag \\ &\; + \Greensfunc{A(x_1)
  (\partial_\nu c(x_2)) \Psi(x_3)} - \Greensfunc{A(x_1)
  (\overline{\lambda}(x_2) \gamma_\nu \epsilon) \Psi(x_3)} \notag \\
  &\; - \ii e \Greensfunc{A(x_1) G_\nu(x_2) c(x_3) \gamma^5 \Psi(x_3)}
  \notag \\ &\; + \ii \Greensfunc{A(x_1) G_\nu(x_2) \left(
  \fmslash{\partial} - \ii e \fmslash{G} (x_3) \gamma^5 \right) \left(
  A(x_3) \gamma^5 + \ii B(x_3) \right) \epsilon} \label{komplitrafo1}
\end{align}
The first and the penultimate Green function do not
contribute. Graphically the second Green function in
(\ref{komplitrafo1}) yields
\begin{equation}
       \parbox{21mm}{%
         \begin{fmfgraph*}(30,20)
           \fmfleft{l1,l2}\fmfright{r1,r2}
           \fmf{plain,label=$-(k_1+k_2) \; \swarrow$,l.side=right}{r2,v2} 
           \fmf{plain,label=$k_1 \; \rightarrow$,l.side=right}{v1,v2}
           \fmf{photon,label=$k_2 \; \nwarrow$,l.side=right}{r1,v2} 
           \fmf{susy_ghost}{l1,v1} 
           \fmf{phantom}{v1,l2}
           \fmfv{decor.shape=square,decor.filled=full,decor.size=2mm}{l1} 
           \fmfv{decor.shape=square,decor.filled=empty,decor.size=1.8mm}{v1}
           \fmfdot{v2}
         \end{fmfgraph*}
         \hfil} 
\end{equation}
and analytically
\begin{equation}
- \dfrac{-\ii\eta_{\nu\beta}}{k_2^2}
  \dfrac{\ii}{\fmslash{k}_1+\fmslash{k}_2} \ii e \gamma^5 \gamma^\beta
  \dfrac{\ii}{\fmslash{k}_1} \gamma^5 \epsilon = \dfrac{e}{k_2^2}
  \dfrac{1}{\fmslash{k}_1 + \fmslash{k}_2} \gamma_\nu
  \dfrac{1}{\fmslash{k}_1} \epsilon 
\end{equation}
As was the case for the 2-point function, a term from the
third Green function in (\ref{komplitrafo1}) does exist with an additional
interaction vertex here, too:
\begin{equation}
       \parbox{21mm}{%
         \begin{fmfgraph*}(30,20)
           \fmfleft{l1,l2}\fmfright{r1,r2}
           \fmf{plain,label=$\swarrow \: -(k_1+k_2)$,l.side=left}{r2,v2} 
           \fmf{plain,label=$k_2 \; \rightarrow$,l.side=left}{v1,v2}
           \fmf{dashes,label=$k_1 \; \nwarrow$,l.side=left}{r1,v2} 
           \fmf{susy_ghost}{l1,v1} 
           \fmf{dots_arrow,label=$k_2 \searrow$}{v1,l2}
           \fmffreeze
           \fmf{photon}{v1,v2}
           \fmfv{decor.shape=square,decor.filled=full,decor.size=2mm}{l1} 
           \fmfv{decor.shape=square,decor.filled=empty,decor.size=1.8mm}{l2}
           \fmfdot{v1,v2}
         \end{fmfgraph*}
         \hfil} 
\end{equation}
It yields
\begin{equation}
- (\ii k_{2,\nu}) \dfrac{\ii}{k_1^2}
  \dfrac{\ii}{\fmslash{k}_1+\fmslash{k}_2} (-\ii e)
  \dfrac{\ii}{\fmslash{k}_2} \ii \fmslash{k}_2 \dfrac{-1}{k_2^2}
  \epsilon = \dfrac{e k_{2,\nu}}{k_1^2 k_2^2} \dfrac{1}{\fmslash{k}_1
  + \fmslash{k}_2} \epsilon  
\end{equation}
The fourth Green function with the diagram 
\begin{equation}
       \parbox{21mm}{%
         \begin{fmfgraph*}(30,20)
           \fmfleft{l1,l2}\fmfright{r1,r2}
           \fmf{plain,label=$-(k_1+k_2) \; \swarrow$,l.side=right}{r2,v2} 
           \fmf{plain,label=$k_2 \; \rightarrow$,l.side=right}{v1,v2}
           \fmf{dashes,label=$k_1 \; \nwarrow$,l.side=right}{r1,v2} 
           \fmf{susy_ghost}{l1,v1} 
           \fmf{phantom}{v1,l2}
           \fmffreeze
           \fmf{photon}{v1,v2}
           \fmfv{decor.shape=square,decor.filled=full,decor.size=2mm}{l1} 
           \fmfv{decor.shape=square,decor.filled=empty,decor.size=1.8mm}{v1}
           \fmfdot{v2}
         \end{fmfgraph*}
         \hfil} 
\end{equation}
gives the result
\begin{equation}
\dfrac{\ii}{k_1^2} \dfrac{\ii}{\fmslash{k}_1+\fmslash{k}_2} (-\ii e)
\dfrac{\ii}{\fmslash{k}_2} \gamma_\nu \epsilon = - \dfrac{e}{k_1^2}
\dfrac{1}{\fmslash{k}_1 + \fmslash{k}_2} \dfrac{1}{\fmslash{k}_2}
\gamma_\nu \epsilon 
\end{equation}
From the last Green function there are two contributions,
\begin{equation}
-\Greensfunc{A(x_1)G_\nu(x_2)\fmslash{\partial} B(x_3)} + e
 \Greensfunc{A(x_1) G_\nu(x_2) \gamma^\lambda (G_\lambda A)(x_3)
 \epsilon} 
\end{equation}
with the two diagrams ($k_{12} \equiv k_1 + k_2$)
\begin{equation}
       \parbox{21mm}{%
         \begin{fmfgraph*}(30,20)
           \fmfleft{l1,l2}\fmfright{r1,r2}
           \fmf{dashes,label=$\swarrow \: k_1$,l.side=left}{r2,v2} 
           \fmf{dbl_dashes,label=$-k_{12} \;
                \rightarrow$,l.side=left}{v1,v2} 
           \fmf{photon,label=$k_2 \; \nwarrow$,l.side=left}{r1,v2} 
           \fmf{susy_ghost}{l1,v1} 
           \fmf{phantom}{v1,l2}
           \fmfv{decor.shape=square,decor.filled=full,decor.size=2mm}{l1} 
           \fmfv{decor.shape=square,decor.filled=empty,decor.size=1.8mm}{v1}
           \fmfdot{v2}
         \end{fmfgraph*}
         \hfil} \qquad\qquad \qquad\qquad 
       \parbox{21mm}{%
         \begin{fmfgraph*}(20,19)
           \fmftop{t1}\fmfbottom{b1,b2}
           \fmf{dashes,label=$k_1 \: \nearrow$,l.side=left}{b1,v}
           \fmf{photon,label=$\nwarrow \: k_2$,l.side=right}{b2,v}
           \fmf{susy_ghost}{t1,v} 
           \fmfv{decor.shape=square,decor.filled=empty,decor.size=1.8mm}{v}
           \fmfv{decor.shape=square,decor.filled=full,decor.size=2mm}{t1}
         \end{fmfgraph*}
         \hfil} \quad \qquad .
\end{equation}
The right diagram yields the analytical expression
\begin{equation}
e \dfrac{\ii}{k_1^2} \dfrac{-\ii\eta_{\nu\lambda}}{k_2^2}
\gamma^\lambda \epsilon = \dfrac{e}{k_1^2 k_2^2} \gamma_\nu \epsilon
\quad , 
\end{equation}
while the left diagram gives the result
\begin{multline}
- \dfrac{\ii}{k_1^2} \dfrac{-\ii\eta_{\nu\beta}}{k_2^2}
  \dfrac{\ii}{(k_1+k_2)^2} e (k_1+(k_1+k_2))^\beta
  (-\ii)(\fmslash{k}_1 + \fmslash{k}_2)\epsilon \\ = \dfrac{-e}{k_1^2
  k_2^2 (k_1+k_2)^2} (2k_1+k_2)_\nu (\fmslash{k}_1+\fmslash{k}_2)
  \epsilon 
\end{multline}

The sum of all four terms vanishes:
\begin{multline}
\dfrac{e}{k_2^2}
  \dfrac{1}{\fmslash{k}_1 + \fmslash{k}_2} \gamma_\nu
  \dfrac{1}{\fmslash{k}_1} \epsilon +
\dfrac{e k_{2,\nu}}{k_1^2 k_2^2} \dfrac{1}{\fmslash{k}_1 
  + \fmslash{k}_2} \epsilon - \dfrac{e}{k_1^2}
\dfrac{1}{\fmslash{k}_1 + \fmslash{k}_2} \dfrac{1}{\fmslash{k}_2} 
\gamma_\nu \epsilon + \dfrac{e}{k_1^2 k_2^2} \gamma_\nu \epsilon \\ -
\dfrac{e}{k_1^2 k_2^2 (k_1+k_2)^2} (2k_1+k_2)_\nu 
(\fmslash{k}_1+\fmslash{k}_2) \epsilon  \\ =
\dfrac{e}{k_1^2 k_2^2 (\fmslash{k}_1+\fmslash{k}_2)} \biggl\{
  \gamma_\nu \fmslash{k}_1 + k_{2,\nu} - \fmslash{k}_2 \gamma_\nu +
  (\fmslash{k}_1 + \fmslash{k}_2) \gamma_\nu - (2k_1 + k_2)_\nu
  \biggr\} \epsilon = 0
\end{multline}
So this STI is fulfilled. Let us check the other example from the
earlier part of the text as well. 

\begin{align}
  0 = &\; \Greensfunc{ \left\{ Q_{\text{BRST}} , A(x_1) B(x_2) \lambda(x_3)
  \right\}} \notag \\
  = & \; - e \Greensfunc{c(x_1)B(x_1) B(x_2)
  \lambda (x_3)} + \Greensfunc{(\overline{\Psi}(x_1) \gamma^5 \epsilon)
  B(x_2) \lambda(x_3)} \notag \\ &\; + e \Greensfunc{A(x_1)
  c(x_2) A(x_2) \Psi(x_3)} + \ii \Greensfunc{A(x_1)
  (\overline{\Psi}(x_2) \epsilon) \lambda(x_3)} \notag \\
  &\; + \dfrac{\ii}{2} e \Greensfunc{A(x_1) B(x_2) \partial_\alpha
  G_\beta(x_3) \lbrack \gamma^\alpha , \gamma^\beta \rbrack \epsilon}
  \notag \\ &\; + \dfrac{e}{2} \Greensfunc{A(x_1) B(x_2) \left(
  A^2(x_3) + B^2(x_3) \right) \gamma^5 \epsilon}
\end{align}
The only vanishing term is the last one with trilinear terms from the
BRST transformations. The nonvanishing contributions are ($k_{12} =
k_1+k_2$): 
\begin{align}
       \parbox{21mm}{%
         \begin{fmfgraph*}(30,20)
           \fmfleft{l1,l2}\fmfright{r1,r2}
           \fmf{plain,label=$-k_{12} \; \swarrow$,l.side=right}{r2,v2} 
           \fmf{plain,label=$k_1 \; \rightarrow$,l.side=right}{v1,v2}
           \fmf{dbl_dashes,label=$\nwarrow\: k_2$,l.side=right}{r1,v2} 
           \fmf{susy_ghost}{l1,v1} 
           \fmf{phantom}{v1,l2}
           \fmffreeze
           \fmf{photon}{r2,v2}
           \fmfv{decor.shape=square,decor.filled=full,decor.size=2mm}{l1} 
           \fmfv{decor.shape=square,decor.filled=empty,decor.size=1.8mm}{v1}
           \fmfdot{v2}
         \end{fmfgraph*}
         \hfil} & \qquad\qquad \qquad = \quad - \dfrac{\ii e}{k_2^2}
         \dfrac{1}{\fmslash{k}_1+\fmslash{k}_2}
         \dfrac{1}{\fmslash{k}_1} \epsilon \\ 
       \parbox{21mm}{%
         \begin{fmfgraph*}(30,20)
           \fmfleft{l1,l2}\fmfright{r1,r2}
           \fmf{plain,label=$-k_{12} \; \swarrow$,l.side=right}{r2,v2} 
           \fmf{plain,label=$k_2 \; \rightarrow$,l.side=right}{v1,v2}
           \fmf{dashes,label=$\nwarrow\: k_1$,l.side=right}{r1,v2} 
           \fmf{susy_ghost}{l1,v1} 
           \fmf{phantom}{v1,l2}
           \fmffreeze
           \fmf{photon}{r2,v2}
           \fmfv{decor.shape=square,decor.filled=full,decor.size=2mm}{l1} 
           \fmfv{decor.shape=square,decor.filled=empty,decor.size=1.8mm}{v1}
           \fmfdot{v2}
         \end{fmfgraph*}
         \hfil} & \qquad\qquad \qquad = \quad \dfrac{\ii e}{k_1^2}
         \dfrac{1}{\fmslash{k}_1+\fmslash{k}_2}
         \dfrac{1}{\fmslash{k}_2} \epsilon \\
       \parbox{21mm}{%
         \begin{fmfgraph*}(30,20)
           \fmfleft{l1,l2}\fmfright{r1,r2}
           \fmf{dashes,label=$k_1 \; \swarrow$,l.side=right}{r2,v2} 
           \fmf{photon,label=$-k_{12} \: \rightarrow$,l.side=right}{v1,v2}
           \fmf{dbl_dashes,label=$\nwarrow\: k_2$,l.side=right}{r1,v2} 
           \fmf{susy_ghost}{l1,v1} 
           \fmf{phantom}{v1,l2}
           \fmfv{decor.shape=square,decor.filled=full,decor.size=2mm}{l1} 
           \fmfv{decor.shape=square,decor.filled=empty,decor.size=1.8mm}{v1}
           \fmfdot{v2}
         \end{fmfgraph*}
         \hfil} & \qquad\qquad \qquad = \quad \dfrac{\ii e}{2 k_1^2
         k_2^2 (k_1+k_2)^2} \lbrack \fmslash{k}_1 + \fmslash{k}_2 ,
         \fmslash{k}_1 - \fmslash{k}_2 \rbrack \epsilon \\
       \parbox{21mm}{%
         \begin{fmfgraph*}(30,20)
           \fmfleft{l1,l2}\fmfright{r1,r2}
           \fmf{plain,label=$\swarrow \: -k_{12}$,l.side=left}{r2,v2} 
           \fmf{susy_ghost}{r1,v2}
           \fmf{dots_arrow,label=$\rightarrow\: k_{12}$,l.side=right}{v2,v1} 
           \fmf{dashes,label=$k_1 \: \nearrow$,l.side=right}{v1,l1} 
           \fmf{phantom}{v1,l2}
           \fmffreeze
           \fmf{photon}{r2,v2}
           \fmfv{decor.shape=square,decor.filled=full,decor.size=2mm}{r1} 
           \fmfv{decor.shape=square,decor.filled=empty,decor.size=1.8mm}{v1}
           \fmfdot{v2}
         \end{fmfgraph*}
         \hfil} & \qquad\qquad \qquad = \quad - \dfrac{\ii e}{k_1^2
         (k_1+k_2)^2} \epsilon \\
       \parbox{21mm}{%
         \begin{fmfgraph*}(30,20)
           \fmfleft{l1,l2}\fmfright{r1,r2}
           \fmf{plain,label=$\swarrow \: -k_{12}$,l.side=left}{r2,v2} 
           \fmf{susy_ghost}{r1,v2}
           \fmf{dots_arrow,label=$\rightarrow\: k_{12}$,l.side=right}{v2,v1} 
           \fmf{dbl_dashes,label=$k_2 \: \nearrow$,l.side=right}{v1,l1} 
           \fmf{phantom}{v1,l2}
           \fmffreeze
           \fmf{photon}{r2,v2}
           \fmfv{decor.shape=square,decor.filled=full,decor.size=2mm}{r1} 
           \fmfv{decor.shape=square,decor.filled=empty,decor.size=1.8mm}{v1}
           \fmfdot{v2}
         \end{fmfgraph*}
         \hfil} & \qquad\qquad \qquad = \quad \dfrac{\ii e}{k_2^2
         (k_1+k_2)^2} \epsilon  
\end{align}
Adding up the five contributions yields:
\begin{multline}
 \dfrac{\ii e}{k_1^2 k_2^2 (k_1+k_2)^2} \biggl\{ - (\fmslash{k}_1 +
 \fmslash{k}_2) \fmslash{k}_1 + (\fmslash{k}_1+\fmslash{k}_2)
 \fmslash{k}_2 - \lbrack \fmslash{k}_1 , \fmslash{k}_2 \rbrack - k_2^2
 + k_1^2 \biggr\} \epsilon = 0
\end{multline}
This STI is valid, too. 

From this examples it can be seen that the formalism of Slavnov-Taylor
identities with the inclusion of the constant SUSY ghosts does indeed
work for supersymmetric gauge theories. Let us give some comments
about that: We only calculated the lowest order of the STI in the
gauge coupling constant {\em and} the SUSY ghost (first order) {\em
and} the translation ghost (zeroth order). This kind of identities can
easily be extended to higher orders in the number of the constant
ghosts due to the filtration property of functionals and Hilbert
space, but {\em not in this model} to higher orders in the gauge
coupling since that special model carries an anomaly and the symmetry
is spoiled when going to the one-loop level or beyond. 

%%%%%%%%%%%%%%%%%%%%%%%%%%%%%%%%%%%%%%%%%%%%%%%%%%%%%%%%%%%%%%%%%%%%%%%%

%%% Local Variables: 
%%% mode: latex
%%% TeX-master: diss
%%% End: 

%% file: chap11.tex
\chapter{Non-Abelian gauge theories: $SU(N)$}

The generalization of the ideas and formalisms for supersymmetric
gauge theories and BRST quantization developed up to now to the non-Abelian
case is the topic of this chapter. For the sake of simplicity we restrict
ourselves to the special unitary groups $SU(N)$; within these gauge
groups the structure of the vertices of four scalar particles is very
simple. For more general gauge groups
cf.~\cite{Cvitanovic:1976:nonabel}.  

We consider an SQCD-like model with two (matter) superfields
$\hat{\Phi}_+$ and $\hat{\Phi}_-$ and their Hermitean adjoints. In this model
the fermion becomes a Dirac fermion by combining the left-handed
components of the $+$-superfield with those of the conjugated
$-$-superfield. Quadratic superpotential terms are allowed here, cubic are
still forbidden by gauge invariance. The Abelian limit of this model will
give SQED, the supersymmetric extension of quantum electrodynamics. 

The main part of the model's details concerning Lagrangean density, equations
of motion, propagators, Feynman rules, etc.~can be found in appendix
\ref{sec:symdetail}. After diagonalizing the mass terms of the fermions, the
model contains two charged scalar fields $\phi_+$ and $\phi_-$ living in the 
fundamental representation and its complex conjugate, respectively, as well as
a charged fermion $\Psi$ in the fundamental representation. Moreover there is
the gauge boson $A_\mu$, the gaugino $\lambda$, the ghost $c$ and the
antighost $\bar{c}$, each in the adjoint representation. 

The BRST transformations of these fields are:
\begin{subequations}
\begin{align}
s\phi_{+,i}(x) &=\; - \ii g c^a(x) T^a_{ij} \phi_{+,j}(x) + \sqrt{2}
\left(\overline{\epsilon} {\cal P}_L \Psi_i(x)\right) - \ii \omega^\nu
\partial_\nu \phi_{+,i}(x) \\
s\phi^\dagger_{+,i}(x) &=\; \ii g c^a(x) \phi^\dagger_{+,j} (x) T^a_{ji} +
\sqrt{2} \left(\overline{\Psi}_i(x){\cal P}_R \epsilon\right) - \ii \omega^\nu
\partial_\nu \phi^\dagger_{+,i} (x) \\
s\phi_{-,i}(x) &=\; \ii g c^a(x) \phi_{-,j}(x) T^a_{ji} - \sqrt{2}
\left( \overline{\Psi}_i(x) {\cal P}_L \epsilon\right) -
\ii\omega^\nu\partial_\nu \phi_{-,i}(x) \\
s\phi^\dagger_{-,i}(x) &=\; -\ii g c^a(x) T^a_{ij}
\phi^\dagger_{-,j}(x) - \sqrt{2} \left( \overline{\epsilon} {\cal P}_R
\Psi_i(x) \right) - \ii\omega^\nu\partial_\nu \phi^\dagger_{-,i} (x)
\\
s \Psi_i(x) &=\; -\ii g c^a(x) T^a_{ij} \Psi_j(x) + \sqrt{2} \Bigl[
(\ii \fmslash{\partial} + m) \phi_{+,i}(x) {\cal P}_R +
(\ii\fmslash{\partial}-m) \phi^\dagger_{-,i}(x) {\cal P}_L \notag\\&
\qquad \qquad + g\fmslash{A}^a (x) T^a_{ij} \left( \phi_{+,j}(x) {\cal
P}_R + \phi^\dagger_{-,j}(x) {\cal P}_L \right) \Bigr] \epsilon - \ii 
\omega^\nu \partial_\nu \Psi_i(x) \\
s \overline{\Psi}_i(x) &=\; \ii g c^a(x) \overline{\Psi}_j(x) T^a_{ji}
+ \sqrt{2} \overline{\epsilon} \Bigl[ {\cal P}_L (\ii
\fmslash{\partial} -m) \phi^\dagger_{+,i}(x) + {\cal P}_R (\ii
\fmslash{\partial} -m)\phi_{-,i}(x)  \notag \\ &\qquad\qquad - g \left(
\phi^\dagger_{+,j}(x) {\cal P}_L + \phi_{-,j}(x) {\cal P}_R \right)
T^a_{ji} \fmslash{A}^a (x) \Bigr] - \ii \omega^\nu \partial_\nu
\overline{\Psi}_i(x) \\ 
sA_\mu^a(x) &=\; (D_\mu c(x))^a - \overline{\epsilon} \gamma_\mu
\lambda^a(x) - \ii \omega^\nu\partial_\nu A_\mu^a(x) \\
s\lambda^a(x) &=\; g f_{abc} c^b(x) \lambda^c(x) + \dfrac{\ii}{2}
F^a_{\alpha\beta}(x) \gamma^\alpha \gamma^\beta \epsilon + g \left(
\phi^\dagger_+ (x) T^a \phi_+ (x) \right) \gamma^5 \epsilon \notag \\
& \qquad\qquad - g \left( \phi_-(x) T^a \phi_-^\dagger(x) \right)
\gamma^5 \epsilon - \ii \omega^\nu \partial_\nu \lambda^a(x) \\
s\overline{\lambda^a}(x) &=\; g f_{abc} c^b(x) \overline{\lambda^c}(x)
- \dfrac{\ii}{2} \overline{\epsilon} \gamma^\alpha \gamma^\beta
F^a_{\alpha\beta} (x) + g \overline{\epsilon} \gamma^5 \left(
\phi_+^\dagger (x) T^a \phi_+(x) \right) \notag \\ & \qquad\qquad - g
\overline{\epsilon} \gamma^5 \left( \phi_-(x) T^a \phi_-^\dagger(x)
\right) - \ii  \omega^\nu \partial_\nu \overline{\lambda^a}(x) \\
s c^a(x) &=\; - \dfrac{g}{2} f_{abc} c^b(x) c^c(x) + \ii
(\overline{\epsilon} \gamma^\mu \epsilon) A_\mu(x) - \ii
\omega^\nu\partial_\nu c^a(x) \\
s\overline{c}^a(x) &=\; \ii B^a(x) - \ii \omega^\nu \partial_\nu
\overline{c}^a(x) \\
sB^a(x) &=\; (\overline{\epsilon}\gamma^\mu\epsilon) \partial_\mu
\overline{c}^a(x) - \ii \omega^\nu \partial_\nu B^a(x) \\
s\epsilon &=\; 0 \\
s\omega^\mu &=\; (\overline{\epsilon} \gamma^\mu \epsilon)
\end{align}
\end{subequations}

The gauge fixing is in complete analogy to the Abelian case
\begin{equation}
  S_{\text{GF+FP}} = - \ii \int d^4 x \; s (\overline{c}^a F^a) = -
  \ii \int d^4 x \left[ (s \overline{c}^a ) F^a - \overline{c}^a (s
  F^a) \right] 
\end{equation}
with the gauge fixing function 
\begin{equation}
  F^a = \partial^\mu A_\mu^a + \dfrac{\xi}{2} B^a .
\end{equation}
$\xi$ is the gauge parameter. For the translational invariance
of the gauge fixing term no translation ghosts $\omega^\mu$ appear,
too, and we get:
\begin{equation}
  \boxed{ 
  \begin{aligned}
  S_{\text{GF+FP}} &=\; \int d^4 x \; \biggl\{ B^a \partial^\mu
  A_\mu^a +  \dfrac{\xi}{2} B^a B^a + \ii \overline{c}^a \partial_\mu
  (D^\mu c)^a \\ & \qquad\qquad\qquad - \ii \overline{c}^a
  (\overline{\epsilon}\fmslash{\partial} \lambda^a) + \ii
  \dfrac{\xi}{2} \overline{c}^a (\overline{\epsilon} \gamma^\mu
  \epsilon) \partial_\mu \overline{c}^a   \biggr\} 
  \end{aligned}} 
\end{equation}
Integrating the Nakanishi-Lautrup field out yields:
\begin{multline}
  S_{\text{GF+FP}} = \int d^4 x \; \biggl\{ - \dfrac{1}{2 \xi} (\partial^\mu
  A_\mu^a) (\partial^\nu A_\nu^a) +  \ii \overline{c}^a \partial_\mu
  (D^\mu c)^a  \\ - \ii \overline{c}^a (\overline{\epsilon}
  \fmslash{\partial} \lambda^a) + \ii \dfrac{\xi}{2} \overline{c}^a
  (\overline{\epsilon} \gamma^\mu \epsilon) \partial_\mu \overline{c}^a
  \biggr\}  
\end{multline}

%%%%%%%%%%%%%%%%%%%%%%%%%%%%%%%%%%%%%%%%%%%%%%%%%%%%%%%%%%%%%%%%%%%%%%%%%%

\section{An example for an STI in SQCD}

We just want to show one example for a Slavnov-Taylor identity in
supersymmetric quantum chromodynamics, SQCD:
\begin{align}
        0 \stackrel{!}{=}&\;  \Greensfunc{ \left\{ Q_{\text{BRST}} ,
        A^a_\mu (x_1) A_\nu^b(x_2) \lambda^c(x_3) \right\} } \notag\\=&\;  
        \Greensfunc{(D_\mu c)^a(x_1) A_\nu^b(x_2) \lambda^c(x_3)}
        - \Greensfunc{ \left( \overline{\epsilon} \gamma_\mu
        \lambda^a(x_1) \right) A_\nu^b(x_2) \lambda^c(x_3) } \notag\\ & \; + 
        \Greensfunc{A_\mu^a(x_1) (D_\nu c)^b(x_2) \lambda^c(x_3)}
        - \Greensfunc{ A_\mu^a(x_1) \left( \overline{\epsilon}
        \gamma_\nu \lambda^b(x_2) \right) \lambda^c(x_3) } \notag\\ &\; +
        \dfrac{\ii}{2} \Greensfunc{ A^a_\mu (x_1) A_\nu^b(x_2)
        \partial_\lambda A^c_\kappa (x_3) \lbrack \gamma^\lambda ,
        \gamma^\kappa \rbrack \epsilon } \notag\\ &\; + \dfrac{\ii g}{4}
        \Greensfunc{ A^a_\mu (x_1) A_\nu^b(x_2) \left( A_\lambda^e
        A_\kappa^f \right) (x_3) \lbrack \gamma^\lambda ,
        \gamma^\kappa \rbrack f^{cef} \epsilon }
\end{align}
(In the sequel $p_{12}$ means $p_1 + p_2$.)
\begin{multline*}
        \qquad\qquad       \parbox{21mm}{%
         \begin{fmfgraph*}(30,20)
           \fmfleft{l1,l2}\fmfright{r1,r2}
           \fmf{plain,label=$\swarrow \: -p_{12}$,l.side=left}{r2,v2} 
           \fmf{plain,label=$p_1 \; \rightarrow$,l.side=left}{v1,v2}
           \fmf{photon,label=$p_2 \; \nwarrow$,l.side=left}{r1,v2} 
           \fmf{susy_ghost}{l1,v1} 
           \fmf{dots_arrow,label=$p_1 \searrow$}{v1,l2}
           \fmffreeze
           \fmf{photon}{v1,v2}
           \fmf{photon}{r2,v2}
           \fmfv{decor.shape=square,decor.filled=full,decor.size=2mm}{l1} 
           \fmfv{decor.shape=square,decor.filled=empty,decor.size=1.8mm}{l2}
           \fmflabel{$a$}{l2}   \fmflabel{$b$}{r1}
           \fmflabel{$c$}{r2}
           \fmfdot{v1,v2}
         \end{fmfgraph*}
         \hfil} \qquad \qquad \qquad +  \qquad \parbox{21mm}{%
         \begin{fmfgraph*}(30,20)
           \fmfleft{l1,l2}\fmfright{r1,r2}
           \fmf{dots_arrow}{v2,v1}
           \fmf{plain,label=$\nwarrow \; -p_{12}$,l.side=right}{r1,v2} 
           \fmf{photon,label=$\nearrow \; p_2$,l.side=right}{l1,v1} 
           \fmf{dots_arrow,label=$p_1 \searrow$}{v1,l2}
           \fmf{susy_ghost}{r2,v2}
           \fmffreeze
           \fmf{photon}{r1,v2}
           \fmfv{decor.shape=square,decor.filled=full,decor.size=2mm}{r2} 
           \fmfv{decor.shape=square,decor.filled=empty,decor.size=1.8mm}{l2}
           \fmflabel{$a$}{l2}   \fmflabel{$b$}{l1}
           \fmflabel{$c$}{r1}
           \fmfdot{v1,v2}
         \end{fmfgraph*}
         \hfil} \\ = - \text{F.T.} \Greensfunc{ \partial_\mu c^a(x_1)
         \bar{c}^d(z) \lambda^c(x_3) \left(\overline{\lambda^d}(z)
         \stackrel{\leftarrow}{\fmslash{\partial}} \epsilon\right)
         A_\nu^b(x_2) } 
\end{multline*}
\begin{align}
        -\dfrac{-1}{p_1^2} \dfrac{-\ii}{p_2^2}
         \dfrac{\ii}{\fmslash{p}_1 + \fmslash{p}_2} g \gamma_\nu
         f^{abc} \dfrac{\ii}{\fmslash{p}_1} (\ii\fmslash{p}_1) \ii
         p_{1,\mu} \epsilon =&\; \dfrac{- \ii g f^{abc}}{p_1^2 p_2^2
         (p_1+p_2)^2} \left( \fmslash{p}_1 + \fmslash{p}_2 \right)
         \gamma_\nu p_{1,\mu} \epsilon  \notag \\ 
        - \dfrac{\ii}{\fmslash{p}_1 + \fmslash{p}_2} \ii\left(
         \fmslash{p}_1 + \fmslash{p}_2 \right) \epsilon
         \dfrac{-1}{(p_1+p_2)^2} \cdot \qquad & \notag \\ (-\ii g
         f^{abc}) p_{1,\nu} 
         \dfrac{-\ii}{p_2^2} \dfrac{-1}{p_1^2} (\ii p_{1,\mu}) =&\;
         \dfrac{-\ii g f^{abc}}{p_1^2 p_2^2 (p_1+p_2)^2} p_{1,\mu}
         p_{1,\nu} \epsilon  \label{contrib1}
\end{align}
\begin{multline*}
        \qquad\qquad       \parbox{21mm}{%
         \begin{fmfgraph*}(30,20)
           \fmfleft{l1,l2}\fmfright{r1,r2}
           \fmf{plain,label=$\swarrow \: -p_{12}$,l.side=left}{r2,v2} 
           \fmf{plain,label=$p_2 \; \rightarrow$,l.side=left}{v1,v2}
           \fmf{photon,label=$p_1 \; \nwarrow$,l.side=left}{r1,v2} 
           \fmf{susy_ghost}{l1,v1} 
           \fmf{dots_arrow,label=$p_2 \searrow$}{v1,l2}
           \fmffreeze
           \fmf{photon}{v1,v2}
           \fmf{photon}{r2,v2}
           \fmfv{decor.shape=square,decor.filled=full,decor.size=2mm}{l1} 
           \fmfv{decor.shape=square,decor.filled=empty,decor.size=1.8mm}{l2}
           \fmflabel{$b$}{l2}   \fmflabel{$a$}{r1}
           \fmflabel{$c$}{r2}
           \fmfdot{v1,v2}
         \end{fmfgraph*}
         \hfil} \qquad \qquad \qquad +  \qquad \parbox{21mm}{%
         \begin{fmfgraph*}(30,20)
           \fmfleft{l1,l2}\fmfright{r1,r2}
           \fmf{dots_arrow}{v2,v1}
           \fmf{plain,label=$\nwarrow \; -p_{12}$,l.side=right}{r1,v2} 
           \fmf{photon,label=$\nearrow \; p_1$,l.side=right}{l1,v1} 
           \fmf{dots_arrow,label=$p_2 \searrow$}{v1,l2}
           \fmf{susy_ghost}{r2,v2}
           \fmffreeze
           \fmf{photon}{r1,v2}
           \fmfv{decor.shape=square,decor.filled=full,decor.size=2mm}{r2} 
           \fmfv{decor.shape=square,decor.filled=empty,decor.size=1.8mm}{l2}
           \fmflabel{$b$}{l2}   \fmflabel{$a$}{l1}
           \fmflabel{$c$}{r1}
           \fmfdot{v1,v2}
         \end{fmfgraph*}
         \hfil} \\ = - \text{F.T.} \Greensfunc{ \partial_\nu c^b(x_2)
         \bar{c}^d(z) \lambda^c(x_3) \left(\overline{\lambda^d}(z)
         \stackrel{\leftarrow}{\fmslash{\partial}} \epsilon\right)
         A_\mu^a(x_1) } 
\end{multline*}
This is just the earlier result with the replacements
$(a\leftrightarrow b)$, $(\mu \leftrightarrow \nu)$, $(p_1
\leftrightarrow p_2)$:
\begin{equation} 
        \dfrac{\ii g f^{abc}}{p_1^2 p_2^2
         (p_1+p_2)^2} \left( \fmslash{p}_1 + \fmslash{p}_2 \right)
         \gamma_\mu p_{2,\nu} \epsilon + 
         \dfrac{\ii g f^{abc}}{p_1^2 p_2^2 (p_1+p_2)^2} p_{2,\mu}
         p_{2,\nu} \epsilon 
\end{equation}
\begin{multline*} 
        \qquad 
       \parbox{21mm}{%
         \begin{fmfgraph*}(30,20)
           \fmfleft{l1,l2}\fmfright{r1,r2}
           \fmf{plain,label=$-p_{12} \; \swarrow$,l.side=right}{r2,v2} 
           \fmf{dots_arrow,label=$p_{12} \; \rightarrow$,l.side=right}{v2,v1}
           \fmf{photon,label=$p_2 \; \nearrow$,l.side=left}{l1,v1} 
           \fmf{susy_ghost}{r1,v2} 
           \fmf{phantom}{v1,l2}
           \fmffreeze
           \fmf{photon}{r2,v2}
           \fmfv{decor.shape=square,decor.filled=full,decor.size=2mm}{r1} 
           \fmfv{decor.shape=square,decor.filled=empty,decor.size=1.8mm}{v1}
           \fmfdot{v2}  \fmflabel{$b$}{l1}
           \fmflabel{$a$}{v1}   \fmflabel{$c$}{r2}              
         \end{fmfgraph*}\hfil} \\ 
         \qquad = - g f^{ade} \text{F.T.} \Greensfunc{
         \left( A^d_\mu c^e \right)(x_1) \bar{c}^f(z) \lambda^c(x_3)
         \left(\overline{\lambda^f}(z)
         \stackrel{\leftarrow}{\fmslash{\partial}} \epsilon\right) 
          A^b_\nu (x_2) }
\end{multline*}
This gives the analytical expression:
\begin{equation}
        -g f^{abc} \dfrac{-\ii}{p_2^2} \eta_{\mu\nu} \dfrac{-1}{(p_1+p_2)^2}
         \dfrac{\ii}{\fmslash{p}_1 + \fmslash{p}_2} \ii \left(
         \fmslash{p}_1 + \fmslash{p}_2 \right) \epsilon = \dfrac{\ii g
         f^{abc}}{p_1^2 p_2^2 (p_1 + p_2)^2} p_1^2 \eta_{\mu\nu}
         \epsilon 
\end{equation}
\begin{multline*} 
        \qquad 
       \parbox{21mm}{%
         \begin{fmfgraph*}(30,20)
           \fmfleft{l1,l2}\fmfright{r1,r2}
           \fmf{plain,label=$-p_{12} \; \swarrow$,l.side=right}{r2,v2} 
           \fmf{dots_arrow,label=$p_{12} \; \rightarrow$,l.side=right}{v2,v1}
           \fmf{photon,label=$p_1 \; \nearrow$,l.side=left}{l1,v1} 
           \fmf{susy_ghost}{r1,v2} 
           \fmf{phantom}{v1,l2}
           \fmffreeze
           \fmf{photon}{r2,v2}
           \fmfv{decor.shape=square,decor.filled=full,decor.size=2mm}{r1} 
           \fmfv{decor.shape=square,decor.filled=empty,decor.size=1.8mm}{v1}
           \fmfdot{v2}  \fmflabel{$a$}{l1}
           \fmflabel{$b$}{v1}   \fmflabel{$c$}{r2}              
         \end{fmfgraph*}\hfil} \\ 
         \qquad = - g f^{bde} \text{F.T.} \Greensfunc{
         \left( A^d_\nu c^e \right)(x_2) \bar{c}^f(z) \lambda^c(x_3)
         \left(\overline{\lambda^f}(z)
         \stackrel{\leftarrow}{\fmslash{\partial}} \epsilon\right) 
          A^a_\mu (x_1) }
\end{multline*}
Again we can use the replacements made above:
\begin{equation}
         \dfrac{-\ii g f^{abc}}{p_1^2 p_2^2 (p_1 + p_2)^2} p_2^2
         \eta_{\mu\nu} \epsilon 
\end{equation}

All of the expressions calculated up to now came from the gauge part
of the BRST transformations, now we discuss the SUSY part. 

\begin{equation*}
       \parbox{21mm}{%
         \begin{fmfgraph*}(30,20)
           \fmfleft{l1,l2}\fmfright{r1,r2}
           \fmf{plain,label=$-p_{12} \; \swarrow$,l.side=right}{r2,v2} 
           \fmf{plain,label=$p_1 \; \rightarrow$,l.side=right}{v1,v2}
           \fmf{photon,label=$\nwarrow \; p_2$,l.side=right}{r1,v2} 
           \fmf{susy_ghost}{l1,v1} 
           \fmf{phantom}{v1,l2}
           \fmffreeze
           \fmf{photon}{v1,v2,r2}
           \fmfv{decor.shape=square,decor.filled=full,decor.size=2mm}{l1} 
           \fmfv{decor.shape=square,decor.filled=empty,decor.size=1.8mm}{v1}
           \fmfdot{v2}          \fmflabel{$a\;$}{v1}
           \fmflabel{$b$}{r1}   \fmflabel{$c$}{r2} 
         \end{fmfgraph*}
         \hfil} \quad \qquad \qquad  = \text{F.T.} \Greensfunc{ A^b_\nu (x_2)
         \lambda^c(x_3) \left( \overline{\lambda^a}(x_1) \gamma_\mu
         \epsilon \right) }
\end{equation*}
We have made use of the two following identities:
\begin{equation}
        \left( \overline{\epsilon} \gamma_\mu \lambda \right) = +
        \left( \overline{\lambda} \gamma_\mu \epsilon \right), \qquad 
        \left( \overline{\lambda_1} \gamma_\mu \epsilon \right)
        \lambda_2 = - \lambda_2 \left( \overline{\lambda_1} \gamma_\mu
        \epsilon \right) . 
\end{equation}
The last diagram yields the analytical expression:
\begin{equation}
        \dfrac{\ii}{\fmslash{p}_1 + \fmslash{p}_2}
        \dfrac{-\ii}{(p_1+p_2)^2} g \gamma_\nu f^{abc}
        \dfrac{\ii}{\fmslash{p}_1} \gamma_\mu \epsilon = \dfrac{\ii g
        f^{abc}}{p_1^2 p_2^2 (p_1+p_2)^2} \left( \fmslash{p}_1 +
        \fmslash{p}_2 \right) \gamma_\nu \fmslash{p}_1 \gamma_\mu
        \epsilon 
\end{equation}
The terms for 
\begin{align*}
       \parbox{21mm}{\hfil\\\hfil\\%
         \begin{fmfgraph*}(30,20)
           \fmfleft{l1,l2}\fmfright{r1,r2}
           \fmf{plain,label=$-p_{12} \; \swarrow$,l.side=right}{r2,v2} 
           \fmf{plain,label=$p_2 \; \rightarrow$,l.side=right}{v1,v2}
           \fmf{photon,label=$\nwarrow \; p_1$,l.side=right}{r1,v2} 
           \fmf{susy_ghost}{l1,v1} 
           \fmf{phantom}{v1,l2}
           \fmffreeze
           \fmf{photon}{v1,v2,r2}
           \fmfv{decor.shape=square,decor.filled=full,decor.size=2mm}{l1} 
           \fmfv{decor.shape=square,decor.filled=empty,decor.size=1.8mm}{v1}
           \fmfdot{v2}          \fmflabel{$b\;$}{v1}
           \fmflabel{$a$}{r1}   \fmflabel{$c$}{r2} 
         \end{fmfgraph*}
         \hfil} \quad \qquad \qquad  = \text{F.T.} \Greensfunc{ A^a_\mu (x_1)
         \lambda^c(x_3) \left( \overline{\lambda^b}(x_2) \gamma_\nu
         \epsilon \right) }
\end{align*}
are again available by the replacements $(a\leftrightarrow b)$, $(\mu
\leftrightarrow \nu)$, $(p_1 \leftrightarrow p_2)$:
\begin{equation}
        \dfrac{- \ii g f^{abc}}{p_1^2 p_2^2 (p_1+p_2)^2} \left( \fmslash{p}_1 +
        \fmslash{p}_2 \right) \gamma_\mu \fmslash{p}_2 \gamma_\nu 
        \epsilon 
\end{equation}
The next step is the transformation of the gluino:
\begin{align*}
       \parbox{21mm}{\hfil\\\hfil\\%
         \begin{fmfgraph*}(30,20)
           \fmfleft{l1,l2}\fmfright{r1,r2}
           \fmf{photon,label=$p_1 \; \swarrow$,l.side=right}{r2,v2} 
           \fmf{photon,label=$\leftarrow \; p_{12}$,l.side=right}{v1,v2}
           \fmf{photon,label=$\nwarrow \; p_2$,l.side=right}{r1,v2} 
           \fmf{susy_ghost}{l1,v1} 
           \fmf{phantom}{v1,l2}
           \fmffreeze
           \fmfv{decor.shape=square,decor.filled=full,decor.size=2mm}{l1} 
           \fmfv{decor.shape=square,decor.filled=empty,decor.size=1.8mm}{v1}
           \fmfdot{v2}          \fmflabel{$c\;$}{v1}
           \fmflabel{$b$}{r1}   \fmflabel{$a$}{r2} 
         \end{fmfgraph*}
         \hfil} \quad \qquad \qquad  = \dfrac{\ii}{2} \text{F.T.}
         \Greensfunc{ A^a_\mu (x_1) 
         A^b_\nu(x_2) \partial_\lambda A^c_\kappa (x_3) \lbrack
         \gamma^\lambda , \gamma^\kappa \rbrack 
         \epsilon }
\end{align*}
From this diagram we get
\begin{multline}
        \dfrac{\ii}{2} \dfrac{-\ii}{p_1^2} \dfrac{-\ii}{p_2^2}
        \dfrac{-\ii}{(p_1+p_2)^2} g f^{abc} (-\ii) (p_1 + p_2)_\lambda \lbrack
        \gamma^\lambda, \gamma^\kappa \rbrack \cdot \\ \biggl[ \eta_{\mu\nu}
        \left( p_1 - p_2 \right)_\kappa + \eta_{\nu\kappa} \left( 2
        p_2 + p_1 \right)_\mu + \eta_{\mu\kappa} \left( - 2 p_1 - p_2
        \right)_\nu \biggr] \epsilon \\ = \dfrac{\ii g f^{abc}}{p_1^2
        p_2^2 (p_1+p_2)^2} \dfrac{1}{2} \biggl[ \eta_{\mu\nu} \lbrack
        \fmslash{p}_1 + \fmslash{p}_2 , \fmslash{p}_1 - \fmslash{p}_2
        \rbrack + \left( 2 p_2 + p_1 \right)_\mu \lbrack \fmslash{p}_1
        + \fmslash{p}_2 , \gamma_\nu \rbrack \\ - \left( 2 p_1 + p_2
        \right)_\nu \lbrack \fmslash{p}_1 + \fmslash{p}_2 , \gamma_\mu
        \rbrack \biggr] \epsilon   
\end{multline}
The final contribution to the STI comes from
\begin{align*}
       \parbox{23mm}{\hfil\\\hfil\\%
         \begin{fmfgraph*}(22,20)
           \fmfleft{l}\fmfright{r1,r2}
           \fmf{photon,label=$p_1 \; \swarrow$,l.side=right}{r2,v} 
           \fmf{photon,label=$p_2 \; \nwarrow$,l.side=left}{r1,v} 
           \fmf{susy_ghost}{l,v} 
           \fmfv{decor.shape=square,decor.filled=full,decor.size=2mm}{l} 
           \fmfv{decor.shape=square,decor.filled=empty,decor.size=1.8mm}{v}
           \fmflabel{$c\;$}{v}
           \fmflabel{$a$}{r2}   \fmflabel{$b$}{r1} 
         \end{fmfgraph*}
         \hfil\\} \quad = \dfrac{\ii g f^{cde}}{4} \text{F.T.}
         \Greensfunc{ A^a_\mu (x_1) 
         A^b_\nu(x_2) \left(A^d_\lambda A^e_\kappa\right) (x_3) \lbrack
         \gamma^\lambda , \gamma^\kappa \rbrack 
         \epsilon }
\end{align*}
It yields (with a symmetry factor two)
\begin{equation}  \label{contrib2} 
        \dfrac{\ii}{2} g f^{abc} \dfrac{-\ii}{p_1^2} \dfrac{-\ii}{p_2^2}
        \lbrack \gamma_\mu , \gamma_\nu \rbrack \epsilon = \dfrac{-\ii g
        f^{abc}}{p_1^2 p_2^2 (p_1+p_2)^2} \dfrac{1}{2} \lbrack \gamma_\mu ,
        \gamma_\nu \rbrack \left( p_1 + p_2 \right)^2 \epsilon 
\end{equation}

Adding up all the contributions (\ref{contrib1})-(\ref{contrib2}) should give
zero; in the following calculation we forget about the common prefactor $\ii g
f^{abc}/(p_1^2 p_2^2 (p_1+p_2)^2)$. First we collect all terms containing a
factor $\fmslash{p}_1 + \fmslash{p}_2$. For the terms of the contribution
with the three-gauge boson vertex we use 
\begin{equation}
  \label{diracalg}
  \dfrac{1}{2} \lbrack \fmslash{p}_1 +\fmslash{p}_2, \fmslash{a} \rbrack = 
  \left( \fmslash{p}_1 + \fmslash{p}_2 \right) \fmslash{a} - \left( p_1 + p_2
  \right) \cdot a \quad ,   
\end{equation}
while for the last contributing term we use $(p_1 + p_2)^2 = (\fmslash{p}_1 +
\fmslash{p}_2) (\fmslash{p}_1 + \fmslash{p}_2)$. We get
\begin{multline}
  \left( \fmslash{p}_1 + \fmslash{p}_2 \right) \biggl\{ - \gamma_\nu p_{1,\mu}
  - \left( 2 p_1 + p_2 \right)_\nu \gamma_\mu + \gamma_\mu p_{2,\nu} +
  \gamma_\nu \fmslash{p}_1 \gamma_\mu - \gamma_\mu \fmslash{p}_2 \gamma_\nu +
  \eta_{\mu\nu} \left( \fmslash{p}_1 - \fmslash{p}_2 \right) \\ + \left( 2 p_2
    + p_1 \right)_\mu \gamma_\nu - \dfrac{1}{2} \left( \fmslash{p}_1 +
    \fmslash{p}_2 \right) \lbrack \gamma_\mu , \gamma_\nu \rbrack \biggr\} \\
  = \left( \fmslash{p}_1 + \fmslash{p}_2 \right) \biggl\{ \eta_{\mu\nu} \left(
    \fmslash{p}_1 - \fmslash{p}_2 \right) - \dfrac{1}{2} \left( \fmslash{p}_1 +
    \fmslash{p}_2 \right) \lbrack \gamma_\mu , \gamma_\nu \rbrack +
  \fmslash{p}_2 \gamma_\mu \gamma_\nu - \fmslash{p}_1 \gamma_\nu \gamma_\mu
  \biggr\} \\ = \dfrac{1}{2} \left( \fmslash{p}_1 + \fmslash{p}_2 \right)
  \left( \fmslash{p}_1 - \fmslash{p}_2 \right) \Bigl[ 2 \eta_{\mu\nu} -
  \gamma_\mu \gamma_\nu - \gamma_\nu \gamma_\mu \Bigr] = 0 .
\end{multline}
In the second equation the Dirac algebra was used for the fourth and fifth
term to cancel out all terms with only one gamma matrix (besides the 
prefactor). 

The terms that do not contain any gamma matrices yield:
\begin{multline}
  \left( 2 p_1 + p_2 \right)_\nu \left( p_1 + p_2 \right)_\mu - \eta_{\mu\nu}
  \left( p_1 + p_2 \right) \cdot \left( p_1 - p_2 \right) + \eta_{\mu\nu}
  \left( p_1^2 - p_2^2 \right) \\ - \left( 2 p_2 + p_1 \right)_\mu \left( p_1 +
  p_2 \right)_\nu - p_{1,\mu} p_{1,\nu} + p_{2,\mu} p_{2,\nu} = 0 .
\end{multline}
So everything cancels, and the STI is fulfilled. Note that this STI contains
almost all elements of the non-Abelian character of the theory, the coupling
of the gluon to the ghosts and to the gluinos, and also the three-gluon
vertex. This identity would be trivial in SQED. 

%%%%%%%%%%%%%%%%%%%%%%%%%%%%%%%%%%%%%%%%%%%%%%%%%%%%%%%%%%%%%%%%%%%%%%%%%%%%%

\section[BRST for spontaneously broken SUSY]{BRST formalism for spontaneously
  broken supersymmetry} 

For spontaneously broken supersymmetry the BRST formalism brings nothing
new. Since supersymmetry is a global symmetry, the Goldstino always is a
physical particle. Only in a supergravity theory with a super-Higgs mechanism
the Goldstino is eaten up by the gravitino to make it massive and add the
missing two $J_z= \pm \frac{1}{2}$ polarizations. But we do not want to
discuss the BRST formalism for supergravity in this thesis.

%%% Local Variables: 
%%% mode: latex
%%% TeX-master: "diss"
%%% End: 

%% file: chap12.tex
\chapter{Implementation in {\em O'Mega}} 

\thispagestyle{headings}

The {\em ``{\bf {\em \underline{O}}}ptimizing {\bf {\em
\underline{M}}}atrix {\bf {\em \underline{E}}}lement {\bf
{\em \underline{G}}}ener{\bf {\em \underline{a}}}tor''}  
{\em O'Mega} invented by Thorsten Ohl  
\cite{Ohl:2000:Omega}, \cite{Moretti/Ohl/Reuter:2001:Omega},
\cite{Moretti/Ohl/Reuter/Schwinn:2001:Omega} provides 
the most efficient way to calculate tree level amplitudes available
today. This is achieved by building up amplitudes recursively by 
fusions of one-particle off-shell wave functions where the redundacies of the
diagrammatic representation are removed by a procedure called {\em common
subexpression elimination}. That is most easily illustrated by an example,
$e^+e^-\to\mu^+\mu^-$ with one additional bremsstrahlung quantum: 
\begin{align*}
        \parbox{18mm}{%
          \begin{fmfgraph*}(30,20)
            \fmfleft{l1,l2,l3,l4} \fmftop{t1,t2,t3,t4,t5} 
            \fmfright{r1,r2}
            \fmfbottom{b1,b2,b3,b4,b5}
            \fmf{phantom}{l1,v1,v,v2,l4}
            \fmf{phantom}{r1,w1,w,w2,r2}
            \fmf{photon}{v,w}
            \fmffreeze
            \fmf{photon}{v1,l2}
            \fmfdot{v,w,v1}
            \fmf{fermion}{l1,v1,v}
            \fmf{fermion}{v,l4}
            \fmf{fermion}{r1,w,r2}
            \fmf{dots}{t3,r2}
            \fmf{dots}{r2,r1}
            \fmf{dots}{r1,b3}            
            \fmf{dots}{b3,t3}
           \end{fmfgraph*}}    &\qquad\qquad + \qquad 
        \parbox{18mm}{%
          \begin{fmfgraph*}(30,20)
            \fmfleft{l1,l2,l3,l4} \fmftop{t1,t2,t3,t4,t5} 
            \fmfright{r1,r2}
            \fmfbottom{b1,b2,b3,b4,b1}
            \fmf{phantom}{l1,v1,v,v2,l4}
            \fmf{phantom}{r1,w1,w,w2,r2}
            \fmf{photon}{v,w}
            \fmffreeze
            \fmf{photon}{v2,l3}
            \fmfdot{v,w,v2}
            \fmf{fermion}{v,v2,l4}
            \fmf{fermion}{l1,v}
            \fmf{fermion}{r1,w,r2}
            \fmf{dots}{t3,r2}
            \fmf{dots}{r2,r1}
            \fmf{dots}{r1,b3}            
            \fmf{dots}{b3,t3}
           \end{fmfgraph*}}
\end{align*}
When we consider the initial state radiation then the muon part (in the dotted
box) remains the same and needs only to be calculated once. If the
result of the one calculation is kept in memory and inserted when
needed instead of being calculated again, a lot of computation time is
saved. That is the way {\em O'Mega} works and how it is able to
reduce the factorial growth of the number of Feynman diagrams
with the number of external particles to an exponential,
cf.~table \ref{processes}. This chapter will be rather formal since we
do not have the space to go into the details of {\em O'Mega} here, they
can be found in the commented source code of the program
\cite{Moretti/Ohl/Reuter/Schwinn:2001:Omega}.   

%%%%%%%%%%%%%%%%%%%%%%%%%%%%%%%%%%%%%%%%%%%%%%%%%%%%%%%%%%%%%%%%%%%%%%%

\section{BRST vertices}

As was briefly mentioned in the foregoing section, {\em O'Mega} constructs
amplitudes by fusing sets of subamplitudes built up recursively from 
one-particle off-shell wave functions (1POWs) fused out of partitions of the
external particles. The fusion works with $k$-ary topologies ($k \geq
2$), so that a new 1POW is fused out of two, three or more 1POWs
depending on subsets of the external momenta:

\begin{align*}
        \parbox{18mm}{\hfil \\%
          \begin{fmfgraph*}(20,20)
            \fmfbottom{b}
            \fmfleft{l1,l2}
            \fmfright{r}
            \fmf{plain}{l2,v,l1}
            \fmf{plain}{v,r}
            \fmf{plain}{v,l4}
            \fmfv{d.sh=circle,d.f=empty,d.si=13pt,l=$\begin{scriptsize}n_1
                \end{scriptsize}$,l.d=0}{l2}
            \fmfv{d.sh=circle,d.f=empty,d.si=13pt,l=$\begin{scriptsize}n_2
                \end{scriptsize}$,l.d=0}{l1}  
            \fmflabel{
              \begin{tabular}{c}   \hfil \\
                Binary topology
              \end{tabular}}{b} 
           \end{fmfgraph*}} \qquad \qquad
         \parbox{18mm}{\hfil \\%
          \begin{fmfgraph*}(20,20)
            \fmfbottom{b}
            \fmfleft{l1,l2,l3}
            \fmfright{r}
            \fmf{plain}{l2,v,l1}
            \fmf{plain}{l3,v,r}
            \fmf{plain}{v,r}
            \fmfv{d.sh=circle,d.f=empty,d.si=13pt,l=$\begin{scriptsize}n_1
                \end{scriptsize}$,l.d=0}{l3}
            \fmfv{d.sh=circle,d.f=empty,d.si=13pt,l=$\begin{scriptsize}n_2
                \end{scriptsize}$,l.d=0}{l2}
            \fmfv{d.sh=circle,d.f=empty,d.si=13pt,l=$\begin{scriptsize}n_3
                \end{scriptsize}$,l.d=0}{l1}  
            \fmflabel{
              \begin{tabular}{c} \hfil \\
                Ternary topology 
              \end{tabular}}{b}
           \end{fmfgraph*}} \qquad\qquad
         \parbox{18mm}{\hfil \\%
          \begin{fmfgraph*}(20,20)
            \fmfcurved
            \fmfbottom{b}
            \fmfleft{l1,l2,l3,l4,l5}
            \fmfright{r}
            \fmf{plain}{l5,v,l1}
            \fmf{plain}{v,r}
            \fmfv{d.sh=circle,d.f=empty,d.si=13pt,l=$\begin{scriptsize}n_1
                \end{scriptsize}$,l.d=0}{l5}
            \fmfv{d.sh=circle,d.f=empty,d.si=13pt,l=$\begin{scriptsize}n_2
                \end{scriptsize}$,l.d=0}{l4}
            \fmfv{d.sh=circle,d.f=empty,d.si=13pt,l=$\begin{scriptsize}n_k
                \end{scriptsize}$,l.d=0}{l1}  
            \fmf{dots,tension=.1}{l2,l3}
            \fmflabel{\begin{tabular}{c} \hfil \\ $k$-ary
          topology\end{tabular}}{b}
            \fmffreeze
            \fmf{plain}{v,l4}
           \end{fmfgraph*}}   \\ \hfil \\      
\end{align*}
The first two cases are needed for the Standard Model's 3- and 4-point
vertices (or other models') while higher topologies are useful for
introducing vertices of a degree higher than four to parameterize
irrelevant operator insertions from an effective theory. But it is {\em
not} possible for {\em O'Mega} to handle 2-point vertices like self
energy insertions, current insertions for operator product expansions,
or also BRST vertices, since every 1POW has to be uniquely labelled by the
external momenta on which it depends. So for {\em O'Mega} it would be 
impossibile to distinguish between a 1POW and one with just an additional
2-point vertex.

The solution to that problem is not difficult: When expressing the
theory with the help of functionals and for deriving Slavnov-Taylor
identities in a functional language, then in the effective
action one has to introduce external sources $K$ for each BRST
transformation \cite{Kugo:Eichtheorie}, \cite{Weinberg:QFTv2:Text},  
\begin{equation}
        \label{brstverteff}
        \ii \int d^4 x \; \sum_{\text{all} \,\Phi} K_\Phi \cdot \left( s \Phi
        \right)  
\end{equation}
sometimes called {\em antifields}. If we add those sources to the
particle content of an {\em O'Mega} model file then we are able to
generate Slavnov-Taylor identities in a manner similar to using
functional derivatives,
\begin{equation*}
        \Greensfunc{(s\Phi) \phi_1 \ldots \phi_n } =
        \Greensfunc{\phi_1 \ldots \phi_n \dfrac{\delta}{\delta
        K_{\Phi}} e^{\ii \Gamma(\phi_i,K_i)}} \biggr|_{K_j=0}  \quad
        . 
\end{equation*}
We also add the BRST vertices in the form
(\ref{brstverteff}) to our model file and generate them in an
amplitude by using the source as an external particle. 
There are always two possibilities to define particles (or better:
fields) in {\em O'Mega}, as propagating, or as being ``only
insertion''. In the first case they can appear as virtual particles in
inner lines while in the latter they are forced to serve exclusively as
external particles. The source $K_\Phi$, which always has the same Lorentz
structure as the transformed field $\Phi$, has to be defined as
nonpropagating, since it just generates a single operator
insertion (that corresponds to setting the sources to zero
above). Graphically, we use again the notation of \cite{Kugo:Eichtheorie}
for the BRST vertex, a small square. For the source we just 
double the line of the transformed field and put a diamond at the end
to demonstrate its nature as a pure insertion (in analogy to the
constant SUSY ghosts). To make this more pictorial we give an example
for the BRST transformation of the matter fermion of our Abelian toy
model from the last part (without the translation ghost, and for
simplicity we also omit the diagrams with the pseudoscalar field $B$)  
\begin{equation*}
    \overline{K_\Psi} \left( s \Psi \right) =\; - \ii e c \, 
    \overline{K_\Psi}  \gamma^5
    \Psi + \ii \overline{K_\Psi} (\fmslash{\partial} - \ii e
    \fmslash{G} \gamma^5) (A \gamma^5 + \ii {\color{blue} B}) \epsilon
\end{equation*}
\begin{align*}
        \Longleftrightarrow \qquad       \parbox{18mm}{%
           \begin{fmfgraph*}(17,16)
                 \fmfleft{t}     \fmfright{b1,b2}
                 \fmf{plain}{b1,v}
                 \fmf{dbl_plain}{t,v}
                 \fmf{dots_arrow}{b2,v}
                 \fmfv{decor.shape=square,decor.filled=empty,
                         decor.size=1.4mm}{v}    
                 \fmfv{decor.shape=diamond,decor.filled=shaded,
                         decor.size=3mm}{t}                 
           \end{fmfgraph*}} \qquad\quad
         \parbox{18mm}{%
           \begin{fmfgraph*}(17,16)
                 \fmfleft{t}     \fmfright{b1,b2}
                 \fmf{dashes}{b1,v}
                 \fmf{dbl_plain}{t,v}
                 \fmf{susy_ghost}{b2,v}
                 \fmfv{decor.shape=square,decor.filled=empty,
                         decor.size=1.4mm}{v}    
                 \fmfv{decor.shape=diamond,decor.filled=shaded,
                         decor.size=3mm}{t}                 
                 \fmfv{decor.shape=square,decor.filled=full,
                         decor.size=2mm}{b2} 
           \end{fmfgraph*}} \qquad\quad 
         \parbox{18mm}{%
           \begin{fmfgraph*}(19,16)
                 \fmfleft{t1}    \fmfright{t2,b1,b2}
                 \fmf{dbl_plain}{t1,v}
                 \fmf{susy_ghost}{b2,v}          
                 \fmf{photon}{t2,v} \fmffreeze
                 \fmf{dashes}{b1,v}
                 \fmfv{decor.shape=square,decor.filled=empty,
                         decor.size=1.4mm}{v}    
                 \fmfv{decor.shape=diamond,decor.filled=shaded,
                         decor.size=3mm}{t1}
                 \fmfv{decor.shape=square,decor.filled=full,
                         decor.size=2mm}{b2}                 
           \end{fmfgraph*}}
\end{align*}

From these diagrams we can immediately see the advantages of this
construction: Slavnov-Taylor identities (STI) for the gauge symmetries as
well as for supersymmetry can be done in the same formalism, one just
has to tell which ghost should be produced, $c$ or $\epsilon$. Without
the BRST transformation source the last of the diagrams written down above 
would have been an ordinary 3-point vertex and 
could have been handled in the usual manner in {\em O'Mega}, but this
vertex arises due to the nonlinearities of supersymmetry (additional
to the nonlinear structure of BRST) while most of the BRST vertices
are of the type of the leftmost diagrams, and furthermore, the
ability to have a unified formalism for managing all kinds of STI is
worth the effort. 

The only objection could lie in the BRST transformation of the gauge
boson into the derivative of the ghost 
\begin{equation*}
        s A_\mu = \partial_\mu c + \ldots , 
\end{equation*}
which -- even when coupled to the antifield -- is still only a 2-point
vertex. There we must use a trick: we establish a dummy field,
another nonpropagating local operator which couples to the BRST source
of the photon and to the ghost, which we denote by a tetragram here:
\begin{equation*}      
        \parbox{19mm}{%
           \begin{fmfgraph*}(18,16)
                 \fmfleft{t}     \fmfright{b2,b1}
                 \fmf{dots_arrow}{b1,v}
                 \fmf{dbl_wiggly}{t,v}
                 \fmf{dbl_dots}{v,b2}
                 \fmfv{decor.shape=square,decor.filled=empty,
                         decor.size=1.4mm}{v}    
                 \fmfv{decor.shape=diamond,decor.filled=shaded,
                         decor.size=3mm}{t}      
                 \fmfv{decor.shape=tetragram,decor.filled=70,
                         decor.size=3mm}{b2}                
           \end{fmfgraph*}} .
\end{equation*}
This means that for generating an STI -- where we have to replace each of
the fields $\phi_i, i=1, \ldots n$ in $\Greensfunc{\left\{
Q_{\text{BRST}} , \phi_1 \phi_2 \ldots \phi_n (\text{ghost})\right\}}$ 
successively by the source of its BRST transformation -- each gauge boson
must be replaced not only by the source of its BRST transformation but
also by the product of the source and that local operator. By the
choice of ``ghost'' we specify whether we want to study a SUSY or a
gauge STI. (Since the BRST charge has ghost number $+1$ we must have
one ghost in our string of fields above to get a nontrivial result.) The
problem with the one-point vertex can be avoided by using physical 
polarization vectors (cf.~the table at the end of that section) for the
external states of the BRST transformed vector bosons, i.e. transversal states
but still off-shell. Since the BRST vertex with the derivative of the ghost
produces the momentum of that gauge boson, the contribution from that term
vanishes. In that formalism the gauge boson can be handled as a matter
particle in the adjoint representation of the gauge group. Of course, a more
stringent test is done by taking the STIs component-wise, where the term
$\partial_\mu c^a$ has been taken accounted for.  

\vspace{2mm}

There is yet another problem which is not so difficult to resolve with
our BRST source construction: {\em O'Mega} produces code for
calculating $S$-matrix elements, i.e. for amputated Green functions,
whereas we want to study STI, relations between off-shell Green
functions usually {\em not} amputated. But we can consider all Green
functions participating in our STIs as being amputated while still
remaining off the mass shell. So all external legs of the Green
functions have been multiplied by their inverse propagators to arrive
at a matrix element-like object, something that can be generated by
{\em O'Mega}. 

For gauge theories one of the fields in the STI must be an
antighost which combines with the ghost from the BRST transformation
to give a ghost propagator. Dividing by that ghost propagator cancels
the ghost propagator in all of the Green functions including the gauge
boson propagator coming from the scalar mode of the gauge boson
produced by BRST transforming the antighost. This is clear for exact
gauge symmetries, whereas for the spontaneously broken case we present
a short proof (we omit the propagator of the Goldstone boson $\phi$ since
it always has the same pole as the ghost, $p^2 - \xi m^2$): 
\begin{multline}
        (-\ii) s \bar{c}^a = \dfrac{1}{\xi} \partial^\mu A_\mu^a + m
        \phi^a \quad \text{produces the propagator with contracted
        momentum:} \\ 
        \dfrac{1}{\xi} \left(\ii p^\mu\right) \dfrac{\ii}{p^2 - m^2}\left(
        -\eta_{\mu\nu} + \dfrac{(1-\xi) p_\mu p_\nu}{p^2 - \xi m^2}
        \right) = \dfrac{-1}{\xi} \dfrac{p_\nu}{p^2-m^2} \left( -1 +
        \dfrac{(1-\xi)p^2}{p^2 - \xi m^2} \right) \\ = \dfrac{1}{\xi}
        \dfrac{p_\nu}{p^2 - m^2} \dfrac{\xi (p^2 - m^2)}{p^2 - \xi
        m^2}  
\end{multline}
The BRST transformed ``physical fields'' (matter fields and gauge
bosons) become internal lines, but so they are in {\em O'Mega} due to
our construction using the sources,
\begin{align*}
        \parbox{17mm}{%
           \begin{fmfgraph*}(30,18)
                 \fmfleft{l1,l2}     \fmfright{r1,r2,r3,r4,r5}
                 \fmf{dots_arrow}{l2,v1}
                 \fmf{dbl_plain}{v1,l1}
                 \fmf{plain}{v1,v2}
                 \fmf{plain}{v2,r1}     \fmf{plain}{v2,r5}      
                 \fmffreeze
                 \fmf{plain}{v2,r2}     \fmf{dots}{r3,r4}
                 \fmfv{decor.shape=square,decor.filled=empty,
                         decor.size=1.4mm}{v1}    
                 \fmfv{decor.shape=diamond,decor.filled=shaded,
                         decor.size=3mm}{l1}
                \fmfv{decor.shape=circle,decor.filled=30,
                         decor.size=6mm}{v2}                  
           \end{fmfgraph*}} \qquad\qquad \qquad ,
\end{align*}
and their propagators are produced automatically in {\em O'Mega}. 
When multiplying with the inverse propagators of all external
particles there is a mismatch according to the formula (for simplicity
we assume to have a gauge, not a supersymmetry here and all particles
to be scalars; the generalization is straightforward):  
\begin{multline}
        \prod_{i=1}^n (-\ii) \left( p_i^2 - m_i^2 \right) \text{F.T.}
        \Greensfunc{ \left\{ Q_{\text{BRST}} , \phi_1 \ldots \phi_n 
        \right\}} \\ = \sum_{j=1}^n \dfrac{p_j^2 - m_j^2}{(p_j + k)^2
        - \tilde{m}_j^2} \cdot 
        \left[ \;\;\; 
        \parbox{18mm}{
        \parbox{17mm}{% 
          \begin{fmfgraph*}(15,15)
                \fmfleft{l1,l2,l3,l4,l5}         \fmfright{r1,r2,r3,r4}
                \fmftop{t}                    \fmfbottom{b}
                \fmf{plain}{l1,v}       \fmf{plain}{v,l5} 
                \fmf{plain}{v,r1}       \fmf{plain}{v,r2}
                \fmf{plain}{v,r3}       \fmf{plain}{v,r4}
                \fmf{plain}{v,t}        \fmf{plain}{v,b}
                \fmf{plain}{v,l4}       \fmf{dots}{l2,l3}
                \fmfv{decor.shape=circle,decor.filled=30,
                         decor.size=6mm}{v}  
          \end{fmfgraph*}}} \right]_{\text{amputated}}^\prime , 
\end{multline}
because the transformed field now has an internal propagator with a
momentum shifted by the ghost's momentum and therefore does not cancel the
inverse propagator (the prime indicates that this internal propagator
has been extracted from the diagram). Besides, in spontaneously broken 
symmetries there is the possibility that field and transformed field
have different masses, e.g.~when connecting electron and neutrino by
an $SU(2)_L$-BRST transformation or the several scalar and fermionic
fields in the O'Raifeartaigh model. This is indicated by a tilde
placed over the mass symbol. In supersymmetric theories the ghost is
constant and brings no momentum into the Green functions, but
obviously the propagators of field and transformed field cannot
cancel since they belong to a fermion and a boson or vice
versa. Thus, generally, one inverse propagator survives which has to
be taken into account. In our formalism this is easily done by
absorbing it into the wavefunction of the BRST source. Consequently,
we associate the following expressions as wavefunctions to our BRST 
sources\footnote{This is based on an idea of Christian
Schwinn. $\epsilon$, $u$ and $v$ could be the ordinary wavefunctions
for external states which were off-shell in that context, or for the
most general tests unit four-vectors and unit four-spinors.} (gauge
boson sources in unitarity gauge): 
\begin{equation} 
     \setlength{\extrarowheight}{2pt}     
        \begin{array}{|c|c||c|c|}\hline
                \text{Particle} & \text{Wavefunction} &
                \text{Particle} & \text{Wavefunction} \\\hline\hline
                K_\phi, \text{incoming} & -\ii(p^2 - m^2) &
                K_\Psi, \text{incoming} & -\ii(\fmslash{p}-m)u(p) \\
                K_\phi, \text{outgoing} & -\ii(p^2 - m^2) &
                K_{\overline{\Psi}}, \text{outgoing} & \ii
                (\fmslash{p} + m) v(p) \\\cline{1-2}
                K_{A_\mu}, \text{incoming} & \ii (p^2 - m^2)
                \epsilon_\mu (p) & 
                K_\Psi, \text{outgoing} & -\ii\overline{u}(p)
                (\fmslash{p} - m) \\ 
                K_{A_\mu}, \text{outgoing} & \ii (p^2 - m^2)
                \epsilon^*_\mu (p) &
                K_{\overline{\Psi}}, \text{incoming} & \ii
                \overline{v}(p) (\fmslash{p} + m) \\\hline 
        \end{array}
\end{equation}
For Majorana fermions there are some minor changes; we will come back
to that point after having discussed another important topic -- the
handling of Fermi statistics with the inclusion of real fermions within
the framework of {\em O'Mega}.  

%%%%%%%%%%%%%%%%%%%%%%%%%%%%%%%%%%%%%%%%%%%%%%%%%%%%%%%%%%%%%%%%%%%%%%%%

\section{Fermi Statistics - Evaluation of Signs}

Fermions are anticommuting objects -- whenever their order is changed
in Green functions or different diagrams contributing to a Green
function they produce sign factors. These signs can be read off from the 
Wick theorem. They arise as relative signs (e.g. for Bhabha scattering between
$s$- and $t$-channel) between whole diagrams, but to avoid the explicit
construction of all diagrams was the strongest motivation for {\em
O'Mega}. How can we cope with Fermi statistics within its framework?
The solution was found by Thorsten Ohl 
\cite{Moretti/Ohl/Reuter/Schwinn:2001:Omega}: When fusing
two (or more) one-particle off-shell wavefunctions (1POWs), then the
sign of the newly produced 1POW is calculated and divided by the signs
of the subamplitudes out of which it was combined. Thus, for every fusion
only the sign relative to 
the subamplitudes is kept in memory, which makes evaluating the signs
compatible with the factorization procedure of {\em O'Mega}. What
remains is to answer the question how to evaluate those signs from
Fermi statistics for a subamplitude.

Each subamplitude depends on a subset of the external particles, part
of which are fermions while the others are not. The solution for
calculating the sign factors lies in a bookkeeping concept for fermion
lines: We ``follow the fermion lines'' through the graphs. At each
fusion we examine whether a fermion line is continued or, together with
another fermion line, fused into a boson line while keeping in mind all
``closed'' fermion lines appearing in the fused 1POWs. The best way 
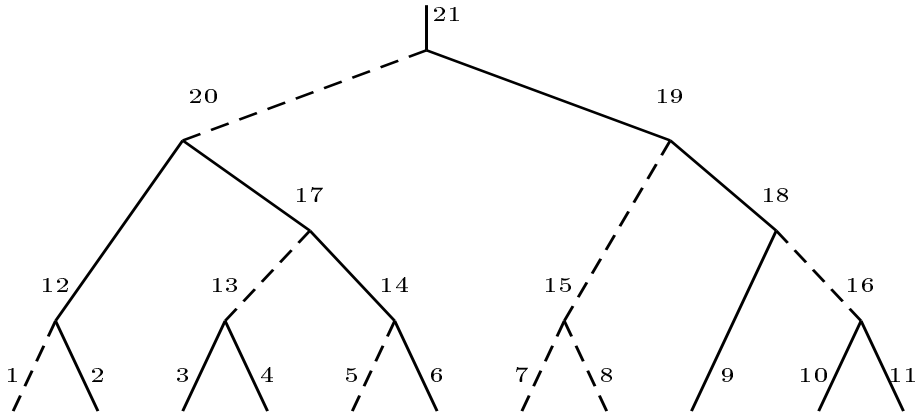
\begin{figure}
\resizebox{12cm}{6cm}{
\begin{pspicture}(0,0)(8.4,5)%  \psset{xunit=.8cm,yunit=.8cm}
        \psline[linestyle=dashed](0,0)(.4,1)    \psline(.8,0)(.4,1)
        \psline(1.6,0)(2,1)     \psline(2.4,0)(2,1)
        \psline[linestyle=dashed](3.2,0)(3.6,1) \psline(4,0)(3.6,1)
        \psline(7.6,0)(8,1)     \psline(8.4,0)(8,1)
        \psline[linestyle=dashed](2,1)(2.8,2)   \psline(3.6,1)(2.8,2)
        \psline[linestyle=dashed](8,1)(7.2,2)   \psline(6.4,0)(7.2,2)   
        \psline(.4,1)(1.6,3)    \psline(2.8,2)(1.6,3)
        \psline[linestyle=dashed](5.6,0)(5.2,1)
        \psline[linestyle=dashed](4.8,0)(5.2,1) 
        \psline[linestyle=dashed](5.2,1)(6.2,3) \psline(7.2,2)(6.2,3)
        \psline(6.2,3)(3.9,4)   \psline[linestyle=dashed](1.6,3)(3.9,4)
        \psline(3.9,4)(3.9,4.5)
        \rput(0,.4){\begin{scriptsize}1\end{scriptsize}}        
        \rput(.8,.4){\begin{scriptsize}2\end{scriptsize}}       
        \rput(1.6,.4){\begin{scriptsize}3\end{scriptsize}}      
        \rput(2.4,.4){\begin{scriptsize}4\end{scriptsize}}      
        \rput(3.2,.4){\begin{scriptsize}5\end{scriptsize}}      
        \rput(4,.4){\begin{scriptsize}6\end{scriptsize}}        
        \rput(4.8,.4){\begin{scriptsize}7\end{scriptsize}}      
        \rput(5.6,.4){\begin{scriptsize}8\end{scriptsize}}      
        \rput(6.75,.4){\begin{scriptsize}9\end{scriptsize}}     
        \rput(7.55,.4){\begin{scriptsize}10\end{scriptsize}}    
        \rput(8.4,.4){\begin{scriptsize}11\end{scriptsize}}     
        \rput(.4,1.4){\begin{scriptsize}12\end{scriptsize}}     
        \rput(2,1.4){\begin{scriptsize}13\end{scriptsize}}      
        \rput(3.6,1.4){\begin{scriptsize}14\end{scriptsize}}    
        \rput(5.15,1.4){\begin{scriptsize}15\end{scriptsize}}   
        \rput(8,1.4){\begin{scriptsize}16\end{scriptsize}}      
        \rput(2.8,2.4){\begin{scriptsize}17\end{scriptsize}}
        \rput(7.2,2.4){\begin{scriptsize}18\end{scriptsize}}     
        \rput(6.2,3.5){\begin{scriptsize}19\end{scriptsize}}    
        \rput(1.8,3.5){\begin{scriptsize}20\end{scriptsize}}    
        \rput(4.1,4.4){\begin{scriptsize}21\end{scriptsize}}    
\end{pspicture}}
\caption{\label{fermigraph} Example for Fermi statistics sign
evaluation. Fermions are denoted by plain lines, bosons by dashed
ones.} 
\end{figure}
to explain this is an example which is provided by
figure \ref{fermigraph}. There the numbers $1$-$11$ denote external
particles on which the 1POW labelled by $21$ depends; this 1POW is a
fermion. (It would be more consistent to label each line by the
external particles on which it depends, e.g. $(1,2)$ for $12$, $(3,4)$
for $13$, $(7,8,9,10,11)$ for $19$ etc.) So, the fermion $21$ is part
of an open fermion line beginning (or ending) with the external
particle $9$ while its 1POW contains the closed fermion lines
$(2\leftrightarrow 6)$, $(3\leftrightarrow 4)$ and $(10\leftrightarrow
11)$. And what about the direction of the lines? 

This is the point where the difference arises between the way Ohl
calculates the sign for theories with Dirac fermions only and
our way for theories with Dirac as well as Majorana fermions. The
Feynman rules for Majorana fermions and their implementation in {\em
O'Mega} rely upon the ideas in \cite{Denner/etal:1992:feynmanrules}. 
When a theory is endowed only with Dirac fermions and is not
supersymmetric, then always a well defined direction can be asserted to each
fermion line -- which is based on the fact that we can always
distinguish between incoming particle and outgoing antiparticle or
vice versa. Out of this concept of directed fermion lines Ohl further
specified, whether a fermionic 1POW is a fermion or an antifermion (it
would be better to say, a spinor or a conjugated spinor, respectively)
-- in the picture of figure \ref{fermigraph} whether the arrow at the
line points upward or downward, respectively. Closed fermion lines
then consist of pairs of numbers indicating the conjugated spinor of
the external particle at which the line begins, and the spinor of the
external particle at which the line ends. This is actually an
abuse of language, since the arrows point from the ``barred'' to the
``unbarred'' spinor, but when writing down an analytical expression we
start with the barred spinor since we write from left to right:
\begin{align*}
\parbox{22mm}{\hfil\\\hfil\\%
          \begin{fmfgraph*}(20,15)
                \fmfstraight
                \fmfleft{l1,l2,l3}         \fmfright{r1,r2,r3}
                \fmftop{t1,t2,t3,t4,t5}
                \fmf{fermion,label=$S_F^{(1)}$,l.side=left}{t2,l2}
                \fmf{fermion}{l2,l1}
                \fmf{fermion}{r1,r2}
                \fmf{fermion,label=$S_F^{(n-1)}$}{r2,t4}
                \fmf{dots}{t2,t4}    
                \fmfdot{l2,t2,t4,r2}
                \fmflabel{$\Gamma^{(1)}$}{l2}
                \fmflabel{$\Gamma^{(2)}$}{t2}
                \fmflabel{$\Gamma^{(n)}$}{r2}
                \fmflabel{$\Gamma^{(n-1)}$}{t4}
                \fmflabel{$\overline{\Psi}$}{l1}
                \fmflabel{$\Psi$}{r1}
          \end{fmfgraph*}\hfil\\}       \qquad \qquad \equiv \;\; 
        \overline{\Psi} \Gamma^{(1)} S_F^{(1)} \Gamma^{(2)} \ldots
        \Gamma^{(n-1)} S_F^{(n-1)} \Gamma^{(n)} \Psi 
\end{align*}

In Ohl's ansatz the sign from Fermi statistics is calculated by
collecting the closed fermion lines (running from the conjugated
spinor to the spinor) and then comparing them with a reference order
of the external fermions and antifermions. The number of
transpositions needed to bring the collected pairs of external
fermionic particles into that reference order gives the relative sign
between the two different orders. As an example we consider Bhabha
scattering (we denote the incoming electron and positron by $1$ and
$2$, respectively, the outgoing electron and positron by $3$ and $4$,
respectively):

\vspace{3mm}

\begin{align*}
        \parbox{18mm}{\hfil\\
        \begin{fmfgraph*}(16,16)
                \fmfstraight
                \fmftop{t1,t2,t3,t4,t5,t6,t7}
                \fmfbottom{b1,b2,b3,b4,b5,b6,b7}
                \fmf{fermion}{b3,v1,b1}
                \fmf{phantom}{v1,t2}
                \fmf{phantom}{b7,v2,b5}
                \fmf{phantom}{v2,t6}
                \fmffreeze
                \fmf{phantom}{v1,v,v2}
                \fmf{fermion}{t4,v}
                \fmffreeze
                \fmf{fermion}{v,b7}
                \fmf{photon}{v,v1}
                \fmflabel{$2$}{b1}
                \fmflabel{$1$}{b3}
                \fmflabel{$3$}{b7}      
                \fmflabel{$4$}{t4}
        \end{fmfgraph*}\hfil\\} \qquad \qquad 
        \parbox{18mm}{\hfil\\
        \begin{fmfgraph*}(16,16)
                \fmfstraight
                \fmftop{t1,t2,t3,t4,t5,t6,t7}
                \fmfbottom{b1,b2,b3,b4,b5,b6,b7}
                \fmf{fermion}{b7,v1,b5}
                \fmf{phantom}{v1,t6}
                \fmf{phantom}{b3,v2,b1}
                \fmf{phantom}{v2,t2}
                \fmffreeze
                \fmf{phantom}{v1,v,v2}
                \fmf{fermion}{t4,v}
                \fmffreeze
                \fmf{fermion}{v,b1}
                \fmf{photon}{v,v1}
                \fmflabel{$2$}{b1}
                \fmflabel{$3$}{b5}
                \fmflabel{$1$}{b7}      
                \fmflabel{$4$}{t4}
        \end{fmfgraph*}\hfil\\}
\end{align*}
The left diagram is the $s$-channel, the right one the $t$-channel. In
our formalism we get $(21)(34) \rightarrow \left\{ 2,1,3,4\right\}$
for the left diagram, and $(31)(24) \rightarrow \left\{ 3,1,2,4
\right\}$ for the right one\footnote{Normally the fusions run up only
to 1POWS depending on nearly half of the external momenta. In our
case, each possible pairing of the fermions would produce a photon, and
in a final step, two off-shell photon wavefunctions are
fused to yield the whole amplitude. Each of those photons would
contain one closed fermion line, but as pairs of numbers $( \,
\underline{\;\;}\, ,\underline{\;\;} \,)$ commute with each other (4
transpositions) the order of the pairs plays no role.}. If we take
$\left\{ 1,2,3,4 \right\}$ as our reference order then the $s$-channel
gets a relative sign, while the $t$-channel not; as the global sign
does not matter, the relative sign between the two channels is
produced\footnote{As discussed above, {\em O'Mega} evaluates the sign(s)
subamplitude-wise, but for human beings the principle is better
understandable when thinking in diagrams.}.  

We therefore have the following ``fusion rules'' for solely Dirac fermions
(again the dashed line is an unspecified boson), by which we mean the
collecting of the fermion lines to evaluate the sign from the number of
transpositions needed to bring the pairs of fermions into the reference order: 
\begin{align}
       \quad 
        \parbox{17mm}{\hfil\\\hfil\\
        \begin{fmfgraph*}(15,15)
                \fmftop{t}
                \fmfbottom{b1,b2,b3,b4,b5}
                \fmf{fermion}{b1,v,b5}
                \fmf{dashes}{v,t}
                \fmflabel{$\phi = \overline{\Psi}_b \Gamma \Psi_a$}{t}
                \fmflabel{$a,l_a$}{b2}
                \fmflabel{$b,l_b$}{b4}
                \fmflabel{$\begin{array}{c} \hfil \\ \boxed{
       \underline{\;\;} \, , \,  \left\{ b, a \right\} \cup l_a \cup l_b
                } \end{array}$}{b3} 
        \end{fmfgraph*}\hfil\\} \qquad \qquad \quad 
        \parbox{17mm}{\hfil\\\hfil\\   
        \begin{fmfgraph*}(15,15)
                \fmftop{t}
                \fmfbottom{b1,b2,b3,b4,b5}
                \fmf{fermion}{b5,v,b1}
                \fmf{dashes}{v,t}
                \fmflabel{$\phi=\overline{\Psi}_a \Gamma \Psi_b$}{t}
                \fmflabel{$a,l_a$}{b2}
                \fmflabel{$b,l_b$}{b4}
                \fmflabel{$\begin{array}{c} \hfil \\ \boxed{
        \underline{\;\;} \, , \,  \left\{ a, b \right\} \cup l_a \cup l_b
                }\end{array}$}{b3} 
        \end{fmfgraph*}\hfil\\} \qquad \qquad \quad 
        \parbox{17mm}{\hfil\\\hfil\\
        \begin{fmfgraph*}(15,15)
                \fmftop{t}
                \fmfbottom{b1,b2,b3,b4,b5}
                \fmf{fermion}{b1,v,t}
                \fmf{dashes}{b5,v}
                \fmflabel{$\Psi_a' = \phi \Gamma \Psi_a$}{t}
                \fmflabel{$a,l_a$}{b2}
                \fmflabel{$\underline{\;\;},l_b$}{b4}
                \fmflabel{$\begin{array}{c} \hfil \\ \boxed{
                    a \, , \,  l_a \cup l_b
                } \end{array}$}{b3} 
        \end{fmfgraph*}\hfil\\}  \qquad \qquad 
        \parbox{17mm}{\hfil\\\hfil\\
        \begin{fmfgraph*}(15,15)
                \fmftop{t}
                \fmfbottom{b1,b2,b3,b4,b5}
                \fmf{fermion}{t,v,b1}
                \fmf{dashes}{v,b5}
                \fmflabel{$\overline{\Psi}'_a = \phi \overline{\Psi}_a \Gamma$}{t}
                \fmflabel{$a,l_a$}{b2}
                \fmflabel{$\underline{\;\;},l_b$}{b4}
                \fmflabel{$\begin{array}{c} \hfil \\ \boxed{
                    a \, , \,  l_a \cup l_b
                } \end{array}$}{b3} 
        \end{fmfgraph*}\hfil\\}   \notag \\ \hfil \notag \\ \hfil \notag \\
\end{align}
Therein $l_a$ and $l_b$ are lists of antifermion--fermion pairs of external
particles belonging to lines already closed, contained in the 1POWs which
take part in that fusion. $a$ and $b$ are fermion labels for the left
and right leg 
in the fusion. In the two left fusions a boson is produced so there 
is no ``open'' fermion index as for the rightmost fusions where a
fermion (or antifermion) is produced again. What {\em
O'Mega} calculates numerically is shown above the diagrams. For the
last two diagrams there also exists a mirror version omitted here,
with the fermion leg on the right. 

The expressions in {\em O'Mega}'s Dirac fermion version are
produced numerically by a final closure of the fermion line where --
via the bilinear product $\overline{\Psi}_1 \Gamma \Psi_2$ -- a
conjugated spinor is multiplied with a spinor to give a non-spinorial
expression. For external particles we assign a spinor to incoming
fermions ($u$) and to outgoing antifermions ($v$), and use conjugated
spinors for outgoing fermions ($\overline{u}$) and incoming
antifermions ($\overline{v}$). Each conjugated spinor is continued
through the subamplitudes representing classes of subdiagrams by right 
multiplication with the vertex and propagator factors which consist of
linear combinations of gamma matrices. In contrast, spinors are left
multiplied by the corresponding factors when following the line,
respectively: 
\begin{equation}
        \overline{\Psi}' = \overline{\Psi} \Gamma \;\; \text{or} \;\;
        \overline{\Psi}'' = \overline{\Psi} S_F \qquad\qquad \Psi' =
        \Gamma \Psi \;\; \text{or} \;\; \Psi'' =  S_F \Psi 
\end{equation}
The bilinear product closing the fermion line could either produce a new
bosonic off-shell wavefunction or means the last keystone in
constructing a part of the whole amplitude. For the latter case it is
indeed a scalar product, e.g.~in the process $e^+e^- \to \gamma\gamma$,
where the positron wavefunction made out of the incoming positron and
one outgoing photon is fused with the electron wavefunction constructed 
from the incoming electron and the other photon. 

\vspace{2mm}

In supersymmetric theories there are two difficulties which make the 
implementation by Thorsten Ohl, described in the last paragraphs
inapplicable: the existence of Majorana fermions, i.e.~real fermions,
to whose lines no specific direction can be assigned since they do not
transport a conserved quantum number, and the possibility of
``clashing arrows'' already seen in our toy model in the first
part. So there are now six types of vertices involving fermions (the
dashed line indicates a boson of whatsoever type): 
\begin{align}
        \parbox{17mm}{\hfil\\
        \begin{fmfgraph*}(15,12)
                \fmftop{t}
                \fmfbottom{b1,b2}
                \fmf{fermion}{b2,v,b1}
                \fmf{dashes}{v,t}
                \fmfdot{v}
        \end{fmfgraph*}\hfil\\}  \quad
        \parbox{17mm}{\hfil\\
        \begin{fmfgraph*}(15,12)
                \fmftop{t}
                \fmfbottom{b1,b2}
                \fmf{fermion}{b2,v}
                \fmf{plain}{v,b1}
                \fmf{dashes}{v,t}
                \fmfdot{v}
        \end{fmfgraph*}\hfil\\}  \quad
        \parbox{17mm}{\hfil\\
        \begin{fmfgraph*}(15,12)
                \fmftop{t}
                \fmfbottom{b1,b2}
                \fmf{fermion}{v,b1}
                \fmf{plain}{v,b2}
                \fmfdot{v}
                \fmf{dashes}{v,t}
        \end{fmfgraph*}\hfil\\}  \quad
        \parbox{17mm}{\hfil\\
        \begin{fmfgraph*}(15,12)
                \fmftop{t}
                \fmfbottom{b1,b2}
                \fmf{plain}{b2,v,b1}
                \fmf{dashes}{v,t}
                \fmfdot{v}
        \end{fmfgraph*}\hfil\\}  \quad
        \parbox{17mm}{\hfil\\
        \begin{fmfgraph*}(15,12)
                \fmftop{t}
                \fmfbottom{b1,b2}
                \fmf{fermion}{b2,v}
                \fmf{fermion}{b1,v}
                \fmf{dashes}{v,t}
                \fmfdot{v}
        \end{fmfgraph*}\hfil\\}  \quad  
        \parbox{17mm}{\hfil\\
        \begin{fmfgraph*}(15,12)
                \fmftop{t}
                \fmfbottom{b1,b2}
                \fmf{fermion}{v,b1}
                \fmf{fermion}{v,b2}
                \fmf{dashes}{v,t}
                \fmfdot{v}
        \end{fmfgraph*}\hfil\\}
\end{align}
Since there no longer is a well-defined direction along the fermion lines
(there may be parts of lines having no direction at all due to propagating
Majorana fermions, or the line's direction changes by one of the rightmost 
vertices) the method presented above is no longer feasible. One possibility is
to artificially assign a direction to a Majorana line and define
different vertices for each case in which the arrows direct to or from
the vertex \cite{Haber/Kane:1985:SUSY}. This would give the correct
analytical expressions for the Feynman diagrams, but the signs between
different diagrams contributing to 
the same amplitude, e.g. for the production of two neutralinos at a linear
collider ($H$ is a shortcut for the three neutral Higgs bosons, $H^0, h^0,
A^0$),
\begin{equation*}
        \parbox{20mm}{\hfil\\
        \begin{fmfgraph*}(19,15)
          \fmfleft{l1,l2}
          \fmfright{r1,r2}
          \fmf{fermion}{l1,v1,l2}
          \fmf{plain}{r1,v2,r2}
          \fmf{dashes,label=$H$,l.side=left}{v1,v2} 
          \fmfdot{v1,v2}
        \end{fmfgraph*}\hfil\\}  \qquad
        \parbox{20mm}{\hfil\\
        \begin{fmfgraph*}(19,15)
          \fmfleft{l1,l2}
          \fmfright{r1,r2}
          \fmf{fermion}{l1,v1,l2}
          \fmf{plain}{r1,v2,r2}
          \fmf{photon,label=$Z^0$,l.side=left}{v1,v2}
          \fmfdot{v1,v2}       
        \end{fmfgraph*}\hfil\\}  \qquad
        \parbox{20mm}{\hfil\\
        \begin{fmfgraph*}(19,15)
          \fmfleft{l1,l2}
          \fmfright{r1,r2}
          \fmf{fermion}{v2,l2}
          \fmf{plain}{v2,r2}
          \fmf{fermion}{l1,v1}
          \fmf{plain}{v1,r1}
          \fmf{dashes,label=$\tilde{e}_{1/2}$}{v1,v2}
          \fmfdot{v1,v2}
        \end{fmfgraph*}\hfil\\} \qquad 
        \parbox{20mm}{\hfil\\
        \begin{fmfgraph*}(19,15)
          \fmfleft{l1,l2}
          \fmfright{r1,r2}
          \fmf{fermion}{v2,l2}
          \fmf{phantom}{v2,r2}
          \fmf{fermion}{l1,v1}
          \fmf{phantom}{v1,r1}
          \fmf{dashes,label=$\tilde{e}_{1/2}$,l.side=left}{v1,v2}
          \fmffreeze
          \fmf{plain}{v1,r2}
          \fmf{plain}{v2,r1}
          \fmfdot{v1,v2}
        \end{fmfgraph*}\hfil\\} \qquad , 
\end{equation*}
have to be derived by the Wick theorem, which is no good
solution for implementing such rules in a computer program. Instead we follow
the Feynman rules invented by Ansgar Denner et
al.~\cite{Denner/etal:1992:feynmanrules}. As definite directions along the
fermion lines are only partially available they give up the concept of fermion
number conservation and use {\em fermion conservation} as an alternative,
which is simply the statement that fermion lines must still run
through the diagrams without being
interrupted. \cite{Denner/etal:1992:feynmanrules} describe in detail
how the contractions of the fermionic field operators can be disentangled
with the help of that concept. The ingredients of Denner's rules are the
``ordinary'' vertices, propagators and external states as well as their
charge conjugated versions:
\begin{align}
  \mathcal{C} \overline{v}^T (p, \sigma) &=\; u (p, \sigma) \\
  \mathcal{C} \overline{u}^T (p, \sigma) &=\; v (p, \sigma) \\
  S'_F \equiv \mathcal{C} S_F^T (p) \mathcal{C}^{-1}  &=\; S_F (-p) =
  \dfrac{\ii}{-\fmslash{p} - m}  \\
  \Gamma' \equiv \mathcal{C} \Gamma^T \mathcal{C}^{-1} &=\; \left\{ 
    \begin{array}{ll}
      + \Gamma & \quad \text{for} \;\; \Gamma \equiv \mathbb{I}, \gamma^5
      \gamma^\mu, \gamma^\mu \\ 
      - \Gamma & \quad \text{for} \;\; \Gamma \equiv \gamma^\mu,
      \sigma^{\mu\nu} 
    \end{array} \right. , \label{gammagammastrich} 
\end{align}
where $\mathcal{C}$ is the charge conjugation matrix. (The basics of Denner's
idea for disentangling the contractions are to replace an interaction operator
bilinear in the fermionic field operators by its transpose, which should leave
everything unchanged as the bilinear is a number (or an array of
numbers), $\overline{\Psi}\Gamma\Psi = \left( \overline{\Psi} \Gamma
\Psi \right)^T$: 
\begin{equation}
        \label{entangle}
  \contracted{}{\stackrel{\stackrel{\hfil}{\hfil}}{\ldots}}{\;}{\contracted{}{\overline{\Psi}}{\Gamma\Psi\;\;\:}{\ldots}}{}
  \qquad \longrightarrow \qquad (-1) \cdot
  \contracted{}{\stackrel{\stackrel{\hfil}{\hfil}}{\ldots}}{}{\Psi^T\:}
  \mathcal{C}^{-1} \left(\mathcal{C} \Gamma^T \mathcal{C}^{-1} \right)
  \mathcal{C} \contracted{}{\;\;\overline{\Psi}^T}{}{\ldots} 
\end{equation}
The sign comes from anticommuting the field operators. The
contractions could have been got tangled up in that way, since all
four possible contractions between field operators and conjugated field
operators are allowed for Majorana fermions, and also due to the appearance of
explicitly charged conjugated fermions as in the chargino--lepton--slepton
vertex. For
Dirac fermions this happens only for the contractions of the field operators
with the asymptotic creation and annihilation operators for incoming or
outgoing antifermions, which have produced the global signs mentioned in the
first part of the text, and for closed fermion loops, of course. On
the level of the analytical expressions for the fermion lines within
Feynman diagrams the relation (\ref{entangle}) for complete lines reads
($w \in \left\{ u, v \right\}$, $w^c \in \left\{ u^c = v, v^c = u
\right\}$ with $w^c \equiv \mathcal{C} \overline{w}^T$)
\begin{equation}
        \label{entangle2}
        \overline{w}_1 \Gamma^{(1)} S_F^{(1)} \Gamma^{(2)} \ldots
        \Gamma^{(n)} w_2 = (-1) \cdot \overline{w^c_2} \, 
        \Gamma^{\prime\,(n)} \ldots S_F^{\prime\,(1)} \Gamma^{\prime\,
        (1)} w^c_1 , 
\end{equation}
where the sign now does not come from fermion anticommutation but from
the antisymmetry of the charge conjugation matrix: $w^T =
\overline{w^c} \, \mathcal{C}^T = - \overline{w^c} \,
\mathcal{C}$. This was also the basis for the ``reversing of fermion
lines'' for the asymptotic STI in part one. Both ways of calculating a
fermion line produce the same result as the sign in (\ref{entangle2}) is
cancelled by the one from fermion anticommutation in
(\ref{entangle}). The key ingredient of Denner's rules is to write down the
Feynman diagram(s), to choose a calculational direction for each
fermion line (i.e.~to decide where to start and where to end at a
given fermion line) and to write down the primed expressions when
the calculational direction is opposite to the arrows of a Dirac
fermion. Disentanglement of the contractions is thereby achieved 
automatically, and there is no need to explicitly use the Wick
theorem. The relative sign of different diagrams can be evaluated by
the method of permutations with respect to a reference order where the
pairs now are of the form ({\em endpoint}, {\em startpoint}) instead
of ({\em conjugated spinor}, {\em spinor}). For details
cf.~\cite{Denner/etal:1992:feynmanrules}.)
\begin{figure}
\[
     \setlength{\extrarowheight}{2pt}     
     \setlength{\fboxsep}{3mm}
        \begin{array}{cccc}
        \text{Fermion} & \;\;\;\text{Antifermion} & \;\;\text{Majorana
     fermion} & \text{Assignment} \\ & & & \\ 
          \parbox{18mm}{\hfil\\        
        \begin{fmfgraph*}(17,5)
          \fmfleft{l1,l2}
          \fmfright{r1,r2}
          \fmfbottomn{b}{8}
          \fmf{fermion}{l2,r2}
          \fmf{dots}{b2,b6}
          \fmf{phantom_arrow}{b6,b7}
          \fmfv{decor.shape=circle,decor.filled=30,
                         decor.size=3mm}{l2}  
        \end{fmfgraph*}} &  \qquad \parbox{18mm}{\hfil\\
        \begin{fmfgraph*}(17,5)
          \fmfleft{l1,l2}
          \fmfright{r1,r2}
          \fmfbottomn{b}{8}
          \fmf{fermion}{r2,l2}
          \fmf{dots}{b2,b6}
          \fmf{phantom_arrow}{b6,b7}
          \fmfv{decor.shape=circle,decor.filled=30,
                         decor.size=3mm}{l2}  
        \end{fmfgraph*}}        &  \qquad \parbox{18mm}{\hfil\\
        \begin{fmfgraph*}(17,5)
          \fmfleft{l1,l2}
          \fmfright{r1,r2}
          \fmfbottomn{b}{8}
          \fmf{plain}{l2,r2}
          \fmf{dots}{b2,b6}
          \fmf{phantom_arrow}{b6,b7}
          \fmfv{decor.shape=circle,decor.filled=30,
                         decor.size=3mm}{l2}  
        \end{fmfgraph*}}        & \qquad\overline{u}(p,\sigma) \\ & & & \\  
          \framebox{\parbox{18mm}{\hfil\\              
        \begin{fmfgraph*}(17,5)
          \fmfleft{l1,l2}
          \fmfright{r1,r2}
          \fmfbottomn{b}{8}
          \fmf{fermion}{l2,r2}
          \fmf{dots}{b3,b7}
          \fmf{phantom_arrow}{b3,b2}
          \fmfv{decor.shape=circle,decor.filled=30,
                         decor.size=3mm}{l2}  
        \end{fmfgraph*}}} &  \qquad \framebox{\parbox{18mm}{\hfil\\
        \begin{fmfgraph*}(17,5)
          \fmfleft{l1,l2}
          \fmfright{r1,r2}
          \fmfbottomn{b}{8}
          \fmf{fermion}{r2,l2}
          \fmf{dots}{b3,b7}
          \fmf{phantom_arrow}{b3,b2}
          \fmfv{decor.shape=circle,decor.filled=30,
                         decor.size=3mm}{l2}  
        \end{fmfgraph*}}}       &  \qquad \framebox{\parbox{18mm}{\hfil\\
        \begin{fmfgraph*}(17,5)
          \fmfleft{l1,l2}
          \fmfright{r1,r2}
          \fmfbottomn{b}{8}
          \fmf{plain}{l2,r2}
          \fmf{dots}{b3,b7}
          \fmf{phantom_arrow}{b3,b2}
          \fmfv{decor.shape=circle,decor.filled=30,
                         decor.size=3mm}{l2}  
        \end{fmfgraph*}}}       & \qquad v(p,\sigma) \\ & & & \\ 
          \framebox{\parbox{18mm}{\hfil\\              
        \begin{fmfgraph*}(17,5)
          \fmfleft{l1,l2}
          \fmfright{r1,r2}
          \fmfbottomn{b}{8}
          \fmf{fermion}{l2,r2}
          \fmf{dots}{b2,b6}
          \fmf{phantom_arrow}{b6,b7}
          \fmfv{decor.shape=circle,decor.filled=30,
                         decor.size=3mm}{r2}  
        \end{fmfgraph*}}} &  \qquad \framebox{\parbox{18mm}{\hfil\\
        \begin{fmfgraph*}(17,5)
          \fmfleft{l1,l2}
          \fmfright{r1,r2}
          \fmfbottomn{b}{8}
          \fmf{fermion}{r2,l2}
          \fmf{dots}{b2,b6}
          \fmf{phantom_arrow}{b6,b7}
          \fmfv{decor.shape=circle,decor.filled=30,
                         decor.size=3mm}{r2}  
        \end{fmfgraph*}}}       &  \qquad \framebox{\parbox{18mm}{\hfil\\
        \begin{fmfgraph*}(17,5)
          \fmfleft{l1,l2}
          \fmfright{r1,r2}
          \fmfbottomn{b}{8}
          \fmf{plain}{l2,r2}
          \fmf{dots}{b2,b6}
          \fmf{phantom_arrow}{b6,b7}
          \fmfv{decor.shape=circle,decor.filled=30,
                         decor.size=3mm}{r2}  
        \end{fmfgraph*}}}       & \qquad u (p,\sigma) \\ & & & \\ 
          \parbox{18mm}{\hfil\\        
        \begin{fmfgraph*}(17,5)
          \fmfleft{l1,l2}
          \fmfright{r1,r2}
          \fmfbottomn{b}{8}
          \fmf{fermion}{l2,r2}
          \fmf{dots}{b3,b7}
          \fmf{phantom_arrow}{b3,b2}
          \fmfv{decor.shape=circle,decor.filled=30,
                         decor.size=3mm}{r2}  
        \end{fmfgraph*}} &  \qquad \parbox{18mm}{\hfil\\
        \begin{fmfgraph*}(17,5)
          \fmfleft{l1,l2}
          \fmfright{r1,r2}
          \fmfbottomn{b}{8}
          \fmf{fermion}{r2,l2}
          \fmf{dots}{b3,b7}
          \fmf{phantom_arrow}{b3,b2}
          \fmfv{decor.shape=circle,decor.filled=30,
                         decor.size=3mm}{r2}  
        \end{fmfgraph*}}        &  \qquad \parbox{18mm}{\hfil\\
        \begin{fmfgraph*}(17,5)
          \fmfleft{l1,l2}
          \fmfright{r1,r2}
          \fmfbottomn{b}{8}
          \fmf{plain}{l2,r2}
          \fmf{dots}{b3,b7}
          \fmf{phantom_arrow}{b3,b2}
          \fmfv{decor.shape=circle,decor.filled=30,
                         decor.size=3mm}{r2}  
        \end{fmfgraph*}}        & \qquad \overline{v}(p,\sigma)
        \end{array} \]
\caption{\label{dennerext} Denner's rules for external fermionic
particles depending upon the direction along which the line is
calculated. Only those in boxes are used for the implementation.} 
\end{figure}

The biggest problem of the incorporation of Denner's rules 
is the factorization procedure used in {\em O'Mega}:
We treat only subamplitudes by successively building up 1POWs, so we do
not know where a beginning fermion line ends (or, where the beginning
of an ending line is. If we knew this, we could choose a calculational 
direction and tell the program how to calculate the line
numerically). At first we arbitrarily assume the calculational
direction to point from the external fermion inwards into the
(sub-)amplitude. Therefore we must assign a spinor instead of a
conjugated spinor to each external fermion. According to Denner the
assignments in table \ref{dennerext} for external fermionic particles
have to be made, wherein the dotted line indicates the chosen calculational
direction. As we use only
spinors for the external particles in {\em O'Mega}, just the
cases in boxes are relevant for us. Consequently every incoming
fermion of whatsoever type is represented by a $u$ spinor while to every
outgoing fermion a $v$ spinor is assigned. But this means that we can
totally forget about conjugated spinors now, since the lines beginning
with a spinor are continued by left multiplication with gamma matrices,
by which a spinor is produced again. We simply have to take $\Psi^T_1
\mathcal{C} \Gamma \Psi_2$ as a bilinear product instead of
$\overline{\Psi}_1 \Gamma \Psi_2$, and so conjugated spinors and right
multiplication with gamma matrices are completely eliminated when
dealing with fermions of mixed types. This choice for our bilinear
product solves another problem: When fusing open ends of two fermion
lines to a bosonic wavefunction (or for the final keystone) there is a 
mismatch in the calculational directions as for both legs they point
from bottom to top:        
\begin{equation}
\parbox{13mm}{
\begin{pspicture}(0,0)(1.2,1)%  \psset{xunit=.8cm,yunit=.8cm}
        \psline(0.1,0)(0.6,0.5) \psline(1.1,0)(0.6,0.5)
        \psline[linestyle=dashed](0.6,0.5)(0.6,1)
        \psline[linestyle=dotted]{->}(0.0,0.2)(0.4,0.6)
        \psline[linestyle=dotted]{->}(1.2,0.2)(0.8,0.6)
\end{pspicture}} \quad \Longrightarrow \quad 
\parbox{13mm}{
\begin{pspicture}(0,0)(1.2,1)%  \psset{xunit=.8cm,yunit=.8cm}
        \psline(0.1,0)(0.6,0.5) \psline(1.1,0)(0.6,0.5)
        \psline[linestyle=dashed](0.6,0.5)(0.6,1)
        \pscurve[linestyle=dotted]{->}(0.0,0.2)(0.4,0.6)(0.8,0.6)(1.2,0.2)
\end{pspicture}}
\quad \text{or} \quad
\fbox{
\parbox{13mm}{
\begin{pspicture}(0,0)(1.2,1)%  \psset{xunit=.8cm,yunit=.8cm}
        \psline(0.1,0)(0.6,0.5) \psline(1.1,0)(0.6,0.5)
        \psline[linestyle=dashed](0.6,0.5)(0.6,1)
        \pscurve[linestyle=dotted]{->}(1.2,0.2)(0.8,0.6)(0.4,0.6)(0.0,0.2)
\end{pspicture}}}
\end{equation}
As indicated in the figure there are now two equal alternatives,
to reverse either the right or the left calculational direction. In
{\em O'Mega} we have chosen the second alternative so that the
calculational direction goes from the right to the left. The
conjugated spinor $\overline{\Psi}$ is the same as ${\Psi^c}^T
\mathcal{C}$, but we take -- as mentioned above -- the product $\Psi^T
\mathcal{C} \Gamma \Psi$ and not ${\Psi^c}^T \mathcal{C} \Gamma
\Psi$. The solution to that obstacle is the following: The part of the
line coming from the left into the fusion has been calculated starting
from the external particle against the calculational direction chosen to
hold after the fusion of two fermion lines. Assume e.g. the left fermion
line simply to be an external incoming fermion; due to the rules in
table \ref{dennerext} {\em O'Mega} takes a $u$ spinor for that
particle but when {\em O'Mega} performs the fusion, it chooses the
calculational direction of the whole line -- now closed by the fusion
-- to go from the right leg of the fusion to the left. Hence, according to
this direction, we actually would have had to take $\overline{v} = u^T
\mathcal{C}$ as external wavefunction, so that performing the fusion like
$\Psi_{\text{left}}^T \mathcal{C} \Gamma \Psi_{\text{right}}$ with
$\Psi_{\text{left}} \equiv u$ is completely correct. The inclusion of
propagators and vertex factors for the left leg will be discussed
below. 

The ``fusion rules'' for fermions of mixed types are:
\begin{align}
\label{fusionrules}
\quad 
\parbox{26mm}{
\begin{pspicture}(0,0)(2.4,3.4)%  \psset{xunit=.8cm,yunit=.8cm}
        \psline(0.2,0.5)(1.2,1.5)       \psline(2.2,0.5)(1.2,1.5)
        \psline[linestyle=dashed](1.2,1.5)(1.2,2.5)
        \pscurve[linestyle=dotted]{->}(2.4,0.9)(1.6,1.7)(0.8,1.7)(0.0,0.9)
        \rput(.2,.2){$a\,,\,l_a$}
        \rput(2.2,.2){$b\,,\,l_b$}
        \rput(0,2.2){$\underline{\;\;} \, , \,  \left\{ a, b \right\}
        \cup l_a \cup l_b \quad$} 
        \rput(1.2,3.1){$\boxed{\phi = \Psi_a^T \mathcal{C} \Gamma^{(\prime)} \Psi_b}$}
\end{pspicture}} \qquad \quad
\parbox{26mm}{
\begin{pspicture}(0,0)(2.4,3.4)%  \psset{xunit=.8cm,yunit=.8cm}
        \psline[linestyle=dashed](0.2,0.5)(1.2,1.5)     \psline(2.2,0.5)(1.2,1.5)
        \psline(1.2,1.5)(1.2,2.5)
        \pscurve[linestyle=dotted]{->}(2.4,0.9)(1.7,1.6)(1.6,1.7)(1.5,2.4)
        \rput(.2,.2){$\underline{\;\;} \,,\,l_a$}
        \rput(2.2,.2){$b\,,\,l_b$}
        \rput(.4,2.2){$b \, , \,  l_a \cup l_b \quad$} 
        \rput(1.2,3.1){$\boxed{\Psi'_b = \phi \Gamma^{(\prime)} \Psi_b}$}
\end{pspicture}} \qquad \quad 
\parbox{26mm}{
\begin{pspicture}(0,0)(2.4,3.4)%  \psset{xunit=.8cm,yunit=.8cm}
        \psline(0.2,0.5)(1.2,1.5)       \psline[linestyle=dashed](2.2,0.5)(1.2,1.5)
        \psline(1.2,1.5)(1.2,2.5)
        \pscurve[linestyle=dotted]{->}(0.0,0.9)(0.7,1.6)(0.8,1.7)(0.9,2.4)
        \rput(.2,.2){$a\,,\,l_a$}
        \rput(2.2,.2){$\underline{\;\;} \,,\,l_b$}
        \rput(0,2.2){$a \, , \, l_a \cup l_b \quad$} 
        \rput(1.2,3.1){$\boxed{\Psi'_a = \phi \Gamma^{(\prime)} \Psi_a}$}
\end{pspicture}} \notag\\ 
\end{align}
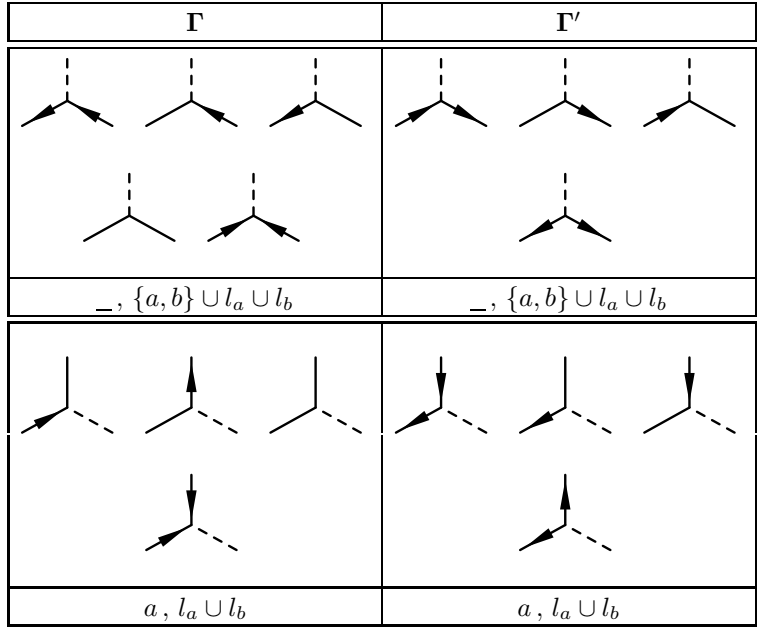
\begin{figure}
\[  
\setlength{\extrarowheight}{2pt}    
\begin{array}{|c|c|}\hline
        {\bf \Gamma} & {\bf \Gamma'} \\\hline\hline     
        \parbox{13mm}{\hfil\\
        \begin{fmfgraph*}(12,10)
          \fmftop{t}
          \fmfbottom{b1,b2}
          \fmf{fermion}{v,b1}   \fmf{fermion}{b2,v}
          \fmf{dashes}{v,t}       
        \end{fmfgraph*}} \quad 
        \parbox{13mm}{\hfil\\
        \begin{fmfgraph*}(12,10)
          \fmftop{t}
          \fmfbottom{b1,b2}
          \fmf{plain}{v,b1}     \fmf{fermion}{b2,v}
          \fmf{dashes}{v,t}       
        \end{fmfgraph*}} \quad 
        \parbox{13mm}{\hfil\\
        \begin{fmfgraph*}(12,10)
          \fmftop{t}
          \fmfbottom{b1,b2}
          \fmf{plain}{v,b2}     \fmf{fermion}{v,b1}
          \fmf{dashes}{v,t}       
        \end{fmfgraph*}}        
 & 
        \parbox{13mm}{\hfil\\
        \begin{fmfgraph*}(12,10)
          \fmftop{t}
          \fmfbottom{b1,b2}
          \fmf{fermion}{b1,v}   \fmf{fermion}{v,b2}
          \fmf{dashes}{v,t}       
        \end{fmfgraph*}} \quad 
        \parbox{13mm}{\hfil\\
        \begin{fmfgraph*}(12,10)
          \fmftop{t}
          \fmfbottom{b1,b2}
          \fmf{plain}{v,b1}     \fmf{fermion}{v,b2}
          \fmf{dashes}{v,t}       
        \end{fmfgraph*}} \quad 
        \parbox{13mm}{\hfil\\
        \begin{fmfgraph*}(12,10)
          \fmftop{t}
          \fmfbottom{b1,b2}
          \fmf{plain}{v,b2}     \fmf{fermion}{b1,v}
          \fmf{dashes}{v,t}       
        \end{fmfgraph*}} 
\\ & \\
        \parbox{13mm}{\hfil\\
        \begin{fmfgraph*}(12,10)
          \fmftop{t}
          \fmfbottom{b2,b1}
          \fmf{plain}{v,b1}     \fmf{plain}{b2,v}
          \fmf{dashes}{v,t}       
        \end{fmfgraph*}} \quad 
        \parbox{13mm}{\hfil\\
        \begin{fmfgraph*}(12,10)
          \fmftop{t}
          \fmfbottom{b2,b1}
          \fmf{fermion}{b1,v}   \fmf{fermion}{b2,v}
          \fmf{dashes}{v,t}       
        \end{fmfgraph*}} 
  &
        \parbox{13mm}{\hfil\\
        \begin{fmfgraph*}(12,10)
          \fmftop{t}
          \fmfbottom{b2,b1}
          \fmf{fermion}{v,b2}   \fmf{fermion}{v,b1}
          \fmf{dashes}{v,t}       
        \end{fmfgraph*}} 
         \\ & \\\hline 
        \underline{\;\;} \, , \, \left\{ a , b \right\} \cup l_a \cup l_b & 
        \underline{\;\;} \, , \, \left\{ a , b \right\} \cup l_a \cup
        l_b \\\hline\hline \vspace{1pt}
        \parbox{13mm}{\hfil\\\hfil\\
        \begin{fmfgraph*}(12,10)
          \fmftop{t}
          \fmfbottom{b2,b1}
          \fmf{plain}{v,t}      \fmf{fermion}{b2,v}
          \fmf{dashes}{b1,v}      
        \end{fmfgraph*}} \quad 
        \parbox{13mm}{\hfil\\\hfil\\
        \begin{fmfgraph*}(12,10)
          \fmftop{t}
          \fmfbottom{b2,b1}
          \fmf{plain}{v,b2}     \fmf{fermion}{v,t}
          \fmf{dashes}{b1,v}      
        \end{fmfgraph*}} \quad
        \parbox{13mm}{\hfil\\\hfil\\
        \begin{fmfgraph*}(12,10)
          \fmftop{t}
          \fmfbottom{b2,b1}
          \fmf{plain}{b2,v}     \fmf{plain}{v,t}
          \fmf{dashes}{b1,v}      
        \end{fmfgraph*}} & 
        \parbox{13mm}{\hfil\\\hfil\\
        \begin{fmfgraph*}(12,10)
          \fmftop{t}
          \fmfbottom{b2,b1}
          \fmf{fermion}{t,v}    \fmf{fermion}{v,b2}
          \fmf{dashes}{b1,v}      
        \end{fmfgraph*}} \quad 
                \parbox{13mm}{\hfil\\\hfil\\
        \begin{fmfgraph*}(12,10)
          \fmftop{t}
          \fmfbottom{b2,b1}
          \fmf{plain}{v,t}      \fmf{fermion}{v,b2}
          \fmf{dashes}{b1,v}      
        \end{fmfgraph*}} \quad 
        \parbox{13mm}{\hfil\\\hfil\\
        \begin{fmfgraph*}(12,10)
          \fmftop{t}
          \fmfbottom{b2,b1}
          \fmf{plain}{b2,v}     \fmf{fermion}{t,v}
          \fmf{dashes}{b1,v}      
        \end{fmfgraph*}}
         \\ & \\ 
        \parbox{13mm}{\hfil\\
        \begin{fmfgraph*}(12,10)
          \fmftop{t}
          \fmfbottom{b2,b1}
          \fmf{fermion}{b2,v}   \fmf{fermion}{t,v}
          \fmf{dashes}{b1,v}      
        \end{fmfgraph*}} 
        & 
        \parbox{13mm}{\hfil\\
        \begin{fmfgraph*}(12,10)
          \fmftop{t}
          \fmfbottom{b2,b1}
          \fmf{fermion}{v,t}    \fmf{fermion}{v,b2}
          \fmf{dashes}{b1,v}      
        \end{fmfgraph*}}        
        \\ & \\\hline 
        a \, , \,  l_a \cup l_b  & a \, , \,  l_a \cup l_b \\\hline
   \end{array}
\]
\caption{\label{fusionmaj} ``Fusion rules'' for fermions of Dirac as well
as Majorana type. For the fusions where a fermion is produced, there
are also the mirror diagrams again. $a$ and $b$ are the fermion labels
for the left fermion and the right fermion, respectively, and $l_a$
and $l_b$ are the corresponding closed fermion lines. For the
conventions concerning the fusions with clashing arrows and with two
Majorana fermions cf.~the text.}  
\end{figure}
As was discussed in the last but one paragraph, the pairings are now
({\em endpoint}, {\em startpoint}) instead of ({\em conjugated
spinor}, {\em spinor}). One part of the Fermi statistics' sign is
calculated as in the Dirac case from the number of transpositions
needed to bring this collection of pairs into a reference order. In
{\em O'Mega} the calculational direction always goes from the right
leg of the fusion to the left, so the pair added to the list of closed
fermion lines is $\left\{ a, b \right\}$ each time, where $a$ is the
fermion index of the left leg and $b$ that of the right one. When a
line is continued as in the rightmost fusions in (\ref{fusionrules}),
then simply the gamma matrix from the vertex is multiplied from the
left to the child spinor. The remaining part of the Fermi statistics'
sign is produced from those gamma matrix vertex factors (and the
propagators, cf.~below), which also answers the question about the
meaning of the primes at the $\Gamma$ in
(\ref{fusionrules}): According to Denner's rules in
(\ref{gammagammastrich}) and the discussion in the text thereafter we
must assign a primed vertex function whenever the calculational 
direction is opposite to the direction of an arrow at a vertex. As the
calculational direction in {\em O'Mega} always points from right to
left or from bottom to top, respectively, we must make the
assignments shown in table \ref{fusionmaj} for the ``fusion rules'' at
the vertices. $\Gamma$ and $\Gamma'$ refer to the property in
(\ref{gammagammastrich}) according to which in the left column always
the ordinary 
vertex factor has to be taken with no additional sign, while for the
right column there is a sign if the vertex $\Gamma$ represents a
vectorial or tensorial coupling. Some remarks about the Majorana lines
and the clashing arrows: In the case of the clashing
arrows we must define the vertices in the {\em O'Mega} model files by
$\overline{\Psi^c_1} \Gamma \Psi_2$ instead of $\overline{\Psi_1}
\Gamma \Psi_2^c$; if we had defined them the other way round, the diagrams with
clashing arrows would have to be exchanged between the two columns in
table \ref{fusionmaj}. There are no ambiguities for vertices with one
Majorana and one Dirac line (e.g. the electron--selectron--neutralino
vertex) because the Dirac fermion automatically gives a direction. For
vertices with two Majorana fermions the case is more complex: If the
two Majorana fermions are identical, the coupling has to be scalar,
pseudoscalar or axial-vectorial, hence there is no problem with signs
(otherwise that part of the interaction Lagrangean vanishes
identically). Consider now the case where the Majorana fermions are
different, e.g. in the vertex between different 
MSSM neutralinos and the $Z$ boson. Here one has to decide whether to 
write the vertex as
\begin{equation*}
        \overline{\tilde{\chi}^0_i} \left( g_V + g_A \gamma^5 \right)
        \fmslash{Z} \tilde{\chi}^0_j \qquad \text{or} \qquad
        \overline{\tilde{\chi}^0_j} \left( - g_V + g_A \gamma^5 \right)
        \fmslash{Z} \tilde{\chi}^0_i \;\; , \quad i \neq j \;\;. 
\end{equation*}    
In the same way we have to choose which of the two possibilities
should be included in the {\em O'Mega} model file. By taking one of
the two versions, we implicitly introduce an arrow: 
\begin{equation*}
        \parbox{18mm}{\hfil\\\hfil\\
        \begin{fmfgraph*}(17,15)
          \fmfleft{l1,l2}
          \fmfright{r}
          \fmfbottom{b}
          \fmf{fermion,label=$\tilde{\chi}^0_j$,l.side=left}{l1,v}      
          \fmf{fermion,label=$\overline{\tilde{\chi}^0_i}$,l.side=right}{v,l2}
          \fmf{photon}{v,r}     
          \fmfdot{v}
          \fmflabel{$\boxed{\overline{\tilde{\chi}^0_i} \left( g_V + g_A
                        \gamma^5 \right) \fmslash{Z} \tilde{\chi}^0_j}$}{b}
        \end{fmfgraph*}\hfil\\\hfil\\} \qquad \qquad    \qquad \qquad 
        \parbox{18mm}{\hfil\\\hfil\\
        \begin{fmfgraph*}(17,15)
          \fmfleft{l1,l2}
          \fmfright{r}
          \fmfbottom{b}
          \fmf{fermion,label=$\tilde{\chi}^0_i$,l.side=left}{l1,v}      
          \fmf{fermion,label=$\overline{\tilde{\chi}^0_j}$,l.side=right}{v,l2}
          \fmf{photon}{v,r}
          \fmfdot{v}
          \fmflabel{$\boxed{\overline{\tilde{\chi}^0_j} \left( - g_V + g_A
                        \gamma^5 \right) \fmslash{Z} \tilde{\chi}^0_i}$}{b}
        \end{fmfgraph*}\hfil\\\hfil\\}
\end{equation*}
Consider the left of the possibilities above: Here the neutralino $i$ becomes
the particle with the arrow pointing away from the vertex whereas the
neutralino $j$ gets the arrow pointing to the vertex. For the right
possibility the arrows are exchanged. That ``pseudo-assigning of
arrows'' means that when contracting the field operators of that
interaction vertex with external states or other interaction vertices,
then we had to write down a conjugated spinor for the first neutralino
and a spinor for the second in the left case, and vice versa for the
right possibility. 

In table \ref{fusionmaj} it is assumed that the vertex for two
Majorana fermions is always written down in such a manner that in the case
of fusioning two fermions 
the left one is the conjugated while in the case of the fermion line
being continued, the fermion fused from the children is the
conjugated. Henceforth no primed vertex factors have to be used. (In
practice, there is a unique representation in {\em O'Mega} for such
vertices, so when the fusion does not match that
representation, then there {\em do} appear signs in front of the
vertex factors. E.g.~when we denote that neutralino neutral current by
the left possiblity above, but the second neutralino appears as a left leg
at a fusion and the first as a right leg, then the vector coupling
constant has to be endowed with an extra minus~\footnote{Do not get
confused when reading \cite{Denner/etal:1992:feynmanrules}: There
the abbreviaton $\overline{\chi} \Gamma \chi$ means $g^a_{ij}
\overline{\chi}_i \Gamma^a \chi_j$ so when reversing the order of the
field operators in that interaction vertex for two Majorana fermions
yields $g^a_{ij} \overline{\chi}_j \Gamma'^a \chi_i$. If $\Gamma'^a$ has
a relative sign with respect to $\Gamma^a$ then we also have $g^a_{ji} =
- g_{ij}^a$ and in Denner's notation we get $\overline{\chi} \Gamma
\chi$ again. But in {\em O'Mega} the coupling constant has
a pre-defined value and we must get an additional sign when reversing
the order of field operators for vectorial and tensorial couplings
even for Majorana fermions.}.)  

The last open point for the handling of real fermions in {\em O'Mega}
is the question of the propagators. First of all, we must say a word
about the momentum flow in {\em O'Mega}: It is always outgoing
(pointing out of an amplitude), i.e.~in those fusion diagrams the
momentum flows always from top to bottom (for all vertices):   
\begin{equation}
\parbox{26mm}{
\begin{pspicture}(0,0)(2.4,2)%  \psset{xunit=.8cm,yunit=.8cm}
        \psline[linestyle=dashed](0.2,0)(1.2,1) 
        \psline[linestyle=dashed](2.2,0)(1.2,1)
        \psline[linestyle=dashed](1.2,1)(1.2,2)
        \psline{->}(.6,0.8)(0,0.2)
        \psline{->}(1.8,0.8)(2.4,0.2)
        \psline{->}(1.5,1.9)(1.5,1.1)
        \rput(0,0.6){$p_l$}
        \rput(2.4,0.6){$p_r$}   
        \rput(2.2,1.5){$p_l + p_r$}
\end{pspicture}}
\end{equation}
All momenta can be expressed as a sum of the external momenta, so each
momentum can be uniquely labelled by a subset of the external
particles. After a fusion has taken place, a fermionic
wavefunction is multiplied by a propagator. Thus, every wavefunction
appearing as a child (left or right leg) in a fusion is either a
wavefunction of an external fermion or has already been multiplied with a
propagator. An exception occurs if the wavefunction is the final
keystone of a subamplitude; then to only one of the fermionic
wavefunctions the propagator has to be assigned. Hence, a propagator is
inserted in all fusion cases where the fermion line is not closed but
runs through to the top. At first one could believe that more than one
propagator type is needed when handling Dirac and Majorana fermions,
but fortunately this is not true. In connection with Denner's rules
above we mentioned that, if the calculational direction is opposite to
the arrow of the Dirac line, then we must use the primed propagator,
i.e.~the propagator with the negative momentum. Of course, also a
momentum flow opposite to the arrow direction results in a minus sign
for the momentum. Altogether the fermion propagator is of the form 
\begin{equation}
        \parbox{18mm}{\hfil\\\hfil\\
        \begin{fmfgraph*}(17,9)
          \fmfleft{l}
          \fmfright{r}
          \fmfbottomn{b}{9}
          \fmftopn{t}{9}
          \fmf{fermion}{l,r}
          \fmf{dashes}{b3,b7}
          \fmf{dots}{t3,t7}
          \fmf{phantom_arrow}{b3,b2}    \fmf{phantom_arrow}{b7,b8}
          \fmf{phantom_arrow}{t3,t2}    \fmf{phantom_arrow}{t7,t8}
          \fmflabel{$p$}{b5}
        \end{fmfgraph*}\hfil\\} \quad =
        \dfrac{\ii}{\xi \fmslash{p} - m}, \qquad\qquad
        \parbox{18mm}{\hfil\\\hfil\\
        \begin{fmfgraph*}(17,9)
          \fmfleft{l}
          \fmfright{r}
          \fmfbottomn{b}{9}
          \fmftopn{t}{9}
          \fmf{plain}{l,r}
          \fmf{dashes}{b3,b7}
          \fmf{dots}{t3,t7}
          \fmf{phantom_arrow}{b3,b2}    \fmf{phantom_arrow}{b7,b8}
          \fmf{phantom_arrow}{t3,t2}    \fmf{phantom_arrow}{t7,t8}
          \fmflabel{$p$}{b5}
        \end{fmfgraph*}\hfil\\} \quad =
        \dfrac{\ii}{\zeta \fmslash{p} - m}
\end{equation}
where $\xi$ and $\zeta$ are sign factors. $\xi$ is $+1$ if the
calculational direction and the momentum flow are both parallel or
both antiparallel to the fermion's arrow, and $-1$ otherwise. For
Majorana fermions $\zeta$ is $+1$ if calculational direction and
momentum flow are parallel and $-1$ if they are antiparallel. The
following table shows that this always leads to a negative sign for
the propagator's momentum within {\em O'Mega}:

\begin{center}
\begin{tabular}{|l|c|c|c|c|}\hline
  fermion type & fermion arrow & mom. & calc. & sign \\\hline\hline
  Dirac fermion &     $\uparrow$  &  $\downarrow$ &
    $\uparrow$ & negative \\\hline
  Dirac antifermion & $\downarrow$ & $\downarrow$ &
    $\uparrow$ & negative   \\\hline
  Majorana fermion  & - & $\downarrow$ & $\uparrow$ & negative \\\hline
\end{tabular}
\end{center}                 
So the universally used fermion propagator for all types of fermions --
fermions, antifermions and Majorana fermions -- is
\begin{equation}
        \boxed{ S_{F,\text{\em O'Mega}} = \dfrac{\ii}{- \fmslash{p} -
        m} } 
\end{equation}

Now we are able to convince ourselves that everything is alright
with the expressions for the left leg at the fusion. Let us assume
that the spinor from the left child calculated by {\em O'Mega} up to
the moment the fusion takes place, has the form
\begin{equation}
        \label{aufbauvonaussen}
        \Psi = S_F^{(1)} \Gamma^{(1)} S_F^{(2)} \ldots \Gamma^{(n)}
        w  , \qquad w \in \left\{ u, v \right\} \, . 
\end{equation}
When the fusion happens, the expression $\Psi^T \mathcal{C}$ is made
out of the left spinor, which after inserting (\ref{aufbauvonaussen}) 
is equal to 
\begin{align}
        \Psi^T \mathcal{C} =&\; w^T  {\Gamma^{(n)}}^T
        \ldots {S_F^{(2)}}^T {\Gamma^{(1)}}^T {S_F^{(1)}}^T \left( -
        \mathcal{C}^{-1} \right)  \notag \\ =&\; 
        w^T  \left( - \mathcal{C} \right)
        \mathcal{C} {\Gamma^{(n)}}^T \mathcal{C}^{-1}
        \ldots \mathcal{C}  {S_F^{(2)}}^T \mathcal{C}^{-1} \mathcal{C}
         {\Gamma^{(1)}}^T \mathcal{C}^{-1}\mathcal{C}{S_F^{(1)}}^T \left( -
        \mathcal{C}^{-1} \right) \notag \\ 
        =&\; \overline{w^c} \, {\Gamma^{(n)}}' \ldots {S_F^{(2)}}' 
         {\Gamma^{(1)}}' {S_F^{(1)}}' \label{finalreverse}
\end{align}
We have already explained that the assignment of the wavefunction for the
external fermion is correct. Now we get the primed expressions for all
vertex factors and propagators. Of course, that is what we really
want, since, indeed, our calculational direction for the whole fermion
line, after the fermion line has been closed (by the fusion), goes the
opposite way as 
it was originally calculated by {\em O'Mega} for the left part of the
line. There the calculational direction at first went from the
external particle inwards, after the fusion it points outwards.  When
{\em O'Mega} calculated the left leg it took the primed expressions
erroneously from the standpoint after the fusion, and as well had let
other expressions unprimed erroneously. By 
performing (\ref{finalreverse}) each vertex factor and
propagator becomes primed so that the wrongly primed factors become
unprimed again (the priming operation is involutory) while factors
left originally unprimed become now correctly primed.  

\vspace{3pt}

From the above discussion it is clear that for generating
Slavnov-Taylor identities for theories including
Majorana fermions  within {\em O'Mega}, we must ``reverse'' the
expressions for the conjugated spinors and assign the external
wavefunction $-\ii \left( \fmslash{p} - m \right) u(p)$ to every
incoming source of a BRST transformed fermion, while $\ii \left(
\fmslash{p} + m \right) v(p)$ for every outgoing source, not
depending on whether the transformed fermionic particle is a fermion,
an antifermion or a Majorana fermion.    

Of course, the formalism outlined above for theories including real
fermions works as well for theories which contain exclusively Dirac
fermions like the Standard Model. In that case the formalism has also
been tested and shown to produce the same numerical results as with
Ohl's ansatz. Finally, let us mention that the problems faced in section
\ref{sec:plentysigns} concerning signs from external antifermions and
from ``clashing arrows'' have been solved by the construction
presented in this chapter: According to figure
\ref{dennerext} the number of external $v$ spinors is simply given by
the number of all fermionic particles in the final state and thus is
obvious, while the ``clashing arrow''-obstacle has been remedied by
our procedure for handling the vertices in figure \ref{fusionmaj}. 

%%%%%%%%%%%%%%%%%%%%%%%%%%%%%%%%%%%%%%%%%%%%%%%%%%%%%%%%%%%%%%%%%%%%%%%%

\section{Numerical checks}

Numerical tests for Slavnov-Taylor identities (STI) have been made
for gauge symmetries as well as for supersymmetry. Therein we
investigated the ratio
\begin{equation}
        R \, \equiv \, \dfrac{\left| \sum_i \mathcal{G}_i
        \right|}{\sum_i \left| \mathcal{G}_i \right|} 
\end{equation}
of the sum of Green functions contributing to the STI
to the sum of their absolute values. In performing such tests for the
Abelian toy model presented in section \ref{sec:abeliantoy}, we
achieved ratios of better than $10^{-10}$, so the STI are fulfilled
numerically to a high level of accuracy. With their help, several
errors in the model files and also in the numerical implementation of
Majorana couplings in {\em O'Mega} have been discovered, showing these
tests to produce nontrivial results.

%%% Local Variables: 
%%% mode: latex
%%% TeX-master: "diss"
%%% End: 

%% file: chap13.tex
\chapter{Summary and Outlook}

Our original task was to implement the whole Minimal
Supersymmetric Standard Model (MSSM) in the matrix element generator
{\em O'Mega} and to further use the calculating power of that program
to produce cross sections and decay rates as predictions for the
coming generation of colliders, LHC and TESLA. This has be done by
creating a new model file for {\em O'Mega}. A compromise has been made
between the desire to be as general as possible and the constraint not
to blow up the model to a complexity which is not only too difficult
to handle within the framework of {\em O'Mega} but also contradicts
current experimental knowledge (flavour-changing neutral currents,
smallness of CP-violation). It became clear that the complexity of
the model is still immense and that checking procedures had to be
found to control the inner consistency of the theory and the
program's numerical stability as well. Although gauge
symmetries have been used as consistency checks ever since,
supersymmetry, though being only global in supersymmetric
field theories, is as powerful for this purpose as those -- or perhaps
even more.  

The idea arose to develop a method to perform such consistency
checks for supersymmetric field theories using supersymmetry as the
vehicle. First, we picked up a formalism invented by Grisaru and
Pendleton in the 1970s and calculated Ward identities for
supersymmetry diagrammatically, out of which relations between {\em
on-shell} {\em S}-matrix elements could be gained. This has been shown using
the simplest supersymmetric field theoretic, the Wess-Zumino
model, and then been extended to a more complex toy model to clarify
questions concerning Fermi statistics and vertices with clashing
arrows. But since this method relies on the annihilation of the ground
state by the supercharge it is only applicable for theories with exact
supersymmetry. The breakdown of the Ward identities has been
demonstrated in the O'Raifeartaigh model. Generally, this formalism does not 
suit well enough for automatized tests (as well as for realistic
models) but it enables some useful insights into the problems with
fermions in supersymmetric theories. 

Further progress was achieved by the investigation of the conserved current
resulting from the supersymmetry of the action. Here Ward identities
can be constructed by inserting the current operator into a Green
function and then taking the derivative with respect to the spacetime
argument of the current. Since the current is still conserved in the
case of spontaneously broken symmetry, this method is applicable not
only for exact supersymmetry. We derived in detail the supersymmetric
current of general models including supersymmetric Yang-Mills
theories. As the supersymmetric current is a spin-$3/2$ object
to which the gravitino couples in gauged supersymmetry, the
incorporation of that method in {\em O'Mega} provided the
infrastructure for supergravity (propagating gravitinos, higher
dimensional vertices with two bosons and two fermions, etc.). This may
be the basis for one of the possible extensions of the work presented
here. Calculations have been done for a simple toy model with a $U(1)$
gauge group, where examples are shown for the Ward identities
constructed in that manner to be fulfilled only on-shell, but not
off-shell. Hence the mentioned formalism can be used for on-shell tests in all
models with global supersymmetry. Nevertheless we are also interested in
off-shell tests since they are more stringent. The understanding of
why Ward identities with current insertions in supersymmetric gauge
theories are only fulfilled between physical on-shell
states, opened the way to a more elegant formalism for consistency
checks in part III. 

Therein we introduced the BRST formalism for supersymmetric field
theories based on the work of White and Sibold. A nilpotent BRST
operator can only be found by including supersymmetry transformations
and translations and using constant ghosts for them. The
Slavnov-Taylor identities from this generalized BRST invariance is the
desired consistency check working also off-shell. The deep-rooted 
reason that supersymmetry seems to be violated off-shell is that the
supercharge for supersymmetric gauge theories does not commute with
the $S$-operator for arbitrary states of the Hilbert space, but only for
physical states from the cohomology of the BRST charge. Whenever we
leave the mass shell, we have to include several additional diagrams
containing the Faddeev-Popov ghosts, which shows that gauge symmetry and
supersymmetry are inseparably entangled. These facts have been
clarified in the context of algebraic renormalization of
supersymmetric field theories by Sibold and co-workers. Here we
presented analytic calculations in a diagrammatic language of
Slavnov-Taylor identities in a pretty simple Abelian toy model and
also for a general supersymmetric Yang-Mills theory. To our knowledge,
this had not been done so far.  

The Slavnov-Taylor identities (STI) have been used for numerical
checks as well, where they justified their application by detecting several
errors in the program libraries of {\em O'Mega}. Some of the
difficulties and fine points concerning Majorana fermions, special
fermion vertices in supersymmetric field theories and the inclusion of
the STI in {\em O'Mega} have been listed for the sake of completeness
in the last part; they maintain the connection to the amount of
computational work not presented in this thesis. The MSSM, which
served as a motivation for this work by its sheer complexity, is
briefly reviewed in the appendix. All the physical fields with the
abundance of mixing angles and phases have been included into the
model file as has been mentioned above. 

\vspace{1mm}

With this work the problem of how to test supersymmetry in scattering
amplitudes and Green functions perturbatively within arbitrary models
analytically and numerically has been solved. The mechanisms by which
the cancellations in Ward and Slavnov-Taylor identities for
supersymmetry happen have been understood in detail. Hence this thesis
supplies the theoretical basis for testing supersymmetric models by
means of these identities. All the infrastructure has been laid to
perform these tests numerically; however this development has only
been sketched in order not to go beyond the scope of this work. 
Notwithstanding the fact that this thesis -- including the
theoretical foundations and the implementation of these identities and
providing a generally applicable checking tool -- is an integral
whole, we would like to give a brief outlook to further projects and
additional ideas, which will be tackled in the future. 

\vspace{3mm}

Among these points is a further debugging of the MSSM model file 
with the help of all existing symmetries: the gauge symmetry, supersymmetry,
also Bose and Fermi symmetries. Since supersymmetry is explicitly
broken in the MSSM, we will have to restore supersymmetry by a spurion
formalism, where superpotential terms containing one or more new
superfields are added. By spontaneous symmetry breaking these new
superfields then generate the soft breaking terms of the MSSM. The
price we have to pay for that restoring of supersymmetry are
additional couplings to the component fields of the new superfields. 
After that we can start producing data for MSSM
processes on a reliable basis. A further project is to enlarge the structure
of {\em O'Mega} so that it can handle propagating fields violating the
spin-statistics theorem with utmost generality. Strictly, this is only
necessary when 
investigating loop processes or STI in supergravity but it would be satisfying
to have a unified formalism managing all eventualities. The propagating
Faddeev-Popov ghosts can be incorporated while evaluating their Fermi
statistics signs separately from those of ``physical'' fermions. The
further diagrammatical examination of supersymmetric Slavnov-Taylor
identities, on the one hand with more than one SUSY ghost, on the 
other hand on an $N$-loop level, should finally be mentioned as a 
project.

%%% Local Variables: 
%%% mode: latex
%%% TeX-master: t
%%% End: 

%%% Local Variables: 
%%% mode: latex
%%% TeX-master: "diss"
%%% End: 

%% file: appen_bas.tex
\chapter{Basics, notations and conventions}

\section{Basics}

Metric:
\begin{equation}
        \eta_{\mu\nu} = \text{diag}\,\left( 1, -1, -1, -1 \right) 
\end{equation}   
%%%
Super-Poincar{\'e} algebra for $N=1$, without central charges:
\begin{equation}
    \label{eq:poincare}
    \boxed{
    \begin{aligned}
      \left[ P^\mu , P^\nu \right] = & \; 0 \\
      \left[ P^\mu , M^{\rho\sigma} \right] = & \; \ii \left( \eta^{\mu\rho}
      P^\sigma - \eta^{\mu\sigma} P^\rho \right) \\
      [M^{\mu\nu} , M^{\rho\sigma} ] = &  \; - {\rm i} ( \eta^{\mu\rho}
      M^{\nu\sigma} - \eta^{\mu\sigma} M^{\nu\rho} - \eta^{\nu\rho}
      M^{\mu\sigma} + \eta^{\nu\sigma} M^{\mu\rho})  \\
      \left\{ Q , \overline{Q} \right\} = & \; 2 \gamma^\mu P_\mu \\
      \left[ Q , M^{\mu\nu} \right] = & \; S^{\mu\nu} Q \quad
    \text{with} \quad       S^{\mu\nu} = \dfrac{\ii}{4} \left[
    \gamma^\mu, \gamma^\nu \right] \\ 
      \left[ Q, P^\mu \right] = & \; 0
    \end{aligned}}
\end{equation}

\vspace{1cm}

Generalized Jacobi identity for $\mathbb{Z}_2$-graded algebras:
\begin{multline}
  \label{eq:jacobi}
  (-1)^{\eta_C \eta_A} [[T_A,T_B\},T_C\} + (-1)^{\eta_A \eta_B}
  [[T_B,T_C\},T_A\} +  \\ (-1)^{\eta_B \eta_C} [[T_C,T_A\},T_B\} = 0
\end{multline}          
This implies the special cases where $B$ is a bosonic and $F$ a
fermionic operator:
\begin{align}
  [[B_A,B_B],B_C] + [[B_B,B_C],B_A] + [[B_C,B_A],B_B] = 0 \notag \\
  [[F_A,B_B],B_C] + [[B_B,B_C],F_A] + [[B_C,F_A],B_B] = 0 \notag \\
  \{[B_A,F_B],F_C\} + [\{F_B,F_C\},B_A] - \{[F_C,B_A],F_B\} = 0 \notag \\ 
  \{\{F_A,F_B\},F_C\} + \{\{F_B,F_C\},F_A\} + \{\{F_C,F_A\},F_B\} = 0 \notag \\
\end{align}
They can be proven by making all the generators bosonic, multiplying
them with Grassmann numbers, writing down the ordinary Jacobi identity
for bosonic operators and extracting all Grassmann numbers to the left
(or to the right) while giving them the same order in all terms. 

%%%%%%%%%%%%%%%%%%%%%%%%%%%%%%%%%%%%%%%%%%%%%%%%%%%%%%%%%%%%%%%%%%%

\section{Superspace}

Supersymmetric field theories are most easily represented on a
$\mathbb{Z}_2$-graded vector space called superspace, containing the
ordinary four-dimensional space-time and four Grass\-mann-odd coordinates, 
\begin{equation}
  \label{eq:supkoord}
  (x^0 , x^1, x^2, x^3, \theta^1 , \theta^2 , \theta^3 , \theta^4 ) \;
  \equiv \begin{pmatrix} x^\mu \\ \theta \end{pmatrix}
  \quad . 
\end{equation}
with the bispinor (four-component spinor) $\theta$. This is the
special case of simple supersymmetry, for $N=n$-supersymmetry, $n\leq
4$ we have the space $\mathbb{R}^{(1,3)|2\cdot 2 N}$. It is possible
to combine the supercharges into a four-component Majorana
spinor $Q$, \cite{Weinberg:QFTv3:Text}. The 14 generators of simple  
supersymmetry $(P^{\mu}, M^{\rho\sigma}, Q)$ generate superspace
transformations -- spacetime translations by $P^{\mu}$, boosts and
rotations by $M^{\rho\sigma}$ and translations of the Grassmann-odd
coordinates by spinorial increments $\xi$, also combined into a Majorana
spinor. They anticommute component-wise, and as these
parameters are constant, they anticommute with the supercharges, too:  
\begin{equation}
  \{ \xi, \xi \} = \{ \xi, \overline{\xi} \} 
  = \{ \xi , Q \} = \{ \overline{\xi}, Q \} = \left\{ \xi,
  \overline{Q} \right\} = \left\{ \overline{\xi}, \overline{Q}
  \right\} = 0.
\end{equation}
Most general element of the Poincar{\'e} supergroup:
\begin{equation}
  \label{eq:susy-trafo}
  S(b,\omega,\xi) = \exp \left[ {\rm i} \left( b_{\mu}
  P^{\mu} + \frac{1}{2} \omega_{\mu\nu} M^{\mu\nu} + \overline{\xi} Q
  \right) \right] . 
\end{equation}
Action of supercharge on superspace:
\begin{equation}
  \label{eq:superladbi2}
\boxed{
  Q = \frac{\partial}{\partial \overline{\theta}} -
  ({\rm i} \gamma^{\mu} \theta) \partial_{\mu}}
\end{equation}
We check that $Q$ fulfills the anticommutation relations
\begin{equation}
        \left\{ Q, Q \right\} = \left\{ \overline{Q} , \overline{Q}
        \right\} = 0, \qquad \left\{ Q , \overline{Q} \right\} = 2
        \ii \fmslash{\partial}  \;\; . 
\end{equation}

The covariant derivatives with respect to the supergroup structure
$\mathcal{D}$, called {\em superderivatives}, anticommute 
with the supercharges: 
\begin{equation}
  \label{eq:antisupsup}
  \left\{ \mathcal{D} , Q \right\} = \left\{ \mathcal{D} ,
  \overline{Q} \right\} = \left\{ \overline{\mathcal{D}} , Q \right\}
  = \left\{ \overline{\mathcal{D}} , \overline{Q} \right\} = 0 
\end{equation}
On superspace they can be given the representation
\begin{equation}
        \boxed{ \mathcal{D} = \frac{\partial}{\partial \overline{\theta}} +
          ({\rm i} \gamma^{\mu} \theta) \partial_{\mu} } 
\end{equation}
With each other they have the (anti-)commutation relations
\begin{equation}
        \left\{ \mathcal{D}, \mathcal{D} \right\} = \left\{
        \overline{\mathcal{D}} , \overline{\mathcal{D}} 
        \right\} = 0, \qquad \left\{ \mathcal{D} ,
        \overline{\mathcal{D}} \right\} = - 2 
        \ii \fmslash{\partial} \;\; .      
\end{equation}
For constructing chiral superfields right- and left-handed
versions of the superderivative are needed:
\begin{equation}
  \label{eq:chiralsupab}
\boxed{
  \begin{aligned}
  \mathcal{D}_L = \frac{1}{2} \left( 1 - \gamma^5 \right) \mathcal{D} = 
  \dfrac{\partial}{\partial \overline{\theta}_L} - ({\rm i} \gamma^{\mu}
  \theta_R) \partial_{\mu} \\  \mathcal{D}_R = \frac{1}{2} \left( 1 + \gamma^5
  \right) \mathcal{D} = \dfrac{\partial}{\partial \overline{\theta}_R} -
  ({\rm i} \gamma^{\mu} \theta_L) \partial_{\mu} 
\end{aligned}}
\end{equation}
For more details about Lie supergroups cf. \cite{Cornwell:1989:SUSY}. 

%%%%%%%%%%%%%%%%%%%%%%%%%%%%%%%%%%%%%%%%%%%%%%%%%%%%%%%%%%%%%%%%%%%%

\section{Properties of Majorana spinors}\label{majo}

Definition of a Majorana spinor
\begin{equation}
  \label{majoranaquer}
  \overline{\Psi}_M \equiv \Psi_M^{\dagger} \gamma^0 = - \Psi_M^T
  \mathcal{C}    ,
\end{equation}
where $\mathcal{C}$ is the antisymmetric charge conjugation matrix,
usually chosen to be equal to $\mathcal{C} = \ii \gamma^2 \gamma^0$.  
In the sequel $\theta$ always means a Grassmann-odd spinor.
\begin{align}
  \overline{\theta}_1 \Gamma \theta_2 & = \, \left( \overline{\theta}_1 \Gamma
  \theta_2 \right)^T = - (\theta_1^T \mathcal{C} \Gamma \theta_2)^T =
  - (\theta_2^T 
  \Gamma^T \mathcal{C} \theta_1) \notag \\ & = \, \overline{\theta}_2
  \mathcal{C}^{-1} \Gamma^T \mathcal{C} 
  \theta_1
\end{align}
Using the well-known relations about gamma matrices 
\begin{equation}
  \Gamma^T = \left\{ 
    \begin{array}{ll} + \, \mathcal{C} \Gamma \mathcal{C}^{-1} &
    \qquad \Gamma = \mathbb{I}, \gamma^5 
      \gamma^{\mu}, \gamma^5 \\ - \, \mathcal{C} \Gamma
    \mathcal{C}^{-1} & \qquad \Gamma = 
      \gamma^{\mu}, [\gamma^{\mu} , \gamma^{\nu}]      
    \end{array} \right. 
\end{equation}
yields
\begin{equation}
  \overline{\theta}_1 \Gamma \theta_2 = \left\{ \begin{array}{ll} + \;
  \overline{\theta}_2 \Gamma \theta_1 & \qquad \Gamma = \mathbb{I}, \gamma^5
  \gamma^{\mu} , \gamma^5 \\ - \; \overline{\theta}_2 \Gamma \theta_1 & \qquad
  \Gamma = \gamma^{\mu}, [\gamma^{\mu} , \gamma^{\nu}]  \end{array} \right.
  \label{majosymm}
\end{equation}
So the only possible bilinears with a single Grassmann-odd spinor are 
\begin{equation}
  \label{eq:bilin}
  \overline{\theta} \theta, \qquad \overline{\theta} \gamma^5 \gamma^{\mu}
  \theta, \qquad \overline{\theta} \gamma^5 \theta , 
\end{equation}
while the other combinations vanish identically:
\begin{equation}
  \label{eq:keinbilin}
  \overline{\theta} \gamma^{\mu} \theta = \overline{\theta} \, [ \gamma^{\mu} ,
  \gamma^{\nu} ] \, \theta = 0.
\end{equation}
Note that for commuting spinors the signs in (\ref{majosymm}) are the
other way round. 

%%%%%%%%%%%%%%%%%%%%%%%%%%%%%%%%%%%%%%%%%%%%%%%%%%%%%%%%%%%%%%%%%%%%%%

\section{Superfields}\label{appen:appen_superfields}

The irreducible representations of the super-Poincar{\'e} algebra on superspace
are called {\em superfields}. They are the basic ingredients of
supersymmetric quantum field theories. We will denote them by a 
hat over the symbol. All superfields have expansions in the superspace
coordinates which only run up to fourth order due to the latter's 
nilpotency. We do not go into the details here; especially we omit
general superfields since they are not needed in the construction of
supersymmetric field theories. We should only mention that products of
superfields underlying some sort of constraints are general
unconstrained superfields again; the highest component in the
superspace expansion of general superfields is called $D$, being a
scalar field of canonical dimension higher by two than the canonical
dimension of the whole superfield. For details
cf.~\cite{Reuter:2000:SUSY}, \cite{Weinberg:QFTv3:Text}. 

A superfield with the constraint 
\begin{equation}
  \label{eq:scasupli}
  \mathcal{D}_R \hat{\Phi} = 0 
\end{equation}
is called {\em left-chiral superfield}, while correspondingly a {\em
right-chiral superfield}, distinguished by a bar, underlies the
constraint 
\begin{equation}
  \label{eq:scasupre}
  \mathcal{D}_L \hat{\bar{\Phi}} = 0   \;\; .
\end{equation}

The chiral superfields have the superspace expansions
\begin{equation}
\boxed{
\begin{aligned}
  \hat{\Phi}(x,\theta) = & \; \phi(x) + \sqrt{2} \left( \overline{\theta}
  \psi_L(x) \right) + \left( \overline{\theta} \left[\dfrac{1 - \gamma^5}{2}
  \right] \theta \right) F(x) + 
  \frac{{\rm i}}{2} \left( \overline{\theta} \gamma^5 \gamma^{\mu} \theta
  \right) 
  \partial_{\mu} \phi(x) \\ & \;\; + \frac{{\rm i}}{\sqrt{2}} \left(
  \overline{\theta} \gamma^5 \theta \right) \left( \overline{\theta}
  \fmslash{\partial} \psi_L(x)\right) + \frac{1}{8} \left( \overline{\theta}
  \gamma^5 \theta \right)^2 \Box \, \phi(x)  \\ \\ 
  \hat{\bar{\Phi}}(x,\theta) = & \; \bar{\phi}(x) + \sqrt{2} \left(
  \overline{\theta} \psi_R(x) \right) + \left( \overline{\theta} \left[
  \dfrac{1 +  \gamma^5}{2} \right] \theta \right) \bar{F}(x) - \frac{{\rm
  i}}{2} \left( \overline{\theta} \gamma^5 \gamma^{\mu} \theta \right)
  \partial_{\mu} \bar{\phi}(x) \\ & 
  \;\; - \frac{{\rm i}}{\sqrt{2}} \left( \overline{\theta} \gamma^5 \theta
  \right) \left( \overline{\theta} \fmslash{\partial} \psi_R(x)\right) +
  \frac{1}{8} \left( \overline{\theta} \gamma^5 \theta \right)^2 \Box \,
  \bar{\phi}(x) 
\end{aligned}} \;\; . 
\end{equation}
Therein $\phi$ and $\bar{\phi}$ are complex scalar fields, $\psi_L$
and $\psi_R$ are left- and righthanded Weyl spinor fields,
respectively, while $F$ and $\bar{F}$ are again complex scalar fields
of canonical dimension two if the dimension of $\phi, \bar{\phi}$ is
one. From the expansions, one can see that the Hermitean adjoint of a
left-chiral superfield is right-chiral and vice versa. 

Products of left-chiral superfields are left-chiral superfields
again. A function~$f$ consisting only of left-chiral
superfields together with an identical contribution of right-chiral superfields
with the complex conjugated prefactors, but neither containing superderivatives
nor spacetime derivatives, is called {\em superpotential}. The $F$ term
(the highest component in the superspace expansion) of the product of
two or three left-chiral superfields are: 
\begin{equation}
  \Bigl[ \hat{\Phi}_1 \hat{\Phi}_2 \Bigr]_F = 
  \left\{ F_1 \phi_2 + F_2 \phi_1 -
  \left( \overline{\psi}_{L,1} \psi_{L,2} \right) \right\}
\end{equation}
\begin{multline}
  \Bigl[ \hat{\Phi}_1 \hat{\Phi}_2 \hat{\Phi}_3 \Bigr]_F = 
  \biggl\{ F_1 \phi_2 \phi_3 + F_2
  \phi_3 \phi_1 + F_3 \phi_1
  \phi_2 - \left( \overline{\psi}_{L,1}  \psi_{L,2}  \right)
  \phi_3 \\ - \left( \overline{\psi}_{L,2} 
  \psi_{L,3}  \right) \phi_1 - \left( \overline{\psi}_{L,3} 
  \psi_{L,1}  \right) \phi_2 \biggr\}
\end{multline}

A superfield constrained by the reality condition
\begin{equation}
        \hat{V}^\dagger = \hat{V} 
\end{equation}
is called a {\em vector superfield} since such superfields are the
supersymmetric generalizations of the gauge boson fields:
\begin{equation}
  \label{eq:veksup1}
  \begin{aligned}
  \hat{V}^a(x,\theta) = & \;\; C^a(x) - {\rm i} \left(
  \overline{\theta} \gamma^5 
  \omega^a(x) \right) - \frac{{\rm i}}{2} \left( \overline{\theta} \gamma^5
  \theta \right) M^a(x) - \frac{1}{2} \left( \overline{\theta} \theta \right)
  N^a(x) \\ & \; - \frac{1}{2} \left( \overline{\theta} \gamma^5
  \gamma^{\mu}
  \theta \right) V^a_{\mu} (x) - {\rm i} \left( \overline{\theta} \gamma^5
  \theta \right) \left( \overline{\theta} \left[ \lambda^a(x) + \frac{{\rm
  i}}{2} \fmslash{\partial} \omega^a (x) \right] \right) \\ & \; - \frac{1}{4}
  \left( \overline{\theta} \gamma^5 \theta \right)^2 \left( D^a(x) -
  \frac{1}{2} \Box C^a(x) \right)
  \end{aligned}
\end{equation}   
As discussed in \cite{Weinberg:QFTv3:Text} there is an extended gauge
symmetry in supersymmetric gauge theories with the gauge parameter
replaced by a whole superfield. This freedom can be used to gauge away
most of the components in (\ref{eq:veksup1}). The remaining part of
the extended gauge transformations, orthogonal to those used above,
represents ordinary gauge invariance. Most famous is the {\em
Wess-Zumino gauge}, 
\begin{equation}
  \label{eq:wess-zumino}
  C^a(x) = \omega^a(x) = M^a(x) = N^a(x) = 0 , 
\end{equation}
\begin{equation}
  \label{eq:vsupwezu}
  \boxed{
    \begin{aligned}
  \hat{V}^a(x,\theta) = & \;\; - \frac{1}{2} \left( \overline{\theta} \gamma^5
  \gamma^{\mu} \theta \right) V^a_{\mu} (x) - {\rm i} \left( \overline{\theta}
  \gamma^5 \theta \right) \left( \overline{\theta} \lambda^a(x) \right)  -
  \frac{1}{4} \left( \overline{\theta} \gamma^5 \theta \right)^2 D^a(x)
    \end{aligned}} \;\;\; , 
\end{equation}         
in which all power series in the vector superfield breaks off after
the quadratic term.  

%%%%%%%%%%%%%%%%%%%%%%%%%%%%%%%%%%%%%%%%%%%%%%%%%%%%%%%%%%%%%%%%%%%%

\section{SUSY transformations of component fields}\label{dwf-appen}

In the sequel we list the SUSY transformations for chiral and vector
superfields: the ``normal'' ones, in the case of supersymmetric gauge
theories the de~Wit--Freedman transformations, where a mixing between
the matter and the gauge superfields occurs. When inserting the
equations of motion for the auxiliary fields, we can forget about the
transformations of the auxiliary fields. 

\subsection{SUSY transformations for chiral superfields}

SUSY transformation:
\begin{equation}
\setlength{\extrarowheight}{3pt}  
  \begin{array}{rl}
          \delta_\xi \phi =& \sqrt{2} \left( \overline{\xi_R}
          \psi_L \right) \\
          \delta_\xi \psi_L =& - \sqrt{2} \, \ii
          (\fmslash{\partial} \phi) \xi_R + \sqrt{2}
          F \xi_L \\
          \delta_\xi F =& - \sqrt{2} \, \ii \left(
          \overline{\xi_L}
          \fmslash{\partial} \psi_L \right)
  \end{array}
\end{equation}  
De Wit-Freedman transformation:
\begin{equation}
\setlength{\extrarowheight}{3pt}  
  \begin{array}{rl}
          \tilde{\delta}_\xi \phi =& \sqrt{2} \left(
          \overline{\xi_R} \psi_L \right) \\
          \tilde{\delta}_\xi \psi_L =& - \sqrt{2} \, \ii
          (\fmslash{D} \phi) \xi_R + \sqrt{2}
          F \xi_L \\
          \tilde{\delta}_\xi F =& - \sqrt{2}
          \, \ii\left( \overline{\xi_L} \fmslash{D}
          \psi_L \right) - 2\ii T^a
          \phi \left( \overline{\xi_L} \lambda^a_R \right)
  \end{array}
\end{equation}
 
Inserting the equations of motion for the auxiliary fields yields the 
``on-shell'' de Wit-Freedman transformation:
\begin{equation}
\setlength{\extrarowheight}{3pt}  
  \begin{array}{rl}
          \tilde{\delta}'_\xi \phi =& \sqrt{2} \left(
          \overline{\xi_R} \psi_L \right) \\
          \tilde{\delta}'_\xi \psi_L =& - \sqrt{2} \, \ii
          (\fmslash{D} \phi) \xi_R - \sqrt{2}
          \left( \dfrac{\partial f(\phi)}{\partial
          \phi} \right)^* \xi_L
  \end{array}
\end{equation}                               
 
%%%

\subsection{SUSY transformations for vector superfields}

SUSY transformation:
\begin{equation}
\setlength{\extrarowheight}{3pt}  
  \begin{array}{rl}
          \delta_\xi A^a_\mu =& - \left( \overline{\xi}
          \gamma_\mu \gamma^5 \lambda^a \right) \\
          \delta_\xi \lambda^a =& - \dfrac{\ii}{2}
          \lbrack \gamma^\alpha , \gamma^\beta \rbrack \gamma^5 \left(
          \partial_\alpha A^a_\beta - \partial_\beta
          A^a_\alpha \right) + D^a \xi \\
          \delta_\xi D^a =& - \ii \left(
          \overline{\xi} \fmslash{\partial} \lambda^a \right)
  \end{array}
\end{equation}
De Wit-Freedman transformation:
\begin{equation}
\setlength{\extrarowheight}{3pt}  
  \begin{array}{rl}
          \tilde{\delta}_\xi A^a_\mu =& - \left( \overline{\xi}
          \gamma_\mu \gamma^5 \lambda^a \right) \\
          \tilde{\delta}_\xi \lambda^a =& - \dfrac{\ii}{2}
          \lbrack \gamma^\alpha , \gamma^\beta \rbrack \gamma^5
          F^a_{\alpha\beta} + D^a \xi \\
          \tilde{\delta}_\xi D^a =& - \ii \left(
          \overline{\xi} \left(\fmslash{D} \lambda\right)^a  \right)
  \end{array}
\end{equation}                     
``On-shell'' de Wit-Freedman transformation:
\begin{equation}
\setlength{\extrarowheight}{3pt}  
  \begin{array}{rl}
          \tilde{\delta}'_\xi A^a_\mu =& - \left( \overline{\xi}
          \gamma_\mu \gamma^5 \lambda^a \right) \\
          \tilde{\delta}'_\xi \lambda^a =& - \dfrac{\ii}{2}
          \lbrack \gamma^\alpha , \gamma^\beta \rbrack \gamma^5
          F^a_{\alpha\beta} - e \left( 
          \phi^\dagger T^a \phi \right) \xi
  \end{array}
\end{equation} 

%%%%%%%%%%%%%%%%%%%%%%%%%%%%%%%%%%%%%%%%%%%%%%%%%%%%%%%%%%%%%%%%%%%%%

\section{Construction of supersymmetric field theories} 

Kinetic terms for matter fields (scalars and fermions): 
\begin{equation}
        \dfrac{1}{2} \left[ \hat{\Phi}^\dagger \hat{\Phi} \right]_D = 
        \partial_\mu \phi^\dagger \partial^\mu \phi + \dfrac{\ii}{2}
        \overline{\psi_L} \fmslash{\partial} \psi_L - \dfrac{\ii}{2}
        \left( \partial_\mu \overline{\psi_R} \right) \gamma^\mu
        \psi_R + \left| F \right|^2 
\end{equation}

Kinetic terms for matter fields with minimal couplings to gauge boson
fields from vector superfields in Wess-Zumino gauge:
\begin{multline}
        \dfrac{1}{2} \left[ \hat{\Phi}^\dagger \exp\left( - T^a
        \hat{V}^a \right) \hat{\Phi} \right]_D =
        (D_\mu \phi)^\dagger (D^\mu \phi) \\ + \dfrac{\ii}{2}
        \overline{\psi_L} \fmslash{D} \psi_L - \dfrac{\ii}{2}
        \left( D_\mu \overline{\psi_R} \right) \gamma^\mu
        \psi_R + \sum_a \left| F^a \right|^2 
\end{multline}

From a vector superfield we construct a new superfield by acting
triply with the superderivative:
\begin{equation}
        \hat{W} = - \dfrac{1}{4} \left( \overline{\mathcal{D}} \mathcal{D}
        \right) \mathcal{D} \hat{V} 
\end{equation}
The complete superfield $\hat{W}$ has a spinor index and is therefore
called a spinor superfield. Projecting with $\frac{1}{2} \left( 1 
\pm \gamma^5 \right)$ gives a right- and left-chiral superfield,
respectively. By the following construction we get kinetic terms
for the gauge boson and the gaugino as well as gauge boson--gaugino
interaction terms:
\begin{equation}
        \dfrac{1}{2} \Re \left[ \, \overline{\hat{W}_R} \hat{W}_L \right]_F = -
        \dfrac{1}{4} F_{\mu\nu}^a F^{\mu\nu}_a + \dfrac{\ii}{2}
        \overline{\lambda^a} \left( \fmslash{D} \lambda \right)^a +
        \dfrac{1}{2} D^a D^a 
\end{equation}

Superpotentials (products of one, two or three chiral superfields) to
construct scalar self-interactions and Yukawa couplings have already
been discussed in section~\ref{appen:appen_superfields}. 

%%%%%%%%%%%%%%%%%%%%%%%%%%%%%%%%%%%%%%%%%%%%%%%%%%%%%%%%%%%%%%%%%%%%

%%% Local Variables: 
%%% mode: latex
%%% TeX-master: "susy"
%%% End: 

%% file: appen_mssm.tex
\chapter{Some details about the MSSM}

All superfields of the MSSM:

\vspace{3pt}

{\renewcommand{\arraystretch}{1.1}
\setlength{\extrarowheight}{1pt}
\begin{tabular}{|c|c|c|c|c|c|}\hline
Superfield & Bosons & Fermions & $U(1)_Y$ & $SU(2)_L$ & $SU(3)_C$
\\\hline\hline 
$\hat{V}^{U(1)_Y}$ & $B$ & $\tilde{B}$ & 0 & 0 & $1$ \\ \hline
$\hat{V}^{SU(2)_L}$ & $W^i$  & $\tilde{W}^i$ & 0 & 1 & $1$  \\ \hline
$\hat{V}^{SU(3)_C}$ & $G^i$ & $\tilde{G}^i$ & 0 & 0 & $8$  \\
\hline\hline 
$\hat{L}_1$ & $(\tilde{\nu}_e,\tilde{e}^-_L)$ & $(\nu_e,e^-)_L$ & -1 & 
$\frac{1}{2}$ & $1$ \\
$\hat{L}_2$ & $(\tilde{\nu}_{\mu},\tilde{\mu}^-_L)$ &
$(\nu_{\mu},\mu^-)_L$ & -1 &
$\frac{1}{2}$ & $1$ \\
$\hat{L}_3$ & $(\tilde{\nu}_{\tau},\tilde{\tau}^-_L)$ & $(\nu_{\tau},\tau^-)_L$ & 
-1 & $\frac{1}{2}$ & $1$ \\
$\hat{\bar{E}}_1$ & $\tilde{e}_R^+$ & $e^+_L$ & 2 & 0 & $1$ \\
$\hat{\bar{E}}_2$ & $\tilde{\mu}_R^+$ & $\mu^+_L$ & 2 & 0 & $1$ \\
$\hat{\bar{E}}_3$ & $\tilde{\tau}_R^+$ & $\tau^+_L$ & 2 & 0 & $1$
\\\hline
$\hat{Q}_1$ & $(\tilde{u}_L , \tilde{d}_L)$ & $(u, d)_L$ & $\frac{1}{3}$ & $\frac{1}{2}$
& $3$ \\
$\hat{Q}_2$ & $(\tilde{c}_L , \tilde{s}_L)$ & $(c, s)_L$ & $\frac{1}{3}$ & $\frac{1}{2}$
& $3$ \\
$\hat{Q}_3$ & $(\tilde{t}_L , \tilde{b}_L)$ & $(t, b)_L$ & $\frac{1}{3}$ & $\frac{1}{2}$
& $3$ \\
$\hat{\bar{U}}_1$ & $\tilde{u}^*_R$ & $u^c_L$ & $-\frac{4}{3}$ & 0 & $\bar{3}$
\\
$\hat{\bar{U}}_2$ & $\tilde{c}^*_R$ & $c^c_L$ & $-\frac{4}{3}$ & 0 & $\bar{3}$
\\
$\hat{\bar{U}}_3$ & $\tilde{t}^*_R$ & $t^c_L$ & $-\frac{4}{3}$ & 0 & $\bar{3}$
\\
$\hat{\bar{D}}_1$ & $\tilde{d}^*_R$ & $d^c_L$ & $\frac{2}{3}$ & 0 & $\bar{3}$
\\
$\hat{\bar{D}}_2$ & $\tilde{s}^*_R$ & $s^c_L$ & $\frac{2}{3}$ & 0 & $\bar{3}$
\\
$\hat{\bar{D}}_3$ & $\tilde{b}^*_R$ & $b^c_L$ & $\frac{2}{3}$ & 0 & $\bar{3}$
\\\hline
$\hat{H}_1$ & $(H_1^0 , H_1^-)$ & $(\tilde{H}_1^0, \tilde{H}^-_1)_L$ &
-1 & $\frac{1}{2}$ & $1$ \\
$\hat{H}_2$ & $(H_2^+ , H_2^0)$ & $(\tilde{H}_2^+, \tilde{H}^0_2)_L$ & 1 & $\frac{1}{2}$ &
$1$ \\\hline
\end{tabular}}   

\vspace{3mm}

Above we listed the superfields of the MSSM all of which are
left-chiral superfields. Naturally, the total field content also
includes their Hermitean conjugates which are right-chiral
superfields. Since we only wanted to write down left-chiral
superfields, all fermionic component fields are left-handed: Therefore
those of the ``barred'' superfields are the left-handed
parts of the antileptons and antiquarks. We omit right-handed
neutrino fields here and so the neutrinos remain massless, but a
generalization is obvious. A tilde on the component fields indicates a
particle with negative $R$ parity and hence a superpartner of a
Standard Model field. The quantum numbers are given by the
hypercharge, the third component of the weak isospin, while the number
in the last column indicates whether the particle is a colour-singlet,
triplet, antitriplet or octet.  

MSSM - the Lagrangean density:

\begin{equation}
  \label{eq:lamssm}
\boxed{
\begin{aligned}
  &  \Large {\bf \mathcal{L}_{\text{MSSM}}} =
  \\
  & \;\;\; \sum_{i=1}^3 \Bigl( \hat{Q}_i^{\dagger} \exp\left[\hat{{\cal V}}\right]
  \hat{Q}_i \Bigr)_D + \sum_{i=1}^3 \Bigl( \hat{L}_i^{\dagger}
  \exp\left[\hat{{\cal V}}\right] 
  \hat{L}_i \Bigr)_D \\ & + \sum_{i=1}^3 \Bigl( \hat{\bar{U}}_i^{\dagger}
  \exp\left[\hat{{\cal V}}\right] 
  \hat{\bar{U}}_i \Bigr)_D + \sum_{i=1}^3 \Bigl(
  \hat{\bar{D}}_i^{\dagger} \exp\left[\hat{{\cal
  V}}\right] \hat{\bar{D}}_i \Bigr)_D \\ & + \sum_{i=1}^3 \Bigl(
  \hat{\bar{E}}_i^{\dagger} \exp\left[\hat{{\cal
  V}}\right] \hat{\bar{E}}_i \Bigr)_D + \Bigl(
  \hat{H}_1^{\dagger} \exp\left[\hat{{\cal
  V}}\right] \hat{H}_1 \Bigr)_D  + \Bigl(
  \hat{H}_2^{\dagger} \exp\left[\hat{{\cal V}}\right]
  \hat{H}_2 \Bigr)_D \\ & + \frac{1}{2} \Re \biggl[ \, 
  \overline{\hat{W}_{R,a}^{SU(3)_C}} \,
  \hat{W}_{L,a}^{SU(3)_C} \biggr]_F +  \frac{1}{2} \Re \biggl[ \, 
  \overline{\hat{W}_{R,a}^{SU(2)_L}}
  \, \hat{W}_{L,a}^{SU(2)_L} \biggr]_F \\ &
  + \frac{1}{2} \Re \biggl[ \, \overline{\hat{W}_R^{U(1)_Y}} \,
  \hat{W}_L^{U(1)_Y} \biggr]_F - \frac{g_s^2 \theta_{\text{QCD}}}{16
  \pi^2} \Im \biggl[ \, \overline{\hat{W}_{R,a}^{SU(3)_C}} \,
  \hat{W}_{L,a}^{SU(3)_C} \biggr]_F  \\ &
   + [{\cal W}]_F + {\cal L}_{SR}
\end{aligned}}
\end{equation}                 
Summation over generation indices from $1$ to $3$ is explicitly shown
while for gauge indices we use the summation
convention. $\theta_{\text{QCD}}$ is the QCD vacuum angle, this term
being the supersymmetric generalization of the term generating instanton
solutions in QCD \cite{Weinberg:QFTv2:Text}, \cite{Weinberg:QFTv3:Text}. 
The vector superfield of the Standard Model $SU(3)_C \times SU(2)_L
\times U(1)_Y$ gauge group is
\begin{equation}
  \hat{{\cal V}} = - \, \hat{V}^{U(1)_Y}_a \cdot
  \frac{Y}{2} - \sum_{a=1}^3 \hat{V}^{SU(2)_L}_a
  \dfrac{\sigma^a}{2} - \sum_{a=1}^8 \hat{V}^{SU(3)_C}_a \dfrac{\lambda^a}{2} 
\end{equation}
Furthermore $\mathcal{W}$ is the {\em superpotential},
\begin{multline}
  \label{eq:superpot}
  {\cal W} = h^E_{kl} \left( \hat{L}_k^a
  \epsilon_{ab} \hat{H}_1^b \right) \hat{\bar{E}}_l + h^D_{kl} \left( 
  \hat{Q}_k^a \epsilon_{ab} \hat{H}_1^b \right) 
  \hat{\bar{D}}_l \\ + h^U_{kl} \left( \hat{Q}_k^a
  \epsilon_{ab} \hat{H}_2^b \right) \hat{\bar{U}}_l + \mu
  \left( \hat{H}_1^a \epsilon_{ab} \hat{H}_2^b \right) +
  {\rm h.c.} ,
\end{multline}      
and ${\cal L}_{SR}$ are the {\em superrenormalizable terms}
parameterizing the unknown SUSY breaking mechanism:
\begin{align}
  {\cal L}_{SR} =&\; - \sum_{ij} (M^2_{\tilde{Q}})_{ij}
  (\tilde{Q}_i^{\dagger}
  \tilde{Q}_j) - \sum_{ij} (M^2_{\bar{U}})_{ij} (\bar{U}_i^{\dagger}
  \bar{U}_j) - \sum_{ij} (M^2_{\bar{D}})_{ij}
  (\bar{D}_i^{\dagger} \bar{D}_j)  \notag \\ &
  - \sum_{ij} (M^2_{\tilde{L}})_{ij} (\tilde{L}_i^{\dagger}
  \tilde{L}_j) - \sum_{ij} 
  (M^2_{\bar{E}})_{ij} (\bar{E}_i^{\dagger} \bar{E}_j) - \Biggl\{
  \frac{1}{2} m_{\text{Gluino}} (\lambda_s  \lambda_s)  \notag \\ & +
  \frac{1}{2} m_{\text{Wino}} (\lambda \lambda ) + \frac{1}{2} m_{\text{Bino}}
  (\lambda' \lambda ') - \sum_{ij} A^D_{ij} h^D_{ij} (\tilde{Q}^T_i \epsilon H_1)
  \bar{D}_j
  \notag \\ & - \sum_{ij} A^E_{ij} h^E_{ij} (\tilde{L}^T_i \epsilon H_1) \bar{E}_j -
  \sum_{ij} A^U_{ij} h^U_{ij} (\tilde{Q}^T_i \epsilon H_2) \bar{U}_j
  \notag \\ & +
  \sum_{ij} C^D_{ij} h^D_{ij} (\tilde{Q}^T_i H^*_2) \bar{D}_j+ \sum_{ij} C^E_{ij}
  h^E_{ij} (\tilde{L}^T_i H^*_2) \bar{E}_j  \notag \\ & + \sum_{ij} C^U_{ij} h^U_{ij}
  (\tilde{Q}^T_i H^*_1) \bar{U}_j + \frac{1}{2} \left( B \mu \right) (H_1^T
  \epsilon H_2) + {\rm h.c.} \Biggr\} \notag \\ & + m_1^2 |H_1^0|^2 + m_2^2
  |H_2^0|^2
\end{align}

%%%%%%%%%%%%%%%%%%%%%%%%%%%%%%%%%%%%%%%%%%%%%%%%%%%%%%%%%%%%%%%%%%%%%%%%%%

\begin{table}
\begin{tabular}
          {|>{\PBS\raggedright\hspace{0pt}}p{9.5cm}
           |>{\PBS\centering\hspace{0pt}}p{2.5cm}|}\hline
\qquad \qquad {\bf VERTICES} & {\bf $\#$} \\\hline\hline
          {\bf Gauge-IA:}
          $\quad WW\gamma$, $WWZ$, $ggg$
          & $3$ \\\hline
          {\bf Gauge-Lepton-IA:}
          $\ell^+\ell^-\gamma$, $\ell^+\ell^-Z$, $\ell^+ \nu W^-$,
          $\ell^- \bar{\nu} W^+$, $\nu\bar{\nu}Z$
          & $5 \:G \rightarrow~15$ \\\hline
          {\bf Gauge-Quark-IA:}
          $\quad q\bar{q}\gamma$, $q\bar{q}Z$, $u\bar{d}W^-$,
          $d\bar{u}W^+$, $q\bar{q}g$
          &  $2\:G(G~+~3) \newline\rightarrow 36$ \\\hline
          {\bf Higgs-IA:}
          $\quad HHH$
          & $1$\\\hline
          {\bf Higgs-Gauge-IA:}
          $HW^+W^-$, $HZZ$
          &$2$ \\\hline
          {\bf Higgs-Lepton-IA:}
          $\ell^+ \ell^- H$
          & $G \rightarrow~3$ \\\hline
          {\bf Higgs-Quark-IA:}
          $\quad q\bar{q}H$
          & $2\:G \rightarrow 6$  \\ \hline\hline
          {\bf Higgs-Gst.-IA:}
          $H\phi^+ \phi^-$, $H\phi\phi$
          & $2$ \\\hline
          {\bf Gst.-Gauge-IA:}
          $\phi^+\phi^-\gamma$, $\phi^+\phi^-Z$, $\phi^\pm\phi W^\mp$,
          $\phi^\pm H W^\mp$, $\phi HZ$, $\phi^\pm W^\mp Z$, $\phi^\pm
          W^\mp\gamma$ & $11$ \\\hline
          {\bf Gst.-Lepton-IA:}
          $\ell^+\ell^-\phi$, $\ell^-\bar{\nu}\phi^+$, $\ell^+\nu\phi^-$
          & $3\:G\rightarrow 9$ \\\hline
          {\bf Gst.-Quark-IA:}
          $q\bar{q}\phi$, $u\bar{d}\phi^-$, $d\bar{u}\phi^+$ & $2\:G(G+1)
          \rightarrow 24$ \\\hline
\end{tabular}
\caption{\label{smvert1}
{\bf 3-Vertices, SM:}  $\left( 2G^2+14G+6 \right) + \left( 2G^2 + 5G +
13 \right) \rightarrow 66 + 46$, with $G$ being the number of
generations, set to $3$ in the final step.}
\end{table}

\begin{table}
\begin{tabular}
          {|>{\PBS\raggedright\hspace{0pt}}p{9.5cm}
           |>{\PBS\centering\hspace{0pt}}p{2.5cm}|}\hline
\qquad \qquad {\bf VERTICES} & {\bf $\#$} \\\hline\hline
          {\bf Gauge-IA:}
          $WW\gamma\gamma$, $WWZ\gamma$, $WWZZ$, $WWWW$, $gggg$ &
          $5$ \\\hline
          {\bf Higgs-IA:}
          $HHHH$ & $1$ \\\hline
          {\bf Higgs-Gauge-IA:}
          $HHW^+W^-$, $HHZZ$ & $2$ \\\hline\hline
          {\bf Higgs-Gst.-IA:}
          $HH\phi^+\phi^-$, $HH\phi\phi$, $\phi^+\phi^-\phi^+\phi^-$,
          $\phi^+\phi^-\phi\phi$, $\phi\phi\phi\phi$ & $5$ \\\hline
          {\bf Gst.-Gauge-IA:}
          $\phi\phi W^+W^-$, $\phi\phi ZZ$, $\phi^+\phi^-WW$,
          $\phi^+\phi^-ZZ$, $\phi^+\phi^-Z\gamma$,
          $\phi^+\phi^-\gamma\gamma$, $H\phi^\pm W^\mp Z$, $H\phi^\pm
          W^\mp \gamma$, $\phi\phi^\pm W^\mp Z$, $\phi\phi^\pm
          W^\mp \gamma$ & $14$ \\\hline
\end{tabular}
\caption{\label{smvert2} {\bf 4-Vertices, SM:} $\quad8$ vertices and
$19$ additional Goldstone vertices.}      
\end{table}

%%%%%%%%%%%%%%%%%%%%%%%%%%%%%%%%%%%%%%%%%%%%%%%%%%%%%%%%%%%%%%%%%%%%%%
\begin{table}
\setlength{\extrarowheight}{1mm}
\begin{tabular}
         {|>{\PBS\raggedright\hspace{0pt}}p{9.7cm}
          |>{\PBS\centering\hspace{0pt}}p{2.3cm}|}\hline
\qquad \qquad {\bf VERTICES} & {\bf $\#$} \\\hline\hline
 {\bf Gauge-IA:} 
$\quad WW\gamma$, $WWZ$, $ggg$   & $3$
\\\hline 
{\bf Gauge-Lepton-IA:} $\quad 
\ell^+\ell^-\gamma$, $\ell^+\ell^-Z$, $\ell^+ \nu W^-$, $\ell^- \bar{\nu} W^+$,
$\nu\bar{\nu}Z$ & $5 \:G \rightarrow~15$ \\\hline          
 {\bf Gauge-Quark-IA:}
$\quad q\bar{q}\gamma$, $q\bar{q}Z$, $u\bar{d}W^-$, $d\bar{u}W^+$, $q\bar{q}g$ 
              
&  $2\:G(G~+~3) \newline\rightarrow 36$ \\\hline 
{\bf Higgs-IA:}
 $\quad H^+H^-H$, $H^+H^-h$, $HHH$, $HHh$, $Hhh$, $hhh$,
$AAH$, $AAh$   & $8$
\\\hline
{\bf Higgs-Gauge-IA:}           
         $AW^\pm H^\mp$, $HAZ$, $hAZ$, $W^+W^-H$, $W^+W^-h$, $W^\pm
         H^\mp H$, $W^\mp H^\pm h$, $ZZH$, $ZZh$, $H^+H^-\gamma$,
         $H^+H^-Z$
& $14$ \\\hline          
                 {\bf Higgs-Lepton-IA:} 
         $H^+ \bar{\nu}\ell^-$, $H^- \nu \ell^+$, $\ell^+ \ell^- H$,
         $\ell^+ \ell^- h$, $\ell^+ \ell^- A$
& $5\:G \rightarrow~15$ \\\hline 
  {\bf Higgs-Quark-IA:}
 $\quad q\bar{q}H$, $q\bar{q}h$, $q\bar{q}A$,
$u\bar{d}H^-$, $d\bar{u}H^+$  &
$2\:G(G~+~3) \newline\rightarrow 36$  \\\hline  
{\bf Higgs-Chargino-Neutralino-IA:}  
$\tilde{\chi}\tilde{\chi}H$, $\tilde{\chi}\tilde{\chi}h$,
$\tilde{\chi}\tilde{\chi}A$, $\tilde{\chi}^+\tilde{\chi}^-H$,
$\tilde{\chi}^+\tilde{\chi}^-h$, $\tilde{\chi}^+\tilde{\chi}^-A$,
$\tilde{\chi}^\pm\tilde{\chi}H^\mp  
$ & $52$  \\\hline   
 {\bf Slepton-Gauge-IA:} 
$\tilde{\ell}^+\tilde{\ell}^-\gamma$,
$\tilde{\ell}^+\tilde{\ell}^-Z$, $\tilde{\nu}\tilde{\nu}^* Z$,
$\tilde{\ell}^+ \tilde{\nu}W^-$, $\tilde{\ell}^- \tilde{\nu}^*W^+$
 & $11\:G\rightarrow~33$
\\\hline   {\bf
Squark-Gauge-IA:}  $\tilde{q}\tilde{q}^*\gamma$,
$\tilde{q}\tilde{q}^*Z$, $\tilde{d}\tilde{u}^*W^+$,
$\tilde{u}\tilde{d}^*W^-$, $\tilde{q}\tilde{q}^*g$ 
& $8\:G(G~+~2) \newline\rightarrow 120$\\\hline
  {\bf
Chargino-Neutralino-Gluino-Gauge-IA:}  $
\tilde{\chi}\tilde{\chi}Z$, $\tilde{\chi}^+\tilde{\chi}^-Z$,
$\tilde{\chi}^+\tilde{\chi}^-\gamma$, $\tilde{\chi}^+\tilde{\chi}W^-$,
$\tilde{\chi}^-\tilde{\chi}W^+$, $\tilde{g}\tilde{g}g$
& $33$
\\\hline   
{\bf Other Chargino-Neutralino-Gluino-IA:}
 $q\tilde{g}\tilde{q}^*$,
$\bar{q}\tilde{g}\tilde{q}$, $\tilde{\chi}^+\ell^-\tilde{\nu}^*$,
$\tilde{\chi}^-\ell^+\tilde{\nu}$,
$\tilde{\chi}^+\bar{\nu}\tilde{\ell}^-$,
$\tilde{\chi}^-\nu\tilde{\ell}^+$, $\tilde{\chi}^+\bar{u}\tilde{d}$,
$\tilde{\chi}^-u\tilde{d}^*$, $\tilde{\chi}^+d\tilde{u}^*$,
$\tilde{\chi}^-\bar{d}\tilde{u}$, $\tilde{\chi}\nu\tilde{\nu}^*$,
$\tilde{\chi}\bar{\nu}\tilde{\nu}$,
$\tilde{\chi}\ell^\mp\tilde{\ell}^\pm$, $\tilde{\chi}q\tilde{q}^*$,
$\tilde{\chi}\bar{q}\tilde{q}  
$ & $4\:G(4\:G~+~19) \newline\rightarrow 372$ \\\hline
 
{\bf Higgs-Slepton-IA:} 
$H\tilde{\nu}\tilde{\nu}^*$, $h\tilde{\nu}\tilde{\nu}^*$,
$H\tilde{\ell}^+\tilde{\ell}^-$, $h\tilde{\ell}^+\tilde{\ell}^-$,
$A\tilde{\ell}^+\tilde{\ell}^-$, $H^+\tilde{\ell}^-\tilde{\nu}^*, 
H^-\tilde{\ell}^+\tilde{\nu}$ & $18\:G
\rightarrow~54$ \\\hline 
  {\bf
Higgs-Squark-IA:}  $\tilde{q}\tilde{q}^* 
H$, $\tilde{q}\tilde{q}^*h$, $\tilde{q}\tilde{q}^*A$, $\tilde{u}\tilde{d}^*
H^-$, $\tilde{u}^*\tilde{d}H^+$ &
$8\:G(3\:G~+~1) \newline\rightarrow~240$ \\\hline
\end{tabular}
\caption{\label{mssmvert1}{\bf MSSM, 3-Vertices} $\; 52\:G^2 +
151\:G + 110 \rightarrow 1031$, with $G$ being the number of generations, 
set to $3$ in the final step.}
\end{table}

%%%%%%%%%%%%%%%%%%%%%%%%%%%%%%%%%%%%%%%%%%%%%%%%%%%%%%%%%%%%%%%%%%%%

In the last two formulae indices from the middle of the alphabet are
used as generation labels while those from the beginning of the
alphabet are gauge group indices. Note that in the soft breaking terms
only components appear and not the whole superfields. $h^E$, $h^D$ and
$h^U$ are arbitrary complex $3\times 3$-matrices in the space of
generations. There are more arbitrary complex $3\times 3$-matrices in
the soft supersymmetry breaking terms, $A^E$, $A^U$ and $A^D$, and
furthermore $C^E$, $C^U$ and $C^D$, the latter not included in 
most reviews about the
MSSM. $M_{\tilde{Q}/\bar{U}/\bar{D}/\tilde{L}/\bar{E}}$ are five
Hermitean mass square matrices in generation space for the
sparticles. The gaugino masses are allowed to be complex, as well as the
Higgs potential parameters $\mu$ and $(B \mu)$, while the mass squares
$m_{1/2}^2$ must be real. 

We now briefly summarize the mixings of the interaction eigenstates to
the mass eigenstates. All particles with identical colour and
electromagnetic quantum numbers are generally allowed to mix. 

The two Higgs doublets are decomposed into the following mass
eigenstates:
\begin{align}
  \label{eq:higgs3}
  H_1 = \begin{pmatrix} \frac{1}{\sqrt{2}} \bigl( v_1 + H^0 \cos\alpha - h^0
  \sin\alpha + {\rm i} A^0 \sin\beta + {\rm i} \phi^0 \cos\beta \bigr) \\ H^-
  \sin\beta + \phi^- \cos\beta \end{pmatrix},
  \\ \notag \\ H_2 = \begin{pmatrix} H^+ \cos\beta - \phi^+ \sin\beta \\
  \frac{1}{\sqrt{2}}
  \bigl( v_2 + H^0 \sin\alpha + h^0 \cos\alpha + {\rm i} A^0 \cos\beta - {\rm
  i} \phi^0 \sin\beta \bigr)
  \end{pmatrix} \label{eq:higgs4}
\end{align}   
There are now five physical Higgs particles, the scalars $H^0$ and
$h^0$, the pseudoscalar $A^0$ and the charged Higgs' $H^\pm$, thus
called in the case of a non-CP violating Higgs potential. $\phi^\pm$
and $\phi^0$ are the Goldstone bosons attached to the $W$ and $Z$
bosons, respectively. The vacuum expectation values are denoted by
$v_1$ and $v_2$ while $\alpha$ and $\beta$ are two real mixing
angles. 

%%%%%%%%%%%%%%%%%%%%%%%%%%%%%%%%%%%%%%%%%%%%%%%%%%%%%%%%%%%%%%%%%%%%%%
\begin{table}
%\slidesubheading{MSSM: 4-Vertices} 
\setlength{\extrarowheight}{1mm}
\begin{tabular}
         {|>{\PBS\raggedright\hspace{0pt}}p{9.5cm}
          |>{\PBS\centering\hspace{0pt}}p{2.5cm}|}\hline
\qquad \qquad {\bf VERTICES} & {\bf $\#$}
\\\hline\hline 
{\bf Gauge-IA:}  $WW\gamma\gamma$,
$WWZ\gamma$, $WWZZ$, $WWWW$, $gggg$ & $5$
\\\hline 
{\bf Higgs-IA:}  $H^+H^-H^+H^-$, $H^+H^-HH$,
$H^+H^-Hh$, $H^+H^-hh$, $H^+H^-AA$, $HHHH$, $HHHh$, $HHhh$, $Hhhh$,
$hhhh$, $HHAA$, $HhAA$, $hhAA$, $AAAA$ &
$14$ 
\\\hline  {\bf
Higgs-Gauge-IA:}  $HHZZ$, $hhZZ$, $AAZZ$, $H^+H^-ZZ$,
$H^+H^-Z\gamma$, $H^+H^-\gamma\gamma$, $H^\pm H W^\mp \gamma$, $H^\pm
h W^\mp \gamma$, $H^\pm H W^\mp Z$, $H^\pm h W^\mp Z$, $HHW^+W^-$,
$hhW^+W^-$, $AAW^+W^-$, $H^\pm A W^\mp \gamma,H^\pm A W^\mp
Z$  & $21$
\\\hline  
{\bf Slepton-Gauge-IA:} $\tilde{\ell}^+
\tilde{\ell}^- \gamma\gamma $, $\tilde{\ell}^+ \tilde{\ell}^- Z
\gamma$, $\tilde{\ell}^+ \tilde{\ell}^- ZZ$,
$\tilde{\nu}\tilde{\nu}^*ZZ$, $\tilde{\ell}^+ \tilde{\ell}^- W^+ W^-$,
$\tilde{\nu}\tilde{\nu}^* W^+W^-$, $\tilde{\ell}^- \tilde{\nu}^* W^+
\gamma$, $\tilde{\ell}^+ \tilde{\nu} W^- \gamma$, $\tilde{\ell}^-
\tilde{\nu}^* W^+ Z$, $\tilde{\ell}^+ \tilde{\nu} W^- Z$
 & $24 G\rightarrow~72$
\\\hline 
{\bf Squark-Gauge-IA:} $\tilde{q}\tilde{q}^*
\gamma\gamma$, $\tilde{q}\tilde{q}^*Z\gamma$,
$\tilde{q}\tilde{q}^*ZZ$, $\tilde{q}\tilde{q}^*W^+W^-$,
$\tilde{u}\tilde{d}^*W^-\gamma$, $\tilde{u}^*\tilde{d}W^+\gamma$,
$\tilde{u}\tilde{d}^*W^-Z$, $\tilde{u}^*\tilde{d}W^+Z$,
$\tilde{q}\tilde{q}^* gg$, $\tilde{q}\tilde{q}^*g\gamma$,
$\tilde{q}\tilde{q}^*gZ$, $\tilde{u}\tilde{d}^* g W^-$,
$\tilde{u}^*\tilde{d} g W^+$ &
$4G(6G+11)\newline\rightarrow~348$ 
\\\hline
{\bf Slepton-Slepton-IA:}
$\quad\tilde{\nu}\tilde{\nu}^*\tilde{\nu}\tilde{\nu}^*$,
$\tilde{\nu}\tilde{\nu}^*\tilde{\ell}^-\tilde{\ell}^+$,
$\tilde{\ell}^-\tilde{\ell}^+\tilde{\ell}^-\tilde{\ell}^+$ & 
$\frac{25}{2}G^2+\frac{1}{2}G+1\newline\rightarrow~115$
\\\hline 
 {\bf Squark-Squark-IA:}
$\quad\tilde{q}\tilde{q}^*\tilde{q}\tilde{q}^*$,
$\tilde{u}\tilde{u}^*\tilde{d}\tilde{d}^*$ &
$2G(8G^3+8G\newline+1)\rightarrow~1446$ \\\hline 
 {\bf Slepton-Squark-IA:}
$\tilde{q}\tilde{q}^*\tilde{\nu}\tilde{\nu}^*$,
$\tilde{q}\tilde{q}^*\tilde{\ell}^+\tilde{\ell}^-$,
$\tilde{u}\tilde{d}^*\tilde{\ell}^-\tilde{\nu}^*$,
$\tilde{u}^*\tilde{d}\tilde{\ell}^+\tilde{\nu}$ &
$8G^2(2G+5)\newline\rightarrow~792$
\\\hline 
{\bf Higgs-Slepton-IA:}
$HH\tilde{\nu}\tilde{\nu}^*$, $Hh\tilde{\nu}\tilde{\nu}^*$,
$hh\tilde{\nu}\tilde{\nu}^*$, $AA\tilde{\nu}\tilde{\nu}^*$,
$H^+H^-\tilde{\nu}\tilde{\nu}^*$, $HH\tilde{\ell}^+\tilde{\ell}^-$,
$Hh\tilde{\ell}^+\tilde{\ell}^-$, $hh\tilde{\ell}^+\tilde{\ell}^-$,
$AA\tilde{\ell}^+\tilde{\ell}^-$, $H^+H^-\tilde{\ell}^+\tilde{\ell}^-$,
$HH^+\tilde{\ell}^-\tilde{\nu}^*$, $hH^+\tilde{\ell}^-\tilde{\nu}^*$,
$AH^+\tilde{\ell}^-\tilde{\nu}^*$, $HH^-\tilde{\ell}^+\tilde{\nu}$,
$hH^-\tilde{\ell}^+\tilde{\nu}$,
$AH^-\tilde{\ell}^+\tilde{\nu}$  &
$37G\rightarrow~111$ 
\\\hline   
{\bf Higgs-Squark-IA:}  $
\tilde{q}\tilde{q}^*HH$, $\tilde{q}\tilde{q}^*Hh$,
$\tilde{q}\tilde{q}^*hh$, $\tilde{q}\tilde{q}^*AA$,
$\tilde{q}\tilde{q}^*H^+H^-$, $\tilde{u}^*\tilde{d}H^+H$,
$\tilde{u}^*\tilde{d}H^+h$, $\tilde{u}^*\tilde{d}H^+A$,
$\tilde{u}\tilde{d}^*H^-H$, $\tilde{u}\tilde{d}^*H^-h$,
$\tilde{u}\tilde{d}^*H^-A$ &
$8G(3G~+~5)\newline\rightarrow 336$ \\\hline 
\end{tabular}
\caption{\label{mssmvert2}{\bf MSSM, 4-Vertices:}  $\;
16G^4+16G^3+\frac{233}{2}G^2+\frac{295}{2}G+41 \rightarrow
3260$}
\end{table}

%%%%%%%%%%%%%%%%%%%%%%%%%%%%%%%%%%%%%%%%%%%%%%%%%%%%%%%%%%%%%%%%%%%%%%

Most important for the MSSM are the mixings of the charged Higgsinos
(the SUSY partners of the Higgs') and the charged gauginos to mass
eigenstates named {\em charginos}, as well as the neutral Higgsinos
and neutral gauginos are linearly combined to states called {\em
neutralinos}. For the charginos it is justified by the smallness of
observed CP-violating effects to choose the imaginary parts of $\mu$
and $m_{\text{Wino}}$ sufficiently small to define orthogonal instead
of unitary mixing matrices 
\begin{equation}
  \label{eq:charg-u}
  U = \begin{pmatrix} \cos\phi_- & \sin\phi_- \\ - \sin\phi_- & \cos\phi_-
  \end{pmatrix},
\end{equation}
\vspace{5pt}
\begin{align}
  \label{eq:charg-v}
   V = \begin{pmatrix} \cos\phi_+ & \sin\phi_+ \\ - \eta \sin\phi_+ &
   \eta \cos\phi_+ \end{pmatrix}, \qquad \eta =  \text{sgn} \, \left[
   \mu \, m_{\text{Wino}} - m_W^2 \sin(2\beta) \right]  
\end{align}      
with the sign factor $\eta$ guaranteeing that the chargino masses are
positive. The mixing angles are
\begin{equation}
  \label{eq:mischcharg1}
  \tan(2\phi_+) = \frac{- 2\sqrt{2} m_W \left( m_{\text{Wino}} \sin\beta + \mu
  \cos\beta \right)}{m_{\text{Wino}}^2 - \mu^2 + 2 m_W^2 \cos(2\beta)},
\end{equation}
\begin{equation}
  \label{eq:mischcharg2}
  \tan(2\phi_-) = \frac{- 2\sqrt{2} m_W \left( m_{\text{Wino}} \cos\beta + \mu
  \sin\beta \right)}{m_{\text{Wino}}^2 - \mu^2 - 2 m_W^2 \cos(2\beta)} .
\end{equation}     
The charginos $\tilde{\chi}^\pm_i, i=1,2$ are then related to the
charged Winos and Higgsinos as 
\begin{equation}
  \label{eq:relagaugcharg1}
\boxed{
  \begin{array}{r@{}c@{\hspace{1mm}}lcr@{}c@{\hspace{1mm}}l}
    \tilde{W}^+_L & \;=\; & V_{i1}^* \tilde{\chi}_{i,L}^+ & \qquad\qquad &
    \overline{\tilde{W}^+_L} & \;=\; &  \overline{\tilde{\chi}_{i,L}^+} V_{i1} \\
    \tilde{W}^+_R & \;=\; & U_{i1} \tilde{\chi}_{i,R}^+ & \qquad\qquad &
    \overline{\tilde{W}^+_R} & \;=\; &  \overline{\tilde{\chi}_{i,R}^+}
        U^*_{i1} \\ 
    \tilde{H}^+_L & \;=\; & V_{i2}^* \tilde{\chi}_{i,L}^+ & \qquad\qquad &
    \overline{\tilde{H}^+_L} & \;=\; &  \overline{\tilde{\chi}_{i,L}^+} V_{i2} \\
    \tilde{H}^+_R & \;=\; & U_{i2} \tilde{\chi}_{i,R}^+ & \qquad\qquad &
    \overline{\tilde{H}^+_R} & \;=\; &  \overline{\tilde{\chi}_{i,R}^+} U^*_{i2}
  \end{array}}
\end{equation}
\begin{equation}
  \label{eq:relagaugcharg2}
\boxed{
  \begin{array}{r@{}c@{\hspace{1mm}}lcr@{}c@{\hspace{1mm}}l}
    \tilde{W}^-_L & \;=\; & U_{i1}^* \tilde{\chi}_{i,L}^- & \qquad\qquad &
    \overline{\tilde{W}^-_L} & \;=\; & \overline{\tilde{\chi}_{i,L}^-} U_{i1} \\
    \tilde{W}^-_R & \;=\; & V_{i1} \tilde{\chi}_{i,R}^- & \qquad\qquad &
    \overline{\tilde{W}^-_R} & \;=\; & \overline{\tilde{\chi}_{i,R}^-} V^*_{i1} \\
    \tilde{H}^-_L & \;=\; & U_{i2}^* \tilde{\chi}_{i,L}^- & \qquad\qquad &
    \overline{\tilde{H}^-_L} & \;=\; & \overline{\tilde{\chi}_{i,L}^-} U_{i2} \\
    \tilde{H}^-_R & \;=\; & V_{i2} \tilde{\chi}_{i,R}^- & \qquad\qquad &
    \overline{\tilde{H}^-_R} & \;=\; & \overline{\tilde{\chi}_{i,R}^-} V^*_{i2}
  \end{array}}
\end{equation}              

For the neutralinos, we introduce the $4\times 4$-matrices $N$ used to
diagonalize the mass matrix 
\begin{align}
  & Y^{0\,\prime} = \notag \\ &
 \begin{pmatrix} m_{\text{Bino}} & 0 & m_Z
  \sin\theta_W \cos\beta & - m_Z \sin\theta_W \sin\beta \\ 0 &
  m_{\text{Wino}} & - m_Z \cos\theta_W \cos\beta & m_Z
  \cos\theta_W \sin\beta \\ m_Z \sin\theta_W \cos\beta & - m_Z
  \cos\theta_W \cos\beta & 0 & - \mu \\ - m_Z \sin\theta_W \sin\beta &
  m_Z \cos\theta_W \sin\beta & - \mu & 0 \end{pmatrix}
\end{align}     
in the form $N^* Y^{0\,\prime} N^{-1} = N_D$; $\theta_W$ is the
Weinberg angle of the electroweak theory. The neutralinos
$\tilde{\chi}^0_i, i=1,\ldots,4$ are then defined as
\begin{equation}
  \label{eq:relagaugneutr1}
\boxed{
  \begin{array}{r@{}c@{\hspace{1mm}}lcr@{}c@{\hspace{1mm}}l}
  \tilde{B}_L & \; = \; & \eta_i^* N_{i1}^* \tilde{\chi}_{i,L}^0 & \qquad\qquad &
  \overline{\tilde{B}_L} & \; = \; &\overline{\tilde{\chi}_{i,L}^0} \eta_i N_{i1}
  \\
  \tilde{B}_R & \; = \; & \eta_i N_{i1} \tilde{\chi}_{i,R}^0  & \qquad\qquad &
  \overline{\tilde{B}_R} & \; = \; &\overline{\tilde{\chi}_{i,R}^0} \eta_i^*
  N_{i1}^*\\
  \tilde{W}^3_L & \; = \; & \eta_i^* N_{i2}^* \tilde{\chi}_{i,L}^0 & \qquad\qquad &
  \overline{\tilde{W}^3_L} & \; = \; &\overline{\tilde{\chi}_{i,L}^0} \eta_i N_{i2}
  \\
  \tilde{W}^3_R & \; = \; & \eta_i N_{i2} \tilde{\chi}_{i,R}^0  & \qquad\qquad &
  \overline{\tilde{W}^3_R} & \; = \; &\overline{\tilde{\chi}_{i,R}^0} \eta_i^*
  N_{i2}^* \\
  \tilde{H}^0_{1,L} & \; = \; & \eta_i^* N_{i3}^* \tilde{\chi}_{i,L}^0 & \qquad\qquad &
  \overline{\tilde{H}^0_{1,L}} & \; = \; &\overline{\tilde{\chi}_{i,L}^0} \eta_i N_{i3}
  \\
  \tilde{H}^0_{1,R} & \; = \; & \eta_i N_{i3} \tilde{\chi}_{i,R}^0  & \qquad\qquad &
  \overline{\tilde{H}^0_{1,R}} & \; = \; &\overline{\tilde{\chi}_{i,R}^0} \eta_i^*
  N_{i3}^*  \\
  \tilde{H}^0_{2,L} & \; = \; & \eta_i^* N_{i4}^* \tilde{\chi}_{i,L}^0 & \qquad\qquad &
  \overline{\tilde{H}^0_{2,L}} & \; = \; &\overline{\tilde{\chi}_{i,L}^0} \eta_i N_{i4}  \\
  \tilde{H}^0_{2,R} & \; = \; & \eta_i N_{i4} \tilde{\chi}_{i,R}^0 & \qquad\qquad &
  \overline{\tilde{H}^0_{2,R}} & \; = \; &\overline{\tilde{\chi}_{i,R}^0} \eta_i^*
  N_{i4}^*
  \end{array}}
\end{equation}    
The $\eta_i$ are phases to guarantee the positivity of the neutralino
masses. 

%%%%%%%%%%%%%%%%%%%%%%%%%%%%%%%%%%%%%%%%%%%%%%%%%%%%%%%%%%%%%%%%%%%%%%%
\begin{table}
\setlength{\extrarowheight}{1mm}
\begin{tabular}
         {|>{\PBS\raggedright\hspace{0pt}}p{9.5cm}
          |>{\PBS\centering\hspace{0pt}}p{2.5cm}|}\hline
\qquad \qquad {\bf VERTICES} & {\bf $\#$}
\\\hline\hline  {\bf
Higgs-Gst.-IA:}  $HA\phi$, $hA\phi$, $H^\pm
H\phi^\mp$, $H^\pm h\phi^\mp$, $H^\pm A\phi^\mp, H\phi\phi$,
$h\phi\phi$, $H\phi^+\phi^-$, $h \phi^+\phi^-$ &
$12$ \\\hline  
{\bf Higgs-Gst.-Gauge-IA:}  $Z\phi^+\phi^-$,
$\gamma\phi^+\phi^-$, $W^\pm \phi^\mp\phi$, $ZH\phi$, $Zh\phi$,
$W^\pm\phi^\mp H$, $W^\pm \phi^\mp h$, $W^\pm Z \phi^\mp$, $W^\pm
\gamma \phi^\mp$   & $14$ \\\hline
  {\bf
Gst.-Lepton-IA:} $\ell^+\ell^-\phi$,
$\ell^-\bar{\nu}\phi^+$, $\ell^+\nu \phi^-$  &
$3G\rightarrow~9$ \\\hline 
 {\bf Gst.-Quark-IA:}
$q\bar{q}\phi, u\bar{d}\phi^-$, $d\bar{u}\phi^+$ &
$2G(G+1)\newline\rightarrow 24$ \\\hline
  {\bf
Gst.-C/N-ino-IA:}    $\tilde{\chi}\tilde{\chi}\phi$,
$\tilde{\chi}^+\tilde{\chi}^-\phi$,
$\tilde{\chi}^+\tilde{\chi}\phi^-$,
$\tilde{\chi}^-\tilde{\chi}\phi^+$ & ${\color{blue}
30}$ \\\hline 
  {\bf
Gst.-Slepton-IA:}   
$\tilde{\ell}^+\tilde{\ell}^-\phi$,
$\tilde{\ell}^-\tilde{\nu}^*\phi^+$, $\tilde{\ell}^+ \tilde{\nu}
\phi^-$ & $8G\rightarrow{\color{blue} 24}$ \\\hline
  {\bf
Gst.-Squark-IA:}  $\tilde{q}\tilde{q}^*\phi$,
$\tilde{u}\tilde{d}^*\phi^-$,
$\tilde{u}^*\tilde{d}\phi^+$ &
$8G(G+1)\newline\rightarrow {\color{blue} 96}$ \\\hline
\end{tabular}
\caption{\label{mssmvert3}{\bf MSSM, 3-Goldstone-Vertices:}  $\;10G^2+21G+56
  \rightarrow 209$} 
\end{table}

%%%%%%%%%%%%%%%%%%%%%%%%%%%%%%%%%%%%%%%%%%%%%%%%%%%%%%%%%%%%%%%%%%%%%%

For each Standard Model fermion there are two complex scalar
superpartners, which can also mix due to the soft breaking
terms. Under (not too) special circumstances the transformations from
the interaction to the mass eigenstates can be assumed to be
orthogonal. 

The matrices of the mass squares for the sfermions to be diagonalized
will always be denoted in the form 
\begin{equation}
        M^2_{\tilde{f}} = \begin{pmatrix}
        m^2_{\tilde{f}_L} & m^2_{\tilde{f}_{L/R}} \\
        \left(m^2_{\tilde{f}_{L/R}}\right)^* & m^2_{\tilde{f}_R} 
        \end{pmatrix} \quad .
\end{equation}
It is easy to perform the diagonalization leading to the mass square
eigenvalues (usually taken as $m^2_{\tilde{f}_1} \leq
m^2_{\tilde{f}_2}$):
\begin{equation}
        \label{masssfermeigen}
        m^2_{\tilde{f}_{1/2,i}} = \dfrac{1}{2} \left(
        m^2_{\tilde{f}_{L,i}} + m^2_{\tilde{f}_{R,i}} \right) \mp
        \dfrac{1}{2} \sqrt{ \left( m^2_{\tilde{f}_{L,i}} -
        m^2_{\tilde{f}_{R,i}} \right)^2 + 4 \left|
        m^2_{\tilde{f}_{L/R}} \right|^2} \quad ,   
\end{equation}

\begin{table}
\setlength{\extrarowheight}{1mm}
\begin{tabular}
         {|>{\PBS\raggedright\hspace{0pt}}p{9.5cm}
          |>{\PBS\centering\hspace{0pt}}p{2.5cm}|}\hline
\qquad \qquad {\bf VERTIZES} & {\bf $\#$} \\\hline\hline
{\bf Higgs-Gst.-IA:}
$HHA\phi$, $HhA\phi$, $hhA\phi$, $AAA\phi$, $H^+H^-A\phi$, $H^\pm
HH\phi^\mp$, $H^\pm Hh\phi^\mp$, $H^\pm hh\phi^\mp$, $H^\pm
AA\phi^\mp$, $H^\pm HA\phi^\mp$, $H^\pm hA\phi^\mp$, $H^\pm
H^+H^-\phi^\mp$, $HH\phi\phi$, $Hh\phi\phi$, $hh\phi\phi$,
$AA\phi\phi$, $H^+H^-\phi\phi$, $H^\pm H \phi^\mp \phi$, $H^\pm h\phi^\mp
\phi$, $H^\pm A \phi^\mp \phi$, $HH\phi^+\phi^-$, $Hh\phi^+\phi^-$,
$hh\phi^+\phi^-$, $AA\phi^+\phi^-$, $H^+H^-\phi^+\phi^-$, $H^\pm
H^\pm\phi^\mp \phi^\mp$, $A\phi\phi\phi$, $H^\pm\phi^\mp\phi\phi$,
$A\phi^+\phi^-\phi$, $H^\pm\phi^\mp\phi^+\phi^-$, $\phi\phi\phi\phi$,
$\phi^+\phi^-\phi\phi$, $\phi^+\phi^-\phi^+\phi^-$ &
${\color{blue} 46}$ \\\hline 
{\bf Higgs-Gst.-Gauge-IA:} $ZZ\phi\phi$,
$ZZ\phi^+\phi^-$, $Z\gamma\phi^+\phi^-$, $\gamma\gamma\phi^+\phi^-$,
$W^+W^-\phi^+\phi^-$, $W^+W^-\phi\phi$, $W^\pm Z \phi^\mp \phi$,
$W^\pm\gamma \phi^\mp \phi$, $W^\pm Z \phi^\mp H$, $W^\pm\gamma
\phi^\mp H$, $W^\pm Z \phi^\mp h$, $W^\pm\gamma \phi^\mp h $
 & ${\color{blue} 18}$ \\\hline 
 {\bf Slepton-Gst.-IA:} $
\tilde{\nu}\tilde{\nu}^*A\phi$,
$\tilde{\nu}\tilde{\nu}^*H^\pm\phi^\mp$,
$\tilde{\nu}\tilde{\nu}^*\phi^+\phi^-$,
$\tilde{\nu}\tilde{\nu}^*\phi\phi$,
$\tilde{\ell}^+\tilde{\ell}^-A\phi$,
$\tilde{\ell}^+\tilde{\ell}^-H^\pm\phi^\mp$,
$\tilde{\ell}^+\tilde{\nu}h\phi^-$,
$\tilde{\ell}^-\tilde{\nu}^*h\phi^+$,
$\tilde{\ell}^+\tilde{\nu}A\phi^-$,
$\tilde{\ell}^-\tilde{\nu}^*A\phi^+$, $\tilde{\ell}^+\tilde{\nu}
\phi\phi^-$, $\tilde{\ell}^-\tilde{\nu}^*\phi\phi^+$,
$\tilde{\ell}^+\tilde{\ell}^- \phi^+\phi^-$,
$\tilde{\ell}^+\tilde{\ell}^- \phi\phi$,
$\tilde{\ell}^+\tilde{\nu}H^-\phi$,
$\tilde{\ell}^-\tilde{\nu}^*H^+\phi$,
$\tilde{\ell}^+\tilde{\nu}H\phi^-$,
$\tilde{\ell}^-\tilde{\nu}^*H\phi^+$ &
$45G\newline\rightarrow{\color{blue} 135}$ \\\hline 
 {\bf Squark-Gst.-IA:} 
$\tilde{q}\tilde{q}^*A\phi$, $\tilde{q}\tilde{q}^*H^\pm\phi^\mp$,
$\tilde{q}\tilde{q}^* \phi^+\phi^-$, $\tilde{q}\tilde{q}^*\phi\phi$,
$\tilde{u}^*\tilde{d}H^+\phi$, $\tilde{u}\tilde{d}^*H^-\phi$,
$\tilde{u}^*\tilde{d}H\phi^+$, $\tilde{u}\tilde{d}^*H\phi^-$,
$\tilde{u}^*\tilde{d}h\phi^+$, $\tilde{u}\tilde{d}^*h\phi^-$,
$\tilde{u}^*\tilde{d}A\phi^+$, $\tilde{u}\tilde{d}^*A\phi^-$,
$\tilde{u}^*\tilde{d}\phi^+\phi$,
$\tilde{u}\tilde{d}^*\phi^-\phi$  
& $40G\cdot\newline(G+1)\newline\rightarrow {\color{blue} 480}$\\\hline
\end{tabular}
\caption{\label{mssmvert4}{\bf MSSM, 4-Goldstone-Vertices:}  $\; 40G^2 + 85G
+ 64 \rightarrow 679$}
\end{table}

while the mixing angle is the solution of 
\begin{equation}
  \label{eq:sfermmisch}
  \tan \theta_{\tilde{f}_i} = \frac{2
  m_{\tilde{f}_{L/R,i}}^2}{m_{\tilde{f}_{L,i}}^2 - 
  m_{\tilde{f}_{R,i}}^2}. 
\end{equation}
The mixing is given by 
\begin{equation}
  \label{eq:orthosferm}
  \begin{array}{rcl}
   \tilde{f}_{1,i} & = & \tilde{f}_{L,i} \cos\theta_{\tilde{f}_i} +
   \tilde{f}_{R,i} \sin\theta_{\tilde{f}_i} \\
   \tilde{f}_{2,i} & = & - \tilde{f}_{L,i} \sin\theta_{\tilde{f}_i} +
   \tilde{f}_{R,i} \cos\theta_{\tilde{f}_i}
  \end{array} \quad .
\end{equation}
Note that this is only an orthogonal transformation for real symmetric
mass square matrices. If $m^2_{\tilde{f}_{L/R}}$ does have an
imaginary part, then there are additional phases involved in
(\ref{masssfermeigen}) and (\ref{eq:sfermmisch}), but they have to be
drastically small in order not to contradict CP-violation
observations.  

We now list the mass square matrices for the up and down squarks as
well as for the sleptons and sneutrinos. 

\underline{up squarks}:
\begin{subequations}
\begin{align}
        m^2_{\tilde{u}_{L,i}} &=\; m^2_{\tilde{Q}_i} + m^2_{u_i} +
        m_Z^2 \cos(2\beta) \left( \dfrac{1}{2} - \dfrac{2}{3} \sin^2
        \theta_W \right) \\ 
        m^2_{\tilde{u}_{R,i}} &=\; m^2_{\bar{U}_i} + m^2_{u_i} +
        \dfrac{2}{3} m_Z^2 \cos(2\beta) \sin^2 \theta_W  \\
        m^2_{\tilde{u}_{L/R,i}} &=\; m_{u_i} \cdot \left( A_i^U +
        (C_i^U + \mu^*) \cot\beta \right)
\end{align}
\end{subequations}

\vspace{2pt}

\underline{down squarks}:
\begin{subequations}
\end{subequations}
\begin{align}
        m^2_{\tilde{d}_{L,i}} &=\; m^2_{\tilde{Q}_i} + m^2_{d_i} -
        m_Z^2 \cos(2\beta) \left( \dfrac{1}{2} - \dfrac{1}{3} \sin^2
        \theta_W \right) \\ 
        m^2_{\tilde{d}_{R,i}} &=\; m^2_{\bar{D}_i} + m^2_{d_i} -
        \dfrac{1}{3} m_Z^2 \cos(2\beta) \sin^2 \theta_W  \\
        m^2_{\tilde{d}_{L/R,i}} &=\; m_{d_i} \cdot \left( A_i^D -
        (C_i^D + \mu^*) \tan\beta \right)
\end{align}
\vspace{2pt}

\underline{sleptons}:
\begin{subequations}
\begin{align}
        m^2_{\tilde{\ell}_{L,i}} &=\; m^2_{\tilde{L}_i} + m^2_{\ell_i} +
        \dfrac{1}{2} m_Z^2 \cos(2\beta) \sin^2
        \theta_W  \\ 
        m^2_{\tilde{\ell}_{R,i}} &=\; m^2_{\bar{E}_i} + m^2_{\ell_i} -
        m_Z^2 \cos(2\beta) \sin^2 \theta_W  \\
        m^2_{\tilde{\ell}_{L/R,i}} &=\; m_{\ell_i} \cdot \left( A_i^E +
        (C_i^E + \mu^*) \tan\beta \right)
\end{align}
\end{subequations}

\vspace{2pt}
\underline{sneutrinos}:

As long as there are no right-handed neutrino fields there is no
mixing between the left- and righthanded sneutrinos. Their mass square
is 
\begin{equation}
  \label{eq:sneumass}
  m_{L,i}^2 + \frac{1}{4} m_Z^2 \cos(2\beta) \sin^2 \theta_W
\end{equation}     

The discussion of the mass terms and the CKM mixing is more or less
the same as in the Standard Model or the non-supersymmetric two
Higgs-doublet model. 

For more details about the MSSM cf.~\cite{Kuroda:1999:MSSM},
\cite{Reuter:2000:SUSY}.

%%% Local Variables: 
%%% mode: latex
%%% TeX-master: "diss"
%%% End: 

%% file: appen1.tex
%%% Local Variables: 
%%% mode: glatex
%%% TeX-master: "swi"
%%% End: 

\chapter{Some technicalities}

\section{Proof of (\ref{eq:proj})}\label{proofinvfourier}

Here we want to prove the relations (\ref{eq:proj}), which are nothing 
but the inverse Fourier transformations back from the Fourier modes to the
field operators. The Fourier expansions of the field operators are
\begin{align}
    \phi (x) & = \int \dfrac{d^3 \vec{k}}{(2 \pi)^3 2 E} \left( a (k) e^{-
    \ii k x} + a^\dagger (k) e^{+ \ii k x} \right), \notag \\
    \psi (x) & = \int \dfrac{d^3 \vec{k}}{(2 \pi)^3 2 E} \; \sum_\sigma \left(
    b (k, \sigma) u (k, \sigma) e^{- \ii k x} + d^\dagger (k, \sigma) v
    (k, \sigma) e^{+ \ii k x} \right), \\
    \overline{\psi} (x) & = \int \dfrac{d^3 \vec{k}}{(2 \pi)^3 2 E} \;
    \sum_\sigma \left( b^\dagger (k, \sigma) \overline{u} (k, \sigma) e^{+
    \ii k x} + d (k, \sigma) \overline{v} (k, \sigma) e^{- \ii k x} 
    \right).    \notag 
\end{align}

In the Majorana case the last two relations are equivalent and read:
\begin{equation}
  \begin{aligned}
    \psi (x) & = \int \dfrac{d^3 \vec{k}}{(2 \pi)^3 2 E} \; \sum_\sigma \left(
    b (k, \sigma) u (k, \sigma) e^{- \ii k x} + b^\dagger (k, \sigma) v
    (k, \sigma) e^{+ \ii k x} \right), \\
    \overline{\psi} (x) & = \int \dfrac{d^3 \vec{k}}{(2 \pi)^3 2 E} \;
    \sum_\sigma \left( b^\dagger (k, \sigma) \overline{u} (k, \sigma) e^{+
    \ii k x} + b (k, \sigma) \overline{v} (k, \sigma) e^{- \ii k x} 
    \right)       
  \end{aligned}
\end{equation}

The inverse relations (\ref{eq:proj}) can be simply verified by inserting the
Fourier expansions of the field operators: 
\begin{align}
    a (k) \stackrel {!}{=} & \; \ii \int d^3 \vec{x} \, e^{\ii k x}
    \stackrel{\leftrightarrow}{\partial}_t \int \dfrac{d^3 \vec{p}}{(2 \pi)^3
    2 E} \biggl( a (p) e^{-\ii p x} + a^\dagger (p) e^{+ \ii p x}
    \biggr)  
    \notag\\ = & \; \ii \int d^3 \vec{x} e^{\ii k x} \int \dfrac{d^3
    \vec{p}}{(2 \pi)^3 2 E} \biggl( a (p) (- \ii E - \ii k^0 ) e^{-
    \ii p x} + a^\dagger (p) (+ \ii E - \ii k^0 ) e^{+ \ii p
    x} \biggr) \notag\\ =  & \;
    \int \dfrac{d^3 \vec{p}}{2 E} \biggl( a (p) (E + k^0 ) \delta^3 (\vec{k} -
    \vec{p} ) \biggr|_{k^0 = E} + a^\dagger (p) (k^0 - E) \delta^3 (\vec{k} +
    \vec{p})\biggr|_{k^0 = E} e^{2 \ii E x^0} \biggr) \notag\\
    = & \; a (k)  \qquad \surd
\end{align}
In the fermionic case:
\begin{align}
    b (k, \sigma) \stackrel{!}{=} & \; \int d^3 \vec{x} \biggl( \overline{u}
    (k, \sigma) \gamma^0 \int \dfrac{d^3 \vec{p}}{(2 \pi)^3 2 E} \sum_\tau
    \left( b (p, \tau) u (p, \tau) e^{- \ii p x} \right. \notag\\
      & \qquad\qquad\qquad \left. + d^\dagger (p, \tau) v
    (p, \tau) e^{+ \ii p x} \right) \biggr) e^{\ii k x} \notag\\ = & \; 
    \overline{u} (k, \sigma) \gamma^0 \int \dfrac{d^3 \vec{p}}{2 E} \sum_\tau
    \biggl( b (p, \tau) u (p, \tau) \delta^3 (\vec{p} - \vec{k}) \biggr|_{k^0
    = E} \notag\\ & \qquad\qquad\qquad + d^\dagger (p, \tau) v (p, \tau) \delta^3
    (\vec{k} + \vec{p}) \biggr|_{k^0 = E} e^{2 \ii E x^0} \biggr) \notag\\ = & \;
    \dfrac{1}{2 E} \sum_\tau \biggl( b (k, \tau) \overline{u} (\vec{k},
    \sigma) \gamma^0 u (\vec{k}, \tau) + d^\dagger ( k, \tau) \overline{u}
    (\vec{k}, \sigma) \gamma^0 v (- \vec{k}, \tau) \biggr) \notag\\
    = & \; b (k, \sigma) \qquad \surd
\end{align}
In the last step we used the identities
\begin{equation}
  \label{eq:ident}
  \begin{aligned}
    \overline{u} (\vec{k}, \sigma) \gamma^0 u (\vec{k}, \tau) = u^\dagger
    (\vec{k}, \sigma) u (\vec{k}, \tau) & = 2 E \, \delta_{\sigma\tau} \\
    u^\dagger (\vec{k}, \sigma) v (- \vec{k}, \tau) & = 0 ,
  \end{aligned}
\end{equation}
which, for example, can be found in the book of Peskin/Schroeder
\cite{Peskin/Schroeder:QFT:Text} on p. 48.

%%%%%%%%%%%%%%%%%%%%%%%%%%%%%%%%%%%%%%%%%%%%%%%%%%%%%%%%%%%%%%%%%%%%%%

\section{Fierz identities}\label{appen_fierz}

We briefly summarize the Fierz identities as they can be found in
\cite{Itzykson/Zuber:1980:textbook}. In the following, $\theta_i,
i=1,\ldots,4$ are four anticommuting, i.e.~Grassmann-odd
four-component spinors like fermion field operators or superspace
coordinates. The scalar, vectorial, tensorial, axialvectorial and
pseudoscalar combination are
\begin{subequations}
\begin{align}
        s(4,2\,;\,3,1) =&\;  \left( \overline{\theta}_4 \theta_2 \right)
        \left( \overline{\theta}_3 \theta_1 \right) \\ 
        v(4,2\,;\,3,1) =&\;  \left( \overline{\theta}_4 \gamma^\mu
        \theta_2 \right) \left( \overline{\theta}_3 \gamma_\mu
        \theta_1 \right) \\ 
        t(4,2\,;\,3,1) =&\;  \dfrac{1}{2} \left( \overline{\theta}_4
        \sigma^{\mu\nu} \theta_2 \right) \left( \overline{\theta}_3
        \sigma_{\mu\nu} \theta_1 \right) \\ 
        a(4,2\,;\,3,1) =&\;  \left( \overline{\theta}_4 \gamma^5
        \gamma^\mu \theta_2 \right) \left( \overline{\theta}_3
        \gamma_\mu \gamma^5 \theta_1 \right) \\ 
        p(4,2\,;\,3,1) =&\;  \left( \overline{\theta}_4 \gamma^5
        \theta_2 \right) \left( \overline{\theta}_3 \gamma^5 \theta_1
        \right)  
\end{align}
\end{subequations}
Watch carefully the convention with respect to the axial vector
adopted from \cite{Itzykson/Zuber:1980:textbook}. The Fierz identities
provide a possibility to rewrite the spinor products with combinations
$(4,2\,;\,3,1)$ as $(4,1\,;\,3,2)$:
\begin{equation}
        \begin{pmatrix} s \\ v \\ t \\ a \\ p \end{pmatrix}
        (4,2\,;\,3,1) = - \, \dfrac{1}{4} \begin{pmatrix} 1 & 1 & 1 &
        1 & 1 \\ 4 & -2 & 0 & 2 & -4 \\ 6 & 0 & -2 & 0 & 6 \\ 4 & 2 &
        0 & -2 & -4 \\ 1 & -1 & 1 & -1 & 1 \end{pmatrix} \;\; 
        \begin{pmatrix} s \\ v \\ t \\ a \\ p \end{pmatrix}
        (4,1\,;\,3,2) 
\end{equation}

%%%%%%%%%%%%%%%%%%%%%%%%%%%%%%%%%%%%%%%%%%%%%%%%%%%%%%%%%%%%%%%%%%%%%%

\section{Derivation of couplings with momenta}

In this short aside we want to get rid of the confusion with respect to the
signs of momenta in 3-point vertices, e.g.~arising in gauge theories by
coupling two scalar fields to a gauge boson. The term under consideration
is established by the trilinear terms in the kinetic parts in the Lagrangean
density after having substituted the partial by the gauge covariant
derivatives. By splitting the (necessarily complex) fields (charge!) in real
and imaginary part we arrive at couplings with real fields. In the sequel we
consider a 3-point vertex in which the scalar fields $\phi_1$ and $\phi_2$
possess the momenta $p_1$ and $p_2$ flowing into the vertex while the vector
boson $A_\mu$ -- not of interest in the following -- has the incoming momentum
$p_3$. The vertex looks like:

\vspace{0.5cm}

\begin{equation} 
       \parbox{28mm}{\fmfframe(2,2)(2,1){\begin{fmfgraph*}(24,24)
         \Threeexternal{\phi_1,,p_1,,x_1}{\phi_2,,p_2,,x_2}{A,,\mu,,p_3,,x_3}
         \fmf{dashes}{v,e1}
         \fmf{dashes}{v,e2}
         \fmf{photon}{v,e3}
         \fmfv{d.sh=circle,d.si=dot_size,label=$y$}{v}       
         \threeincoming
       \end{fmfgraph*}}} 
\end{equation}

\vspace{0.5cm}

Its analytical form is:
\begin{equation}
  \mathcal{L}_{\text{int}} = e A_\mu \left( \phi_1 \partial^\mu \phi_2 -
    \phi_2 \partial^\mu \phi_1 \right) 
\end{equation}
All prefactors, numerical ones and also factors of $\ii$ are understood
to have been absorbed into the ``coupling constant''. We think about
this vertex as being 
part of an $n$-point Green function, so that the fields of this interaction
term are contracted with other field operators to give the propagators
to be discussed below. This is shown
here only with one term, the other is analogous; furthermore, as mentioned
above, we ignore the vector field. It is not needed in the following
discussion. To be more precise, we add to the derivatives the spacetime
argument they act upon:
\begin{equation}
  \contracted{}{\phi_1}{(x_1)}{\phi_1}{(y)} \partial^\mu_y 
  \contracted{}{\phi_2}{(y)}{\phi_2}{(x_2)} = D_F (x_1 - y) \partial^\mu_y D_F 
  (y-x_2)
\end{equation}
For the Feynman propagators in momentum space we have to perform the
Fourier transformation from coordinate space with the momentum flowing from
$y$ to $x_1$, that is $-p_1$, and with the momentum flowing from $x_2$ to $y$,
that is $p_2$. Note that due to
\begin{equation}
  D_F (x-y) = \contracted{}{\phi}{(x)}{\phi}{(y)} = \Greensfunc{\phi (x) \phi
    (y)} = \int \dfrac{d^4 p}{(2\pi)^4} \dfrac{\ii e^{-\ii p
    (x-y)}}{p^2 - m^2 + \ii \epsilon} 
\end{equation}
the Fourier momentum flows from $y$ to $x$ according to the time
ordering. This yields:
\begin{equation}
  D_F (x_1 - y) \partial^\mu_y D_F (y-x_2)
  \stackrel{\text{F.T.}}{\longrightarrow}  (-\ii p_2^\mu) \text{F.T.}
  D_F (x_1 - y) D_F (y-x_2). 
\end{equation}
Finally, the analytical expression for the vertex (with the
additional factor $\ii$ stemming from the perturbation expansion) becomes:
\begin{equation} 
       \parbox{28mm}{\fmfframe(2,2)(2,1){\begin{fmfgraph*}(24,24)
         \Threeexternal{\phi_1,,p_1,,x_1}{\phi_2,,p_2,,x_2}{A,,\mu,,p_3,,x_3}
         \fmf{dashes}{v,e1}
         \fmf{dashes}{v,e2}
         \fmf{photon}{v,e3}
         \fmfv{d.sh=circle,d.si=dot_size,label=$y$}{v}       
         \threeincoming
       \end{fmfgraph*}}} \qquad \qquad = \quad e \left( p_2 - p_1 \right)_\mu
\end{equation}

This can be stated as the mnemonic:
\begin{equation}
  \boxed{\ii \partial^\mu \longrightarrow + (\text{incoming momentum})^\mu}
\end{equation}

%%% Local Variables: 
%%% mode: latex
%%% TeX-master: t
%%% End: 

%% file: appen_sym.tex
\chapter{Details to the supersymmetric current} 

\section[The current for a general model]{The current for a general model without gauge symmetry}\label{sec:currentgenmodel}\label{appen:generalcurrent}

In this section we want to briefly repeat the derivation of the supersymmetric
current for a general model without gauge symmetries from
\cite{Weinberg:QFTv3:Text} and prove its conservation not shown in the
reference. The Lagrangean density of a general model given by a quantum field
theory endowed with exact supersymmetry can be found in the equations
(26.3.30), (26.4.7) and (26.7.7) in \cite{Weinberg:QFTv3:Text}: 
\begin{multline}
  \label{eq:allgmodell}
  {\cal L} = \sum_n \biggl[ (\partial_\mu \phi_n^*) (\partial^\mu \phi_n) +
  F_n^* F_n + \dfrac{\ii}{2} \left( \overline{\psi_{n,L}} \,
  \fmslash{\partial} \, \psi_{n,L} \right) + \dfrac{\ii}{2} \left(
  \overline{\psi_{n,R}} \, \fmslash{\partial} \, \psi_{n,R} \right) \biggr] \\
  - \dfrac{1}{2} \sum_{n,m} \dfrac{\partial^2 f (\phi)}{\partial \phi_n
  \partial \phi_m} \left( \overline{\psi_{n,R}} \psi_{m,L} \right) -
  \dfrac{1}{2} \sum_{n,m} \left( \dfrac{\partial^2 f (\phi)}{\partial \phi_n
  \partial \phi_m} \right)^* \left( \overline{\psi_{n,L}} \psi_{m,R} \right)
  \\ + \sum_n F_n \dfrac{\partial f (\phi)}{\partial \phi_n} + \sum_n F_n^*
  \left( \dfrac{\partial f (\phi)}{\partial \phi_n} \right)^*    
\end{multline}
In this general model there are $n$ different chiral superfields. 
$f$ is an arbitrary function of these chiral superfields;
when we impose renormalizability as a constraint, it is only allowed to
be a polynom with degree three as an upper bound. The Noether part can be
calculated by (\ref{eq:noetheranteil}) 
\begin{equation}
  \label{eq:noetheranteilallg}
  \begin{aligned}
    N^\mu = & \; - \sum_n \biggl[ \sqrt{2} (\partial^\mu \phi_n^*) \psi_{n,L}
    + \sqrt{2} (\partial^\mu \phi_n) \psi_{n,R} + \dfrac{1}{\sqrt{2}}
    (\fmslash{\partial} \phi_n) \gamma^\mu \psi_{n,R}  \\ & \qquad \qquad
    \qquad + \dfrac{1}{\sqrt{2}} (\fmslash{\partial} \phi_n^*) \gamma^\mu
    \psi_{n,L} - \dfrac{\ii}{\sqrt{2}} F_n \gamma^\mu \psi_{n,R} - 
    \dfrac{\ii}{\sqrt{2}} F_n^* \gamma^\mu \psi_{n,L} \biggr] 
  \end{aligned}
\end{equation}

Next we derive the SUSY transformation of the general Lagrangean density in the
same manner as for the WZ model, writing it in the form 
\begin{equation}
  \label{eq:allmodellkonstrukt}
  S_{\text{general}} = \dfrac{1}{2} \sum_n \int d^4 x \Bigl[
  \hat{\Phi}_n^\dagger \hat{\Phi}_n \Bigr]_D + \int d^4 x \Bigl[ f
  (\hat{\Phi}) \Bigr]_F + \int d^4 x \Bigl[ f (\hat{\Phi}) \Bigr]^*_F .
\end{equation}
The kinetic part can be taken from (\ref{eq:kkinetisch}),
\begin{equation}
  \label{eq:kkinetischallg}
  K^\mu_{\text{kin}} = \dfrac{1}{\sqrt{2}} \gamma^\mu \sum_n \biggl( \left(
  \fmslash{\partial} \phi_n \right) \psi_{n,R} + \left( \fmslash{\partial}
  \phi_n^* \right) \psi_{n,L} - \ii F_n \psi_{n,R} - \ii F^*_n \psi_{n,L}
  \biggr) \quad .   
\end{equation}
For the contribution to the current from the potential we remember that
\begin{equation}
  \label{eq:potallgmodell}
  \delta_\xi {\cal L}_{\text{pot}} = \biggl( \delta_\xi \left[ f (\hat{\Phi})
  \right]_F + \text{h.c.} \biggr) = \biggl( - \ii \sqrt{2} \: \overline{\xi}
  \fmslash{\partial} \left[ f (\hat{\Phi}) \right]_{\psi_L} + \text{h.c.}
  \biggr)  
\end{equation}
for the SUSY transformation of the superpotential. When expanding the
superpotential in a power series of the superfields,
\begin{equation*}
  f (\hat{\Phi}) = \sum_{k = 0}^\infty \sum_{\substack{n_1, n_2, \ldots , n_k
  \\ n_1 + n_2 + \ldots + n_k = k}} f_{n_1 n_2 \ldots n_k}
  \hat{\Phi}_{i_1}^{n_1} \cdot \hat{\Phi}_{i_2}^{n_2} \cdot \ldots \cdot
  \hat{\Phi}_{i_k}^{n_k} \qquad ,    
\end{equation*}
with $i_j \in \left\{ \text{appearing superfields} \right\}$,
the spinor component of the superpotential can easily be read off:
\begin{equation}
  \label{eq:spinorkompsuppot}
  \left[ f(\hat{\Phi}) \right]_{\psi_L} = \sum_n \left( \dfrac{\partial f
  (\phi)}{\partial \phi_n} \right) \psi_{n,L} \quad .  
\end{equation}
By the notation $f (\phi)$ we want to stress that the superfields as arguments
of the function $f$ have been replaced by their scalar components. This
produces the potential part of the supersymmetric current
\begin{equation}
  \label{eq:potbeitragallg}
  K^\mu_{\text{pot}} = - \ii \sqrt{2} \, \sum_n \gamma^\mu \Biggl[ \left(
  \dfrac{\partial f (\phi)}{\partial \phi_n} \right) \psi_{n,L} + \left(
  \dfrac{\partial f (\phi)}{\partial \phi_n} \right)^* \psi_{n,R} \Biggr] .
\end{equation}

Finally the supersymmetric current for a general model (without gauge
interactions which will be studied later on) is:
\begin{equation}
  \label{eq:supstromallg}
  \boxed{
  \begin{aligned}
      {\cal J}^\mu = & \; - \sqrt{2} \sum_n \Biggl[ (\fmslash{\partial}
      \phi_n) \gamma^\mu \psi_{n,R} + (\fmslash{\partial} \phi_n)^* \gamma^\mu
      \psi_{n,L} \\ & \; \qquad \qquad \quad + \ii \gamma^\mu \left(
      \dfrac{\partial f (\phi)}{\partial \phi_n} \right) \psi_{n,L} + \ii
      \gamma^\mu \left( \dfrac{\partial f (\phi)}{\partial \phi_n} \right)^*
      \psi_{n,R} \Biggr]  
  \end{aligned}}
\end{equation}

We check the current conservation:
\begin{equation*}
  \begin{aligned}
   - \dfrac{1}{\sqrt{2}} \partial_\mu {\cal J}^\mu = & \; \sum_n \Biggl[ (\Box
    \phi_n) 
    \psi_{n,R} + (\Box \phi_n^*) \psi_{n,L} + \underline{(\fmslash{\partial}
    \phi_n) (\fmslash{\partial} \psi_{n,R})} + \underline{(\fmslash{\partial}
    \phi_n^*) (\fmslash{\partial} \psi_{n,L})} \\ & \qquad \qquad \qquad + \ii
    \left( \dfrac{\partial f (\phi)}{\partial \phi_n} \right)
    \fmslash{\partial} \psi_{n,L} + \ii \left( \dfrac{\partial f
    (\phi)}{\partial \phi_n} \right)^* \fmslash{\partial} \psi_{n,R} \Biggr]
    \\ & \; + \ii  \sum_{m,n} \Biggl[ \underline{\left( \dfrac{\partial^2
    f(\phi)}{\partial \phi_m \partial \phi_n} \right) (\fmslash{\partial}
    \phi_m) \psi_{n,L}} + \underline{\left( \dfrac{\partial^2
    f(\phi)}{\partial \phi_m \partial \phi_n} \right)^* (\fmslash{\partial}
    \phi_m^*) \psi_{n,R}} \Biggr] \\ = & \; \sum_n \Biggl[ 
    - \dfrac{1}{2} \sum_{k,l} \left( \dfrac{\partial^3 f (\phi)}{\partial
    \phi_n \partial \phi_k \partial \phi_l} \right) \left(
    \overline{\psi_{k,R}} \psi_{l,L} \right) \psi_{n,L} \\ & \qquad -
    \dfrac{1}{2} \sum_{k,l} \left( \dfrac{\partial^3 f(\phi)}{\partial \phi_n
    \partial \phi_k \partial \phi_l} \right)^* \left( \overline{\psi_{k,L}}
    \psi_{l,R} \right) \psi_{n,R} \\ & \qquad + \underline{\underline{\sum_k
    F_k \left( 
    \dfrac{\partial^2 f (\phi)}{\partial \phi_n \partial \phi_k} \right)
    \psi_{n,L}}} + \underline{\underline{\sum_k F_k^* \left(
    \dfrac{\partial^2 \phi}{\partial \phi_n \partial \phi_k} \right)^*
    \psi_{n,R}}} \\ & \qquad - \underline{\underline{\ii F_n
    (\fmslash{\partial} 
    \psi_{n,L})}} - \underline{\underline{\ii F_n^* (\fmslash{\partial}
    \psi_{n,R})}} \Biggr]  
  \end{aligned}
\end{equation*}
In the first identity the underlined terms vanish due to
the equations of motion for the fermions. For the second equality 
the equations of motion for the scalar particles were inserted 
yielding the leading four terms. The two rightmost terms are produced 
using the equation of motion for the auxiliary fields $F_n$ and
$F_n^*$. The doubly underlined terms cancel due to the fermions' equations of 
motion. There still remain the trilinear fermion
terms. The three indices are summed over, so we can split the
terms in three cyclic contributions (it will prove easier to use the
2-spinor formalism here): 
\begin{multline}
  - \dfrac{1}{2} \sum_{n,k,l} \left( \dfrac{\partial^3 f (\phi)}{\partial
    \phi_n \partial \phi_k \partial \phi_l} \right) (\overline{\psi_{k,R}}
    \psi_{l,L}) \psi_{n,L} + \text{h.c.} \\ = - \dfrac{1}{6} \sum_{n,k,l}
    \left( \dfrac{\partial^3 f (\phi)}{\partial \phi_n \partial \phi_k
    \partial \phi_l} \right) \Bigl( (\psi_k \psi_l) \psi_n + (\psi_l \psi_n)
    \psi_k + (\psi_n \psi_k) \psi_l \Bigr) + \text{h.c.} = 0 
\end{multline}
This vanishes due to the Schouten identity, cf.~for instance
\cite{Reuter:1999:dipl}, which is valid for Grassmann odd 2-spinors as
well, for there is always an even number of transpositions in the
cyclic sum. 

%%%%%%%%%%%%%%%%%%%%%%%%%%%%%%%%%%%%%%%%%%%%%%%%%%%%%%%%%%%%%%%%%%%%

\section{Derivation of the SYM current}\label{sec:symcurrent}

Here we present the detailed derivation of the SYM current omitted in
the text. 

We apply the de~Wit--Freedman transformation to the matter
Lagrangean density (\ref{eq:lmateich}) and get:
\begin{align}
  \tilde{\delta}_\xi \Bigl[ \left( D_\mu \phi \right)^\dagger \left( D^\mu
  \phi \right) \Bigr] = & \; \sqrt{2} \left( D_\mu \phi \right)^\dagger \left(
  \overline{\xi} D^\mu \Psi_L \right) + \ii g \left( D_\mu \phi
  \right)^\dagger \left( \overline{\xi} \gamma^\mu \gamma^5 \vec{T} 
  \phi \cdot \vec{\lambda} \right) \notag \\ & \; + \sqrt{2} \left( D_\mu \phi
  \right)^T \left( \overline{\xi} D^\mu \Psi_R \right) - \ii g \phi^\dagger
  \left( \overline{\xi} \gamma_\mu \gamma^5 \vec{T} \vec{\lambda} \right)
  D^\mu \phi   \label{eq:dewit1}  \\
  %%%
  \tilde{\delta}_\xi \Bigl[ F^\dagger F \Bigr] = & \; - \ii \sqrt{2} F^\dagger
  \left( \overline{\xi} \fmslash{D} \Psi_L \right) - \ii \sqrt{2} F^T \left(
  \overline{\xi} \fmslash{D} \Psi_R \right) + 2 g F^\dagger \left(
  \overline{\xi} \vec{T} \phi \cdot \vec{\lambda}_R \right) \notag \\ & \; + 2
  g \phi^\dagger \left( \overline{\xi} \vec{T} F \cdot \vec{\lambda}_R \right)
  \label{eq:dewit2} \\ 
  %%%
  \tilde{\delta}_\xi \Bigl[ \dfrac{\ii}{2} \overline{\Psi} \fmslash{D} \Psi
  \Bigr] = & \; - \dfrac{1}{\sqrt{2}} \left( \overline{\xi} \gamma^\mu \left(
  D_\mu \phi \right)^\dagger \fmslash{D} \Psi_L \right) +
  \dfrac{\ii}{\sqrt{2}} \left( \overline{\xi} F^\dagger \fmslash{D} \Psi_L
  \right) \notag \\ & \; + \dfrac{1}{\sqrt{2}} \left( \overline{\xi}
  \gamma^\mu \gamma^\nu \left( D_\nu D_\mu \phi^T \right) \Psi_R \right) -
  \dfrac{\ii}{\sqrt{2}} \left( \overline{\xi} \left( \fmslash{D} F^T \right)
  \Psi_R \right) \notag \\ & \; - \dfrac{1}{\sqrt{2}} \left( \overline{\xi}
  \gamma^\mu \left( D_\mu \phi \right)^T \fmslash{D} \Psi_R \right) +
  \dfrac{\ii}{\sqrt{2}} \left( \overline{\xi} F^T \fmslash{D} \Psi_R \right)
  \notag \\ & \; + \dfrac{1}{\sqrt{2}} \left( \overline{\xi}
  \gamma^\mu \gamma^\nu 
  \left( D_\nu D_\mu \phi \right)^\dagger \Psi_L \right) -
  \dfrac{\ii}{\sqrt{2}} \left( \overline{\xi} (\fmslash{D} F^\dagger)
  \Psi_L \right) \notag \\ & \; - \dfrac{g}{2} \left( \overline{\xi}
  \gamma_\mu \gamma^5 \vec{\lambda} \right) \cdot \left(
  \overline{\Psi_L} \gamma^\mu \vec{T} \Psi_L \right) + \dfrac{g}{2}
  \left( \overline{\xi} \gamma_\mu \gamma^5 \vec{\lambda} \right)
  \cdot \left( \overline{\Psi_R} \gamma^\mu \vec{T} \Psi_R \right) 
  \label{eq:dewit3} \\ 
  %%%
  \tilde{\delta}_\xi \Bigl[ - \sqrt{2} g \overline{\vec{\lambda}}
  \cdot \phi^\dagger \vec{T} \Psi_L \Bigr] = & \; \dfrac{\ii
  g}{\sqrt{2}} \left( \overline{\xi} \gamma^5 \gamma^\nu \gamma^\mu
  \vec{F}_{\mu\nu} \cdot \phi^\dagger \vec{T} \Psi_L \right) -
  \sqrt{2} g \left( \overline{\xi} \vec{D} \cdot \phi^\dagger \vec{T}
  \Psi_L \right) \notag \\ & \; - 2 g \left( \overline{\xi} \Psi_R
  \right) \left( \overline{\vec{\lambda}} \cdot \vec{T} \Psi_L \right)
  - 2 \ii g \left( \overline{\xi} \phi^\dagger \vec{T} (\fmslash{D}
  \phi) \cdot \vec{\lambda}_L \right) \notag \\ & \; - 2 g \left(
  \overline{\xi} \phi^\dagger \vec{T} F \cdot \vec{\lambda} \right)
  \label{eq:dewit4} \\ 
  %%%
  \tilde{\delta}_\xi \Bigl[ - \sqrt{2} g \overline{\Psi_L} \vec{T}
  \phi \cdot \vec{\lambda} \Bigr] = & \; - 2 \ii g \left(
  \overline{\xi} \gamma^\mu (D_\mu \phi)^\dagger \vec{T} \phi \cdot
  \vec{\lambda}_R \right) - 2 g \left( \overline{\xi} F^\dagger
  \vec{T} \phi \cdot \vec{\lambda}_R \right) \notag \\ & \; - 2 g
  \left( \overline{\xi} \Psi_L \right) \left( \overline{\Psi_L}
  \vec{T} \cdot \vec{\lambda} \right) + \dfrac{\ii g}{\sqrt{2}} \left( 
  \overline{\xi} \gamma^5 \gamma^\nu \gamma^\mu  \vec{F}_{\mu\nu}
  \cdot \phi^T \vec{T} \Psi_R \right) \notag \\ & \; - \sqrt{2} g \left(
  \overline{\xi} \vec{D} \cdot \phi \vec{T} \Psi_R \right)
  \label{eq:dewit5} \\
  %%%
  \tilde{\delta}_\xi \Bigl[ g \phi^\dagger \vec{T} \phi \cdot \vec{D}
  \Bigr] = & \; \sqrt{2} g \left( \overline{\xi} \vec{D} \cdot \phi
  \vec{T} \Psi_R \right) + \sqrt{2} g \left( \overline{\xi} \vec{D}
  \cdot \phi^\dagger \vec{T} \Psi_L \right) \notag \\ & \; - \ii g
  \left( \phi^\dagger \vec{T} \phi \right) \cdot \left( \overline{\xi} 
  \fmslash{D} \vec{\lambda} \right) \label{eq:dewit6} \\
  %%%
  \tilde{\delta}_\xi \mathcal{W} (\phi, \Psi, F) \equiv & \;
  \partial_\mu \overline{\xi} \tilde{K}^\mu (\phi, \Psi, F)
  \label{eq:dewitspezial}
\end{align}
Now we start to examine all the produced terms. At first, all
terms containing four spinors cancel each other -- the two rightmost
terms of (\ref{eq:dewit3}) and the third term of (\ref{eq:dewit4}) and
(\ref{eq:dewit5}), respectively. Using the Fierz identities (note
again the global sign due to the presence of anticommuting spinors)
yields:  
\begin{equation*}
  \begin{aligned}
    & - 2 g \left( \overline{\xi_R} \Psi_L \right) \left( \overline{\Psi_L}
    \vec{T} \cdot \vec{\lambda}_R \right) \\ & \qquad = \; \dfrac{g}{2}
    \underline{\left( \overline{\xi_R} \vec{\lambda}_R \right) \cdot \left( 
    \overline{\Psi_L} \vec{T} \Psi_L \right)} + \dfrac{g}{2} \left(
    \overline{\xi_R} \gamma^\mu \vec{\lambda}_R \right) \cdot \left(
    \overline{\Psi_L} \gamma_\mu \vec{T} \Psi_L \right) \\ & \qquad \; \; \; + 
    \dfrac{g}{4} \underline{ \left( \overline{\xi_R} \sigma^{\mu\nu}
    \vec{\lambda}_R \right) \cdot \left( \overline{\Psi_L} \sigma_{\mu\nu}
    \vec{T} \Psi_L \right)} + \dfrac{g}{2} \left( \overline{\xi_R}
    \gamma^\mu \gamma^5 \vec{\lambda}_R \right) \cdot \left( \overline{\Psi_L}
    \gamma^5 \gamma_\mu \vec{T} \Psi_L \right) \\ & \qquad \; \; \; +
    \dfrac{g}{2} \underline{\left( \overline{\xi_R} \gamma^5 \vec{\lambda}_R
    \right) \cdot \left( \overline{\Psi_L} \gamma^5 \vec{T} \Psi_L \right)}
    \\ & \qquad = - g \left( \overline{\xi_R} \gamma^\mu \gamma^5
    \vec{\lambda}_R \right) \cdot \left( \overline{\Psi_L} \gamma_\mu \gamma^5
    \vec{T} \Psi_L \right) 
  \end{aligned}
\end{equation*}
The underlined terms vanish here and in the following calculation, since
scalar, pseudoscalar and tensor bilinears cannot couple spinors of
like chirality to each other. 
\begin{equation*}
  \begin{aligned}
    & - 2 g \left( \overline{\xi_L} \Psi_R \right) \left(
    \overline{\vec{\lambda}_R} \vec{T} \cdot \Psi_L \right) = - 2 g \left(
    \overline{\xi_L} \Psi_R \right) \left( \overline{\Psi_R} \vec{T}
    \vec{\lambda}_L \right) \\ & \qquad = - g \left( \overline{\xi_L}
    \gamma^\mu \gamma^5 \vec{\lambda}_L \right) \cdot \left( \overline{\Psi_R} 
    \gamma_\mu \gamma^5 \vec{T} \Psi_R \right) 
  \end{aligned}
\end{equation*}
For the other terms we get:
\begin{equation*}
  \begin{aligned}
  & \dfrac{g}{2} \left( \overline{\xi} \gamma_\mu \gamma^5 \vec{\lambda}
  \right) \cdot \left( \overline{\Psi_R} \gamma^\mu \gamma^5 \Psi_R \right) \\
  & \qquad = \dfrac{g}{2} \left( \overline{\xi_L} \gamma_\mu \gamma^5
  \vec{\lambda}_L \right) \cdot \left( \overline{\Psi_R} \gamma^\mu \gamma^5
  \Psi_R \right) + \dfrac{g}{2} \left( \overline{\xi_R} \gamma_\mu \gamma^5
  \vec{\lambda}_R \right) \cdot \left( \overline{\Psi_R} \gamma^\mu \gamma^5
  \Psi_R \right) \\ & \qquad = \dfrac{g}{2} \left( \overline{\xi_L} \gamma_\mu
  \gamma^5 \vec{\lambda}_L \right) \cdot \left( \overline{\Psi_R} \gamma^\mu
  \gamma^5 \Psi_R \right) + \dfrac{g}{2} \left( \overline{\xi_R} \gamma_\mu
  \gamma^5 \vec{\lambda}_R \right) \cdot \left( \overline{\Psi_L} \gamma^\mu
  \gamma^5 \Psi_L \right) 
  \end{aligned} 
\end{equation*}
\begin{equation*}
  \begin{aligned}
  & \dfrac{g}{2} \left( \overline{\xi} \gamma_\mu \gamma^5 \vec{\lambda}
  \right) \cdot \left( \overline{\Psi_L} \gamma^\mu \gamma^5 \Psi_L \right) \\
  & \qquad = \dfrac{g}{2} \left( \overline{\xi_L} \gamma_\mu \gamma^5
  \vec{\lambda}_L \right) \cdot \left( \overline{\Psi_L} \gamma^\mu \gamma^5
  \Psi_L \right) + \dfrac{g}{2} \left( \overline{\xi_R} \gamma_\mu \gamma^5
  \vec{\lambda}_R \right) \cdot \left( \overline{\Psi_L} \gamma^\mu \gamma^5
  \Psi_L \right) \\ & \qquad = \dfrac{g}{2} \left( \overline{\xi_L} \gamma_\mu
  \gamma^5 \vec{\lambda}_L \right) \cdot \left( \overline{\Psi_R} \gamma^\mu
  \gamma^5 \Psi_R \right) + \dfrac{g}{2} \left( \overline{\xi_R} \gamma_\mu
  \gamma^5 \vec{\lambda}_R \right) \cdot \left( \overline{\Psi_R} \gamma^\mu
  \gamma^5 \Psi_R \right) 
  \end{aligned} 
\end{equation*}
Each of the second manipulations for the two latest identities follow
from the Majorana properties of the spinor field $\Psi$. As promised,
all four terms from the last four identities cancel.

It is obvious that all terms containing the auxiliary fields $D^a$ -- 
the second from (\ref{eq:dewit4}), the last from (\ref{eq:dewit5}) and
the first two from (\ref{eq:dewit6}) -- give zero.

Let us consider the terms containing $F^\dagger$ now. There are five of
them: the first and third term in (\ref{eq:dewit2}), the second and
eighth from (\ref{eq:dewit3}) and the second term of
(\ref{eq:dewit5}). As is immediately seen, the latter cancels the third
term from (\ref{eq:dewit2}). What remains is: 
\begin{equation}
  \label{eq:strom_sym1}
  \begin{aligned}
    & \qquad - \ii \sqrt{2} \left( \overline{\xi} F^\dagger \fmslash{D} \Psi_L 
    \right) + \dfrac{\ii}{\sqrt{2}} \left( \overline{\xi} F^\dagger
    \fmslash{D} \Psi_L \right) - \dfrac{\ii}{\sqrt{2}} \left( \overline{\xi}
    (F^\dagger \overleftarrow{\fmslash{D}}') \Psi_L \right) \\ & = \; -
    \dfrac{\ii}{\sqrt{2}} \left( \overline{\xi} F^\dagger \fmslash{D} \Psi_L
    \right) - \dfrac{\ii}{\sqrt{2}} \left( \overline{\xi} (F^\dagger
    \overleftarrow{\fmslash{D}}') \Psi_L \right) \\ & = \; -
    \dfrac{\ii}{\sqrt{2}} \partial_\mu \left( \overline{\xi} \gamma^\mu
    F^\dagger \Psi_L \right)   
  \end{aligned}
\end{equation}
The symbol $D'_\mu$ introduced here is the covariant
derivative originally acting on the righthanded spinor field
$\Psi_R$, so it has the opposite sign compared to the
covariant derivative for the lefthanded fields. Only the contributions
with the partial derivatives survive. 

The calculation for the parts with $F^T$ (or $F$) proceeds
analogously. Here the 
last term of (\ref{eq:dewit4}) and the last one from
(\ref{eq:dewit2}) cancel, leaving the second term of (\ref{eq:dewit2})
as well as the fourth and sixth from (\ref{eq:dewit3}):
\begin{equation}
  \label{eq:strom_sym2}
  \begin{aligned}
    & \qquad - \ii \sqrt{2} \left( \overline{\xi} F^T \fmslash{D} \Psi_R
    \right) - \dfrac{\ii}{\sqrt{2}} \left( \overline{\xi} (F^T
    \overleftarrow{\fmslash{D}}') \Psi_R \right) + \dfrac{\ii}{\sqrt{2}}
    \left( \overline{\xi} F^T \fmslash{D} \Psi_R \right) \\ & = \; -
    \dfrac{\ii}{\sqrt{2}} \partial_\mu \left( \overline{\xi} \gamma^\mu F^T
    \Psi_R \right)  
  \end{aligned}
\end{equation}
Herein $D'_\mu$ is the covariant derivative originally acting on the
lefthanded spinor $\Psi_L$. Again, only the terms with the partial
derivatives remain. 

Next we turn our attention to the contributions containing both $\phi$
and $\phi^\dagger$, i.e.~the second and fourth of (\ref{eq:dewit1}), the
fourth term of (\ref{eq:dewit4}), the first of (\ref{eq:dewit5}) and the
rightmost of (\ref{eq:dewit6}). We split the covariant derivatives in partial
derivatives and the gauge field parts:
\begin{align*}
  & \qquad \ii g \left( \overline{\xi} (\fmslash{D} \phi^\dagger) \gamma^5
    \vec{T} \phi \cdot \vec{\lambda} \right) - \ii g \left( \overline{\xi}
    \gamma_\mu \gamma^5 \phi^\dagger \vec{T} (D^\mu \phi) \cdot \vec{\lambda}
    \right) - 2 \ii g \left( \overline{\xi} \gamma^\mu \phi^\dagger \vec{T}
    (D_\mu \phi) \cdot \vec{\lambda}_L \right) \\ & \qquad - 2 \ii g \left(
    \overline{\xi} \gamma^\mu (D_\mu \phi)^\dagger \vec{T} \phi \cdot
    \vec{\lambda}_R \right) - \ii g \left( \phi^\dagger \vec{T} \phi \right)
    \cdot \left( \overline{\xi} \fmslash{D} \vec{\lambda} \right) \\ & = \;
    \ii g \underline{\left( \overline{\xi} (\fmslash{\partial} \phi^\dagger)
    \gamma^5 \vec{T} \phi \cdot \vec{\lambda} \right)} - \ii g
    \underline{\left( \overline{\xi} \phi^\dagger \vec{T} (\fmslash{\partial}
    \phi) \gamma^5 \cdot \vec{\lambda} \right)} - \ii g \left( \overline{\xi}
    \phi^\dagger \vec{T} (\fmslash{\partial} \phi) \cdot (1 -
    \underline{\gamma^5}) \vec{\lambda} \right) \\ & \qquad - \ii g \left(
    \overline{\xi} (\fmslash{\partial} \phi^\dagger) \vec{T} \phi \cdot (1 +
    \underline{\gamma^5}) \vec{\lambda} \right) - \ii g \left( \overline{\xi}
    \phi^\dagger \vec{T} \phi \cdot \fmslash{\partial} \vec{\lambda} \right)
    \\ & \quad \; \: - g^2 \underline{\left( \overline{\xi} \gamma^\mu
    \gamma^5 \phi^\dagger \vec{T} ( \vec{T} \cdot \vec{A}_\mu) \phi \cdot
    \vec{\lambda} \right)} - g^2 \underline{\left( \overline{\xi} \gamma^\mu
    \gamma^5 \phi^\dagger \vec{T} (\vec{T} \cdot \vec{A}_\mu) \phi \cdot
    \vec{\lambda} \right)} \\ & \qquad - g^2 \left( \overline{\xi} \gamma^\mu
    \phi^\dagger \vec{T} (\vec{T} \cdot \vec{A}_\mu) \phi \cdot (1 -
    \underline{\gamma^5}) \vec{\lambda} \right) + g^2 \left( \overline{\xi}
    \gamma^\mu \phi^\dagger (\vec{T} \cdot \vec{A}_\mu) \vec{T} \phi \cdot (1
    + \underline{\gamma^5}) \vec{\lambda} \right) \\ & \qquad - \ii g^2 \left(
    \phi^\dagger \vec{T} \phi \right) \cdot \left( \overline{\xi} f^a_{\;bc}
    A_\mu^b \gamma^\mu \lambda^c \right) \\ & = \; - \ii g \partial_\mu \left(
    \overline{\xi} \gamma^\mu \phi^\dagger \vec{T} \phi \cdot \vec{\lambda}
    \right) \\ & \qquad + g^2 \underline{\left( \overline{\xi} \gamma^\mu
    \phi^\dagger \left[ T^a , T^b \right] \phi A_\mu^a \lambda^b \right)} -
    \ii g^2 \underline{\left( \overline{\xi} \gamma^\mu \phi^\dagger T^a \phi
    f^a_{\;bc} A_\mu^b \lambda^c \right)}   
\end{align*}
Underlined contributions cancel each other. The terms in the last line
vanish due to the Lie algebra of the gauge group. We get the gradient
\begin{equation}
  \label{eq:strom_sym3}
  - \ii g \partial_\mu \left( \overline{\xi} \gamma^\mu \phi^\dagger \vec{T}
    \phi \cdot \vec{\lambda} \right)
\end{equation}

With the help of the algebra of covariant derivatives
\begin{equation}
  D_\mu D_\nu - D_\nu D_\mu = - \ii g \vec{T} \cdot \vec{F}_{\mu\nu} 
\end{equation}
we can rewrite the two terms with the field strength tensor of the
gauge field, the first term of (\ref{eq:dewit4}) and the fourth in
(\ref{eq:dewit5}):  
\begin{align*}
  & \qquad \dfrac{\ii g}{\sqrt{2}} \left( \overline{\xi} \gamma^5 \gamma^\nu
  \gamma^\mu \vec{F}_{\mu\nu} \cdot \phi^\dagger \vec{T} \Psi_L \right) +
  \dfrac{\ii g}{\sqrt{2}} \left( \overline{\xi} \gamma^5 \gamma^\nu \gamma^\mu 
  \vec{F}_{\mu\nu} \cdot \phi^T \vec{T} \Psi_R \right) \\ & = \; - \dfrac{\ii 
  g}{\sqrt{2}} \left( \overline{\xi} \gamma^\nu \gamma^\mu \vec{F}_{\mu\nu}
  \cdot \phi^\dagger \vec{T} \Psi_L \right) + \dfrac{\ii g}{\sqrt{2}} \left(
  \overline{\xi} \gamma^\nu \gamma^\mu \vec{F}_{\mu\nu} \cdot \phi^T
  \vec{T} \Psi_R \right) \\ & = \; + \dfrac{1}{\sqrt{2}} \left( \overline{\xi}
  \gamma^\nu \gamma^\mu \left[ (D^\dagger_\nu D^\dagger_\mu - D^\dagger_\mu
  D^\dagger_\nu) \phi^\dagger \right] \Psi_L \right) - \dfrac{1}{\sqrt{2}}
  \left( \overline{\xi} \gamma^\nu \gamma^\mu \left[ (D_\mu D_\nu - D_\nu
  D_\mu) \phi^T \right] \Psi_R \right) 
\end{align*}
The order of spacetime indices in both terms can be understood from
the fact that in the first one the commutator of covariant derivatives
acts upon the Hermitean adjoint scalar field and therefore to the left, so
we have to replace the operators by their Hermitean adjoints and to
revert their order. For the right term there is no such effect
since we simply have the transposed representation of the gauge group
there. Consider all missing terms containing the scalar field or its
adjoint together with the spinor field $\Psi$ (i.e. the first and third
from (\ref{eq:dewit1}) and the first, third, fifth and seventh in
(\ref{eq:dewit3})), and use the Dirac algebra
\begin{equation}
  \eta^{\mu\nu} = \dfrac{1}{2} \left( \gamma^\mu \gamma^\nu + \gamma^\nu
  \gamma^\mu \right) 
\end{equation}
for the following manipulations (underlined terms cancel after summing
up all terms)
\begin{align*}
  \sqrt{2} \left( \overline{\xi} (D_\mu \phi)^\dagger D^\mu \Psi_L \right) = &
  \; \dfrac{1}{\sqrt{2}} \left( \overline{\xi} (\underline{\gamma^\mu
  \gamma^\nu} + \gamma^\nu \gamma^\mu) (D_\mu \phi)^\dagger D_\nu \Psi_L
  \right) \\ 
  \sqrt{2} \left( \overline{\xi} (D_\mu \phi)^T D^\mu \Psi_R \right) = &
  \; \dfrac{1}{\sqrt{2}} \left( \overline{\xi} (\underline{\gamma^\mu
  \gamma^\nu} + \gamma^\nu \gamma^\mu) (D_\mu \phi)^T D_\nu \Psi_R
  \right)  \\
  - \dfrac{1}{\sqrt{2}} \left( \overline{\xi} \gamma^\mu (D_\mu \phi)^\dagger
  \fmslash{D} \Psi_L \right) = & \; - \dfrac{1}{\sqrt{2}} \underline{\left(
  \gamma^\mu \gamma^\nu (D_\mu \phi)^\dagger D_\nu \Psi_L \right)} \\
  \dfrac{1}{\sqrt{2}} \left( \overline{\xi} (\phi^T
  \overleftarrow{\fmslash{D}} \overleftarrow{\fmslash{D}}) \Psi_R \right) = &
  \; \dfrac{1}{\sqrt{2}} \left( \overline{\xi} \gamma^\mu \gamma^\nu (D_\nu
  D_\mu \phi^T) \Psi_R \right) \\ 
  - \dfrac{1}{\sqrt{2}} \left( \overline{\xi} \gamma^\mu (D_\mu \phi)^T
  \fmslash{D} \Psi_R \right) = & - \dfrac{1}{\sqrt{2}} \underline{\left(
  \overline{\xi} \gamma^\mu \gamma^\nu (D_\mu \phi)^T D_\nu \Psi_R \right)} \\ 
  \dfrac{1}{\sqrt{2}} \left( \overline{\xi} (\phi
  \overleftarrow{D}_\mu)^\dagger \gamma^\mu \overleftarrow{\fmslash{D}}'
  \Psi_L \right) = & \; \dfrac{1}{\sqrt{2}} \left( \overline{\xi} \gamma^\mu
  \gamma^\nu (D'_\nu (D_\mu \phi)^\dagger) \Psi_L \right) 
\end{align*}
After adding the contributions from the field strength tensors we
arrive at:
\begin{align*}
  & \qquad \dfrac{1}{\sqrt{2}} \overline{\xi} \gamma^\nu \gamma^\mu \biggl\{
  (D_\mu \phi)^\dagger D_\nu \Psi_L + (D_\mu \phi)^T D_\nu \Psi_R +
  \underline{(D_\mu D_\nu \phi^T) \Psi_R} + (D'_\mu (D_\nu \phi)^\dagger)
  \Psi_L \\ & \qquad\qquad\qquad\quad + (D^\dagger_\nu D^\dagger_\mu
  \phi^\dagger) \Psi_L - (D^\dagger_\mu D^\dagger_\nu \phi^\dagger) \Psi_L -
  \underline{(D_\mu D_\nu \phi^T) \Psi_R} + (D_\nu D_\mu \phi^T) \Psi_R
  \biggr\} \\ & = \; \dfrac{1}{\sqrt{2}} \overline{\xi} \gamma^\nu \gamma^\mu
  \biggl\{ (D_\mu \phi)^\dagger D_\nu \Psi_L + (D_\mu \phi)^T D_\nu \Psi_R +
  \underline{(D^\dagger_\mu D^\dagger_\nu \phi^\dagger) \Psi_L} +
  (D^\dagger_\nu D^\dagger_\mu \phi^\dagger) \Psi_L \\ &
  \qquad\qquad\qquad\quad - \underline{(D^\dagger_\mu D^\dagger_\nu
  \phi^\dagger) \Psi_L} + (D_\nu D_\mu \phi^T) \Psi_R \biggr\} \\ & = \;
  \dfrac{1}{\sqrt{2}} \partial_\mu \biggl\{ \left( \overline{\xi} \gamma^\mu 
  \gamma^\nu (D_\nu \phi)^\dagger \Psi_L \right) + \left( \overline{\xi}
  \gamma^\mu \gamma^\nu (D_\nu \phi)^T \Psi_R \right) \biggr\}
\end{align*}
For the second identity it has been used, that the primed covariant
derivative originally acting on the righthanded fermion field and
hence endowed with a positive sign in front of the gauge field, is
identical to the Hermitean adjoint of the ``ordinary'' covariant
derivative (in the fundamental representation). In the last equation
all gauge field contributions from the second covariant derivatives in
each term cancel, hence the final result is a total derivative. We also
relabelled the indices $(\mu \leftrightarrow \nu)$.

Altogether, for the de~Wit-Freedman transformation of the matter
Lagrangean density, we get the total derivative
\begin{equation}
\label{eq:deltaeich1}
\boxed{
\begin{aligned}
  \tilde{\delta}_\xi \mathcal{L}_{\text{mat}} = & \; - \dfrac{\ii}{\sqrt{2}}
  \partial_\mu \left( \overline{\xi} F^\dagger \gamma^\mu \Psi_L \right) -
  \dfrac{\ii}{\sqrt{2}} \partial_\mu \left( \overline{\xi} F^T \gamma^\mu
  \Psi_R \right) + \dfrac{1}{\sqrt{2}} \partial_\mu \left( \overline{\xi}
  \gamma^\mu \gamma^\nu (D_\nu \phi)^\dagger \Psi_L \right) \\ & \; +
  \dfrac{1}{\sqrt{2}} \partial_\mu \left( \overline{\xi} \gamma^\mu \gamma^\nu
  (D_\nu \phi)^T \Psi_R \right) - \ii g \partial_\mu \left( \overline{\xi}
  \gamma^\mu \phi^\dagger \vec{T} \phi \cdot \vec{\lambda} \right) +
  \partial_\mu \overline{\xi} \tilde{K}^\mu (\phi, \Psi, F) 
\end{aligned}}
\end{equation}

We now turn to the gauge part of the Lagrangean density,
(\ref{eq:leicheich}). The transformations of the various parts are
\begin{align}
  \tilde{\delta}_\xi \biggl[ \dfrac{\ii}{2} \left(\overline{\lambda^a}
  \gamma^\mu (D_\mu \lambda)^a \right) \biggr] = & \; \dfrac{1}{4} \left(
  \overline{\xi} \gamma^5 \gamma^\nu \gamma^\mu \gamma^\alpha F^a_{\mu\nu}
  (D_\alpha \lambda)^a \right) - \dfrac{1}{4} \left( \overline{\xi}
  \gamma^5 \gamma^\nu \gamma^\mu \gamma^\alpha (D_\alpha F_{\mu\nu})^a
  \lambda^a \right) \notag \\ & \; + \dfrac{\ii}{2} \left( \overline{\xi}
  \gamma^\mu D^a (D_\mu \lambda)^a \right) - \dfrac{\ii}{2} \left(
  \overline{\xi} \gamma^\mu (D_\mu D)^a \lambda^a \right) \label{eq:dewit7} 
  \\ 
  %%%
  \tilde{\delta}_\xi \biggl[ - \dfrac{1}{4} F^a_{\mu\nu} F^{\mu\nu}_a
  \biggr] = & \; F_a^{\mu\nu} \partial_\mu \left( \overline{\xi} \gamma_\nu
  \gamma^5 \lambda^a \right) + g F_a^{\mu\nu} f^a_{\;bc} \left(
  \overline{\xi} \gamma_\mu \gamma^5 \lambda^b \right) A_\nu^c
  \label{eq:dewit8} \\ 
  %%%
  \tilde{\delta}_\xi \biggl[ \dfrac{1}{2} D^a D^a \biggr] = & \; - \ii D^a
  \left( \overline{\xi} \fmslash{D} \lambda \right)^a \label{eq:dewit9} 
\end{align}

The term  $- (\ii g/ 2) f^a_{\;bc}
\left( \overline{\lambda^a} \gamma^\mu \lambda^c \right) \left(
\overline{\xi} \gamma_\mu \gamma^5 \lambda^b \right)$ produced by
the transformation of the gauge field in the covariant derivative
vanishes due to the Fierz identities: The scalar, pseudoscalar and 
vector parts vanish because the bilinears contracted with the totally
antisymmetric structure constants $f^a_{\;bc}$ are symmetric in the
gauge group indices $(ab)$ since they are built from Majorana
spinors. Note that since we are using $\gamma^5 \lambda^b$ as spinor
in the Fierz identity the vector part becomes the axial vector and
vice versa. Only the pseudovector is antisymmetric and survives. After
relabelling the indices we get
\begin{equation}
\label{dreigauginossindnull}
f^a_{\;bc} \left( \overline{\lambda^a} \gamma^\mu \lambda^c \right)
\left( \overline{\xi} \gamma_\mu \gamma^5 \lambda^b \right) = + 
\dfrac{1}{2} f^a_{\;bc} \left( \overline{\lambda^a} \gamma^\mu
\lambda^c \right) \left( \overline{\xi} \gamma_\mu \gamma^5 \lambda^b
\right) = 0 , 
\end{equation}
which is seen to vanish after a second Fierz transformation. 

The terms containing the auxiliary field $D^a$ (the rightmost from
(\ref{eq:dewit7}) and (\ref{eq:dewit9})) together yield:
\begin{align}
  & \qquad - \dfrac{\ii}{2} \left( \overline{\xi} D^a (\fmslash{D}
  \lambda)^a \right) - \dfrac{\ii}{2} \left( \overline{\xi} (\fmslash{D}
  D)^a \lambda^a \right) \notag \\ & = \; - \dfrac{\ii}{2} \left(
  \overline{\xi} D^a (\fmslash{\partial} \lambda^a) \right) -
  \dfrac{\ii}{2} \left( \overline{\xi} (\fmslash{\partial} D^a) \lambda^a
  \right) - \dfrac{\ii g}{2} \underline{\left( \overline{\xi} \gamma^\mu D^a
  f^a_{\;bc} A_\mu^b \lambda^c \right)} - \dfrac{\ii g}{2} \underline{\left(
  \overline{\xi} \gamma^\mu f^a_{\;bc} A_\mu^b D^c \lambda^a \right)}
  \notag \\ & = \; - \dfrac{\ii}{2} \partial_\mu \left( \overline{\xi}
  \gamma^\mu D^a \lambda^a \right) \label{eq:strom_sym4}    
\end{align}
Underlined terms cancel each other. 

To calculate the remaining terms with the field strength tensors we
need the following identity for gamma matrices,
\begin{equation}
 \label{eq:gammaidentitaet}
 \left[ \gamma^\mu , \gamma^\nu \right] \gamma^\rho = - 2 \eta^{\mu\rho}
 \gamma^\nu + 2 \eta^{\nu\rho} \gamma^\mu - 2 \ii
 \epsilon^{\mu\nu\rho\sigma} \gamma_\sigma \gamma^5,
\end{equation}
which can be easily derived by expanding a general $4\times4$ matrix
as a linear combination of the 16 gamma matrices $\mathbb{I}$,
$\gamma^5$, $\sigma^{\mu\nu}$, $\gamma^\mu$, $\gamma^\mu\gamma^5$. As
the only available Lorentz invariant tensor coefficients are the
metric $\eta^{\mu\nu}$ and the Levi-Civit\`a tensor
$\epsilon^{\mu\nu\rho\sigma}$, the three leftmost combinations are not
possible. Considering the properties under parity transformation shows
that only the product of metric and tensor as well as the product of
the pseudovector with the Epsilon-tensor are allowed. The explicit
prefactors can be calculated by inserting $(121)$ and $(123)$ for
$(\mu\nu\rho)$. 

With the identity (\ref{eq:gammaidentitaet}) we are able to rewrite
one of the appearing terms: 
\begin{align}
\left( \overline{\xi} \gamma^5 \gamma^\beta \gamma^\alpha \gamma^\mu (D_\mu
F_{\alpha\beta})^a \lambda^a \right) = & \; \dfrac{1}{2} \left(
\overline{\xi} \gamma^5 \left[ \gamma^\beta , \gamma^\alpha \right]
\gamma^\mu (D_\mu  F_{\alpha\beta})^a \lambda^a \right) \notag \\ = & \; - 
\left( \overline{\xi} \gamma^5 \gamma^\alpha (D^\beta F_{\alpha\beta})^a
\lambda^a \right) + \left( \overline{\xi} \gamma^5 \gamma^\beta (D^\alpha
F_{\alpha\beta})^a \lambda^a \right) \notag \\ & \; - \ii \left(
\overline{\xi} \gamma^5 \gamma_\sigma \gamma^5
\epsilon^{\sigma\beta\alpha\mu} (D_\mu F_{\alpha\beta})^a 
\lambda^a \right) \notag \\ = & \; - 2 \left( \overline{\xi} \gamma^5
\gamma^\alpha (D^\beta F_{\alpha\beta})^a \lambda^a \right) 
\end{align}
The first identity as well as the equality of the first two terms in
the middle line hold due to the antisymmetry of the field
strength tensor in the spacetime indices. The Bianchi identity of
non-Abelian gauge theories causes the third term in the middle line
to vanish. 

We collect the remaining terms:
\begin{align}
& \qquad \dfrac{1}{4} \left( \overline{\xi} \gamma^5 \gamma^\beta
\gamma^\alpha \gamma^\mu F_{\alpha\beta}^a (D_\mu \lambda)^a \right) -
\dfrac{1}{4} \left( \overline{\xi} \gamma^5 \gamma^\beta \gamma^\alpha
\gamma^\mu (D_\mu F_{\alpha\beta})^a \lambda^a \right) + F_a^{\mu\nu}
\partial_\mu \left( \overline{\xi} \gamma_\nu \gamma^5 \lambda^a \right)
\notag \\ & \qquad + g F_a^{\mu\nu} f^a_{\;bc} \left( \overline{\xi}
\gamma_\mu \gamma^5 \lambda^b \right) A_\nu^c  \notag \\ & = \;
\dfrac{1}{4} \left( \overline{\xi} \gamma^5 \gamma^\beta \gamma^\alpha
\gamma^\mu F_{\alpha\beta}^a (\partial_\mu \lambda^a) \right) +
\dfrac{g}{4} \underline{\left( \overline{\xi} \gamma^5 \gamma^\beta
\gamma^\alpha \gamma^\mu F_{\alpha\beta}^a f^a_{\;bc} A_\mu^b \lambda^c
\right)} \notag \\ & \qquad + \dfrac{1}{4} \left( \overline{\xi} \gamma^5
\gamma^\beta \gamma^\alpha \gamma^\mu (\partial_\mu F_{\alpha\beta}^a)
\lambda^a \right) + \dfrac{g}{4} \underline{\left( \overline{\xi} \gamma^5
\gamma^\beta \gamma^\alpha \gamma^\mu f^a_{\;bc} A_\mu^b F_{\alpha\beta}^c
\lambda^a \right)} \notag \\ & \qquad - \dfrac{1}{2} \left( \overline{\xi}
\gamma^5 \gamma^\beta \gamma^\alpha \gamma^\mu (D_\mu F_{\alpha\beta})^a
\lambda^a \right) + F_a^{\mu\nu} \partial_\mu \left( \overline{\xi}
\gamma_\nu \gamma^5 \lambda^a \right) + g F_a^{\mu\nu} f^a_{\;bc} \left(
\overline{\xi} \gamma_\mu \gamma^5 \lambda^b \right) A_\nu^c \notag \\ &
\stackrel{(\ref{eq:gammaidentitaet})}{=} \; \dfrac{1}{4} \partial_\mu
\left( \overline{\xi} \gamma^\alpha \gamma^\beta \gamma^\mu \gamma^5
F_{\alpha\beta}^a \lambda^a \right) + F_a^{\mu\nu} \partial_\mu \left(
\overline{\xi} \gamma_\nu \gamma^5 \lambda^a \right) + \underline{g
F_a^{\mu\nu} f^a_{\;bc} \left( \overline{\xi} \gamma_\mu \gamma^5 \lambda^b
\right) A_\nu^c} \notag \\ & \qquad + \left( \overline{\xi} (\partial_\beta
F^{\beta\alpha}_a) \gamma_\alpha \gamma^5 \lambda^a \right) + \underline{g
F_c^{\alpha\beta} f^a_{\;bc} \left( \overline{\xi} \gamma^5 \gamma_\alpha
\lambda^a \right) A_\beta^b} \notag \\ & = \; \dfrac{1}{4} \partial_\mu
\left( \overline{\xi} \gamma^\alpha \gamma^\beta \gamma^\mu \gamma^5
F_{\alpha\beta}^a \lambda^a \right) + \partial_\mu \left( \overline{\xi}
F_a^{\mu\nu} \gamma_\nu \gamma^5 \lambda^a \right) \label{eq:strom_sym5} 
\end{align}

The divergence to which the gauge part of the Lagrangean density of an
SYM theory is transformed under a de~Wit\--Freedman transformation finally
is: 
\begin{equation}
\label{eq:deltaeich2}
\boxed{
\begin{aligned}
 \tilde{\delta}_\xi \mathcal{L}_{\text{gauge}} = \partial_\mu \left(
\overline{\xi} F_a^{\mu\nu} \gamma_\nu \gamma^5 \lambda^a \right) +
\dfrac{1}{4} \partial_\mu \left( \overline{\xi} \gamma^\alpha \gamma^\beta
\gamma^\mu \gamma^5  F_{\alpha\beta}^a \lambda^a \right) - \dfrac{\ii}{2}
\partial_\mu \left( \overline{\xi} \gamma^\mu D^a \lambda^a \right) 
\end{aligned}}
\end{equation}

The last point is a possible Fayet\--Iliopoulos contribution. Its
transformation can be written down immediately:
\begin{equation}
  \label{eq:deltaeich3}
  \boxed{ \tilde{\delta}_\xi \mathcal{L}_{\text{FI}} = - \ii \partial_\mu
  \left( \overline{\xi} \gamma^\mu \zeta^a \lambda^a \right) } 
\end{equation}

In the long run we have to construct the current from the contribution
of the SUSY transformation of the Lagrangean density and the
``Noether'' part. The terms produced by the SUSY transformation of the
superpotential are identical to those found for theories without gauge
symmetry, (\ref{eq:potbeitragallg}). We can think of the problem as
having gauged the global 
symmetry in the models discussed earlier. As the superpotential
contains no derivatives of the fields, this contribution to the
Lagrangean density is then of course also locally invariant. The
difference between the ordinary SUSY transformations and the
de~Wit--Freedman transformations can be written as a local gauge
transformation with special scalar and spinor fields as gauge
parameters (cf.~\cite{deWit/Freedman:1975:susyeich}). Since the
superpotential is invariant under SUSY transformations as well as
under gauge transformations and also under the above mentioned special
gauge transformations, it is invariant under de~Wit--Freedman
transformations. As the superpotential contains no derivatives it does
not contribute to the current,
\begin{equation}
\tilde{K}^\mu_{\text{local}} = \tilde{K}^\mu_{\text{global}} . 
\end{equation}

The Noether part of the supersymmetric current is:
\begin{equation}
  \sum_{\text{all fields}} \dfrac{\partial_R \mathcal{L}}{\partial
  (\partial_\mu \Lambda)} \tilde{\delta}_\xi \Lambda = - \overline{\xi} N^\mu 
\end{equation}
Finally, we have the contributions:
\begin{align}
  \dfrac{\partial \mathcal{L}}{\partial (\partial_\mu \phi)} \tilde{\delta}_\xi
  \phi = & \; \sqrt{2} (D^\mu \phi)^\dagger \left( \overline{\xi} \Psi_L
  \right) \\
  \dfrac{\partial \mathcal{L}}{\partial (\partial_\mu \phi^\dagger)}
  \tilde{\delta}_\xi \phi^\dagger = & \; \sqrt{2} (D^\mu \phi) \left(
  \overline{\xi} \Psi_R \right) \\ 
  \dfrac{\partial \mathcal{L}}{\partial (\partial_\mu \Psi)} \tilde{\delta}_\xi
  \Psi = & \; \dfrac{1}{\sqrt{2}} \Bigl( \overline{\xi} \gamma^\nu \gamma^\mu
  \left( (D_\nu \phi)^T {\cal P}_R + (D_\nu \phi)^\dagger {\cal P}_L \right)
  \Psi \Bigr) \notag \\ & \qquad - \dfrac{\ii}{\sqrt{2}} \left(
  \overline{\xi} \gamma^\mu \left( F^T {\cal P}_R + F^\dagger {\cal P}_L
  \right) \Psi \right) \\ 
  \dfrac{\partial \mathcal{L}}{\partial (\partial_\mu \lambda^a)}
  \tilde{\delta}_\xi \lambda^a = & \; - \dfrac{1}{4} \left( \overline{\xi}
  \gamma^\alpha \gamma^\beta \gamma^\mu \gamma^5 F_{\alpha\beta}^a \lambda^a
  \right) - \dfrac{\ii}{2} \left( \overline{\xi} \gamma^\mu D^a \lambda^a
  \right) \\
  \dfrac{\partial \mathcal{L}}{\partial (\partial_\mu A_\nu^a)}
  \tilde{\delta}_\xi  
  A_\nu^a = & \; F_a^{\mu\nu} \left( \overline{\xi} \gamma_\nu \gamma^5
  \lambda^a \right) 
\end{align}
Adding these terms to the contributions from the variation of the
Lagrangean density (\ref{eq:deltaeich1}), (\ref{eq:deltaeich2}),
(\ref{eq:deltaeich3}), we arrive at the supersymmetric current for
supersymmetric Yang-Mills theories, which has the form 
\begin{equation}
  \label{eq:strom_sym_appen}
\boxed{
\begin{aligned}
  \mathcal{J}^\mu = & \; - \sqrt{2} \gamma^\nu \gamma^\mu (D_\nu \phi)^T
  \Psi_R - \sqrt{2} \gamma^\nu \gamma^\mu (D_\nu \phi)^\dagger \Psi_L - \ii
  \gamma^\mu \zeta^a \lambda^a \\ & \; + \dfrac{1}{2} \gamma^\alpha
  \gamma^\beta \gamma^\mu \gamma^5 F_{\alpha\beta}^a \lambda^a - \ii g
  \gamma^\mu \left( \phi^\dagger \vec{T} \phi \right) \cdot
  \vec{\lambda} \\ & \; - 
  \ii \sqrt{2} \gamma^\mu \left( \dfrac{\partial f (\phi)}{\partial \phi}
  \right)^T \Psi_L - \ii \sqrt{2} \gamma^\mu \left( \dfrac{\partial f
  (\phi)}{\partial \phi} \right)^\dagger \Psi_R 
\end{aligned}} \quad .
\end{equation}

%%%%%%%%%%%%%%%%%%%%%%%%%%%%%%%%%%%%%%%%%%%%%%%%%%%%%%%%%%%%%%%%%%%%%

\section{Proof of SYM current conservation}\label{sec:symcurrentcons}

In this section we want to check the conservation of the
supersymmetric current for super-Yang--Mills theories given in
(\ref{eq:strom_sym_appen}). For this purpose, we list the equations of motion
of all participating fields:
\begin{align}
  (D_\mu D^\mu) \phi = & \; - \sqrt{2} g \overline{\vec{\lambda}} \cdot 
  \vec{T} \Psi_L + g \vec{T} \phi \cdot \vec{D} - \dfrac{1}{2}
  \left(\dfrac{\partial^3 f (\phi)}{\partial \phi^3} \right)^\dagger \left(
  \overline{\Psi_L} \Psi_R \right) + F^\dagger \left(
  \dfrac{\partial^2 f (\phi)}{\partial \phi^2} \right)^\dagger
  \label{bewegphi} \\
  (D_\mu D^\mu \phi)^\dagger = & \; - \sqrt{2} g \overline{\Psi_L}  
  \vec{T} \cdot \vec{\lambda} + g \phi^\dagger \vec{T} \cdot \vec{D} -
  \dfrac{1}{2} \left(\dfrac{\partial^3 f (\phi)}{\partial \phi^3} \right)
  \left( \overline{\Psi_R} \Psi_L \right) + F \left( \dfrac{\partial^2 f
  (\phi)}{\partial \phi^2} \right) \label{bewegphid} \\
  \ii \fmslash{D} \Psi_L = & \; \sqrt{2} g \vec{T} \phi \cdot
  \vec{\lambda}_R + \left( \dfrac{\partial^2 f(\phi)}{\partial \phi^2}
  \right)^\dagger \Psi_R \label{bewegpsil} \\
  \ii \fmslash{D} \Psi_R = & \; \sqrt{2} g \phi^\dagger \vec{T}
  \cdot \vec{\lambda}_L + \left( \dfrac{\partial^2 f(\phi)}{\partial \phi^2}
  \right) \Psi_L \label{bewegpsir} \\
  \ii (\fmslash{D} \lambda)^a = & \; \sqrt{2} g \phi^\dagger T^a
  \Psi_L + \sqrt{2} g T^a \phi \Psi_R \label{beweglambda} \\
  D^a = & \; - \zeta^a - g \left( \phi^\dagger T^a \phi \right) \label{bewegd}
  \\
  F = & \; - \left( \dfrac{\partial f (\phi)}{\partial \phi} \right)^\dagger
  \label{bewegf} \\
  F^\dagger = & \; - \left( \dfrac{\partial f (\phi)}{\partial \phi}
  \right) \label{bewegfd} \\
  \partial_\mu F_a^{\mu\nu} = & \; g f_{abc} F_c^{\nu\rho} A_\rho^b - \left(
  \ii g \phi^\dagger T^a \right) (D^\nu \phi) + \ii g (D^\nu \phi)^\dagger
  (T^a \phi) \notag \\ & \; - \dfrac{g}{2} \left( \overline{\Psi_L} \gamma^\nu
  T^a \Psi_L \right) + \dfrac{g}{2} \left( \overline{\Psi_R} \gamma^\nu T^a
  \Psi_R \right) - \dfrac{\ii g}{2} \left( \overline{\lambda^b} f_{abc}
  \gamma^\nu \lambda^c \right) \label{bewegb}   
\end{align}
The divergence of the current is:
\begin{equation}
\label{eq:stromerhaltunganhang}
  \begin{aligned}
    \partial_\mu \mathcal{J}^\mu = & \; - \sqrt{2} \gamma^\nu \gamma^\mu
    \Bigl[ \partial_\mu (D_\nu \phi)^T \Bigr] \Psi_R - \sqrt{2} \gamma^\nu
    \gamma^\mu \Bigl[ \partial_\mu (D_\nu \phi)^\dagger \Bigr] \Psi_L -
    \sqrt{2} \gamma^\nu (D_\nu \phi)^T \fmslash{\partial} \Psi_R \\ & \; -
    \sqrt{2} \gamma^\nu (D_\nu \phi)^\dagger \fmslash{\partial} \Psi_L - \ii
    \zeta^a \fmslash{\partial} \lambda^a + \dfrac{1}{2} \gamma^\alpha
    \gamma^\beta \gamma^\mu \gamma^5 (\partial_\mu F_{\alpha\beta}^a)
    \lambda^a - \dfrac{1}{2} \gamma^\alpha \gamma^\beta \gamma^5
    F_{\alpha\beta}^a \fmslash{\partial} \lambda^a \\ & \; - \ii g
    (\fmslash{\partial} \phi^\dagger) \vec{T} \phi \cdot \vec{\lambda} - \ii g
    \phi^\dagger \vec{T} (\fmslash{\partial} \phi) \cdot \vec{\lambda} - \ii g
    \left( \phi^\dagger \vec{T} \phi \right) \cdot \fmslash{\partial}
    \vec{\lambda} \\ & \; - \sqrt{2} \, \ii \left( \dfrac{\partial^2
    f(\phi)}{\partial 
    \phi^2} \right)^T (\fmslash{\partial} \phi) \Psi_L - \sqrt{2} \,
    \ii \left( \dfrac{\partial^2 f(\phi)}{\partial \phi^2} \right)^\dagger
    (\fmslash{\partial} \phi^\dagger) \Psi_R \\ & \; - \sqrt{2} \, \ii \left(
    \dfrac{\partial f(\phi)}{\partial \phi} \right)^T \fmslash{\partial}
    \Psi_L  - \sqrt{2} \, \ii \left( \dfrac{\partial
    f(\phi)}{\partial \phi} 
    \right)^\dagger \fmslash{\partial} \Psi_R 
  \end{aligned}
\end{equation}

For the first two terms we calculate:
\begin{equation*}
\begin{aligned}
  \partial_\mu (D_\nu \phi)^T = & \; \partial_\mu \partial_\nu \phi^T - \ii g
  \vec{T}^T \vec{A}_\nu \partial_\mu \phi^T - \ii g \vec{T}^T (\partial_\mu
  \vec{A}_\nu) \phi^T \\
  \partial_\mu (D_\nu \phi)^\dagger = & \; \partial_\mu \partial_\nu
  \phi^\dagger  + \ii g \vec{T} \vec{A}_\nu \partial_\mu \phi^\dagger + \ii g
  \vec{T} (\partial_\mu \vec{A}_\nu) \phi^\dagger 
\end{aligned}
\end{equation*}
This is split up into a symmetric and an antisymmetric part; note that
the first term, which only contains partial derivatives, has no
antisymmetric contribution. The symmetrization and antisymmetrization
respectively can be transferred to the gamma matrices; using the Dirac
algebra, the first two terms of the current's divergence read: 
\begin{equation}
\label{dummy1}
  \begin{aligned}
    & - \sqrt{2} \left( \partial^\mu D_\mu \phi \right)^T \Psi_R - \sqrt{2}
      \left( \partial^\mu D_\mu \phi \right)^\dagger \Psi_L +
      \dfrac{\ii g}{\sqrt{2}} \left[ \gamma^\nu , \gamma^\mu \right] \vec{T}^T
      \vec{A}_\nu \partial_\mu \phi^T \\ & - \dfrac{\ii g}{\sqrt{2}} \left[
      \gamma^\nu , \gamma^\mu \right] \vec{T} \vec{A}_\nu \partial_\mu
      \phi^\dagger + \dfrac{\ii g}{2 \sqrt{2}} \left[ \gamma^\nu , \gamma^\mu
      \right] \vec{T}^T (\partial_\mu \vec{A}_\nu - \partial_\nu \vec{A}_\mu)
      \phi^T  \\ & \; -  \dfrac{\ii g}{2 \sqrt{2}} \left[ \gamma^\nu ,
      \gamma^\mu \right] \vec{T} (\partial_\mu \vec{A}_\nu - \partial_\nu
      \vec{A}_\mu) \phi^\dagger  
  \end{aligned}
\end{equation}
In the first two terms of (\ref{dummy1}) we complete the squares of
the covariant derivatives, insert the equations of motion
(\ref{bewegphi}) and (\ref{bewegphid}) for the scalar field and get
\begin{align}
    \bf{I} = & + 2 g \left( \overline{\vec{\lambda}} \cdot \vec{T}^T \Psi_L
    \right) \Psi_R - \sqrt{2} g \phi^T \vec{T}^T \cdot \vec{D} \Psi_R -
    \sqrt{2} F^\dagger \left( \dfrac{\partial^2 f(\phi)}{\partial \phi^2}
    \right)^\dagger \Psi_R \notag\\ & 
    + 2 g \left( \overline{\Psi_L} \vec{T} \cdot \vec{\lambda} \right)
    \Psi_L - \sqrt{2} g \phi^\dagger \vec{T} \cdot \vec{D} \Psi_L - \sqrt{2}
    F^T \left( \dfrac{\partial^2 f(\phi)}{\partial \phi^2} \right) \Psi_L
    \notag\\ 
    & - \sqrt{2} \, \ii g \vec{T}^T \vec{A}^\mu (D_\mu \phi)^T \Psi_R +
    \sqrt{2} \, \ii g \vec{T} \vec{A}^\mu (D_\mu \phi)^\dagger \Psi_L \notag\\
    & + \dfrac{\ii g}{\sqrt{2}} \left[ \gamma^\nu , \gamma^\mu \right]
    \vec{T}^T \vec{A}_\nu \partial_\mu \phi^T \Psi_R - \dfrac{\ii g}{\sqrt{2}}
    \left[ \gamma^\nu , \gamma^\mu \right] \vec{T} \vec{A}_\nu \partial_\mu
    \phi^\dagger \Psi_L \notag\\ & + \dfrac{\ii g}{2 \sqrt{2}} \left[
    \gamma^\nu , 
    \gamma^\mu \right] \vec{T}^T (\partial_\mu \vec{A}_\nu - \partial_\nu
    \vec{A}_\mu) \phi^T \Psi_R - \dfrac{\ii g}{2 \sqrt{2}} \left[ \gamma^\nu ,
    \gamma^\mu \right] \vec{T} (\partial_\mu \vec{A}_\nu - \partial_\nu
    \vec{A}_\mu) \phi^\dagger \Psi_L   \label{bruchteil1}
\end{align}
Here we already used the vanishing of 
$(\overline{\Psi_L} \Psi_R) \Psi_R$ and 
$(\overline{\Psi_R} \Psi_L) \Psi_L$  
as third powers of a Grassmann odd spinor. In the case of a general
(nonrenormalizable) superpotential with more than three matter
superfields, this contribution vanishes due to the Schouten identity. 

For all terms in (\ref{eq:stromerhaltunganhang}) containing derivatives
of the matter fermions (i.e.~the third and fourth and the two
rightmost), we insert the equations of motion for the fermions
(\ref{bewegpsil}) and (\ref{bewegpsir}), in which we must bring the
gauge field term 
from the covariant derivative to the right hand side. Furthermore we
replace the derivatives of the superpotential with respect to the
scalar fields by the equations of motion for the auxiliary fields
(\ref{bewegf}) and (\ref{bewegfd}). This results in:
\begin{equation}
  \label{dummy2}
  \begin{aligned}
  & + \sqrt{2} \, \ii g \gamma^\nu (D_\nu \phi)^T \vec{T}^T \fmslash{\vec{A}} 
  \Psi_R + 2 \ii g \gamma^\nu (D_\nu \phi)^T \phi^\dagger \vec{T} \cdot
  \vec{\lambda}_L + \sqrt{2} \, \ii \gamma^\nu (D_\nu \phi)^T \left(
  \dfrac{\partial^2 f(\phi)}{\partial \phi^2} \right) \Psi_L  \\ &
  - \sqrt{2} \, \ii g \gamma^\nu (D_\nu \phi)^\dagger \vec{T}
  \fmslash{\vec{A}} \Psi_L + 2 \ii g \gamma^\nu (D_\nu \phi)^\dagger \vec{T}
  \phi^T \cdot \vec{\lambda}_R + \sqrt{2} \, \ii \gamma^\nu (D_\nu
  \phi)^\dagger \left( \dfrac{\partial^2 f(\phi)}{\partial \phi^2}
  \right)^\dagger \Psi_R \\ & 
  - \sqrt{2} g F^\dagger \vec{T} \fmslash{\vec{A}} \Psi_L + \sqrt{2} g
  F^\dagger \vec{T} \phi^T \cdot \vec{\lambda}_R + \sqrt{2} F^\dagger \left(
  \dfrac{\partial^2 f(\phi)}{\partial \phi^2} \right)^\dagger \Psi_R  \\
  & + \sqrt{2} g F^T \vec{T}^T \fmslash{\vec{A}} \Psi_R + \sqrt{2} g
  F^T  \phi^\dagger \vec{T} \cdot \vec{\lambda}_L + \sqrt{2} F^T \left(
  \dfrac{\partial^2 f(\phi)}{\partial \phi^2} \right) \Psi_L 
  \end{aligned}
\end{equation}
The two terms $- \sqrt{2} \, \ii \left( \partial^2 f(\phi) /
\partial \phi^2 \right)^T (\fmslash{\partial} \phi) \Psi_L$ and $- \sqrt{2} \,
\ii \left( \partial^2 f(\phi) / \partial \phi^2 \right)^\dagger
(\fmslash{\partial} \phi^\dagger) \Psi_R$, i.e. the terms in the fourth line 
of (\ref{eq:stromerhaltunganhang}), 
cancel the partial derivatives of the scalar fields from the covariant
ones in (\ref{dummy2}) so that only the gauge field terms remain. We
yield as a second contribution to the current's divergence: 
\begin{align}
  \bf{II} = & + \sqrt{2} \, \ii g \gamma^\nu (D_\nu \phi)^T \vec{T}^T
  \fmslash{\vec{A}} \Psi_R + 2 \ii g \gamma^\nu (D_\nu \phi)^T \phi^\dagger
  \vec{T} \cdot \vec{\lambda}_L + \sqrt{2} g \fmslash{\vec{A}} \vec{T}^T \phi^T
  \left( \dfrac{\partial^2 f(\phi)}{\partial \phi^2} \right) \Psi_L  \notag\\ &
  - \sqrt{2} \, \ii g \gamma^\nu (D_\nu \phi)^\dagger \vec{T}
  \fmslash{\vec{A}} \Psi_L + 2 \ii g \gamma^\nu (D_\nu \phi)^\dagger \vec{T}
  \phi^T \cdot \vec{\lambda}_R - \sqrt{2} g \fmslash{\vec{A}} \vec{T}
  \phi^\dagger \left( \dfrac{\partial^2 f(\phi)}{\partial \phi^2}
  \right)^\dagger \Psi_R \notag\\ & 
  - \sqrt{2} g F^\dagger \vec{T} \fmslash{\vec{A}} \Psi_L + \sqrt{2} g
  F^\dagger \vec{T} \phi^T \cdot \vec{\lambda}_R + \sqrt{2} F^\dagger \left(
  \dfrac{\partial^2 f(\phi)}{\partial \phi^2} \right)^\dagger \Psi_R  \notag\\
  & + \sqrt{2} g F^T \vec{T}^T \fmslash{\vec{A}} \Psi_R + \sqrt{2} g
  F^T  \phi^\dagger \vec{T} \cdot \vec{\lambda}_L + \sqrt{2} F^T \left(
  \dfrac{\partial^2 f(\phi)}{\partial \phi^2} \right) \Psi_L
  \label{bruchteil2} 
\end{align}   
For the fifth, seventh and tenth term in
(\ref{eq:stromerhaltunganhang}) we simply have
to insert the gaugino equation of motion (\ref{beweglambda}). Again we
must transfer the gauge field term to the right hand side, though it
cancels in combination with the Fayet--Iliopoulos constant due to the
condition (\ref{lfieich}). This immediately gives a third contribution
to the divergence:
\begin{equation}
  \label{bruchteil3}
  \begin{aligned}
    \bf{III} = &  - \sqrt{2} g \zeta^a \phi^\dagger T^a \Psi_L - \sqrt{2} g
    \zeta^a T^a \phi^T \Psi_R + \ii g^2 f^a_{\;bc} \left( \phi^\dagger T^a
    \phi \right) \fmslash{A}^b \lambda^c \\ & - \sqrt{2} g^2 \left(
    \phi^\dagger T^a \phi \right) \phi^\dagger T^a \Psi_L - \sqrt{2} g^2
    \left( \phi^\dagger T^a \phi \right) \phi^T T^a \Psi_R + \dfrac{g}{2}
    \gamma^\alpha \gamma^\beta \gamma^5 F_{\alpha\beta}^a f^a_{\;bc}
    \fmslash{A}^b \lambda^c \\ & + \dfrac{\ii g}{\sqrt{2}} \gamma^\alpha
    \gamma^\beta \gamma^5 F_{\alpha\beta}^a \phi^\dagger T^a \Psi_L + 
    \dfrac{\ii g}{\sqrt{2}} \gamma^\alpha \gamma^\beta \gamma^5
    F_{\alpha\beta}^a \phi^T T^a \Psi_R  
  \end{aligned}
\end{equation}

At first, we leave the eighth and ninth term from the divergence of
the current unchanged:
\begin{align}
  \label{bruchteil4}
  \bf{IV} = & - \ii g (\fmslash{\partial} \phi^\dagger) \vec{T} \phi \cdot
    \vec{\lambda} - \ii g \phi^\dagger \vec{T} (\fmslash{\partial} \phi) \cdot
    \vec{\lambda}  
\end{align}

For further manipulations on the last remaining term with the
derivative of the field strength tensor for the gauge field we use
again the gamma matrix identity (\ref{eq:gammaidentitaet}). This
happens in the second step of the following calculation:
\begin{equation*}
\begin{aligned}
\dfrac{1}{2} \gamma^\alpha \gamma^\beta \gamma^\mu \gamma^5 \left(
\partial_\mu F_{\alpha\beta}^a \right) \lambda^a = & \; \dfrac{1}{4} \left[
\gamma^\alpha , \gamma^\beta \right] \gamma^\mu \gamma^5 \left( \partial_\mu
F_{\alpha\beta}^a \right) \lambda^a \\ = & \; 
\dfrac{1}{2} \gamma^\alpha \gamma^5 \left( \partial^\beta
F_{\alpha\beta}^a \right) \lambda^a - \dfrac{1}{2} \gamma^\beta
\gamma^5 \left( \partial^\alpha F_{\alpha\beta}^a \right) \lambda^a \\
& \qquad \qquad -
\dfrac{\ii}{2} \epsilon^{\alpha\beta\mu\sigma} \gamma_\sigma \left(
\partial_\mu F_{\alpha\beta}^a \right) \lambda^a \\
= & \; \gamma^\alpha \gamma^5 \left( \partial^\beta F_{\alpha\beta}^a
\right) \lambda^a - \dfrac{\ii}{2} \epsilon^{\alpha\beta\mu\sigma}
\gamma_\sigma \left( \partial_\mu F_{\alpha\beta}^a \right) \lambda^a
\end{aligned}
\end{equation*}
The insertion of the gauge field's equation of motion in the form
(\ref{bewegb}) for the first term yields a fifth contribution to the
divergence of the current:
\begin{equation}
\label{bruchteil5}
\begin{aligned}
\bf{V} = & - g \gamma^\alpha \gamma^5 f^a_{\;bc} F^c_{\alpha\rho} \lambda^a
A_b^\rho + \ii g \gamma^\alpha \gamma^5 \phi^\dagger T^a (D_\alpha
\phi) \lambda^a - \ii g \gamma^\alpha \gamma^5 (D_\alpha \phi)^\dagger
T^a \phi \lambda^a \\ & + \dfrac{g}{2} \left( \overline{\Psi_L}
\gamma_\alpha T^a \Psi_L \right) \gamma^\alpha \gamma^5 \lambda^a -
\dfrac{g}{2} \left( \overline{\Psi_R} \gamma_\alpha T^a \Psi_R \right)
\gamma^\alpha \gamma^5 \lambda^a + \dfrac{\ii g}{2} \left(
\overline{\lambda^b} f_{abc} \gamma_\alpha \lambda^c \right)
\gamma^\alpha \gamma^5 \lambda^a \\ & - \dfrac{\ii}{2}
\epsilon^{\sigma\alpha\beta\mu} \gamma_\sigma \left( \partial_\mu
F_{\alpha\beta}^a \right) \lambda^a 
\end{aligned}
\end{equation}

As a next step we use the equation of motion for the auxiliary field
$D^a$, (\ref{bewegd}), to see that the second and fifth term of 
(\ref{bruchteil1}) cancel the first, the second, the fourth and the
fifth term from (\ref{bruchteil3}). 

With the help of relation (\ref{dreigauginossindnull}) and the
discussion above this equation, we can set the term with three
gauginos, the penultimate in (\ref{bruchteil5}), equal to zero. 

Now we consider the contributions with two matter fermions and one
gaugino. These are the first and fourth term in (\ref{bruchteil1}) as
well as the fourth and fifth from (\ref{bruchteil5}). By the Fierz
identity these four terms add up to zero in the same manner as
discussed below (\ref{eq:dewitspezial}).

We may rewrite the last term of (\ref{bruchteil5}) as follows
\begin{equation}
 - \dfrac{\ii}{2} \epsilon^{\alpha\beta\mu\sigma} \gamma_\sigma \left(
   \partial_\mu F_{\alpha\beta}^a \right) \lambda^a = + \dfrac{\ii
   g}{2} \epsilon^{\alpha\beta\mu\sigma} \gamma_\sigma f^a_{\;bc}   A_\mu^b F_{\alpha\beta}^c \lambda^a ,
\end{equation}
as the antisymmetrized covariant derivative of the field strength
tensor vanishes due to the Bianchi identity. We manipulate the sixth
term from (\ref{bruchteil3}) with help of the identity
(\ref{eq:gammaidentitaet}):  
\begin{equation*}
- \dfrac{g}{4} \left[ \gamma^\alpha , \gamma^\beta \right] \gamma^\mu
  \gamma^5 F_{\alpha\beta}^a f^a_{\;bc} A_\mu^b \lambda^c =  
  g \gamma^\beta \gamma^5 F_{\alpha\beta}^a f^a_{\;bc}
  A^\alpha_b \lambda^c + \dfrac{\ii g}{2} \gamma_\sigma
  \epsilon^{\alpha\beta\mu\sigma} F_{\alpha\beta}^a f^a_{\;bc} A_\mu^b
  \lambda^c 
\end{equation*}
It is obvious that this cancels the term in the last equation as well
as the first term of (\ref{bruchteil5}).  

It is also clear that the sum of the third and sixth term out of
(\ref{bruchteil1}) and the ninth and last one from (\ref{bruchteil2})
is zero.  

To rewrite the third and sixth term in (\ref{bruchteil2}) we remember
that the superpotential in an SYM theory must be a gauge invariant
function  of the fields. Hence it is constrained by the condition 
\begin{equation}
\label{suppoteichinv}
\left( \dfrac{\partial f (\phi)}{\partial \phi} \right) T^a \phi = 0  
\end{equation}
Differentiating with respect to $\phi$ and using the equation of
motion for $F$ yields the relation (similarly for $\phi^\dagger$):
\begin{align}
 \left( \dfrac{\partial^2 f(\phi)}{\partial \phi^2} \right) T^a \phi &
 = \; - \left( \dfrac{\partial f(\phi)}{\partial \phi} \right) T^a  =
 F^\dagger T^a  \\ 
 \left( \dfrac{\partial^2 f(\phi)}{\partial \phi^2} \right)^\dagger
 T^a \phi^\dagger & = \; - \left( \dfrac{\partial f(\phi)}{\partial
 \phi} \right)^\dagger T^a  = F T^a
\end{align}
Now we see that these two terms cancel against the seventh and tenth of
(\ref{bruchteil2}).

By (\ref{suppoteichinv}) the eighth and penultimate term out of
(\ref{bruchteil2}) vanish, too. 

We want to inspect the terms containing derivatives of the scalar
fields now. These are the first, second, fourth and fifth term from
(\ref{bruchteil2}), the two remaining ones out of (\ref{bruchteil5}) --
the second and the third -- as well as the seventh up to the tenth term
of (\ref{bruchteil1}). Consider first all of these terms containing
gauginos. We calculate: 
\begin{equation*}
  \begin{aligned}
    & \; 2 \ii g \gamma^\nu (D_\nu \phi)^T \phi^\dagger \vec{T} \cdot
    \vec{\lambda}_L - \ii g \phi^\dagger \vec{T} (\fmslash{\partial} \phi^T)
    \cdot \vec{\lambda} + \ii g \gamma^\nu \gamma^5 \phi^\dagger \vec{T}
    (D_\nu \phi)^T \vec{\lambda} \\ = & \; 
    \underline{2 \ii g (\fmslash{\partial} \phi^T) \phi^\dagger \vec{T} \cdot
    \vec{\lambda}_L} + 2 g^2 \gamma^\nu \phi^\dagger T^b T^a A_\nu^b \phi^T
    \lambda_L^a + \underline{\ii g (\fmslash{\partial} \phi^T) \phi^\dagger
    \vec{T} \cdot \gamma^5 \vec{\lambda}} \\ & \; + g^2 \gamma^\nu
    \phi^\dagger T^b T^a A_\nu^b \phi^T \gamma^5 \lambda^a - \underline{\ii g
    \phi^\dagger \vec{T} (\fmslash{\partial} \phi^T) \cdot \vec{\lambda}} \\ =
    & \; g^2 \gamma^\nu \phi^\dagger T^a T^b A_\nu^b \phi^T \lambda^a 
  \end{aligned}
\end{equation*}
The underlined terms with the spacetime derivatives all vanish, solely
the gauge field parts are left. An analogous calculation works for the
complex conjugated scalar fields:
\begin{equation*}
  \begin{aligned}
    & \; 2 \ii g \gamma^\nu (D_\nu \phi)^\dagger \vec{T} \phi^T \cdot
    \vec{\lambda}_R - \ii g (\fmslash{\partial} \phi^\dagger) \vec{T} \phi^T
    \cdot \vec{\lambda} - \ii g \gamma^\nu \gamma^5 (D_\nu \phi)^\dagger
    \vec{T} \phi^T \vec{\lambda} \\ = & \; 
    - g^2 \gamma^\nu \phi^\dagger T^b T^a \phi^T A_\nu^b \lambda^a 
  \end{aligned}
\end{equation*}
Summing up the last two resulting expressions yields
\begin{equation}
  g^2 \gamma^\nu \phi^\dagger \left[ T^a , T^b \right] \phi^T A_\nu^b
    \lambda^a = \ii g^2 \gamma^\nu \phi^\dagger f_{abc} T^c \phi^T A_\nu^b
    \lambda^a
\end{equation}
and hence cancel the third term of (\ref{bruchteil3}).

The first and fourth term of (\ref{bruchteil2}) can be rewritten with
the help of the Dirac algebra:
\begin{equation*}
  \begin{aligned}
    \sqrt{2} \, \ii g \gamma^\nu \gamma^\mu (D_\nu \phi)^T \vec{T}^T
    \vec{A}_\mu \Psi_R = & \; \sqrt{2} \, \ii g \Bigl( \dfrac{1}{2} \left\{
    \gamma^\nu , \gamma^\mu \right\} + \dfrac{1}{2} \left[ \gamma^\nu ,
    \gamma^\mu \right] \Bigr) (D_\nu \phi)^T \vec{T}^T \vec{A}_\mu \Psi_R \\
    = & \; + \sqrt{2} \, \ii g \vec{T}^T \vec{A}^\mu (D_\mu \phi)^T \Psi_R \\
    & \; + \dfrac{\ii g}{\sqrt{2}} \left[ \gamma^\nu , \gamma^\mu \right]
    (\partial_\nu \phi)^T \vec{T}^T \vec{A}_\mu \Psi_R \\ & \; +
    \dfrac{g^2}{\sqrt{2}} \left[ \gamma^\nu , \gamma^\mu \right] A_\nu^a T^b
    T^a A_\mu^b \phi^T \Psi_R  
  \end{aligned}
\end{equation*}
Of the newly established terms the first cancels the seventh term in
(\ref{bruchteil1}) while the second one eliminates the ninth from
(\ref{bruchteil1}). For the complex conjugated fields we get
analogously  
\begin{equation*}
  \begin{aligned}
    - \sqrt{2} \, \ii g \gamma^\nu \gamma^\mu (D_\nu \phi)^\dagger \vec{T}
    \vec{A}_\mu \Psi_L = & \; - \sqrt{2} \, \ii g \vec{T} \vec{A}^\mu (D_\mu
    \phi)^\dagger \Psi_L \\ & \; - \dfrac{\ii g}{\sqrt{2}} \left[ \gamma^\nu ,
    \gamma^\mu \right] (\partial_\nu \phi^\dagger) \vec{T} \vec{A}_\mu \Psi_L
    \\ & \; + \dfrac{g^2}{\sqrt{2}} \phi^\dagger \left[ \gamma^\nu ,
    \gamma^\mu \right] A_\nu^a T^a T^b A_\mu^b \Psi_L 
  \end{aligned}
\end{equation*}
A similar cancellation takes place here: the first term cancels the
eighth of (\ref{bruchteil1}), the second one the tenth out of
(\ref{bruchteil1}). The third terms from the last and the penultimate
identity are further manipulated with taking attention to their
antisymmetrization: 
\begin{equation}
  \begin{aligned}
    \dfrac{g^2}{\sqrt{2}} \left[ \gamma^\nu , \gamma^\mu \right] A_\nu^a
    A_\mu^b T^b T^a \phi^T \Psi_R = & \; \dfrac{g^2}{2 \sqrt{2}} \left[
    \gamma^\nu , \gamma^\mu \right] \left( A_\nu^a A_\mu^b - A_\mu^a A_\nu^b
    \right) T^b T^a \phi^T \Psi_R \\ = & \; \dfrac{g^2}{2 \sqrt{2}} \left[
    \gamma^\nu , \gamma^\mu \right] A_\nu^a A_\mu^b \left[ T^b , T^a \right]
    \phi^T \Psi_R \\ = & \; \dfrac{\ii g^2}{2 \sqrt{2}} \left[
    \gamma^\nu , \gamma^\mu \right] A_\nu^a A_\mu^b f_{bac} T^c \phi^T \Psi_R
  \end{aligned}
\end{equation}
Analogously:
\begin{equation}
    \dfrac{g^2}{\sqrt{2}} \phi^\dagger \left[ \gamma^\nu , \gamma^\mu \right]
    A_\nu^a T^a T^b A_\mu^b \Psi_L = \dfrac{\ii g^2}{2 \sqrt{2}} \left[
    \gamma^\nu , \gamma^\mu \right] \phi^\dagger A_\nu^a A_\mu^b f_{abc} T^c
    \Psi_L  
\end{equation}
Together with the last two terms from (\ref{bruchteil1}) this yields
\begin{equation}
   + \dfrac{\ii g}{2 \sqrt{2}} \left[ \gamma^\nu , \gamma^\mu \right] T^a 
    F^a_{\mu\nu} \phi^T \Psi_R - \dfrac{\ii g}{2 \sqrt{2}} \left[ \gamma^\nu ,
    \gamma^\mu \right] T^a F^a_{\mu\nu} \phi^\dagger \Psi_L   \quad ,
\end{equation}
which cancels the last remaining terms, the two rightmost ones
in (\ref{bruchteil3}), so after all we get the desired result
\begin{equation}
  \label{eq:stromerhaltung_sym}
  \boxed{ \partial_\mu \mathcal{J}^\mu = 0 . }
\end{equation}

%%% Local Variables: 
%%% mode: latex
%%% TeX-master: diss
%%% End: 

%% file: appen_wz.tex
\section{The Wess-Zumino model}\label{appen_wz}

Action: 
\begin{equation}
  \label{eq:lag_wz}
  S_{WZ} =  \int d^4 x \, \biggl\{ \dfrac{1}{2} \left[ \hat{\Phi}^\dagger
  \hat{\Phi} \right]_D + \left[ \mu \hat{\Phi} + \frac{m}{2} \hat{\Phi}^2 +
  \frac{\lambda}{3!} \hat{\Phi}^3  \, \text{+ h.c.} \right]_F \biggr\}
\end{equation}
As is shown in \cite{Wess/Bagger:SUSY:text}, the contribution to the
superpotential linear in the superfields can always be eliminated by a
redefinition of the superfields. So in the sequel we set $\mu \equiv 0$. 

The quadratic and cubic part of the superpotential yield:
\begin{align}
  \frac{m}{2} \left[ \hat{\Phi}^2 \, \text{+ h.c.} \right]_F & = m \phi F
 + m \phi^* F^* - \frac{m}{2} \left( \psi \psi + \bar{\psi} \bar{\psi} \right)
 , \\ \frac{\lambda}{3!} \left[ \hat{\Phi}^3 \text{+ h.c.} \right]_F & =
 \frac{\lambda}{2} \phi^2 F + \frac{\lambda}{2} (\phi^*)^2 F^* -
 \frac{\lambda}{2} \phi \psi \psi - \frac{\lambda}{2} \phi^* \bar{\psi}
 \bar{\psi}   
\end{align}
The kinetic part is:
\begin{equation}
  \label{eq:kin_WZ}
  \dfrac{1}{2} \left[ \hat{\Phi}^\dagger \hat{\Phi} \right]_D = \left(
  \partial_\mu \phi \right)^* \left( \partial^\mu \phi \right) + \frac{\ii}{2}
  \bar{\psi} \bar{\sigma}^\mu \partial_\mu \psi + \frac{\ii}{2} \psi
  \sigma^\mu \partial_\mu \bar{\psi} + \left| F \right|^2 
\end{equation}
The equation of motion for the auxiliary field is:
\begin{equation}
  \label{eq:beweghilf}
  F^* = - m \phi - \frac{\lambda}{2} \phi^2 \quad .
\end{equation}
After introducing the bispinor notation
\begin{equation}
  \label{eq:majobisp}
  \Psi = 
  \begin{pmatrix}
    \psi \\ \bar{\psi}
  \end{pmatrix}
\end{equation}
for the fermionic degrees of freedom (the Yukawa coupling terms will soon be
brought into bispinor form) we get the Lagrangean density
\begin{equation}
  \label{eq:lag_WZ_2}
  {\cal L}_{WZ} = \left( \partial_\mu \phi \right)^* \left( \partial^\mu \phi
  \right) + \frac{1}{2} \overline{\Psi} \left( \ii \fmslash{\partial} - m
  \right) \Psi - \frac{\lambda}{2} \phi \psi \psi - \frac{\lambda}{2} \phi^*
  \bar{\psi} \bar{\psi} - \left| F \right|^2 
\end{equation}
The scalar part of the superpotential reads
\begin{equation}
  \label{eq:sup_WZ}
  \left| F \right|^2 = m^2 \left| \phi \right|^2 + \frac{\lambda^2}{4} \left(
  \left| \phi \right|^2 \right)^2 + \frac{1}{2} m \lambda \left| \phi
  \right|^2 \left( \phi + \phi^* \right) 
\end{equation}
After splitting the complex scalar field into real and imaginary part
(cf. chapter 2), \[ \phi = \frac{1}{\sqrt{2}} \left( A + \ii B \right), \] 
it is easier to write the Yukawa interactions in the bispinor language. 
\begin{equation}
  \begin{aligned}
    \phi \psi \psi + \phi^* \bar{\psi} \bar{\psi} = & \; \dfrac{1}{2 \sqrt{2}}
    \overline{\Psi} \left( 1 - \gamma^5 \right) \Psi \: \left( A + \ii B
    \right) + \dfrac{1}{2 \sqrt{2}} \overline{\Psi} \left( 1 + \gamma^5
    \right) \Psi \: \left( A - \ii B \right) \\ = & \; \dfrac{1}{\sqrt{2}}
    \overline{\Psi} \Psi A - \dfrac{\ii}{\sqrt{2}} \overline{\Psi}
    \gamma^5 \Psi B \quad . 
  \end{aligned}
\end{equation}
Finally, the whole Lagrangean density for the Wess-Zumino model reads:
\begin{equation}
  \label{eq:kinvoll}
  \begin{aligned}
  {\cal L}_{WZ} = & \; \frac{1}{2} \left( \partial_\mu A \partial^\mu A - m^2
  A^2 \right) + \frac{1}{2} \left( \partial_\mu B \partial^\mu B - m^2 B^2
  \right) + \frac{1}{2} \overline{\Psi} \left( \ii \fmslash{\partial} - m
  \right) \Psi \\ & \; - \dfrac{\lambda}{2 \sqrt{2}} \overline{\Psi} \Psi A +
  \dfrac{\ii\lambda}{2 \sqrt{2}} \overline{\Psi} \gamma^5 \Psi B -
  \frac{\lambda^2}{16} A^4 - \frac{\lambda^2}{16} B^4 - \frac{\lambda^2}{8}
  A^2 B^2 \\ & \; - \frac{1}{2 \sqrt{2}} m \lambda A^3  - \frac{1}{2 \sqrt{2}}
  m \lambda A B^2  
  \end{aligned}
\end{equation}

We also state the SUSY transformations for the component fields:
\begin{subequations}
\begin{align}
  \lbrack \ii \overline{\xi} Q, A \rbrack &=  \left( \overline{\xi} \Psi 
  \right) \\ \lbrack \ii \overline{\xi} Q,B \rbrack  &= \left( \ii 
  \overline{\xi} \gamma^5 \Psi \right) \\
  \lbrack \ii \overline{\xi} Q , \Psi \rbrack &= - \left( \ii 
  \fmslash{\partial} + m\right) (A + \ii \gamma^5 B) \xi - 
  \dfrac{\lambda}{2\sqrt{2}} \left( A^2 - B^2 \right) \xi - 
  \dfrac{\ii \lambda}{\sqrt{2}} A B \gamma^5 \xi 
\end{align}
\end{subequations}

{\bf The vertices of the WZ model:}  
%%%%%%%%%%%%%%%%%%%%%%%%%%%%%%%%%%%%%%%%%%%%%%%%
\begin{equation*}     %%%  A-A-A
 \begin{aligned}
  \parbox{5cm}{
  \begin{fmfchar*}(25,25)
    \fmfleft{i1,i2}
    \fmfright{o}
    \fmf{dashes,label=$A$}{i2,v}
    \fmf{dashes,label=$A$}{i1,v}
    \fmf{dashes,label=$A$}{o,v}
    \fmfdot{v}
  \end{fmfchar*}} & \hfill &
  \parbox{5cm}{
  \begin{math}
   \qquad
   \begin{array}{l}
   - \ii \dfrac{3}{\sqrt{2}} m \lambda
     \end{array} 
  \end{math}}
 \end{aligned}
\end{equation*}
\vspace{0.8cm}
%%%%%%%%%%%%%%%%%%%%%%%%%%%%%%%%%%%%%%%%%%%%%%%%
\begin{equation*}     %%%  A-B-B
 \begin{aligned}
  \parbox{5cm}{
  \begin{fmfchar*}(25,25)
    \fmfleft{i1,i2}
    \fmfright{o}
    \fmf{dbl_dashes,label=$B$}{i2,v}
    \fmf{dbl_dashes,label=$B$}{i1,v}
    \fmf{dashes,label=$A$}{o,v}
    \fmfdot{v}
  \end{fmfchar*}} & \hfill &
  \parbox{5cm}{
  \begin{math}
   \qquad
   \begin{array}{l}
   - \ii \dfrac{1}{\sqrt{2}} m \lambda
     \end{array} 
  \end{math}}
 \end{aligned}
\end{equation*}
\vspace{0.8cm}  
%%%%%%%%%%%%%%%%%%%%%%%%%%%%%%%%%%%%%%%%%%%%%%%%%%%
\begin{equation*}      %%%  A-A-A-A
 \begin{aligned}
 \parbox{5cm}{
  \begin{fmfchar*}(25,25)
  \fmfleft{i1,i2}
  \fmfright{o1,o2}
  \fmf{dashes,label=$A$,l.side=left}{i2,v}
  \fmf{dashes,label=$A$,l.side=left}{o2,v}
  \fmf{dashes,label=$A$}{v,o1}
  \fmf{dashes,label=$A$}{v,i1}
  \fmfdot{v}
  \end{fmfchar*}} & \hfill &
 \parbox{5cm}{
  \begin{math} \qquad
   \begin{array}{c}
     - \dfrac{3}{2} \ii \lambda^2
   \end{array}
 \end{math}}
 \end{aligned}
\end{equation*}
\vspace{0.8cm} 
%%%%%%%%%%%%%%%%%%%%%%%%%%%%%%%%%%%%%%%%%%%%%%%%%%%
\begin{equation*}      %%%  B-B-B-B
 \begin{aligned}
 \parbox{5cm}{
  \begin{fmfchar*}(25,25)
  \fmfleft{i1,i2}
  \fmfright{o1,o2}
  \fmf{dbl_dashes,label=$B$,l.side=left}{i2,v}
  \fmf{dbl_dashes,label=$B$,l.side=left}{o2,v}
  \fmf{dbl_dashes,label=$B$}{v,o1}
  \fmf{dbl_dashes,label=$B$}{v,i1}
  \fmfdot{v}
  \end{fmfchar*}} & \hfill &
 \parbox{5cm}{
  \begin{math} \qquad
   \begin{array}{c}
     - \dfrac{3}{2} \ii \lambda^2
   \end{array}
 \end{math}}
 \end{aligned}
\end{equation*}
\vspace{0.8cm}  
%%%%%%%%%%%%%%%%%%%%%%%%%%%%%%%%%%%%%%%%%%%%%%%%%%%
\begin{equation*}      %%%  A-A-B-B
 \begin{aligned}
 \parbox{5cm}{
  \begin{fmfchar*}(25,25)
  \fmfleft{i1,i2}
  \fmfright{o1,o2}
  \fmf{dashes,label=$A$,l.side=left}{i2,v}
  \fmf{dashes,label=$A$,l.side=left}{o2,v}
  \fmf{dbl_dashes,label=$B$}{v,o1}
  \fmf{dbl_dashes,label=$B$}{v,i1}
  \fmfdot{v}
  \end{fmfchar*}} & \hfill &
 \parbox{5cm}{
  \begin{math} \qquad
   \begin{array}{c}
     - \dfrac{1}{2} \ii \lambda^2
   \end{array}
 \end{math}}
 \end{aligned}
\end{equation*}
\vspace{0.8cm}
%%%%%%%%%%%%%%%%%%%%%%%%%%%%%%%%%%%%%%%%%%%%%%%%%%%%%%%%%
\begin{equation*}  %%%  Psi-A-Psi
 \begin{aligned}
  \parbox{5cm}{
  \begin{fmfchar*}(25,25)
    \fmfleft{i1,i2}
    \fmfright{o}
    \fmf{plain,label=$\Psi$}{v,i2}
    \fmf{plain,label=$\Psi$,l.side=left}{i1,v}
    \fmf{dashes,label=$A$}{o,v}
    \fmfdot{v}
  \end{fmfchar*}} & \hfill &
  \parbox{5cm}{
  \begin{math}
   \qquad
   \begin{array}{l}
     - \dfrac{\ii \lambda}{\sqrt{2}}
  \end{array}
  \end{math}}
 \end{aligned}
\end{equation*}
\vspace{0.8cm}                                                                
%%%%%%%%%%%%%%%%%%%%%%%%%%%%%%%%%%%%%%%%%%%%%%%%%%%%%%%%%
\begin{equation*}  %%%  Psi-B-Psi
 \begin{aligned}
  \parbox{5cm}{
  \begin{fmfchar*}(25,25)
    \fmfleft{i1,i2}
    \fmfright{o}
    \fmf{plain,label=$\Psi$}{v,i2}
    \fmf{plain,label=$\Psi$,l.side=left}{i1,v}
    \fmf{dbl_dashes,label=$B$}{o,v}
    \fmfdot{v}
  \end{fmfchar*}} & \hfill &
  \parbox{5cm}{
  \begin{math}
   \qquad
   \begin{array}{l}
     - \dfrac{\lambda}{\sqrt{2}} \cdot \gamma^5
  \end{array}
  \end{math}}
 \end{aligned}
\end{equation*}
\vspace{0.8cm}       

%%% Local Variables: 
%%% mode: latex
%%% TeX-master: t
%%% End: 

%% file: appen_toy1.tex
\section{A toy model}\label{appen_toy1} 

Action:
\begin{equation}
  \label{eq:lagrangespiel}
  {\cal S}_{\text{toy}} = \int d^4 x \Biggl\{ \dfrac{1}{2} \left[
   \hat{\Phi}_1^\dagger \hat{\Phi}_1 \right]_D + \dfrac{1}{2} \left[
   \hat{\Phi}_2^\dagger \hat{\Phi}_2 \right]_D + \biggl[ m
  \hat{\Phi}_1 \hat{\Phi}_2 + g \hat{\Phi}_1
  \hat{\Phi}_1 \hat{\Phi}_2   \, \text{+ h.c.} \biggr]_F \Biggr\} 
\end{equation}
Equations of motion for the auxiliary fields:
\begin{equation}
  \label{eq:beweghilfsspiel}
  F_1^* = - m \phi_2 - 2 g \phi_1 \phi_2 , \qquad \qquad F_2^* = - m \phi_1 -
  g \phi_1^2  
\end{equation}
Lagrangean density in components,
\begin{equation}
  \label{eq:lagrangespiel2}
  \begin{aligned} 
  {\cal L}_{\text{toy}} = & \quad (\partial_\mu \phi_1^*) (\partial^\mu \phi_1) +
  (\partial_\mu \phi_2^*) (\partial^\mu \phi_2) + \ii \partial_\mu \bar{\psi}_1
  \bar{\sigma}^\mu \psi_1 \\ & \; + \ii \psi_2 \sigma^\mu \partial_\mu
  \bar{\psi}_2 + \ii \partial_\mu \bar{\psi}_1 \bar{\sigma}^\mu \psi_1 + \ii
  \psi_2 \sigma^\mu \partial_\mu \bar{\psi}_2 \\ & \; - \left| F_1 \right|^2 -
  \left| F_2 \right|^2 - \Bigl( m \psi_1 \psi_2 + g \psi_1 \psi_1 \phi_2 + 2 g
  \psi_1 \psi_2 \phi_1 \text{+ h.c.} \Bigr) .
  \end{aligned}
\end{equation}
Superpotential:
\begin{equation}
  \label{eq:suppotspiel}
  \begin{aligned}
  |F|^2 = & \; m^2 |\phi_1|^2 + m^2 |\phi_2|^2 + m g |\phi_1|^2 \left( \phi_1 + \phi_1^*
  \right) \\ & \qquad + 2 m g |\phi_2|^2 \left( \phi_1 + \phi_1^* \right) + 4 g^2
  |\phi_1|^2 |\phi_2|^2 + g^2 \left( |\phi_1|^2 \right)^2   
  \end{aligned}
\end{equation}
The structure of the scalar interaction terms makes the following
redefinitions reasonable:
\begin{equation}
  \label{eq:umdef}
  \begin{aligned}
    \phi_1 & \; \longrightarrow \quad \dfrac{1}{\sqrt{2}} \left( A + \ii B
    \right) \\ 
    \phi_1^* & \; \longrightarrow \quad \dfrac{1}{\sqrt{2}} \left( A - \ii B
    \right) \\ 
    \phi_2 & \; \longrightarrow \quad \phi \\
    \phi_2^* & \; \longrightarrow \quad \phi^* 
  \end{aligned}
\end{equation}
To diagonalize the mass terms of the fermions, we introduce a Dirac bispinor 
\begin{equation}
  \label{eq:diracspiel}
  \Psi \equiv \begin{pmatrix} \psi_1 \\ \bar{\psi}_2 \end{pmatrix} \quad .
\end{equation}
So our Lagrangean density now looks like
\begin{equation}
  \label{eq:lagrangespiel3}
  {\cal L}_{\text{toy}} = {\cal L}_{\text{kin}} + {\cal L}_{\text{pot}} +
  {\cal L}_{\text{Yukawa}} \qquad , 
\end{equation}
with
\begin{align}
  {\cal L}_{\text{kin}} = & \; \dfrac{1}{2} \Bigl[ (\partial_\mu A)
  (\partial^\mu A) - m^2 A^2 \Bigr] + \dfrac{1}{2} \Bigl[ (\partial_\mu B)
  (\partial^\mu B) - m^2 B^2 \Bigr] \notag \\ & \quad + (\partial_\mu \phi^*)
  (\partial^\mu \phi) - m^2 |\phi|^2 + \overline{\Psi} \left( \ii
  \fmslash{\partial} - m \right) \Psi \\ & \notag \\
  {\cal L}_{\text{pot}} = & \; - \dfrac{1}{\sqrt{2}} m g \left( A^2 + B^2
  \right) A - 2 \sqrt{2} m g |\phi|^2 A - 2 g^2 |\phi|^2 \left( A^2 + B^2
  \right) \notag \\ & \qquad - \dfrac{g^2}{4} \left( A^4 + 2 A^2 B^2 + B^4
  \right) \\ & \notag 
  \\ 
  {\cal L}_{\text{Yukawa}} = & - \sqrt{2} g \overline{\Psi} \Psi A + \sqrt{2}
  \ii g \overline{\Psi} \gamma^5 \Psi B - g \overline{\Psi^c} {\cal P}_L \Psi
  \phi  - g \overline{\Psi} {\cal P}_R \Psi^c \phi^*
\end{align}

The Feynman rules seem to be obvious, but there are some delicacies, so we
write the vertices down in a graphical notation as in the Wess-Zumino
model. The point that is (in the bispinor formalism
(\ref{eq:lagrangespiel2})) easily overlooked is the crucial symmetry
factor two stemming from the Weyl spinors $\psi$ and $\bar{\psi}$ appearing
quadratically in the couplings to the scalar $\phi$, which can be seen by
applying functional derivatives for deriving the Feynman rules.  

{\bf Feynman rules of the toy model:}
%%%%%%%%%%%%%%%%%%%%%%%%%%%%%%%%%%%%%%%%%%%%%%%%
\begin{equation*}     %%%  A-A-A
 \begin{aligned}
  \parbox{5cm}{
  \begin{fmfchar*}(25,25)
    \fmfleft{i1,i2}
    \fmfright{o}
    \fmf{dashes,label=$A$}{i2,v}
    \fmf{dashes,label=$A$}{i1,v}
    \fmf{dashes,label=$A$}{o,v}
    \fmfdot{v}
  \end{fmfchar*}} & \hfill &
  \parbox{5cm}{
  \begin{math}
   \qquad
   \begin{array}{l}
     - 3 \sqrt{2} \ii m g 
     \end{array} 
  \end{math}}
 \end{aligned}
\end{equation*}
\vspace{0.8cm}
%%%%%%%%%%%%%%%%%%%%%%%%%%%%%%%%%%%%%%%%%%%%%%%%
\begin{equation*}     %%%  A-B-B
 \begin{aligned}
  \parbox{5cm}{
  \begin{fmfchar*}(25,25)
    \fmfleft{i1,i2}
    \fmfright{o}
    \fmf{dbl_dashes,label=$B$}{i2,v}
    \fmf{dbl_dashes,label=$B$}{i1,v}
    \fmf{dashes,label=$A$}{o,v}
    \fmfdot{v}
  \end{fmfchar*}} & \hfill &
  \parbox{5cm}{
  \begin{math}
   \qquad
   \begin{array}{l}
     -  \sqrt{2} \ii m g 
     \end{array} 
  \end{math}}
 \end{aligned}
\end{equation*}
\vspace{0.8cm}
%%%%%%%%%%%%%%%%%%%%%%%%%%%%%%%%%%%%%%%%%%%%%%%%
\begin{equation*}     %%%  A-phi-phi*
 \begin{aligned}
  \parbox{5cm}{
  \begin{fmfchar*}(25,25)
    \fmfleft{i1,i2}
    \fmfright{o}
    \fmf{dots,label=$\phi$}{i2,v}
    \fmf{dots,label=$\phi^*$}{i1,v}
    \fmf{dashes,label=$A$}{o,v}
    \fmfdot{v}
  \end{fmfchar*}} & \hfill &
  \parbox{5cm}{
  \begin{math}
   \qquad
   \begin{array}{l}
     - 2 \sqrt{2} \ii m g 
     \end{array} 
  \end{math}}
 \end{aligned}
\end{equation*}
\vspace{0.8cm}
%%%%%%%%%%%%%%%%%%%%%%%%%%%%%%%%%%%%%%%%%%%%%%%%%%%
\begin{equation*}      %%%  A-A-phi-phi*
 \begin{aligned}
 \parbox{5cm}{
  \begin{fmfchar*}(25,25)
  \fmfleft{i1,i2}
  \fmfright{o1,o2}
  \fmf{dashes,label=$A$,l.side=left}{i2,v}
  \fmf{dots,label=$\phi$,l.side=left}{o2,v}
  \fmf{dots,label=$\phi^*$}{v,o1}
  \fmf{dashes,label=$A$}{v,i1}
  \fmfdot{v}
  \end{fmfchar*}} & \hfill &
 \parbox{5cm}{
  \begin{math} \qquad
   \begin{array}{c}
     - 4 \ii g^2
   \end{array}
 \end{math}}
 \end{aligned}
\end{equation*}
\vspace{0.8cm}
%%%%%%%%%%%%%%%%%%%%%%%%%%%%%%%%%%%%%%%%%%%%%%%%%%%
\begin{equation*}      %%%  B-B-phi-phi*
 \begin{aligned}
 \parbox{5cm}{
  \begin{fmfchar*}(25,25)
  \fmfleft{i1,i2}
  \fmfright{o1,o2}
  \fmf{dbl_dashes,label=$B$,l.side=left}{i2,v}
  \fmf{dots,label=$\phi$,l.side=left}{o2,v}
  \fmf{dots,label=$\phi^*$}{v,o1}
  \fmf{dbl_dashes,label=$B$}{v,i1}
  \fmfdot{v}
  \end{fmfchar*}} & \hfill &
 \parbox{5cm}{
  \begin{math} \qquad
   \begin{array}{c}
     - 4 \ii g^2
   \end{array}
 \end{math}}
 \end{aligned}
\end{equation*}
\vspace{0.8cm}
%%%%%%%%%%%%%%%%%%%%%%%%%%%%%%%%%%%%%%%%%%%%%%%%%%%
\begin{equation*}      %%%  A-A-A-A
 \begin{aligned}
 \parbox{5cm}{
  \begin{fmfchar*}(25,25)
  \fmfleft{i1,i2}
  \fmfright{o1,o2}
  \fmf{dashes,label=$A$,l.side=left}{i2,v}
  \fmf{dashes,label=$A$,l.side=left}{o2,v}
  \fmf{dashes,label=$A$}{v,o1}
  \fmf{dashes,label=$A$}{v,i1}
  \fmfdot{v}
  \end{fmfchar*}} & \hfill &
 \parbox{5cm}{
  \begin{math} \qquad
   \begin{array}{c}
     - 6 \ii g^2
   \end{array}
 \end{math}}
 \end{aligned}
\end{equation*}
\vspace{0.8cm}
%%%%%%%%%%%%%%%%%%%%%%%%%%%%%%%%%%%%%%%%%%%%%%%%%%%
\begin{equation*}      %%%  A-A-B-B
 \begin{aligned}
 \parbox{5cm}{
  \begin{fmfchar*}(25,25)
  \fmfleft{i1,i2}
  \fmfright{o1,o2}
  \fmf{dashes,label=$A$,l.side=left}{i2,v}
  \fmf{dbl_dashes,label=$B$,l.side=left}{o2,v}
  \fmf{dbl_dashes,label=$B$}{v,o1}
  \fmf{dashes,label=$A$}{v,i1}
  \fmfdot{v}
  \end{fmfchar*}} & \hfill &
 \parbox{5cm}{
  \begin{math} \qquad
   \begin{array}{c}
     - 2 \ii g^2
   \end{array}
 \end{math}}
 \end{aligned}
\end{equation*}
\vspace{0.8cm}
%%%%%%%%%%%%%%%%%%%%%%%%%%%%%%%%%%%%%%%%%%%%%%%%%%%
\begin{equation*}      %%%  B-B-B-B
 \begin{aligned}
 \parbox{5cm}{
  \begin{fmfchar*}(25,25)
  \fmfleft{i1,i2}
  \fmfright{o1,o2}
  \fmf{dbl_dashes,label=$B$,l.side=left}{i2,v}
  \fmf{dbl_dashes,label=$B$,l.side=left}{o2,v}
  \fmf{dbl_dashes,label=$B$}{v,o1}
  \fmf{dbl_dashes,label=$B$}{v,i1}
  \fmfdot{v}
  \end{fmfchar*}} & \hfill &
 \parbox{5cm}{
  \begin{math} \qquad
   \begin{array}{c}
     - 6 \ii g^2
   \end{array}
 \end{math}}
 \end{aligned}
\end{equation*}
\vspace{0.8cm}
%%%%%%%%%%%%%%%%%%%%%%%%%%%%%%%%%%%%%%%%%%%%%%%%%%%%%%%%%
\begin{equation*}  %%%  Psi-A-Psi
 \begin{aligned}
  \parbox{5cm}{
  \begin{fmfchar*}(25,25)
    \fmfleft{i1,i2}
    \fmfright{o}
    \fmf{fermion,label=$\overline{\Psi}$}{v,i2}
    \fmf{fermion,label=$\Psi$,l.side=left}{i1,v}
    \fmf{dashes,label=$A$}{o,v}
    \fmfdot{v}
  \end{fmfchar*}} & \hfill &
  \parbox{5cm}{
  \begin{math}
   \qquad
   \begin{array}{l}
     - \sqrt{2} \ii g
  \end{array}
  \end{math}}
 \end{aligned}
\end{equation*}
\vspace{0.8cm}  
%%%%%%%%%%%%%%%%%%%%%%%%%%%%%%%%%%%%%%%%%%%%%%%%%%%%%%%%%
\begin{equation*}  %%%  Psi-B-Psi
 \begin{aligned}
  \parbox{5cm}{
  \begin{fmfchar*}(25,25)
    \fmfleft{i1,i2}
    \fmfright{o}
    \fmf{fermion,label=$\overline{\Psi}$}{v,i2}
    \fmf{fermion,label=$\Psi$,l.side=left}{i1,v}
    \fmf{dbl_dashes,label=$B$}{o,v}
    \fmfdot{v}
  \end{fmfchar*}} & \hfill &
  \parbox{5cm}{
  \begin{math}
   \qquad
   \begin{array}{l}
     - \sqrt{2} g \gamma^5
  \end{array}
  \end{math}}
 \end{aligned}
\end{equation*}
\vspace{0.8cm}  
%%%%%%%%%%%%%%%%%%%%%%%%%%%%%%%%%%%%%%%%%%%%%%%%%%%%%%%%%
\begin{equation*}  %%%  Psi-phi-Psi
 \begin{aligned}
  \parbox{5cm}{
  \begin{fmfchar*}(25,25)
    \fmfleft{i1,i2}
    \fmfright{o}
    \fmf{fermion,label=$\Psi$}{i2,v}
    \fmf{fermion,label=$\Psi$,l.side=left}{i1,v}
    \fmf{dots,label=$\phi$}{o,v}
    \fmfdot{v}
  \end{fmfchar*}} & \hfill &
  \parbox{5cm}{
  \begin{math}
   \qquad
   \begin{array}{l}
     - 2 \ii g {\cal P}_L
  \end{array}
  \end{math}}
 \end{aligned}
\end{equation*}
\vspace{0.8cm}  
%%%%%%%%%%%%%%%%%%%%%%%%%%%%%%%%%%%%%%%%%%%%%%%%%%%%%%%%%
\begin{equation*}  %%%  Psi-phi*-Psi
 \begin{aligned}
  \parbox{5cm}{
  \begin{fmfchar*}(25,25)
    \fmfleft{i1,i2}
    \fmfright{o}
    \fmf{fermion,label=$\overline{\Psi}$}{v,i2}
    \fmf{fermion,label=$\overline{\Psi}$,l.side=left}{v,i1}
    \fmf{dots,label=$\phi^*$}{o,v}
    \fmfdot{v}
  \end{fmfchar*}} & \hfill &
  \parbox{5cm}{
  \begin{math}
   \qquad
   \begin{array}{l}
     - 2 \ii g {\cal P}_R
  \end{array}
  \end{math}}
 \end{aligned}
\end{equation*}
\vspace{0.8cm}  

For the last two vertices compare the remark at the end of the next section. 

%%% Local Variables: 
%%% mode: latex
%%% TeX-master: "diss"
%%% End: 

%% file: appen_or.tex
\section{The O'Raifeartaigh model}

Action:
\begin{equation}
  \label{eq:oraif}
  S_{OR} = \int d^4 x \: \Biggl\{ \dfrac{1}{2} \sum_{i=1}^3 \left[
  \hat{\Phi}_i^\dagger \hat{\Phi}_i \right]_D 
  +  \left[ \lambda \hat{\Phi}_1 + m \hat{\Phi}_2
  \hat{\Phi}_3 + g \hat{\Phi}_1 \hat{\Phi}_2 \hat{\Phi}_2  \; \text{+
  h.c.} \right]_F \Biggr\} 
\end{equation}  
The equations of motion for the auxiliary fields:
\begin{align}
  F_1^* = - \lambda - g \phi_2^2 , \\
  F_2^* = - m \phi_3 - 2 g \phi_1 \phi_2 , \\
  F_3^* = - m \phi_2 .
\end{align}  
The kinetic part produces the massless kinetic terms for the fields
$\phi_i, \psi_i, i = 1,2,3$ and their complex conjugates. Again the
superpotential (after inserting the equations of motion for the
auxiliary fields) consists of $- \left| F \right|^2$ and the Yukawa
couplings. To diagonalize the mass terms of the fermions, we introduce
the Dirac bispinor 
\begin{equation}
  \label{eq:or_bispi}
  \Psi = \begin{pmatrix} \psi_2 \\ \bar{\psi}_3 \end{pmatrix} \; \text{with}
  \;\; 
  \ii \psi_3  \sigma^\mu \partial_\mu \bar{\psi}_3 + \ii \psi_2
  \sigma^\mu \partial_\mu \bar{\psi}_3 - m \left( \psi_2 \psi_3 + \bar{\psi}_2
  \bar{\psi}_3 \right) = \overline{\Psi} \left( \ii \fmslash{\partial} - m
  \right) \Psi .
\end{equation}
The spinor from the first superfield remains massless and can be
extended to a Majorana spinor field $\chi$. The Lagrangean
density then reads 
\begin{equation}
  \begin{aligned}
  {\cal L}_{OR} = & \; \overline{\Psi} \left( \ii \fmslash{\partial} - m
  \right) \Psi + \dfrac{\ii}{2} \overline{\chi} \fmslash{\partial} \chi +
  \sum_{i = 1}^3 \left( \partial_\mu \phi^*_i \right) \left( \partial^\mu
  \phi_i \right) - \sum_{i = 1}^3 \left| F_i \right|^2  \\ & \; - 2 g \left(
  \psi_1 \psi_2 \phi_2 + \bar{\psi}_1 \bar{\psi}_2 \phi_2^* \right) - g \left(
  \psi_2 \psi_2 \phi_1 + \bar{\psi}_2 \bar{\psi}_2 \phi_1^* \right)
  \end{aligned}
\end{equation}
In the next step we calculate the superpotential:
\begin{equation}
  \label{eq:superpotor}
  \begin{aligned}
    \mathcal{W} = & \; \left| F_1 \right|^2 + \left| F_2 \right|^2 + \left| F_3
    \right|^2  \\ = & \; \left| \lambda + g \phi_2^2 \right|^2 + \left| m
    \phi_3 + 2 g \phi_1 \phi_2 \right|^2 + m^2 \left| \phi_2 \right|^2
    \\ = &
    \; \lambda^2 + \lambda g \left( \phi_2^2 + (\phi_2^*)^2 \right) + g^2
    \left( \left| \phi_2 \right|^2 \right)^2 + m^2 \left| \phi_3
    \right|^2 + 2 g m \left( \phi_1 \phi_2 \phi_3^* + \phi_1^* \phi_2^* \phi_3
    \right) \\ & \qquad \qquad + 4 g^2 \left| \phi_1 \right|^2 \left| \phi_2
    \right|^2 + m^2 \left| \phi_2 \right|^2
  \end{aligned}
\end{equation}
As for the WZ model the complex scalar field $\phi_2$ is split up into
real and imaginary part , $\phi_2 = \frac{1}{\sqrt{2}} \left( A + \ii B
\right)$, which changes its kinetic parts to
\begin{equation}
  ( \partial_\mu \phi_2^* ) ( \partial^\mu \phi_2 ) = \dfrac{1}{2} (
  \partial_\mu A ) (\partial^\mu A) + \dfrac{1}{2} ( \partial_\mu B ) (
  \partial^\mu B ) .
\end{equation}
The terms quadratic in the scalar fields in the superpotential yield:
\begin{multline}
    \lambda g \left( \phi_2^2 + (\phi_2^*)^2 \right) + m^2 \left| \phi_3
    \right|^2 + m^2 \left| \phi_2 \right|^2 \\ = \dfrac{1}{2} \left( m^2 + 2
    \lambda g \right) A^2 + \dfrac{1}{2} \left( m^2 - 2 \lambda g \right) B^2 +
    m^2 \left| \phi_3 \right|^2
\end{multline}   
Altogether, the scalar part consists of a complex scalar field with
zero mass, $\phi_1$, a complex scalar field with mass $m$, $\phi_3$,
as well as two real (scalar and pseudoscalar) fields,$A$ and $B$, with
masses $\sqrt{m^2 + 2 \lambda g}$ and $\sqrt{m^2 - 2 \lambda g}$. In
the following we rename the field $\phi_3$ by $\phi$ and substitute
the notation $\Phi$ for the massive field $\phi_3$. The final form of
the kinetic part then looks like: 
\begin{multline}
  {\cal L}_{\text{kin}} = \overline{\Psi} \left( \ii \fmslash{\partial} -
  m \right) \Psi + \dfrac{\ii}{2} \overline{\chi} \fmslash{\partial} \chi
  + \dfrac{1}{2} \left[ (\partial_\mu A) (\partial^\mu A) - (m^2 + 2 \lambda
  g) A^2 \right] \\ + \dfrac{1}{2} \left[ (\partial_\mu B) (\partial^\mu B) - 
  (m^2 - 2 \lambda g) B^2 \right] + \left( \partial_\mu \phi^* \right) \left(
  \partial^\mu \phi \right) \\ + \left( \partial_\mu \Phi^* \right) \left(
  \partial^\mu \Phi \right) - m^2 \left| \Phi \right|^2
\end{multline}
The remaining part of the scalar potential rewritten with the physical
fields: 
\begin{multline}
  {\cal L}_{\text{scalar}} = - \lambda^2 - \dfrac{g^2}{4} \left( A^4 + 2 A^2
  B^2 + B^4 \right) - 2 g^2 \left| \phi \right|^2 \left( A^2 + B^2 \right) \\
  - \sqrt{2} g m A \left( \phi \Phi^* + \phi^* \Phi \right) - \sqrt{2} \ii
  g m B \left( \phi \Phi^* - \phi^* \Phi \right) 
\end{multline}
The term $- \lambda^2$ provides a contribution to the cosmological constant. 

Finally we have to transform the Yukawa couplings; again we meet the
complication of vertices with fermions whose arrows both point to the
vertex or both away from it. This, as mentioned for our toy
model, can be handled within the general formalism of
\cite{Denner/etal:1992:feynmanrules}. The Yukawa terms have the structure: 
\begin{multline}
  {\cal L}_{\text{Yukawa}} = - \frac{g}{\sqrt{2}} A \Bigl( \overline{\chi} (1
  - \gamma^5) \Psi + \overline{\Psi} (1 + \gamma^5) \chi \Bigr) \\ -
  \dfrac{\ii g}{\sqrt{2}} B \Bigl( \overline{\chi} ( 1 - \gamma^5) \Psi -
  \overline{\Psi} ( 1 + \gamma^5 ) \chi \Bigr) \\ - \dfrac{g}{2}
  \overline{\Psi^c} (1 - \gamma^5) \Psi \phi - \dfrac{g}{2} \overline{\Psi} (1
  + \gamma^5) \Psi^c \phi^*
\end{multline}

\begin{table}[htbp]
  \begin{center}
    \setlength{\extrarowheight}{2pt}    
    \begin{tabular}{c|c|l}
      Particle & Mass & Description \\ \hline
      $A$ & $\sqrt{m^2 + 2 \lambda g}$ & neutral scalar \\
      $B$ & $\sqrt{m^2 - 2 \lambda g}$ & neutral pseudoscalar \\
      $\chi$ & $0$ & neutral fermion (Goldstino) \\
      $\Phi^{(*)}$ & $m$ & charged scalar \\
      $\Psi, \overline{\Psi}$ & $m$ & charged fermion \\
      $\phi^{(*)}$ & $0$ & charged scalar
    \end{tabular}
    \caption{Physical particle content of the OR model.}
    \label{tab:teilchen_OR}
  \end{center}
\end{table}

For further discussion of the OR model in the text we write down the
equations of motion and the SUSY transformations of all fields:
\begin{equation}
  \label{eq:beweggleich_or}
  \begin{aligned}
    \left( \ii \fmslash{\partial} - m \right) \Psi = & \; \sqrt{2} g \left( A
    - \ii B \right) {\cal P}_R \chi + 2 g \phi^* {\cal P}_R \Psi^c \\
    \ii \fmslash{\partial} \chi = & \; \sqrt{2} g \left( A + \ii B \right)
    {\cal P}_L \Psi + \sqrt{2} g \left( A - \ii B \right) \Psi^c \\
    \left( \Box + m^2 + 2 \lambda g \right) A = & \; - \sqrt{2} g \left(
      \overline{\chi} {\cal P}_L \Psi \right) - \sqrt{2} g \left(
      \overline{\Psi} {\cal P}_R \chi \right) \\ & \qquad - g^2 A^3 - g^2 A
      B^2 - 4 g^2 A \left| \phi \right|^2 - \sqrt{2} g m \left( \phi \Phi^* +
      \phi^* \Phi \right) \\
    \left( \Box + m^2 - 2 \lambda g \right) B = & \; - \sqrt{2} \ii g \left(
      \overline{\chi} {\cal P}_L \Psi \right) + \sqrt{2} \ii g \left(
      \overline{\Psi} {\cal P}_R \chi \right) \\ & \qquad - g^2 B^3 - g^2 A^2
    B - 4 g^2 B \left| \phi \right|^2 - \sqrt{2} \ii g m \left( \phi \Phi^* -
      \phi^* \Phi \right) \\
    \Box \, \phi = &  \; - \sqrt{2} g m \left( A - \ii B \right) \Phi - g
    \left( \overline{\Psi} {\cal P}_R \Psi^c \right) - 2 g^2 \phi \left( A^2 +
      B^2 \right) \\
    \left( \Box + m^2 \right) \Phi = & \; - \sqrt{2} g m \left( A + \ii B
    \right) \phi 
  \end{aligned}
\end{equation}
The missing equations can be obtained by complex or by charge
conjugation. For the fermionic fields the SUSY transformation laws are 
\begin{align}
  \delta_\xi \chi = & \; - \sqrt{2} \ii \fmslash{\partial} \left( \phi {\cal
  P}_R + \phi^* {\cal P}_L \right) \xi - \sqrt{2} \lambda \xi -
  \dfrac{g}{\sqrt{2}} \left( A^2 - B^2 \right) \xi - \sqrt{2} \ii g A B
  \gamma^5 \xi \\
  \delta_\xi \Psi = & \; - \left( \ii \fmslash{\partial} + m \right) \left( A
  + \ii B \right) {\cal P}_R \xi - \sqrt{2} \left( \ii \fmslash{\partial} + m
  \right) \Phi^* {\cal P}_L \xi - 2 g \phi^* \left( A - \ii B \right)
  {\cal P}_L \xi \\
  \delta_\xi \Psi^c = & \; - \left( \ii \fmslash{\partial} + m \right)
  \left( A - \ii B \right) {\cal P}_L \xi - \sqrt{2} \left( \ii
  \fmslash{\partial} + m \right) \Phi {\cal P}_R \xi - 2 g \phi \left(
  A + \ii B \right) {\cal P}_R \xi ,
\end{align}
while for the scalar fields we have
\begin{align}
  \delta_\xi A = & \; \left( \overline{\xi} {\cal P}_L \Psi \right) +
  \left( \overline{\xi} {\cal P}_R \Psi^c \right) \\
  \delta_\xi B = & \; - \ii \left( \overline{\xi} {\cal P}_L \Psi
  \right) + \ii \left( \overline{\xi} {\cal P}_R \Psi^c \right) \\
  \delta_\xi \phi = & \; \sqrt{2} \left( \overline{\xi} {\cal P}_L
  \chi \right) \\
  \delta_\xi \phi^* = & \; \sqrt{2} \left( \overline{\xi} {\cal P}_R
  \chi \right) \\
  \delta_\xi \Phi = & \; \sqrt{2} \left( \overline{\xi} {\cal P}_L
  \Psi^c \right) \\
   \delta_\xi \Phi^* = & \; \sqrt{2} \left( \overline{\xi} {\cal P}_R
  \Psi \right) \quad .
\end{align}

We denote the Feynman rules for the OR model graphically:

%%%%%%%%%%%%%%%%%%%%%%%%%%%%%%%%%%%%%%%%%%%%%%%%
\begin{equation*}     %%%  A-phi-Phi
 \begin{aligned}
  \parbox{5cm}{
  \begin{fmfchar*}(25,25)
    \fmfleft{i1,i2}
    \fmfright{o}
    \fmf{dashes,label=$A$}{i2,v}
    \fmf{dots,label=$\phi (\phi^*)$}{i1,v}
    \fmf{dbl_dots,label=$\Phi^* (\Phi)$}{o,v}
    \fmfdot{v}
  \end{fmfchar*}} & \hfill &
  \parbox{5cm}{
  \begin{math}
   \qquad
   \begin{array}{l}
      - \sqrt{2} \ii g m 
     \end{array} 
  \end{math}}
 \end{aligned}
\end{equation*}
\vspace{0.8cm}
%%%%%%%%%%%%%%%%%%%%%%%%%%%%%%%%%%%%%%%%%%%%%%%%
\begin{equation*}     %%%  B-phi-Phi
 \begin{aligned}
  \parbox{5cm}{
  \begin{fmfchar*}(25,25)
    \fmfleft{i1,i2}
    \fmfright{o}
    \fmf{dbl_dashes,label=$B$}{i2,v}
    \fmf{dots,label=$\phi (\phi^*)$}{i1,v}
    \fmf{dbl_dots,label=$\Phi^* (\Phi)$}{o,v}
    \fmfdot{v}
  \end{fmfchar*}} & \hfill &
  \parbox{5cm}{
  \begin{math}
   \qquad
   \begin{array}{l}
      \pm \sqrt{2} g m 
     \end{array} 
  \end{math}}
 \end{aligned}
\end{equation*}
\vspace{0.8cm}
%%%%%%%%%%%%%%%%%%%%%%%%%%%%%%%%%%%%%%%%%%%%%%%%%%%
\begin{equation*}      %%%  A-A-A-A
 \begin{aligned}
 \parbox{5cm}{
  \begin{fmfchar*}(25,25)
  \fmfleft{i1,i2}
  \fmfright{o1,o2}
  \fmf{dashes,label=$A$,l.side=left}{i2,v}
  \fmf{dashes,label=$A$,l.side=left}{o2,v}
  \fmf{dashes,label=$A$}{v,o1}
  \fmf{dashes,label=$A$}{v,i1}
  \fmfdot{v}
  \end{fmfchar*}} & \hfill &
 \parbox{5cm}{
  \begin{math} \qquad
   \begin{array}{c}
     - 6 \ii g^2
   \end{array}
 \end{math}}
 \end{aligned}
\end{equation*}
\vspace{0.8cm}
%%%%%%%%%%%%%%%%%%%%%%%%%%%%%%%%%%%%%%%%%%%%%%%%%%%
\begin{equation*}      %%%  B-B-B-B
 \begin{aligned}
 \parbox{5cm}{
  \begin{fmfchar*}(25,25)
  \fmfleft{i1,i2}
  \fmfright{o1,o2}
  \fmf{dbl_dashes,label=$B$,l.side=left}{i2,v}
  \fmf{dbl_dashes,label=$B$,l.side=left}{o2,v}
  \fmf{dbl_dashes,label=$B$}{v,o1}
  \fmf{dbl_dashes,label=$B$}{v,i1}
  \fmfdot{v}
  \end{fmfchar*}} & \hfill &
 \parbox{5cm}{
  \begin{math} \qquad
   \begin{array}{c}
     - 6 \ii g^2
   \end{array}
 \end{math}}
 \end{aligned}
\end{equation*}
\vspace{0.8cm}  
%%%%%%%%%%%%%%%%%%%%%%%%%%%%%%%%%%%%%%%%%%%%%%%%%%%
\begin{equation*}      %%%  A-A-B-B
 \begin{aligned}
 \parbox{5cm}{
  \begin{fmfchar*}(25,25)
  \fmfleft{i1,i2}
  \fmfright{o1,o2}
  \fmf{dashes,label=$A$,l.side=left}{i2,v}
  \fmf{dashes,label=$A$,l.side=left}{o2,v}
  \fmf{dbl_dashes,label=$B$}{v,o1}
  \fmf{dbl_dashes,label=$B$}{v,i1}
  \fmfdot{v}
  \end{fmfchar*}} & \hfill &
 \parbox{5cm}{
  \begin{math} \qquad
   \begin{array}{c}
     - 2 \ii g^2
   \end{array}
 \end{math}}
 \end{aligned}
\end{equation*}
\vspace{0.8cm}
%%%%%%%%%%%%%%%%%%%%%%%%%%%%%%%%%%%%%%%%%%%%%%%%%%%
\begin{equation*}      %%%  A-A-phi-phi
 \begin{aligned}
 \parbox{5cm}{
  \begin{fmfchar*}(25,25)
  \fmfleft{i1,i2}
  \fmfright{o1,o2}
  \fmf{dashes,label=$A$,l.side=left}{i2,v}
  \fmf{dashes,label=$A$,l.side=left}{o2,v}
  \fmf{dots,label=$\phi$}{v,o1}
  \fmf{dots,label=$\phi^*$}{v,i1}
  \fmfdot{v}
  \end{fmfchar*}} & \hfill &
 \parbox{5cm}{
  \begin{math} \qquad
   \begin{array}{c}
     - 4 \ii g^2
   \end{array}
 \end{math}}
 \end{aligned}
\end{equation*}
\vspace{0.8cm}
%%%%%%%%%%%%%%%%%%%%%%%%%%%%%%%%%%%%%%%%%%%%%%%%%%%
\begin{equation*}      %%%  B-B-phi-phi
 \begin{aligned}
 \parbox{5cm}{
  \begin{fmfchar*}(25,25)
  \fmfleft{i1,i2}
  \fmfright{o1,o2}
  \fmf{dbl_dashes,label=$B$,l.side=left}{i2,v}
  \fmf{dbl_dashes,label=$B$,l.side=left}{o2,v}
  \fmf{dots,label=$\phi$}{v,o1}
  \fmf{dots,label=$\phi^*$}{v,i1}
  \fmfdot{v}
  \end{fmfchar*}} & \hfill &
 \parbox{5cm}{
  \begin{math} \qquad
   \begin{array}{c}
     - 4 \ii g^2
   \end{array}
 \end{math}}
 \end{aligned}
\end{equation*}
\vspace{0.8cm}
%%%%%%%%%%%%%%%%%%%%%%%%%%%%%%%%%%%%%%%%%%%%%%%%%%%%%%%%%
\begin{equation*}  %%%  Chi-A-Psi
 \begin{aligned}
  \parbox{5cm}{
  \begin{fmfchar*}(25,25)
    \fmfleft{i1,i2}
    \fmfright{o}
    \fmf{plain,label=$\chi$}{v,i2}
    \fmf{fermion,label=$\Psi$,l.side=left}{i1,v}
    \fmf{dashes,label=$A$}{o,v}
    \fmfdot{v}
  \end{fmfchar*}} & \hfill &
  \parbox{5cm}{
  \begin{math}
   \qquad
   \begin{array}{l}
     - \sqrt{2} \ii g {\cal P}_L
  \end{array}
  \end{math}}
 \end{aligned}
\end{equation*}
\vspace{0.8cm}  
%%%%%%%%%%%%%%%%%%%%%%%%%%%%%%%%%%%%%%%%%%%%%%%%%%%%%%%%%
\begin{equation*}  %%%  Psi-A-Chi
 \begin{aligned}
  \parbox{5cm}{
  \begin{fmfchar*}(25,25)
    \fmfleft{i1,i2}
    \fmfright{o}
    \fmf{fermion,label=$\Psi$}{v,i2}
    \fmf{plain,label=$\chi$,l.side=left}{i1,v}
    \fmf{dashes,label=$A$}{o,v}
    \fmfdot{v}
  \end{fmfchar*}} & \hfill &
  \parbox{5cm}{
  \begin{math}
   \qquad
   \begin{array}{l}
     - \sqrt{2} \ii g {\cal P}_R
  \end{array}
  \end{math}}
 \end{aligned}
\end{equation*}
\vspace{0.8cm}  
%%%%%%%%%%%%%%%%%%%%%%%%%%%%%%%%%%%%%%%%%%%%%%%%%%%%%%%%%
\begin{equation*}  %%%  Chi-B-Psi
 \begin{aligned}
  \parbox{5cm}{
  \begin{fmfchar*}(25,25)
    \fmfleft{i1,i2}
    \fmfright{o}
    \fmf{plain,label=$\chi$}{v,i2}
    \fmf{fermion,label=$\Psi$,l.side=left}{i1,v}
    \fmf{dbl_dashes,label=$B$}{o,v}
    \fmfdot{v}
  \end{fmfchar*}} & \hfill &
  \parbox{5cm}{
  \begin{math}
   \qquad
   \begin{array}{l}
     + \sqrt{2} g {\cal P}_L
  \end{array}
  \end{math}}
 \end{aligned}
\end{equation*}
\vspace{0.8cm}  
%%%%%%%%%%%%%%%%%%%%%%%%%%%%%%%%%%%%%%%%%%%%%%%%%%%%%%%%%
\begin{equation*}  %%%  Psi-B-Chi
 \begin{aligned}
  \parbox{5cm}{
  \begin{fmfchar*}(25,25)
    \fmfleft{i1,i2}
    \fmfright{o}
    \fmf{fermion,label=$\Psi$}{v,i2}
    \fmf{plain,label=$\chi$,l.side=left}{i1,v}
    \fmf{dbl_dashes,label=$B$}{o,v}
    \fmfdot{v}
  \end{fmfchar*}} & \hfill &
  \parbox{5cm}{
  \begin{math}
   \qquad
   \begin{array}{l}
     - \sqrt{2} g {\cal P}_R
  \end{array}
  \end{math}}
 \end{aligned}
\end{equation*}
\vspace{0.8cm} 
%%%%%%%%%%%%%%%%%%%%%%%%%%%%%%%%%%%%%%%%%%%%%%%%%%%%%%%%%
\begin{equation*}  %%%  Psic-phi-Psi
 \begin{aligned}
  \parbox{5cm}{
  \begin{fmfchar*}(25,25)
    \fmfleft{i1,i2}
    \fmfright{o}
    \fmf{fermion,label=$\Psi$}{i2,v}
    \fmf{fermion,label=$\Psi$,l.side=left}{i1,v}
    \fmf{dots,label=$\phi$}{o,v}
    \fmfdot{v}
  \end{fmfchar*}} & \hfill &
  \parbox{5cm}{
  \begin{math}
   \qquad
   \begin{array}{l}
     - 2 \ii g {\cal P}_L
  \end{array}
  \end{math}}
 \end{aligned}
\end{equation*}
\vspace{0.8cm} 
%%%%%%%%%%%%%%%%%%%%%%%%%%%%%%%%%%%%%%%%%%%%%%%%%%%%%%%%%
\begin{equation*}  %%%  Psi-phi*-Psic
 \begin{aligned}
  \parbox{5cm}{
  \begin{fmfchar*}(25,25)
    \fmfleft{i1,i2}
    \fmfright{o}
    \fmf{fermion,label=$\overline{\Psi}$}{v,i2}
    \fmf{fermion,label=$\overline{\Psi}$,l.side=left}{v,i1}
    \fmf{dots,label=$\phi^*$}{o,v}
    \fmfdot{v}
  \end{fmfchar*}} & \hfill &
  \parbox{5cm}{
  \begin{math}
   \qquad
   \begin{array}{l}
     - 2 \ii g {\cal P}_R
  \end{array}
  \end{math}}
 \end{aligned}
\end{equation*}
\vspace{0.8cm}

\vspace{20pt}

{\bf Remark:}

\vspace{5pt}
Again there is the phenomenon known from the discussion of the toy
model in the foregoing section, which always appears in supersymmetric
models with mixings between the fermionic components of more than one
superfield, called the problem of ``clashing arrows''. 
The examples in the MSSM where charged gauginos and Higgsinos are
combined into the charginos are more prominent. Of course, in the last
two vertices we 
can reverse the arrow of one of the Dirac fermions, and
label the lines by the charge conjugated field
$\Psi^c$ instead of $\Psi$. For the structure of the bispinor products
to be intact this would indeed be the better solution. We again have a
symmetry factor two in the last two vertices as in the toy model, which
is perhaps most clearly seen in the formulation with two-component
spinors, but can also be understood when looking at the two factors of
$\Psi$ or $\overline{\Psi}$ appearing in the vertex -- here this description
is better suited than the one with the charge conjugated field. 

%% file: appen_sagt.tex
\section{An Abelian toy model}\label{sec:sagt}

Action:
\begin{equation}
     S = \int d^4 x \biggl\{ \dfrac{1}{2} \begin{bmatrix}
     \hat{\Phi}^\dagger \exp(\mathcal{-V}) \hat{\Phi} \end{bmatrix}_D
     + \dfrac{1}{2} \Re \begin{bmatrix} \overline{W_R} W_L
     \end{bmatrix}_F \biggr\} 
\end{equation}  
Equations of motion for the auxiliary fields:
\begin{subequations}
\begin{align}
        F \equiv &\; 0 \\ 
        D =&\; - e |\phi|^2 
\end{align}
\end{subequations}
By introducing the bispinor notation for matter fermion and gaugino 
\footnote{The strange prefactor of $\ii\gamma^5$ is due to the
conventions originally introduced by Wess and Bagger to define
two-component gaugino fields, cf.~the appendix
of \cite{Haber/Kane:1985:SUSY}}
\begin{equation}
\Psi = \begin{pmatrix} \psi \\ \bar{\psi}
\end{pmatrix} , \qquad \qquad \lambda = \ii \gamma^5 \begin{pmatrix}
\lambda \\ \bar{\lambda}
\end{pmatrix},
\end{equation}
splitting the complex scalar field in real and imaginary part $A$ and
$B$ as well as integrating out the auxiliary fields $D$ and $F$ (the
latter vanishing identically), we find the Lagrangean density in
components as well as the Feynman rules:
\begin{multline}
  \mathcal{L} = \dfrac{1}{2} (\partial_\mu A) (\partial^\mu A) +
    \dfrac{1}{2} (\partial_\mu B) (\partial^\mu B) + \dfrac{\ii}{2}
    \overline{\Psi} \fmslash{\partial} \Psi - \dfrac{1}{4} F_{\mu\nu}
    F^{\mu\nu} + \dfrac{\ii}{2} \overline{\lambda} \fmslash{\partial}
    \lambda \\
    + e G_\mu \left( B \partial^\mu A - A \partial^\mu B
    \right) + \dfrac{e^2}{2} G_\mu G^\mu \left( A^2 + B^2 \right) - e
    \left( \overline{\Psi} \lambda \right) A \\
    - \ii e \left( \overline{\Psi} \gamma^5 \lambda \right) B -
    \dfrac{e}{2} \overline{\Psi} \fmslash{G} \gamma^5 \Psi - \dfrac{e^2}{8}
    \left( A^4 + B^4 + 2 A^2 B^2 \right)
\end{multline}
Following the derivation from the foregoing appendix we can
immediately write down the current: 
\begin{multline}
  \mathcal{J}^\mu = - (\fmslash{\partial} A) \gamma^\mu \Psi - \ii
  (\fmslash{\partial} B) \gamma^\mu \gamma^5 \Psi + \ii e A \fmslash{G}
  \gamma^\mu \gamma^5 \Psi \\  - e B \fmslash{G}  \gamma^\mu \Psi +
  \dfrac{1}{2} \lbrack \gamma^\alpha , \gamma^\beta \rbrack \gamma^\mu
  \gamma^5 (\partial_\alpha G_\beta) \lambda - \dfrac{\ii e}{2} \left( A^2
  + B^2 \right) \gamma^\mu \lambda
\end{multline}      
Equations of motion for all fields:
\begin{subequations}
\begin{align}
  \Box A &= - 2 e G_\mu \partial^\mu B - e B \partial_\mu G^\mu +
  e^2 G_\mu G^\mu A - e \overline{\Psi} \lambda - \dfrac{e^2}{2} \left(
  A^3 + A B^2 \right) \\
  \Box B &= 2 e G_\mu \partial^\mu A + e A \partial_\mu G^\mu
  + e^2 G_\mu G^\mu B - \ii e \overline{\Psi} \gamma^5 \lambda -
  \dfrac{e^2}{2} \left( B^3 + B A^2 \right) \\
  \ii \fmslash{\partial} \Psi &= e A \lambda + \ii e B \gamma^5 \lambda
  + e \fmslash{G} \gamma^5 \Psi \\
  \ii \fmslash{\partial} \lambda &= e A \Psi + \ii e B \gamma^5 \Psi \\
  \partial^\nu F_{\nu\mu} &= e \left( A \partial_\mu B - B
  \partial_\mu A \right) - e^2 G_\mu \left( A^2 + B^2 \right) +
  \dfrac{e}{2} \overline{\Psi} \gamma_\mu \gamma^5 \Psi
\end{align}
\end{subequations}      

The charge generates the SUSY transformations of the fields: 
\begin{subequations}
  \begin{align}
    \lbrack \ii \overline{\xi} Q, A \rbrack &= \left( \overline{\xi} \Psi
    \right) \\ \lbrack \ii \overline{\xi} Q,B \rbrack  &= \left( \ii
      \overline{\xi} \gamma^5 \Psi \right) \\
    \lbrack \ii \overline{\xi} Q , \Psi \rbrack &= - \ii
      (\fmslash{\partial} - \ii e \fmslash{G} \gamma^5) (A + \ii \gamma^5 B)
      \xi \\
    \lbrack \ii \overline{\xi} Q , \lambda \rbrack &=  - \dfrac{\ii}{2}
      F_{\alpha\beta} \gamma^\alpha \gamma^\beta \gamma^5 \xi - \dfrac{e}{2}
      \left( A^2 + B^2 \right) \xi 
    \\
    \lbrack \ii \overline{\xi} Q , G_\mu \rbrack &= - \left( \overline{\xi}
      \gamma_\mu \gamma^5 \lambda \right)
  \end{align}
\end{subequations}
The gauge transformations with transformation parameter $\theta =
\theta(x)$ are (note that this model has a chiral
charge, the gauge boson couples to the axial vector of the fermion, and
scalar and pseudoscalar are interchanged by gauge transformations):
\begin{subequations}
\begin{align}
        \delta A =&\; - e \theta B \\
        \delta B =&\; + e \theta A \\
        \delta \Psi = &\; - \ii e \gamma^5 \theta \Psi \\
        \delta G_\mu = &\; \partial_\mu \theta \\
        \delta \lambda =&\; 0 
\end{align}
\end{subequations}

The ``covariance of the equations of motions'' is important for the
Green functions to have the same poles. We just show one
example: 
\begin{align}
  \lbrack \ii \overline{\xi} Q , \Box A \rbrack = & \; 2 \ii e \left(
    \overline{\xi} (\fmslash{\partial} B) \gamma^5 \lambda \right) + 2 e
  \left( \overline{\xi} \gamma^5 G_\mu \partial^\mu \Psi \right) - \ii e
  \left( \overline{\xi} B \gamma^5 \fmslash{\partial} \lambda \right)
  \notag\\ & \;
  + e (\partial_\mu G^\mu) \left( \overline{\xi} \gamma^5 \Psi \right) -
  2 \ii e^2 A \left( \overline{\xi} \fmslash{G} \gamma^5 \lambda \right) +
  \ii e^2 G_\mu G^\mu \left( \overline{\xi} \Psi \right) \notag\\ & \;
  + e \overline{\xi} \Bigl( \left( \fmslash{\partial} - \ii e \fmslash{G}
    \gamma^5 \right) \left( A - \ii \gamma^5 B \right) \Bigr) \lambda +
  \dfrac{e^2}{2} \left( \overline{\xi} \lbrack \gamma^\alpha , \gamma^\beta
    \rbrack \gamma^5 (\partial_\alpha G_\beta) \Psi \right) \notag \\ & \; +
  \dfrac{\ii e^2}{2} \left( A^2 + B^2 \right) \left( \overline{\xi} \Psi
  \right) - \dfrac{3 \ii e^2}{2} A^2 \left( \overline{\xi} \Psi \right) -
  \dfrac{\ii e^2}{2} B^2 \left( \overline{\xi} \Psi \right) \notag \\ & \;
  + e^2 A B \left( \overline{\xi} \gamma^5 \Psi \right) \notag
  \\ = & \;    
  \ii e \left(
    \overline{\xi} (\fmslash{\partial} B) \gamma^5 \lambda \right) + 2 e
  \left( \overline{\xi} \gamma^5 G_\mu \partial^\mu \Psi \right) - \ii e
  \left( \overline{\xi} B \gamma^5 \fmslash{\partial} \lambda \right)
  \notag\\ & \;
  + e (\partial_\mu G^\mu) \left( \overline{\xi} \gamma^5 \Psi \right) -
  \ii e^2 A \left( \overline{\xi} \fmslash{G} \gamma^5 \lambda \right) +
  \ii e^2 G_\mu G^\mu \left( \overline{\xi} \Psi \right) \notag\\ & \;
  + e \left( \overline{\xi} (\fmslash{\partial} A) \lambda \right)
  + e^2 B \left( \overline{\xi} \fmslash{G} \lambda \right) +
  \dfrac{e^2}{2} \left( \overline{\xi} \lbrack \gamma^\alpha , \gamma^\beta
    \rbrack \gamma^5 (\partial_\alpha G_\beta) \Psi \right) \notag \\ & \;
  - \ii e^2 A^2 \left( \overline{\xi} \Psi \right) + e^2 A B
  \left( \overline{\xi} \gamma^5 \Psi \right)
\end{align}
\begin{align}
  \Box \: \lbrack \ii \overline{\xi} Q , A \rbrack = & \; \ii \left(
    \overline{\xi} \fmslash{\partial} \fmslash{\partial} \Psi \right) \notag
  \\ = & \; e A \left( \overline{\xi} \fmslash{\partial} \lambda \right)
  + e \left( \overline{\xi} (\fmslash{\partial} A) \lambda \right) - \ii e
  B \left( \overline{\xi} \gamma^5 \fmslash{\partial} \lambda \right) +
  \ii e \left( \overline{\xi} (\fmslash{\partial} B) \gamma^5 \lambda
  \right) \notag \\ & \; + e \left( \overline{\xi} (\fmslash{\partial}
    \fmslash{G}) \gamma^5 \Psi \right) + e \left( \overline{\xi} \gamma^\mu
    \fmslash{G} \gamma^5 \partial_\mu \Psi \right) \notag \\ = & \;
  - \ii e^2 A^2 \left( \overline{\xi} \Psi \right) + e^2 A B \left(
    \overline{\xi} \gamma^5 \Psi \right) + e \left( \overline{\xi}
    (\fmslash{\partial} A) \lambda \right) - \ii e \left( \overline{\xi}
    B \gamma^5 \fmslash{\partial} \lambda \right) \notag \\ & \; + \ii e
  \left( \overline{\xi} (\fmslash{\partial} B) \gamma^5 \lambda \right) +
  \dfrac{e}{2} \left( \overline{\xi} \lbrack \gamma^\alpha , \gamma^\beta
    \rbrack \gamma^5 (\partial_\alpha G_\beta) \Psi \right) + e (\partial_\mu
  G^\mu) \left( \overline{\xi} \gamma^5 \Psi \right) \notag \\ & \; + 2 e
  \left( \overline{\xi} \gamma^5 G_\mu \partial^\mu \Psi \right)
  - \ii e^2 A \left( \overline{\xi} \fmslash{G} \gamma^5 \lambda \right) +
  e^2 B \left( \overline{\xi} \fmslash{G} \lambda \right) \notag \\ & \; +
  \ii e^2 (G_\mu G^\mu) \left( \overline{\xi} \Psi \right)
\end{align}
\begin{equation}
  \Longrightarrow \lbrack \overline{\xi} Q , \Box A \rbrack \, = \,
  \Box \: \lbrack \overline{\xi} Q , A \rbrack
\end{equation}     

To clear our notation with respect to the propagators -- especially in
comparison with the ghost propagators in the third part of the text --
we write down the propagators for the particles of the model:
\begin{subequations}
  \begin{align}
    \parbox{21mm}{%
      \begin{fmfgraph*}(20,5)
        \fmfleft{i}\fmfright{o}
        \fmflabel{$A(-p)$}{i}
        \fmflabel{$A(p)$}{o}
        \fmf{dashes}{i,o}
        \fmfdot{i,o}
      \end{fmfgraph*}}\qquad\quad
        &= \frac{\mathrm{i}}{p^2+\mathrm{i}\epsilon} \\
    \parbox{21mm}{%
      \begin{fmfgraph*}(20,5)
        \fmfleft{i}\fmfright{o}
        \fmflabel{$B(-p)$}{i}
        \fmflabel{$B(p)$}{o}
        \fmf{dbl_dashes}{i,o}
        \fmfdot{i,o}
      \end{fmfgraph*}}\qquad\quad
        &= \frac{\mathrm{i}}{p^2+\mathrm{i}\epsilon} \\
    \parbox{21mm}{%
      \begin{fmfgraph*}(20,5)
        \fmfleft{i}\fmfright{o}
        \fmflabel{$G_\mu(-p)$}{i}
        \fmflabel{$G_\nu(p)$}{o}           \fmf{photon}{i,o}
        \fmfdot{i,o}
      \end{fmfgraph*}}\qquad\quad
        &= \frac{- \mathrm{i} \eta_{\mu\nu}}{p^2+\mathrm{i}\epsilon} \\
    \parbox{21mm}{%
      \begin{fmfgraph*}(20,5)
        \fmfleft{i}\fmfright{o}
        \fmflabel{$\Psi(-p)$}{i}
        \fmflabel{$\overline{\Psi}(p)$}{o}
        \fmf{plain}{i,o}
        \fmfdot{i,o}
      \end{fmfgraph*}}\qquad\quad
        &= \frac{\mathrm{i} \fmslash{p}}{p^2+\mathrm{i}\epsilon} \\
    \parbox{21mm}{%
      \begin{fmfgraph*}(20,5)
        \fmfleft{i}\fmfright{o}
        \fmflabel{$\lambda(-p)$}{i}
        \fmflabel{$\overline{\lambda}(p)$}{o}
        \fmf{plain}{i,o} \fmf{photon,wiggly_len=1mm}{i,o}
        \fmfdot{i,o}
      \end{fmfgraph*}}\qquad\quad
        &= \frac{\mathrm{i} \fmslash{p}}{p^2+\mathrm{i}\epsilon}
  \end{align}
\end{subequations}                   

Vertices of our Abelian toy model (all momenta incoming):
\begin{align*}
  \parbox{5cm}{%
    \hfil\\\hfil\\
    \begin{fmfgraph*}(25,25)
      \fmfleft{p1}\fmfright{p2,p3}
      \fmf{photon,label=$G_\mu$}{p1,v}
      \fmf{dashes,label=$A(p_A)$,l.side=left}{p3,v} 
      \fmf{dbl_dashes,label=$B(p_B)$,l.side=right}{p2,v}
      \fmfdot{v}
    \end{fmfgraph*}\\
    \hfil}\qquad\quad
      &e (p_A - p_B)_\mu \\
  \parbox{5cm}{%
    \hfil\\\hfil\\
    \begin{fmfgraph*}(25,25)
      \fmfleft{p1}\fmfright{p2,p3}
      \fmf{photon,label=$G_\mu$}{p1,v}
      \fmf{plain,label=$\Psi$}{p2,v}
      \fmf{plain,label=$\Psi$}{p3,v}
      \fmfdot{v}
    \end{fmfgraph*}\\
    \hfil}\qquad\quad
      &\ii e \gamma^5 \gamma_\mu    \\
  \parbox{5cm}{%
    \hfil\\\hfil\\
    \begin{fmfgraph*}(25,25)
      \fmfleft{p1}\fmfright{p2,p3}
      \fmf{dashes,label=$A$}{p1,v}
      \fmf{plain,label=$\lambda$}{p2,v}
      \fmf{plain,label=$\Psi$}{p3,v} \fmffreeze
      \fmf{photon}{v,p2}
      \fmfdot{v}        
   \end{fmfgraph*}\\
    \hfil}\qquad\quad
      &- \ii e \\
  \parbox{5cm}{%
    \hfil\\\hfil\\
    \begin{fmfgraph*}(25,25)
      \fmfleft{p1}\fmfright{p2,p3}
      \fmf{dbl_dashes,label=$B$}{p1,v}
      \fmf{plain,label=$\lambda$}{p2,v}
      \fmf{plain,label=$\Psi$}{p3,v} \fmffreeze
      \fmf{photon}{v,p2}
      \fmfdot{v}
    \end{fmfgraph*}\\
    \hfil}\qquad\quad
      &e \gamma^5 \\
  \parbox{5cm}{%
    \hfil\\\hfil\\
    \begin{fmfgraph*}(25,25)
      \fmfleft{p1,p2}\fmfright{p3,p4}
      \fmf{dashes,label=$A$}{v,p1}
      \fmf{dashes,label=$A$}{v,p2}
      \fmf{dashes,label=$A$}{v,p3}
      \fmf{dashes,label=$A$}{v,p4}
      \fmfdot{v}
    \end{fmfgraph*}\\
    \hfil}\qquad\quad
      &- 3 \ii e^2  \\
  \parbox{5cm}{%
    \hfil\\\hfil\\
    \begin{fmfgraph*}(25,25)
      \fmfleft{p1,p2}\fmfright{p3,p4}
      \fmf{dbl_dashes,label=$B$}{v,p1}
      \fmf{dbl_dashes,label=$B$}{v,p2}
      \fmf{dbl_dashes,label=$B$}{v,p3}
      \fmf{dbl_dashes,label=$B$}{v,p4}
      \fmfdot{v}
    \end{fmfgraph*}\\
    \hfil}\qquad\quad
      &- 3 \ii e^2   \\
  \parbox{5cm}{%
    \hfil\\\hfil\\
    \begin{fmfgraph*}(25,25)
      \fmfleft{p1,p2}\fmfright{p3,p4}
      \fmf{dashes,label=$A$}{v,p1}
      \fmf{dashes,label=$A$}{v,p2}         
     \fmf{dbl_dashes,label=$B$}{v,p3}
     \fmf{dbl_dashes,label=$B$}{v,p4}
      \fmfdot{v}
    \end{fmfgraph*}\\
    \hfil}\qquad\quad
      &- \ii e^2 \\
  \parbox{5cm}{%
    \hfil\\\hfil\\
    \begin{fmfgraph*}(25,25)
      \fmfleft{p1,p2}\fmfright{p3,p4}
      \fmf{dashes,label=$A$}{v,p1}
      \fmf{dashes,label=$A$}{v,p2}
      \fmf{photon,label=$G_\mu$}{v,p3}
      \fmf{photon,label=$G_\nu$}{v,p4}
      \fmfdot{v}
    \end{fmfgraph*}\\
    \hfil}\qquad\quad
      &2 \ii e^2 \eta_{\mu\nu} \\
  \parbox{5cm}{%
    \hfil\\\hfil\\
    \begin{fmfgraph*}(25,25)
      \fmfleft{p1,p2}\fmfright{p3,p4}
      \fmf{dbl_dashes,label=$B$}{v,p1}
      \fmf{dbl_dashes,label=$B$}{v,p2}
      \fmf{photon,label=$G_\mu$}{v,p3}
      \fmf{photon,label=$G_\nu$}{v,p4}
      \fmfdot{v}
    \end{fmfgraph*}\\
    \hfil}\qquad\quad
      &2 \ii e^2 \eta_{\mu\nu}
\end{align*}

%%% Local Variables: 
%%% mode: latex
%%% TeX-master: "diss"
%%% End: 

%% file: appen_modelsym.tex
%%%%%%%%%%%%%%%%%%%%%%%%%%%%%%%%%%%%%%%%%%%%%%%%%%%%%%%%%%%%%%%%%%%

\section{Supersymmetric Yang--Mills theory (SYM)}\label{sec:symdetail}

The Lagrangean density without gauge fixing is:
\begin{multline}
\mathcal{L} = (D_\mu \phi_+)^\dagger (D^\mu \phi_+) + (D_\mu
\phi_-)^\dagger (D^\mu \phi_-) + \dfrac{\ii}{2} \overline{\psi}_{+,i}
(\fmslash{D} \psi_+)_i + \dfrac{\ii}{2} \overline{\psi}_{-,i}
(\fmslash{D} \psi_-)_i \\ + |F_+|^2 + |F_-|^2 - \sqrt{2} g
(\overline{\lambda^a} \phi^\dagger_{+,i} T^a_{ij} {\cal P}_L
\psi_{+,j}) - \sqrt{2} g (\overline{\psi}_{+,i} T^a_{ij} \phi_{+,j}
{\cal P}_R \lambda^a) \\ - \sqrt{2} g (\overline{\lambda^a}
\phi_{-,i}^\dagger (-T^{a\:*})_{ij} {\cal P}_L \psi_{-,j}) - \sqrt{2}
g (\overline{\psi}_{-,i} (-T^{a\:*})_{ij} \phi_{-,j} {\cal P}_R
\lambda^a) \\ + g (\phi^\dagger_{+,i} T^a_{ij} \phi_{+,j}) D^a + g
(\phi^\dagger_{-,i} (-T^{a\:*})_{ij} \phi_{-,j}) D^a - \dfrac{1}{4}
F^a_{\mu\nu} F^{\mu\nu}_a \\ + \dfrac{\ii}{2} \overline{\lambda^a}
(\fmslash{D} \lambda)^a + \dfrac{1}{2} (D^a D^a) + m \phi_{+,i}
F_{-,i} + m \phi^\dagger_{+,i} F^\dagger_{+,i} \\ + m \phi_{-,i} F_{-,i}
+ m \phi^\dagger_{-,i} F^\dagger_{-,i} - m \left( \psi_+ \psi_- +
\bar{\psi}_+ \bar{\psi}_- \right)  
\end{multline}
The generators of the gauge group fulfill the Lie algebra
\begin{equation}
 \lbrack T^a , T^b \rbrack = \ii f_{abc} T^c , \qquad\qquad \qquad 
 \lbrack (-T^a)^* , (-T^b)^* \rbrack = \ii f_{abc} (-T^c)^* 
\end{equation}
For the auxiliary fields the equations of motion are
\begin{subequations}
\begin{align}
 F_{+,i} = - m \phi^\dagger_{-,i} & \qquad\qquad F_{+,i}^\dagger = - m
 \phi_{-,i} \\
 F_{-,i} = - m \phi^\dagger_{+,i} & \qquad\qquad F_{-,i}^\dagger = - m
 \phi_{+,i} 
\end{align}
\begin{equation}
D^a = - g (\phi^\dagger_+ T^a \phi_+) + g (\phi^\dagger_- T^{a\:*}
\phi_-) 
\end{equation}
\end{subequations}
We diagonalize the mass terms of the fermions by introducing the bispinors
\begin{equation}
 \Psi_i = \begin{pmatrix} \psi_{+,i} \\ \bar{\psi}_{i,-}
 \end{pmatrix}, \qquad\qquad \overline{\Psi}_i = \left( \psi_{-,i} ,
 \bar{\psi}_{+,i} \right) \quad . 
\end{equation}
By the redefinitions of the fermion fields and after integrating out
all auxiliary fields we get the Lagrangean density (with gauge-fixing and
Faddeev-Popov terms)
\begin{multline}
     \mathcal{L} = (D_\mu \phi_+)^\dagger (D^\mu \phi_+) - m^2
     |\phi_+|^2 + (D_\mu \phi_-)^\dagger (D^\mu \phi_-) - m^2
     |\phi_-|^2 \\ + \overline{\Psi} (\ii\fmslash{D} - m) \Psi +
     \dfrac{\ii}{2} \overline{\lambda^a} (\fmslash{D} \lambda)^a -
     \dfrac{1}{4} F_{\mu\nu}^a F^{\mu\nu}_a - \sqrt{2} \, \ii g
     \phi^\dagger_{+,i} T^a_{ij} (\overline{\lambda^a} {\cal P}_L \Psi_j) \\
     - \sqrt{2} \, \ii g \phi_{-,i} T^a_{ij} (\overline{\lambda^a} {\cal P}_R
     \Psi_j) + \sqrt{2} \, \ii g (\overline{\Psi}_i {\cal P}_R \lambda^a)
     T^a_{ij} \phi_{+,j} \\ + \sqrt{2} \, \ii g (\overline{\Psi}_i {\cal P}_L
     \lambda^a) T^a_{ij} \phi^\dagger_{-,j} - \dfrac{g^2}{2} \left(
     \phi^\dagger_{+,i} T^a_{ij} \phi_{+,j} \right)\left(
     \phi^\dagger_{+,k} T^a_{kl} \phi_{+,l} \right) \\ - \dfrac{g^2}{2} \left(
     \phi_{-,i} T^a_{ij} \phi^\dagger_{-,j} \right)\left(
     \phi_{-,k} T^a_{kl} \phi^\dagger_{-,l} \right) + g^2 \left(
     \phi^\dagger_{+,i} T^a_{ij} \phi_{+,j} \right)\left(
     \phi_{-,k} T^a_{kl} \phi^\dagger_{-,l} \right) \\
     - \dfrac{1}{2 \xi} (\partial^\mu A_\mu^a) (\partial^\nu A_\nu^a)
     + \ii \overline{c}^a 
     \partial_\mu (D^\mu c)^a - \ii \overline{c}^a (\overline{\epsilon}
     \fmslash{\partial} \lambda^a) + \dfrac{\ii\xi}{2} \overline{c}^a (
     \overline{\epsilon} \gamma^\mu \epsilon) \partial_\mu \overline{c}^a
\end{multline}  
The propagators for the particles of the model are
\begin{subequations}
     \begin{align}
       \parbox{25mm}{%
         \begin{fmfgraph*}(20,5)
           \fmfleft{i}\fmfright{o}
           \fmflabel{$\phi_{+,i}(-p)$}{i}
           \fmflabel{$\phi_{+,j}^\dagger(p)$}{o}
           \fmf{dashes_arrow}{o,i}
           \fmfdot{i,o}
         \end{fmfgraph*}}\qquad\quad
           &= \frac{\mathrm{i}\delta_{ij}}{p^2-m^2+\mathrm{i}\epsilon} \\
       \parbox{25mm}{%
         \begin{fmfgraph*}(20,5)
           \fmfleft{i}\fmfright{o}
           \fmflabel{$\phi_{-,i}^\dagger(-p)$}{i}
           \fmflabel{$\phi_{-,j}(p)$}{o}
           \fmf{dbl_dashes_arrow}{o,i}
           \fmfdot{i,o}
         \end{fmfgraph*}}\qquad\quad
           &= \frac{\mathrm{i}\delta_{ij}}{p^2-m^2+\mathrm{i}\epsilon} \\
       \parbox{25mm}{%
         \begin{fmfgraph*}(20,5)
           \fmfleft{i}\fmfright{o}
           \fmflabel{$A^a_\mu(-p)$}{i}
           \fmflabel{$A^b_\nu(p)$}{o}
           \fmf{photon}{i,o}
           \fmfdot{i,o}
         \end{fmfgraph*}}\qquad\quad
           &= \dfrac{-\ii\delta_{ab}}{p^2+\ii\epsilon} \left(\eta_{\mu\nu}
              - (1-\xi) \dfrac{p_\mu p_\nu}{p^2} \right) \notag\\
           & \qquad \stackrel{\xi\to 1}{\longrightarrow}
               \frac{- \mathrm{i} \eta_{\mu\nu} \delta_{ab}}{p^2+\mathrm{i}
               \epsilon} \\                            
       \parbox{25mm}{%
         \begin{fmfgraph*}(20,5)
           \fmfleft{i}\fmfright{o}
           \fmflabel{$\Psi_i(-p)$}{i}
           \fmflabel{$\overline{\Psi}_j(p)$}{o}
           \fmf{fermion}{o,i}
           \fmfdot{i,o}
         \end{fmfgraph*}}\qquad\quad
           &= \frac{\mathrm{i} \delta_{ij}(\fmslash{p}+m)}{p^2-m^2+
              \mathrm{i}\epsilon} \\
       \parbox{25mm}{%
         \begin{fmfgraph*}(20,5)
           \fmfleft{i}\fmfright{o}
           \fmflabel{$\lambda^a(-p)$}{i}
           \fmflabel{$\overline{\lambda^b}(p)$}{o}
           \fmf{plain}{i,o} \fmf{photon,wiggly_len=1mm}{i,o}
           \fmfdot{i,o}
         \end{fmfgraph*}}\qquad\quad
           &= \frac{\mathrm{i} \delta_{ab}\fmslash{p}}{p^2+\mathrm{i}
              \epsilon}  \\
       \parbox{25mm}{%
         \begin{fmfgraph*}(20,5)
           \fmfleft{i}\fmfright{o}
           \fmflabel{$c^a(-p)$}{i}
           \fmflabel{$\overline{c}^b(p)$}{o}
           \fmf{dots_arrow}{o,i}
           \fmfdot{i,o}
         \end{fmfgraph*}}\qquad\quad
           &= \frac{- \delta_{ab}}{p^2+\mathrm{i}
              \epsilon}
     \end{align}
\end{subequations}                       
The 3-vertices are (all momenta incoming):
\begin{subequations}            
     \begin{align}
       \parbox{21mm}{%
         \hfil\\\hfil\\
         \begin{fmfgraph*}(20,20)
           \fmfleft{p1}\fmfright{p2,p3}
           \fmflabel{$A^a_\mu(p_1)$}{p1}
           \fmflabel{$A^b_\nu(p_2)$}{p2}
           \fmflabel{$A^c_\rho(p_3)$}{p3}
           \fmf{photon}{p1,v}
           \fmf{photon}{p3,v} \fmf{photon}{p2,v}
           \fmfdot{v}
         \end{fmfgraph*}\\
         \hfil}\qquad\quad
           &\begin{matrix} = g f_{abc} \lbrack \eta_{\mu\nu} \left( p_1 - p_2
              \right)_\rho + \eta_{\nu\rho} \left( p_2 - p_3 \right)_\mu \\
              + \eta_{\rho\mu} \left( p_3 - p_1 \right)_\nu \rbrack
              \end{matrix} \\ \notag \\
       \parbox{21mm}{%
         \hfil\\\hfil\\
         \begin{fmfgraph*}(20,20)
           \fmfleft{p1}\fmfright{p2,p3}
           \fmflabel{$A^a_\mu(p_1)$}{p1}
           \fmflabel{$\phi_{+,j}(p_3)$}{p2}
           \fmflabel{$\phi^\dagger_{+,i}(p_2)$}{p3}
           \fmf{photon}{p1,v}
           \fmf{dashes_arrow}{p2,v,p3}
           \fmfdot{v}
         \end{fmfgraph*}\\
         \hfil}\qquad\quad
           &= \ii g \left( p_3 - p_2 \right)_\mu T^a_{ij} \\ \notag \\                \parbox{21mm}{%
         \hfil\\\hfil\\
         \begin{fmfgraph*}(20,20)
           \fmfleft{p1}\fmfright{p2,p3}
           \fmflabel{$A^a_\mu(p_1)$}{p1}
           \fmflabel{$\phi^\dagger_{-,j}(p_3)$}{p2}
           \fmflabel{$\phi_{-,i}(p_2)$}{p3}
           \fmf{photon}{p1,v}
           \fmf{dbl_dashes_arrow}{p2,v,p3}
           \fmfdot{v}
         \end{fmfgraph*}\\
         \hfil}\qquad\quad
           &= \ii g \left( p_3 - p_2 \right)_\mu T^a_{ij} \\ \notag \\
       \parbox{21mm}{%
         \hfil\\\hfil\\
         \begin{fmfgraph*}(20,20)
           \fmfleft{p1}\fmfright{p2,p3}
           \fmflabel{$A^a_\mu(p_1)$}{p1}
           \fmflabel{$\overline{\Psi}_i(p_2)$}{p3}
           \fmflabel{$\Psi_j(p_3)$}{p2}
           \fmf{photon}{p1,v}
           \fmf{fermion}{p2,v,p3}
           \fmfdot{v}
         \end{fmfgraph*}\\
         \hfil}\qquad\quad
           &= \ii g \gamma_\mu T^a_{ij} \\ \notag \\                      
       \parbox{21mm}{%
         \hfil\\\hfil\\
         \begin{fmfgraph*}(20,20)
           \fmfleft{p1}\fmfright{p2,p3}
           \fmflabel{$A^a_\mu(p_1)$}{p1}
           \fmflabel{$\lambda^c(p_3)$}{p2}
           \fmflabel{$\overline{\lambda^b}(p_2)$}{p3}
           \fmf{photon}{p1,v}
           \fmf{plain}{p2,v,p3} \fmffreeze
           \fmf{photon}{p2,v,p3}
           \fmfdot{v}
         \end{fmfgraph*}\\
         \hfil}\qquad\quad
           &= g \gamma_\mu f_{abc} \\ \notag \\
       \parbox{21mm}{%
        \hfil\\\hfil\\
         \begin{fmfgraph*}(20,20)
           \fmfleft{p1}\fmfright{p2,p3}
           \fmflabel{$\phi^\dagger_{+,i}(p_1)$}{p1}
           \fmflabel{$\Psi_j(p_3)$}{p2}
           \fmflabel{$\overline{\lambda^a}(p_2)$}{p3}
           \fmf{dashes_arrow}{v,p1}
           \fmf{fermion}{p2,v}
           \fmf{plain}{v,p3} \fmffreeze
           \fmf{photon}{v,p3}
           \fmfdot{v}
         \end{fmfgraph*}\\
        \hfil}\qquad\quad
           &= \dfrac{g}{\sqrt{2}} \left( 1 - \gamma^5 \right) T^a_{ij} \\
                                                                   \notag \\  
     \parbox{21mm}{%
        \hfil\\\hfil\\
         \begin{fmfgraph*}(20,20)
           \fmfleft{p1}\fmfright{p2,p3}
           \fmflabel{$\phi_{-,i}(p_1)$}{p1}
           \fmflabel{$\Psi_j(p_3)$}{p2}
           \fmflabel{$\overline{\lambda^a}(p_2)$}{p3}
           \fmf{dbl_dashes_arrow}{v,p1}
           \fmf{fermion}{p2,v}
           \fmf{plain}{v,p3} \fmffreeze
           \fmf{photon}{v,p3}
           \fmfdot{v}
         \end{fmfgraph*}\\
        \hfil}\qquad\quad
           &= \dfrac{g}{\sqrt{2}} \left( 1 + \gamma^5 \right) T^a_{ij} \\
                                                                   \notag \\
       \parbox{21mm}{%
        \hfil\\\hfil\\
         \begin{fmfgraph*}(20,20)
           \fmfleft{p1}\fmfright{p2,p3}
           \fmflabel{$\phi_{+,j}(p_1)$}{p1}
           \fmflabel{$\lambda^a(p_3)$}{p2}
           \fmflabel{$\overline{\Psi}_i(p_2)$}{p3}
           \fmf{dashes_arrow}{p1,v}
           \fmf{fermion}{v,p3}
           \fmf{plain}{v,p2} \fmffreeze
           \fmf{photon}{v,p2}
           \fmfdot{v}
         \end{fmfgraph*}\\
        \hfil}\qquad\quad
           &= - \dfrac{g}{\sqrt{2}} \left( 1 + \gamma^5 \right)
       T^a_{ij} \\ \notag \\ 
        \parbox{21mm}{%
        \hfil\\\hfil\\
         \begin{fmfgraph*}(20,20)
           \fmfleft{p1}\fmfright{p2,p3}
           \fmflabel{$\phi^\dagger_{-,j}(p_1)$}{p1}
           \fmflabel{$\lambda^a(p_3)$}{p2}
           \fmflabel{$\overline{\Psi}_i(p_2)$}{p3}
           \fmf{dbl_dashes_arrow}{p1,v}
           \fmf{fermion}{v,p3}
           \fmf{plain}{v,p2} \fmffreeze
           \fmf{photon}{v,p2}
           \fmfdot{v}
         \end{fmfgraph*}\\
        \hfil}\qquad\quad
           &= - \dfrac{g}{\sqrt{2}} \left( 1 - \gamma^5 \right) T^a_{ij} \\
                                                                     \notag \\
        \parbox{21mm}{%
        \hfil\\\hfil\\
         \begin{fmfgraph*}(20,20)
           \fmfleft{p1}\fmfright{p2,p3}
           \fmflabel{$A^b_\mu(p_1)$}{p1}
           \fmflabel{$c^c(p_3)$}{p2}
           \fmflabel{$\overline{c}^a(p_2)$}{p3}
           \fmf{photon}{p1,v}
           \fmf{dots_arrow}{p2,v,p3}
           \fmfdot{v}
         \end{fmfgraph*}\\
        \hfil}\qquad\quad
           &= - \ii g f_{abc} p_{2,\mu} \\ \notag \\    
        \parbox{21mm}{%
        \hfil\\\hfil\\
         \begin{fmfgraph*}(20,20)
           \fmfleft{p1}\fmfright{p2,p3}
           \fmflabel{$\overline{c}^a(-p)$}{p1}
           \fmflabel{$\lambda^b(p)$}{p2}
           \fmflabel{$\overline{\epsilon}$}{p3}
           \fmf{dots_arrow}{v,p1}
           \fmf{plain}{p2,v}
           \fmf{susy_ghost}{p3,v} \fmffreeze
           \fmf{photon}{p2,v}
           \fmfv{decor.shape=square,decor.filled=full,decor.size=2mm}{p3}
           \fmfdot{v}
         \end{fmfgraph*}\\
        \hfil}\qquad\quad
           &= - \ii \fmslash{p} \delta_{ab}
     \end{align}
\end{subequations}    
We have the following 4-vertices

\vspace{3mm}

\begin{subequations}
     \begin{align}
       \parbox{21mm}{%\hfil\\\hfil\\\hfil\\\hfil\\
         \begin{fmfgraph*}(20,20)
           \fmfleft{p1,p2}\fmfright{p3,p4}
           \fmflabel{$A_\mu^a(p_1)$}{p1}
           \fmflabel{$A_\nu^b(p_2)$}{p2}
           \fmflabel{$A_\rho^c(p_3)$}{p3}
           \fmflabel{$A_\sigma^d(p_4)$}{p4}
           \fmf{photon}{p1,v,p2}
           \fmf{photon}{p3,v,p4}
           \fmfdot{v}
         \end{fmfgraph*}\\
         \hfil}\qquad\quad
           &\begin{matrix} = - \ii g^2 \lbrack f_{abe} f_{cde} \left(
            \eta_{\mu\rho} \eta_{\nu\sigma} - \eta_{\mu\sigma} \eta_{\nu\rho}
            \right)  \\ \qquad + f_{ace} f_{bde} \left(
            \eta_{\mu\nu} \eta_{\rho\sigma} - \eta_{\mu\sigma} \eta_{\nu\rho}
            \right)  \\ \qquad + f_{ade} f_{bce} \left(
            \eta_{\mu\nu} \eta_{\rho\sigma} - \eta_{\mu\rho} \eta_{\nu\sigma}
            \right)  \rbrack \end{matrix} \\ \notag \\
       \parbox{21mm}{%
         \hfil\\\hfil\\
         \begin{fmfgraph*}(20,20)
           \fmfleft{p1,p2}\fmfright{p3,p4}
           \fmflabel{$\phi_{+,i}(p_1)$}{p1}
           \fmflabel{$\phi^\dagger_{+,j}(p_2)$}{p2}
           \fmflabel{$A_\mu^a(p_3)$}{p3}
           \fmflabel{$A_\nu^b(p_4)$}{p4}
           \fmf{dashes_arrow}{p1,v,p2}
           \fmf{photon}{p3,v,p4}
           \fmfdot{v}
         \end{fmfgraph*}\\
         \hfil}\qquad\quad
           &= \ii g^2 \eta_{\mu\nu} \left\{ T^a , T^b \right\}_{ij} \\
       \notag \\ 
       \parbox{21mm}{%
         \hfil\\\hfil\\
         \begin{fmfgraph*}(20,20)
           \fmfleft{p1,p2}\fmfright{p3,p4}
           \fmflabel{$\phi^\dagger_{-,i}(p_1)$}{p1}
           \fmflabel{$\phi_{-,j}(p_2)$}{p2}
           \fmflabel{$A_\mu^a(p_3)$}{p3}
           \fmflabel{$A_\nu^b(p_4)$}{p4}
           \fmf{dbl_dashes_arrow}{p1,v,p2}
           \fmf{photon}{p3,v,p4}
           \fmfdot{v}
         \end{fmfgraph*}\\
         \hfil}\qquad\quad
           &= \ii g^2 \eta_{\mu\nu} \left\{ T^a , T^b \right\}_{ij} \\\notag\\
       \parbox{21mm}{%
         \hfil\\\hfil\\
         \begin{fmfgraph*}(20,20)
           \fmfleft{p1,p2}\fmfright{p3,p4}
           \fmflabel{$\phi_{+,j}(p_1)$}{p1}
           \fmflabel{$\phi^\dagger_{+,i}(p_2)$}{p2}
           \fmflabel{$\phi_{+,l}(p_3)$}{p3}
           \fmflabel{$\phi^\dagger_{+,k}(p_4)$}{p4}
           \fmf{dashes_arrow}{p1,v,p2}
           \fmf{dashes_arrow}{p3,v,p4}
           \fmfdot{v}
         \end{fmfgraph*}\\
         \hfil}\qquad\quad
           &= - \dfrac{\ii g^2}{4} \left( \delta_{il} \delta_{jk} -
                \dfrac{1}{N} \delta_{ij} \delta_{kl} \right)  \\
       \notag \\  
       \parbox{21mm}{%
         \hfil\\\hfil\\
         \begin{fmfgraph*}(20,20)
           \fmfleft{p1,p2}\fmfright{p3,p4}
           \fmflabel{$\phi^\dagger_{-,j}(p_1)$}{p1}
           \fmflabel{$\phi_{-,i}(p_2)$}{p2}
           \fmflabel{$\phi^\dagger_{-,l}(p_3)$}{p3}
           \fmflabel{$\phi_{-,k}(p_4)$}{p4}
           \fmf{dbl_dashes_arrow}{p1,v,p2}
           \fmf{dbl_dashes_arrow}{p3,v,p4}
           \fmfdot{v}
         \end{fmfgraph*}\\
         \hfil}\qquad\quad
           &= - \dfrac{\ii g^2}{4} \left( \delta_{il} \delta_{jk} -
                \dfrac{1}{N} \delta_{ij} \delta_{kl} \right)  \\ \notag \\
       \parbox{21mm}{%
         \hfil\\\hfil\\
         \begin{fmfgraph*}(20,20)
           \fmfleft{p1,p2}\fmfright{p3,p4}
           \fmflabel{$\phi_{+,j}(p_1)$}{p1}
           \fmflabel{$\phi^\dagger_{+,i}(p_2)$}{p2}
           \fmflabel{$\phi^\dagger_{-,l}(p_3)$}{p3}
           \fmflabel{$\phi_{-,k}(p_4)$}{p4}
           \fmf{dashes_arrow}{p1,v,p2}
           \fmf{dbl_dashes_arrow}{p3,v,p4}
           \fmfdot{v}
         \end{fmfgraph*}\\
         \hfil}\qquad\quad
           &= \dfrac{\ii g^2}{2} \left( \delta_{il} \delta_{jk} - \dfrac{1}{N}
                               \delta_{ij} \delta_{kl} \right)
       \\\notag\\   
       \parbox{21mm}{%
         \hfil\\\hfil\\
         \begin{fmfgraph*}(20,20)
           \fmfleft{p1,p2}\fmfright{p3,p4}
           \fmflabel{$\epsilon$}{p1}
           \fmflabel{$\overline{c}^a(-p)$}{p2}
           \fmflabel{$\overline{\epsilon}$}{p3}
           \fmflabel{$\overline{c}^b(p)$}{p4}
           \fmf{dots_arrow}{v,p2}
           \fmf{dots_arrow}{v,p4}
           \fmf{susy_ghost}{p1,v} 
           \fmf{susy_ghost}{p3,v}
           \fmfv{decor.shape=square,decor.filled=full,decor.size=2mm}{p1,p3}
           \fmfdot{v}
         \end{fmfgraph*}\\
         \hfil}\qquad\quad
           &= \xi \fmslash{p} \delta_{ab}
     \end{align}
\end{subequations} 

%%% Local Variables: 
%%% mode: latex
%%% TeX-master: "diss"
%%% End: 

%% file: dank.tex
\chapter*{} 

\pagestyle{empty}

\newpage
\begin{Large}
\begin{center}
\bf{DANKSAGUNG}
\end{center}
\end{Large}
\vspace{10pt}
Mein Dank gilt in erster Linie Prof.~Dr.~Panagiotis
Manakos f\"ur die Erm\"oglichung meiner Promotion in seiner
Arbeitsgruppe und seine weitere Unterst\"utzung. Ganz besonders danken
m\"ochte ich Dr.~Thorsten Ohl, der mir durch seine Betreuung
in zahllosen Diskussionen immer wieder Anregungen gegeben hat, die
wesentlich zum Gelingen dieser Arbeit beigetragen haben; vor allem bin
ich f\"ur die geduldige Hilfestellung bei den Unwegsamkeiten mit Soft-
und Hardware dankbar. 

Gefreut habe ich mich in den vergangenen Jahren stets \"uber die
fruchtbaren Diskussionen innerhalb der Hochenergie--Arbeitsgruppe der
TU Darmstadt, neben den oben genannten Dr.~Harald Anlauf,
Dipl.-Phys.~David Ondreka, Dipl.-Phys. Heike Pieschel und
Dipl.-Phys. Christian Schwinn. Letztem m\"ochte ich f\"ur die gute,
gemeinsame Zusammenarbeit an dem {\em O'Mega}-Projekt und seine Hilfe
bei der Einarbeitung in {\em O'Caml} danken. 

Schlie{\ss}lich gilt mein besonderer Dank Heike f\"ur ihre 
Unterst\"utzung und Liebe. Zu guter letzt danke ich meinen Eltern, die
mich in all den Jahren stets unterst\"utzt haben.